%% file: main.tex
\documentclass[10pt]{article}

\input{macro}

\begin{document}

\title{An invitation to topological orders and category theory}
\author[a,b,c]{Liang Kong \thanks{Email: \href{mailto:kongl@sustech.edu.cn}{\tt kongl@sustech.edu.cn}}}
\author[d,a]{Zhi-Hao Zhang \thanks{Email: \href{mailto:zzh31416@mail.ustc.edu.cn}{\tt zzh31416@mail.ustc.edu.cn}}}
\affil[a]{Shenzhen Institute for Quantum Science and Engineering, \authorcr
Southern University of Science and Technology, Shenzhen, 518055, China}
\affil[b]{International Quantum Academy, Shenzhen 518048, China}
\affil[c]{Guangdong Provincial Key Laboratory of Quantum Science and Engineering, \authorcr
Southern University of Science and Technology, Shenzhen, 518055, China}
\affil[d]{Wu Wen-Tsun Key Laboratory of Mathematics of Chinese Academy of Sciences, \authorcr
School of Mathematical Sciences, \authorcr
University of Science and Technology of China, Hefei, 230026, China}
\date{\vspace{-5ex}}

\maketitle

\begin{abstract}
Although it has been a well-known fact, for more than two decades, that category theory is needed for the study of topological orders, it is still a non-trivial challenge for students and working physicists to master the abstract language of category theory. In this work, for those who have no background in category theory, we explain in great details how the structure of a (braided) fusion category naturally emerges from lattice models and physical intuitions. Moreover, we show that nearly all mathematical notions and constructions in fusion categories and its representation theory, such as (monoidal) functors, Drinfeld center, module categories, Morita equivalence, condensation completion and fusion 2-categories, naturally emerge from lattice models and physical intuitions. In this process, we also introduce some basic notions and important results of topological orders. 
\end{abstract}

\tableofcontents

\include{introduction}

\include{basic}

\include{2d}

\include{1d}



\include{adv-topics}

\bibliography{Top}

\end{document}

%% file: macro.tex
\usepackage{amsfonts,amsmath,amssymb}
\usepackage{mathtools}
\usepackage{stackrel}
\usepackage{leftidx}
\usepackage[amsmath,amsthm,thmmarks]{ntheorem}

\usepackage{authblk}

\usepackage{euscript}
\usepackage{pxfonts}
\usepackage{dsfont}
\usepackage{bbold}

\usepackage[shortlabels]{enumitem}

\usepackage{hyperref}
\usepackage{cleveref}

\usepackage{xcolor}
\usepackage{tcolorbox}
\usepackage{graphicx}
\usepackage{subfigure}

\usepackage[all,2cell,arrow,matrix]{xy} \UseAllTwocells \SilentMatrices
\usepackage{tikz-cd}
\usepackage{tikz}
\usetikzlibrary{arrows,arrows.meta,decorations.markings,decorations.pathreplacing,calc}

\usepackage{verbatim}
\usepackage{comment}

\usepackage{epstopdf} 

\usepackage[nottoc]{tocbibind} 
\usepackage{appendix}

\usepackage{geometry}
\newcommand\void[1]       {}

\tikzset{->-/.style={decoration={markings,mark=at position #1 with {\arrow{stealth}}},postaction={decorate}},->-/.default=0.55}
\colorlet{e_ext}{red}
\colorlet{m_ext}{blue!30}
\tikzset{e_str/.style={very thick,red!80}}
\tikzset{m_str/.style={very thick,blue!80}}
\tikzset{m_dual_str/.style={thick,dashed,blue}}
\tikzset{link_label/.style={scale=0.8,black}}

\theoremstyle{definition}

\newtheorem{thm}{Theorem}[subsection]

\newtheorem{pthm}[thm]{Theorem$^{\text{ph}}$}

\newtheorem{notation}[thm]{Notation}
\newtheorem{exercise}[thm]{Exercise}
\newtheorem{principle}[thm]{Principle}
{
\theoremsymbol{\mbox{${\blacksquare}$}}
\newtheorem{defn}[thm]{Definition}
}
{
\theoremsymbol{\mbox{$\vardiamondsuit$}}
\newtheorem{expl}[thm]{Example}
}
{
\theoremsymbol{\mbox{$\diamondsuit$}}
\newtheorem{rem}[thm]{Remark}
}
\qedsymbol{\mbox{$\square$}}

\numberwithin{equation}{subsection}

\newcommand\be            {\begin{equation}}
\newcommand\ee            {\end{equation}}
\newcommand\bea           {\begin{eqnarray}}
\newcommand\eea         {\end{eqnarray}}
\newcommand\bnu          {\begin{enumerate}}
\newcommand\enu          {\end{enumerate}}
\newcommand\bit          {\begin{itemize}}
\newcommand\eit          {\end{itemize}}

\newcommand{\pf}{\begin{proof}}
\newcommand{\epf}{\qed\end{proof}}

\DeclareMathOperator*{\medotimes}{\vcenter{\hbox{{\scalebox{0.8}{$\bigotimes$}}}}}
\DeclareMathOperator*{\medoplus}{\vcenter{\hbox{{\scalebox{0.8}{$\bigoplus$}}}}}

\makeatletter
\providecommand{\leftsquigarrow}{%
  \mathrel{\mathpalette\reflect@squig\relax}%
}
\newcommand{\reflect@squig}[2]{%
  \reflectbox{$\m@th#1\rightsquigarrow$}%
}
\makeatother

\DeclareMathAlphabet{\mathcal}{OMS}{cmsy}{m}{n}	
\DeclareMathAlphabet{\mathsf}{OT1}{cmss}{m}{n}	

\newcommand\Cb			{\mathbb{C}}
\newcommand\Nb			{\mathbb{N}}

\newcommand\Qb			{\mathbb{Q}}
\newcommand\Rb			{\mathbb{R}}
\newcommand\Zb			{\mathbb{Z}}

\newcommand\CA			{\EuScript{A}}
\newcommand\CB			{\EuScript{B}}
\newcommand\CC			{\EuScript{C}}
\newcommand\CD			{\EuScript{D}}
\newcommand\CE			{\EuScript{E}}

\newcommand\CM			{\EuScript{M}}
\newcommand\CN			{\EuScript{N}}

\newcommand\CP			{\EuScript{P}}
\newcommand\CQ			{\EuScript{Q}}

\newcommand\CX			{\EuScript{X}}

\newcommand\CXs			{{\EuScript{X}^\sharp}}

\newcommand{\FZ}			{\text{\usefont{U}{euf}{m}{n}Z}}

\newcommand\BB			{\mathbf{B}}
\newcommand\BC			{\mathbf{C}}
\newcommand\BD			{\mathbf{D}}
\newcommand\BE			{\mathbf{E}}

\newcommand\BM			{\mathbf{M}}

\newcommand\SA			{\mathsf{A}}
\newcommand\SB			{\mathsf{B}}
\newcommand\SC			{\mathsf{C}}
\newcommand\SD			{\mathsf{D}}
\newcommand\SE			{\mathsf{E}}

\newcommand\SL			{\mathsf{L}}
\newcommand\SM			{\mathsf{M}}
\newcommand\SN			{\mathsf{N}}

\newcommand\SP			{\mathsf{P}}
\newcommand\SQ			{\mathsf{Q}}

\newcommand\SX			{\mathsf{X}}
\newcommand\SY			{\mathsf{Y}}

\newcommand{\toric}{\EuScript{TC}}

\newcommand{\Toric}{\EuScript{TC}}

\DeclareMathOperator{\Hom}{Hom}
\DeclareMathOperator{\End}{End}
\DeclareMathOperator{\Aut}{Aut}

\DeclareMathOperator{\coker}{coker}
\DeclareMathOperator{\im}{im}

\DeclareMathOperator{\id}{id}

\DeclareMathOperator{\tr}{tr}
\DeclareMathOperator{\Ind}{Ind}

\DeclareMathOperator{\ev}{ev}

\DeclareMathOperator{\fun}{Fun}

\DeclareMathOperator{\LMod}{LMod}
\DeclareMathOperator{\RMod}{RMod}
\DeclareMathOperator{\BMod}{BMod}

\DeclareMathOperator{\Gr}{Gr}

\newcommand{\op}			{\mathrm{op}}
\newcommand{\rev}			{\mathrm{rev}}

\newcommand{\one}			{\mathbb{1}}

\newcommand{\ob}          {\mathrm{ob}}
\newcommand{\tTO}          {\mathsf{1}}

\newcommand\fpdim       {\mathrm{FPdim}}

\newcommand\Cat			{\EuScript{C}\mathrm{at}}
\newcommand\cat			{\mathrm{Cat}}

\newcommand\set			{\mathrm{Set}}

\newcommand\vect			{\mathrm{Vec}}
\newcommand\svect			{\mathrm{sVec}}
\newcommand\hilb			{\mathrm{Hilb}}

\newcommand\rep			{\mathrm{Rep}}
\newcommand\grp			{\mathrm{Grp}}

\newcommand\nao			{\mbox{$n$+1}}

\newcommand\Irr			{\mathrm{Irr}}

\newcommand{\TO}{\mathsf{TO}}

\newcommand{\Bulk}{\mbox{$\mathsf{Bulk}$}}

%% file: introduction.tex

\section{Introduction}

In the last two decades, we have witnessed unprecedented interactions between topics in topological phases of matter (such as topological orders) and those in pure mathematics. It is not surprising to see the application of geometry, topology and representation theory in physics because it is a long and old tradition. What is surprising is that category theory, one of the most abstract subjects in mathematics, has deep application in the study of topological phases. This application is not accidental but rooted in the very heart of quantum many-body theory or quantum field theory. As we will show in this work, category theory is becoming an indispensable language and tool in the study of all quantum many-body systems.

Although there are many excellent textbooks on category theory (see for example \cite{Mac78,Awo10,Rie17}) and a modern textbook on tensor categories \cite{EGNO15}, unfortunately, none of them are written for physicists. Many years have passed since the first application of category theory in the study of topological orders. It remains a daunting task for working physicists and students to master or even to appreciate the language of category theory by first reading these books. In fact, there are still many mathematicians regarding category theory as a boring and useless subject or finding it difficult to adapt themselves to the abstractness of category theory. Taking into account the fact that all mathematicians were educated in the abstract language of modern mathematics from their first days of professional training, we can then imagine how difficult it can be for working physicists, who accustomed themselves to a concrete way of thinking (or computing) based on physical intuitions, to find category theory inspiring or useful. 

Therefore, in order to help students and working physicists to enter this fast developing field and to catch up with the latest developments, it is highly desirable to have some introductory articles or books focusing on the application of category theory in physics. The goal of this work is to partially fill the gap. For those who have no background in category theory, we explain in great details how categorical structures naturally emerge from simple lattice models and physical intuitions. In other words, instead of first introducing categorical notions then giving their physical application, we reverse the order by first analyzing structures hidden in a lattice model then summarizing them as categorical notions. As we proceed, almost all ingredients in the mathematical theory of monoidal categories, such as (braided) fusion categories, (monoidal) functors, natural transformations, module categories, module functors, Drinfeld centers, Morita equivalence, condensation completion and fusion 2-categories, etc., naturally emerge. Although the mathematical theory that we can cover in this work seems quite rich already, it is still just the tip of the iceberg. At the end of this work (see Section \ref{sec:adv-topics}), we provide a brief outline of a few advanced topics. 

Although the field of topological orders is more than 30 years old, our understanding of subject is still quite limited and unsatisfying. One of the manifestation of this fact is the constant evolution of the basic notions and the occasional radical changes of the point of view. This phenomenon causes a lot of difficulties and confusion in reading the literature. In this work, we try our best to distinguish notions with subtle differences and clarify misleading statements that are common in the literature. In a few places, we adopt some modern points of view available only recently \cite{KZ20,KZ22}. At the same time, we try to be open to future evolution, an attitude which is reflected in many of our remarks. For this reason, we believe that this work can be useful to experts in the fields. It might also be useful to mathematicians who want to get some ideas on how category theory is used in physics.

\medskip
The popularity of topological orders in recent years is not a scientific reason why we are interested in this subject. 
In Section \ref{sec:motivations} and \ref{sec:impacts}, we explain the theoretical motivations for and the theoretical significance of the study of topological orders, some of which are either becoming apparent only recently or not so well-known. In Section \ref{sec:new-language}, we briefly explain the significance and impacts of category theory in both mathematics and physics. In Section \ref{sec:contents}, we briefly explain the content and the layout of this paper, and provide some suggestions on how to read the paper. 

\medskip
Throughout this work, we use $n$d to represent the spatial dimension and $n$D to represent the spacetime dimension.

\subsection{Motivations} \label{sec:motivations}

We start from introducing some terminologies used in this work. A quantum matter is a state of matter at zero temperature. A quantum phase is a universal class of quantum matters. There are two types of quantum phases: gapped and gapless. In this work, we are interested in a special family of gapped quantum phases called \emph{gapped quantum liquids} \cite{ZW15} (see Section \ref{sec:GQL}). A gapped quantum liquid without symmetry is called a topological order \cite{Wen89, Wen90}. Gapped quantum liquids with symmetries include gapped spontaneous symmetry breaking orders, symmetry protected trivial (SPT) orders \cite{GW09, CLW11, CGLW13} and symmetry enriched topological (SET) orders \cite{CGW10}. In this work, we focus on 2d and 1d topological orders and only briefly comment on other quantum phases in Remarks and in Section \ref{sec:adv-topics}. 

At the zero temperature, due to the finite energy gap, a gapped quantum phase hardly responds to any external probes. It seems that it must be featureless and trivial. It was one of the greatest discoveries in 1980's that this seemingly `trivial' phase is actually non-trivial. There are three sources of this great discovery. The first one is in the study of the experimentally discovered 2d fractional quantum Hall (FQH) phases \cite{TSG82}. It was realized that different FQH phases have the same symmetry but different in other features, such as non-trivial ground state degeneracies (GSD), fractionalized excitations and chiral gapless boundaries (see \cite{Sto92} and references therein). The second source is chiral spin liquids emerged from the study of superconductors at high critical temperatures \cite{KL87}. It was realized that different chiral spin liquids can have the same symmetry but different numbers of heat-conducting edge modes \cite{WWZ89}. The third source is Witten's construction of 2+1D Chern-Simons topological quantum field theories (TQFT) \cite{Wit89}. In these TQFT's, the spaces of states are those of ground states on different closed 2d manifolds, and are different for different TQFT's; and they are non-trivial also in that there are non-trivial Wilson line defects leading to infinitely many new topological invariants \cite{RT91}. These TQFT's should be viewed as the low energy effective quantum field theories of exotic gapped quantum phases. Therefore, these new exotic gapped phases are highly non-trivial, and are beyond Landau's paradigm of spontaneous symmetry breaking theory \cite{Wen89, Wen90}. They provide a gateway to an entirely new world. This motivated Xiao-Gang Wen to introduced the notion of a `topological order' in 1989 \cite{Wen89, Wen90} (see recent reviews \cite{Wen17,Wen19} and references therein). Another commonly stated motivation is that it provides the physical foundation of the so-called fault-tolerant quantum computing \cite{Kit03,FLW02,Fre01}. These motivations have already attracted many condensed matter physicists, mathematical physicists, computer scientists and mathematicians working jointly in this field.

Interestingly, as the study of topological orders or gapped quantum liquids deepens, we have encountered more surprises. The significances and the impacts of this study have gone far beyond condensed matter physics and have already reached out to the many other fields, such as particle physics, quantum gravity, quantum computing and even pure mathematics. In the next subsection, we discuss some of the significances and the impacts of this study.  

\subsection{Significances and impacts} \label{sec:impacts}

The first significance of the study of gapped quantum liquids is that in order to go beyond Landau's paradigm, this study demands us to return to perhaps the most fundamental question in condensed matter physics:
\[
\textit{What is a phase or phase transition?}?
\]
Landau's theory was developed from the study of concrete phase transitions. The tools and the language developed for this study automatically provide a way to distinguish different phases, more precisely, a partial characterization of a phase. This approach  does not provide a prior reason for the completeness of this characterization. Landau's theory is not developed from the first principle, by which we mean first defining the notion of a phase precisely, then finding a mathematical characterization of a phase transition. The discovery of new exotic phases beyond Landau's paradigm provides us with a unique chance and motivation to study the notion of a quantum phase from the first principle. Indeed, it has already motivated many attempts to define the notion of a gapped quantum liquid precisely from both the microscopic perspective \cite{CGW10,ZW15} and the macroscopic perspective \cite{Kit06,KW14,KWZ15,JF20,KLWZZ20,KZ20a,KZ22}. Microscopically, a gapped quantum liquid is an equivalence class of gapped lattice models. The physical description of the equivalence relation between two models is essentially known as a path connecting two models in the space of `models' without closing the gap nor changing the ground state degeneracy anywhere on the path. Moving along the path is defined by properly enlarging the Hilbert space and adding local perturbations allowed by symmetries \cite{ZW15}. The real challenge lies in how to formulate this equivalence relation mathematically precise and prove its compatibility with the macroscopic description of gapped quantum liquids. In this work, we focus our attention on the macroscopic description. Note that a macroscopic description of a topological order is possible simply because the notion of a quantum phase is itself a macroscopic notion. Therefore, it is reasonable that the precise characterization of a quantum phase can be obtained by summarizing the complete\footnote{We want to emphasize that the mathematical description of the complete set of  `observables' is fundamentally different from that of a proper subset in that certain structures and results, such as the boundary-bulk relation of topological excitations (see Section \ref{sec:boundary_bulk} and \ref{sec:bbr}) and the mathematical theory of anyon condensation (see Section \ref{sec:anyon-condensation}), are possible only when the description is complete (at least to certain extent), and are impossible to see within a partial description.}  set of `observables' in the thermodynamic limit and in the long wave length limit. This point of view is both fundamental and natural, and has already played an important role in the study of topological orders. Interestingly, it also provided a new way of characterizing the symmetry-breaking phases in old Landau's paradigm \cite{KZ20a,KWZ22}, and led to a grand unification of topological orders, SPT/SET orders, gapped symmetry-breaking phases and CFT-type gapless phases \cite{KZ20,KZ21,KZ20a,KWZ22}. 


In retrospect, to be able to study topological orders from the first principle is because topological orders are the simplest quantum phases. Indeed, the categorical description of a topological order is much simpler than that of a spontaneous symmetry breaking order \cite{KWZ22}. This simplicity makes the precise and complete description and other precise results possible. We believe that the role played by topological orders in the study of quantum many-body physics is somewhat similar to that of the two-body problem in Newtonian mechanics and that of the Hydrogen atom in quantum mechanics. As many new results of quantum phases are gradually emerging, it seems that we are only at the beginning stage of revealing the true nature of quantum phases and quantum phase transitions.

\medskip
The second significance of the study of gapped quantum liquids is that it leads to a new approach towards the study of gapless quantum phases (see the discussion in \cite[Section 7]{KZ21}). It is both surprising and ironically obvious in retrospect. It is in everyone's mind that gapless quantum phases are significantly richer and harder than gapped quantum phases. This richness lead people to a natural impression that the intuitions and methods developed in the study of the gapped cases might not be useful, if not misleading, to that of the gapless cases. From this perspective, it was somewhat surprising to realize that the mathematical characterization of a gapped quantum liquid is an indispensable ingredient of that of a gapless quantum phase, and the theory of gapped liquids might provide a systematic approach to the study of gapless phases. On the other hand, it is also completely obvious. A generic gapless phase can be obtained by stacking a gapless phase $\SX$ with a gapped quantum liquid $\SA$ as illustrated in the first picture in \eqref{eq:gapless-phase}. Therefore, the mathematical characterization of a gapped quantum liquid, such as the higher category of topological defects, is also an indispensable ingredient of that of a gapless quantum phase. This categorical structure might be changed if we introduce coupling (or the glue $\SC$) between two layers as illustrated in the second picture in \eqref{eq:gapless-phase}, but its categorical nature remains intact. 
\be \label{eq:gapless-phase}
\SA \boxtimes \SX \coloneqq
\begin{array}{c}
\begin{tikzpicture}
\useasboundingbox (-0.1,-0.2) rectangle (3.5,0.8) ;
\draw[ultra thick,->-] (0,0)--(3,0) node[right] {$\SX$} node[near start,below,scale=0.6] {gapless} ;
\draw[ultra thick,->-] (0,0.5)--(3,0.5) node[right] {$\SA$} node[near start,above,scale=0.6] {gapped} ;
\draw[thick,decorate,decoration=brace] (-0.1,0)--(-0.1,0.5) ;
\end{tikzpicture}
\end{array}
\qquad
\SB \boxtimes_\SC \SY \coloneqq
\begin{array}{c}
\begin{tikzpicture}
\useasboundingbox (-0.1,-0.2) rectangle (3.5,0.8) ;
\fill[gray!20] (0,0) rectangle (3,0.5) node[midway,xshift=1cm,black] {$\SC$} ;
\draw[ultra thick,->-] (0,0)--(3,0) node[right] {$\SY$} node[near start,below,scale=0.6] {gapless} ;
\draw[ultra thick,->-] (0,0.5)--(3,0.5) node[right] {$\SB$} node[near start,above,scale=0.6] {gapped} ;
\draw[thick,decorate,decoration=brace] (-0.1,0)--(-0.1,0.5) ;
\end{tikzpicture}
\end{array}
\ee
Instead of using stacking, we can also describe this categorical structure intrinsically. In a gapless quantum phase, there are, in general, gapped excitations, which are topological (or superselection) sectors in the Hilbert space. These topological sectors should form a higher categorical structure similar to those in a gapped quantum liquid. We call this categorical structure the `topological skeleton' of the gapless quantum phase \cite[Section 7]{KZ21}. The first examples of topological skeletons of gapless quantum phases are explicitly constructed for the chiral/non-chiral gapless edges of 2d topological orders \cite{KZ20,KZ21}. Although the topological skeleton is not enough to determine a gapless quantum phase, it has already provided sufficient amount of information to study certain types of questions that are irrelevant to the missing data, such as the boundary-bulk relation, Morita equivalence of two boundaries, computing dimensional reduction processes and condensations \cite{KZ21}. Since all the missing data must be compatible with the topological skeleton, this compatibility provides non-trivial clues to the construction of the missing data. Moreover, it provides new tools to study topological phase transitions \cite{CJKYZ20}. All of these seems to suggest that the theory of gapped quantum liquids might provide a brand new and systematic approach towards the study of gapless quantum phases \cite[Section 7]{KZ21}\cite{KZ20a,KWZ22}. Indeed, it leads to a new mathematical theory of all gapped/gapless quantum liquids  (with/without symmetries) \cite{KZ22}.

\medskip
The third significance is its impact on the study of problems that are topics in the field of high energy physics. Indeed, it has stimulated an unexpected and unprecedented interaction between high energy theory and condensed matter theory. The new results and methods developed in study of gapped quantum liquids, such as many-body entanglement, string-net condensations \cite{LW05} and emergent gauge fields and gapless modes on the boundaries, turn out to be useful and inspirational to the study of many other fundamental problems in physics, such as emergent chiral fermions in standard model \cite{Wen03,Wen13,YX15,AC16,CS17}, beyond standard model \cite{Wan21} (and references therein), string theory \cite{KPMT20,TY21}, emergent gravitational modes \cite{Xu06,GW12,Gu17}, applications to loop quantum gravity \cite{DG17,Zuo17}, many of the recent developments of dualities among 2+1D CFT's (see for example \cite{SSWW16,SSWX19} and references therein), and holography through the study of entanglement entropy (see for example \cite{RT06,Swi12,Qi18} and references therein). Moreover, important phenomena found in the study of gapped phases, such as bulk-boundary correspondence \cite{LH08,QKL12} and boundary-bulk relation of topological excitations \cite{KK12,KWZ17}, also echo with holographic principle in quantum gravity \cite{Hoo93,Mal99} and the open-closed duality in string theory. It seems that we are only at a beginning stage to grasp some universal features of quantum many-body systems. 
In retrospect, perhaps the reason for its influence and connection to quantum gravity is the fact that the what-is-a-phase question cannot be completely separated from the question:
\[
\textit{What are the space, the time and elementary particles?}
\]
if they are not the same question (are they?). After all, they are all quantum many-body problems.

\medskip
The fourth significance is its impact on pure mathematics. It was known at the dawn of modern science that mathematics is efficient in solving problems in physics \cite{Wig60}. There is no exception in our case. The mathematical notions of a 1+1D conformal field theory (CFT) \cite{Seg88} and a TQFT \cite{Ati88} were introduced earlier than that of a topological order \cite{Wen89}. So it is not surprising to see that many mathematical results about TQFT's, CFT's and related subjects, such as quantum groups, vertex operator algebras and tensor categories, have played important roles in the development of the theory of topological orders. In recent years, more advanced mathematical topics, such as higher algebras \cite{Lur17} and factorization homology \cite{AF20}, also joined the party \cite{AKZ17,JF20}. A natural question is: ``Can physics return the favor?''. Although it has already been a few decades since physics returned the favor through string theory (see Remark \ref{rem:return-favor}), it is still not obvious if the favor can be return through the study of topological orders. But it did happen. The physical problems and intuitions of topological orders have already directly produced or inspired many new mathematical results in topology (see for example \cite{FH20,FH21,FT21}) and in category theory (see for examples \cite{DMNO13,BNRW15,KZ18,LKW16a,AKZ17,KZ18a,KTZ20,GJF19,VR19,KZ21,KYZ21,JF20,KLWZZ20,KZ20a,JF20a,JFY21a,JFR21,KZ21a}) and even theories of new mathematical theories (see for example \cite{DR18,GJF19,JF20,KYZZ21,KZ22}). 

A direct reason behind this return of favor is that the study of topological orders has raised many important new questions and provided surprisingly powerful physical intuitions that are not so obvious from the mathematical perspective. For example, questions related to topological phase transitions between topological orders are completely new to mathematicians who are experts in the mathematical theory of TQFT's. These new questions demand and have inspired new mathematical theories. As the new approach towards the study of gapless quantum phases just starts, it is only reasonable to expect more impacts on pure mathematics in the coming future. 

\begin{rem} \label{rem:return-favor}
Why can physics help to discover new mathematical structures and solve mathematical problems? Let us venture out to give a partial answer to this philosophical question. An infinite dimensional mathematical object, viewed from generators and relations, is very complicated and lacks of global intuition. However, if it describes a quantum many-body system, say a solid material, our physical intuition, including the ordinary visual effect of a solid material in the sight of human eyes, is the consequence of a highly non-trivial RG computation, i.e. integrating out all microscopic degrees of freedom. Similarly, all phenomenological theories, Hamiltonian or Lagrangian descriptions and even physical measurement carried out in physics labs can all be regarded as powerful computations of systems with infinite number of degrees of freedom. Although we are still far from understanding the true nature of these transcendental computations, it is not hard to imagine that they, as gifts from mother nature, can be surprisingly powerful in solving mathematical problems. Moreover, quantum field theories have revealed an entirely new world of infinite dimensional mathematics. A lot of new mathematical theories developed in mathematics were inspired by quantum field theories. 
\end{rem}

\subsection{Category theory as a new language} \label{sec:new-language}

The mathematical theory of symmetries (in the the classical sense) is that of groups. The new exotic phases of matter challenge us to find radically new mathematical language to characterize gapped quantum liquids. The enormous effort has been made to meet this challenge (see for example \cite{Kit06,LW05,KK12,BBCW19,KWZ15,LKW16a,JF20,KLWZZ20,KZ20a,JF20a,JFY21a} and references therein). As a result, up to invertible gapped quantum liquids, the mathematical structure that characterizes a gapped quantum liquid is found to be a (higher) category, which describes topological defects in the phase. 

\smallskip
This success is, however, somewhat bittersweet to many working physicists. The mathematical theory of higher categories, or category theory, is far more exotic than the group theory, and is notorious for its abstractness even in mathematics. As a consequence, even after 30 years of development, it is still a big challenge for a working physicist to master the abstract language of category theory. Although it is obvious to some experts in this field that category theory is indispensable (also powerful) in the study of topological orders, ironically, if a physical paper on topological orders uses a lot of category theory, it is often considered by many physicists as mathematics or perhaps non-physics, and is not accepted by physics journals. The main goal of this paper is to bridge the gap between physical/geometrical intuitions and mathematical notions. More precisely, for physicists with no background in category theory, we explain in great details how the mathematical notions of a unitary modular tensor category (UMTC), a fusion category, a monoidal functor, Drinfeld center, a module category, Morita equivalence, condensation completion and fusion 2-categories, etc. naturally emerge from physical/geometrical intuitions or explicit lattice models calculations. We also clarify some subtle issues, misleading statements and confusions that are abundant in literature along the way. Although all results in this work were known, many parts of our explanation are hard to find in literature. Some of them should be interesting and useful even to experts. 

From a physical perspective, we have already argued in Section  \ref{sec:impacts} that the theory of gapped quantum liquids is not just a corner in physics. It leads us to a new approach towards the study of the gapless quantum phases, and links to many fundamental questions in physics, including particle physics and quantum gravity. From a mathematical perspective, category theory is also far more than just a generalization of group theory. It is as basic as the set theory, and provides a new foundation of mathematics. It is capable of supplying many different kinds of new `calculus' and `linear algebras'. It has already become the basic languages or powerful tools in a variety of fields, such as logic, functional analysis, algebraic number theory, algebraic geometry, algebraic topology and representation theory. Moreover, it unifies `continuous' with `discrete', `algebraic' with `analytic or geometric' and `finite' with `infinite'. Perhaps, the most important fact is that there are a lot of new mathematical truths that can only be expressed in the categorical language. In other words, these mathematical truths are beyond set theory. There has been a new wave in mathematics of replacing set theory by category theory as the new foundation of mathematics since the rise of the new paradigm of algebraic geometry developed in 1960's by Alexander Grothendieck and his schools. Therefore, the real question is: can category theory become the language of a new calculus for quantum many-body physics?

\medskip
The application of category theory in physics start as early as 1980's in the study of $1+$1D conformal field theories (CFT) and $n+$1D TQFT's. Mathematical definition of a CFT (or a TQFT) was originated from the idea of formalizing the factorization property of path integrals as certain monoidal functor defined on a cobordism category \cite{Seg88,Ati88}. Soon after, Moore and Seiberg discovered the notion of a modular tensor category\footnote{The name was suggested by Igor Frenkel and first appeared in \cite{MS90}. But Moore-Seiberg's notion was not formulated in the categorical language and was called `Moore-Seiberg data' in \cite{BK01}.} in their study of $1+$1D rational CFT's as the category of modules over the so-called chiral algebras \cite{MS89,MS90} (see \cite{Hua08a,Hua08}) for a mathematical proof). In 1991, Reshetikhin and Turaev reformulated Moore-Seiberg's notion categorically as what we see now, and used it to give a construction of the so-called 2+1D Reshetikhin-Turaev TQFT's \cite{RT91,TV92,Tur20}. It eventually entered the study of topological order as the category of topological excitations (or anyons) in a 2d topological order (see \cite{Kit06} for a review). Through the construction of Levin-Wen models \cite{LW05} and gapped boundaries and domain walls \cite{KK12} and through the anyon condensation theory \cite{Kon14}, nearly all modern developments of tensor categories (see a book \cite{EGNO15}) have entered the field of topological orders in their full strength. After that, the applications of the tensor category theory to the study of gapped quantum liquids (with symmetries) has entered a golden age (see for example \cite{BBCW19,KZ18,LKW16a,AKZ17,KZ18a,CJKYZ20,KZ20,KZ21}). Recenlty, the new wave of applying higher category theory to the study of gapped quantum liquids is picking up momentum \cite{KW14,KWZ15,BHW17,DR18,KTZ20,BD19,GJF19,BD20,XZGC21,JF20,KLWZZ20,BD21,KTZ20a,JFY21,KZ20a,BD21a,JFY21a,JFR21,KZ21a}. Category theory also played a crucial role in the study of homological mirror symmetry \cite{Kon94} as the category of D-branes in certain superstring theory. This application also planted the seed for the later development of higher algebras \cite{Lur09}. We believe that the mathematical theory of higher algebras and higher representation theories \cite{Lur08,Lur09,Lur17,AF20} will eventually solve important physics problems.

It is beneficial to reflect upon the reason why the category theory is so useful in the study of quantum many-body physics. In general, a physical system with infinite degrees of freedom is very complicated if you measure it via the inefficient device of the 1-dimensional ground field $\Cb$. It is often expressed in terms of certain complicated infinite dimensional $\Cb$-linear algebraic structures, such as vertex operator algebras and algebras over operads or PROPs. However, if we change our measuring device to more efficient one: an infinite dimensional creature $A$, then above algebraic structures turn themselves into familiar (even `finite dimensional') notions of `$A$-linear' algebras in certain highly non-trivial monoidal higher categories (e.g. $E_k$-monoidal higher categories). This fact is deeply related to or perhaps not different from the fact that all gapped defects in a quantum many-body system or a quantum field theory automatically form a higher category.

The applications of category theory have also reached out to computer science, linguistics, philosophy and economics, etc. Just as the calculus is once the universal language of all sciences, this is not hard to imagine that the category theory, which is capable of generating many new calculus (including the old one), will eventually become a universal language of all sciences. Our humble hope is that this paper might help to speed up this process. 

\begin{rem}
We would like to give a remark that might help some readers to overcome a psychological barrier in the study of category theory. We have heard from some physicists and students that they have spent a lot of time studying category theory, but only found it very formal, dry, uncomputable and useless. This feeling and reaction is very common and natural. It reflects a rather deep truth. In fact, category theory is as basic as set theory. Both of them can serve as the foundations of mathematics. Physicists' reaction to set theory is likely to be the same: formal, dry, uncomputable and useless. But physicists really do not need set theory because all they need are calculus and linear algebra. Similarly, it is impossible to see the power of category theory until you see `categorical calculus' and `categorical linear algebras'. Physicists can study calculus directly without go through set theory because all they need from set theory is automatically covered by the theory of functions in calculus. But for the `categorical calculus or linear algebra', as far as we can tell, it is impossible to bypass the basic notions of a category, a functor, a natural transformation, Yoneda Lemma, adjoint functors, etc. This work provides a friendly way to get a glimpse of the `categorical linear algebra' through concrete physics models. We hope that it can help physicists to overcome the language barrier. 
\end{rem}

\subsection{Layout and how to read} \label{sec:contents}

We explain the layout of this paper: in Section \ref{sec:basic}, we briefly review some basic notions related to topological orders; in Section \ref{sec:2d}, we explain in details how the structure of a unitary modular tensor category emerges from physical and geometric intuitions of topological excitations; in Section \ref{sec:1d}, we explain the emergence of many mathematical notions associated to (anomalous) 1d topological orders and the boundary-bulk relation, such as unitary fusion categories and Drinfeld center, etc; in Section \ref{sec:adv-topics}, we give a brief guideline of some advanced topics.  

\medskip
In the main text of this paper, we try to be self-contained, slowly paced and not technical. We hide many technical parts in Remarks. One of the purposes of this paper to let readers to get a glimpse of the incredible richness of category theory and its application in the study of topological orders. Therefore, we choose to comment on some advanced physical or mathematical topics only in remarks, where intrigued readers can find relevant references for further reading. That is also to say that if you are a beginner and do not understand those technical and advanced Remarks, you can simply ignore them.

\medskip
Throughout this work, we use ``Theorem$^{\text{ph}}$'' to highlight a physical result, and use ``Theorem'' to represent a mathematically rigorous result.

\bigskip
\noindent {\bf Acknowledgements}: We would like to thank Meng Cheng, Arthur Jaffe, Xiao-Liang Qi, Yang Qi, Tian Lan, Chenjie Wang, Xiao-Gang Wen and Cen-Ke Xu for useful comments. We would also like to thank An-Si Bai, Chun-Yu Bai, Rongge Xu and Holiverse Yang for finding typos. We are supported by NSFC (Grant No. 11971219) and by Guangdong Provincial Key Laboratory (Grant No.2019B121203002) and by Guangdong Basic and Applied Basic Research Foundation (Grant No. 2020B1515120100). ZHZ is also supported by Wu Wen-Tsun Key Laboratory of Mathematics at USTC of Chinese Academy of Sciences.

%% file: basic.tex

\section{Basic notions for topological orders} \label{sec:basic}

In this section, we briefly review a few basic notions associated to topological orders and introduce some notations along the line. 

\subsection{Gapped quantum phases} \label{sec:GQL}

A quantum phase is a universality class of quantum many-body systems defined in the thermodynamic limit and at zero temperature. Since a \emph{quantum phase} can be realized by lattice models, it can be defined microscopically as an equivalence class of lattice models. In this subsection, we briefly sketch this microscopic perspective of gapped quantum phases without giving details. 

\medskip
By a \emph{lattice model} we mean a collection of the following data:
\bit
\item There are some discrete sites in the space, and the number of sites is called the size of the lattice model.
\item For each site $i$ we associate to it a Hilbert space $\mathcal H_i$ (called a \emph{local Hilbert space}), which represents the local degrees of freedom.
\item The \emph{total Hilbert space} is defined by $\mathcal H \coloneqq \medotimes_i \mathcal H_i$.
\item The \emph{Hamiltonian} $H$ acting on $\mathcal H$ is the sum of local interaction terms. A term in $H$ is called local if it acts as identities on all sites except some spatially local sites.
\eit
For convenience, we also use the total Hilbert space $\mathcal H$ to denote a lattice model. The thermodynamic limit is defined for a quantum (many-body) system\footnote{When a lattice model $\mathcal H$ is translation invariant, it induces a canonical sequence $\{\mathcal H^{(k)}\}$ of lattice models. In this case, `a lattice model' and `a quantum system' have the same meaning.}, which is a sequence $\{\mathcal H^{(k)}\}$ of lattice models, where the size of $\mathcal H^{(k)}$ tends to infinity as $k \to \infty$. In order for the limit to be well-defined, additional data and conditions are needed. The literature on this subject is gigantic. We avoid to go into detail here but recommend a review \cite{GJ87} and references therein. In principle, a \emph{quantum phase} could be defined microscopically as an equivalence class of quantum systems. However, it is hard to define the equivalence relation in this generality.

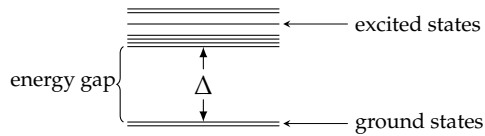
\begin{figure}[htbp]
\centering
\begin{tikzpicture}[scale=1.0]
\useasboundingbox (0,-1.05) rectangle (3,0.5) ;
\foreach \y in {-1.05,-1,0,0.05,...,0.2,0.3,0.45,0.5}
	\draw (0,\y)--(2,\y) ;
\draw[-latex] (1,-0.3)--(1,0) ;
\draw[-latex] (1,-0.7)--(1,-1) ;
\node at (1,-0.5) {$\Delta$} ;
\draw[-stealth,very thin] (2.9,-1.02) node[right,scale=0.8] {ground states}--(2.05,-1.02) ;
\draw[-stealth,very thin] (2.9,0.3) node[right,scale=0.8] {excited states}--(2.05,0.3) ;
\draw[decorate,decoration=brace] (-0.05,-1)--(-0.05,0) node[midway,left,scale=0.8] {energy gap} ;
\end{tikzpicture}
\caption{the energy spectrum of a typical gapped system}
\label{fig:gapped_system}
\end{figure}

In this work, we are only interested in a special kind of quantum systems, which satisfy the following conditions (see Figure \ref{fig:gapped_system}): 
\bit 
\item there exists a fixed $\Delta > 0$ independent of $k$ such that
\bnu
\item there is no eigenvalue of $H^{(k)}$ in an energy window of size $\Delta$ as $k\to \infty$;
\item the number of eigenstates below the energy window is finite and independent of $k$ as $k\to \infty$;
\item the energy splitting of those eigenstates below the energy window tends to zero as $k \to \infty$.
\enu 
\eit
Such a quantum system is called a {\it gapped liquid\footnote{Gapped {\it non-liquid} quantum systems are known (see for example \cite{Cha05,Haa11}).} quantum system}. The subspace spanned by those eigenstates below the energy window is called the \emph{ground state subspace}, and its dimension is called the {\it ground state degeneracy} (GSD). The quantum phase determined by a gapped liquid quantum system is called a {\it gapped quantum liquid} (or a {\it gapped liquid} for short), and is defined by an equivalence class of gapped liquid quantum systems. In principle, the equivalence relation could be defined by the (symmetry allowed) perturbations of the quantum system without closing the gap and without changing GSD (see Remark \ref{rem:LU}).

\begin{rem} \label{rem:LU}
To define the equivalence relation precisely is a very subtle problem. Note that we have not yet made the notion of a quantum system very precise. Since only the ground state subspace is physically relevant at zero temperature, one way is to define a gapped liquid directly by an equivalence class of the ground state wave functions. The proposal of such an equivalence relation can be found in \cite{CGW10,ZW15,SM16}. 
\end{rem}

\begin{expl} \label{expl:trivial_quantum_phase}
Suppose $\mathcal V$ is a finite dimensional Hilbert space and $P$ is a Hermitian operator on $\mathcal V$ with a unique ground state $\lvert \psi \rangle$. For each $n \in \Nb$, there is an $n$d lattice model defined as follows:
\bit
\item Each local Hilbert space is $\mathcal H_i \coloneqq \mathcal V$.
\item The Hamiltonian is $H \coloneqq \sum_i P_i$, where $P_i = \cdots \otimes 1 \otimes P \otimes 1 \otimes \cdots$ is the operator acting on $\mathcal H_i$ as $P$ and acting on other local Hilbert spaces as identity.
\eit
This defines a gapped liquid quantum system with a unique the ground state given by the \emph{product state} $\lvert \psi \rangle \otimes \lvert \psi \rangle \otimes \cdots$. It realizes the \emph{trivial quantum phase} or the \emph{trivial gapped liquid}, denoted by $1_n$.
\end{expl}

\begin{expl} \label{expl:stacking_quantum_phase}

The stacking of two $n$d lattice models (or quantum systems or quantum phases) is a two-layer system without coupling between two layers. For two $n$d quantum phases $\SC_n$ and $\SD_n$, their stacking is illustrated in the following picture:  
\[ 
\SC_n \boxtimes \SD_n \coloneqq
\begin{array}{c}
\begin{tikzpicture}
\useasboundingbox (-0.1,-0.2) rectangle (3.5,0.8) ;
\draw[ultra thick,->-] (0,0)--(3,0) node[right] {$\SD_n$} node[near start,below,scale=0.6] {} ;
\draw[ultra thick,->-] (0,0.5)--(3,0.5) node[right] {$\SC_n$} node[near start,above,scale=0.6] {} ;
\draw[thick,decorate,decoration=brace] (-0.1,0)--(-0.1,0.5) ;
\end{tikzpicture}
\end{array}
\]
This stacking defines a commutative multiplication on the set of all quantum phases, and the trivial phase $1_n$ is the unit under this multiplication, i.e. $1_n \boxtimes C_n=C_n$. 
\end{expl}





A gapped liquid without symmetry is called a \emph{topological order}. We denote the set of all $n$d topological orders by $\TO_n$. By Example \ref{expl:trivial_quantum_phase}, there is at least one topological order in each dimension: the \emph{trivial topological order}, i.e. $\tTO_n \in \TO_n$. The set $\TO_n$, together with the multiplication $\boxtimes$ and the unit $1_n$, defines a commutative monoid. 


\begin{rem} \label{rem:TO_EFT_TQFT}
The notion of a topological order is closely related to that of a topological quantum field theory (TQFT) but with a subtle difference (see Remark \ref{rem:TO-TQFT}). It was generally believed that the low energy effective theory of a topological order is a TQFT \cite{Wit89}. Indeed, the low energy effective theories of chiral spin liquids \cite{WWZ89} and fraction quantum Hall liquids \cite{ZHK89} are Chern-Simons TQFT's; those of Kitaev's quantum double models \cite{Kit03} are 3+1D Dijkgraaf-Witten TQFT's; those of Levin-Wen models \cite{LW05} are Turaev-Viro-Barrett-Westbury TQFT's \cite{TV92,BW99}.   
\end{rem}

\begin{rem}
Gapped liquids with symmetries include gapped spontaneous symmetry breaking orders, symmetry enriched topological (SET) orders \cite{CGW10} and symmetry protected trivial (SPT) orders \cite{GW09, CLW11, CGLW13} (see also Section \ref{sec:SPT/SET_2d}).
\end{rem}



\subsection{\texorpdfstring{$\Rb^n$}{Rn}-observables} \label{sec:Rn_observable}

At zero temperature, regardless gapped or gapless, only physically relevant data are those survived in the low energy limit, or equivalently, in the long wave length limit. In other words, the notion of a quantum phase is actually a macroscopic notion defined in the long wave length limit. Therefore, 
\begin{quote}
in principle, a quantum phase should be characterized by all possible `observables' in the long wave length limit. 
\end{quote}
When the quantum phase is gapless, typical examples of such `observables' are the correlation functions as those in quantum field theories (QFT), which should be viewed as the low energy effective field theories of quantum many-body systems. When the quantum phase is gapped, all the correlation functions decay exponentially. They are not observables in the long wave length limit. In this case, there are other `observables' which are discussed in Section \ref{sec:top_skeleton}.

\begin{figure}[htbp]
\centering
\begin{tikzpicture}
\begin{scope}[xshift=-3cm,rotate=-20]
\draw[even odd rule,fill=gray!20] (0,0) ellipse (1.5 and 1) (0.5,0) .. controls (0.2,0.3) and (-0.2,0.3) .. (-0.5,0) .. controls (-0.2,-0.3) and (0.2,-0.3) .. cycle ;
\draw (-0.6,0.1)--(-0.5,0) (0.5,0)--(0.6,0.1) ;
\coordinate (X) at (1,0.2) ;
\draw[densely dashed] (X) circle (0.25) ;
\coordinate (A) at ([shift=(135:0.25cm)] X) ;
\coordinate (B) at ([shift=(-85:0.25cm)] X) ;
\end{scope}
\begin{scope}[xshift=4cm,rotate=10]
\draw[even odd rule,fill=gray!20] (2.5,0) .. controls (2.5,1) and (1,1) .. (0,1) .. controls (-1,1) and (-2.5,1) .. (-2.5,0) .. controls (-2.5,-1) and (-1,-1) .. (0,-1) .. controls (1,-1) and (2.5,-1) .. cycle (-0.5,0) .. controls (-0.8,0.3) and (-1.2,0.3) .. (-1.5,0) .. controls (-1.2,-0.3) and (-0.8,-0.3) .. cycle (1.5,0) .. controls (1.2,0.3) and (0.8,0.3) .. (0.5,0) .. controls (0.8,-0.3) and (1.2,-0.3) .. cycle ;
\draw (-1.6,0.1)--(-1.5,0) (-0.5,0)--(-0.4,0.1) ;
\draw (0.4,0.1)--(0.5,0) (1.5,0)--(1.6,0.1) ;
\coordinate (Y) at (-1.9,-0.3) ;
\draw[densely dashed] (Y) circle (0.25) ;
\coordinate (C) at ([shift=(50:0.25cm)] Y) ;
\coordinate (D) at ([shift=(-100:0.25cm)] Y) ;
\end{scope}
\begin{scope}
\draw[clip] (0,0) circle (0.8) ;
\draw[help lines,step=0.3cm,xshift=0.1cm,yshift=0.05cm] (-1.5,-1.5) grid (1.5,1.5) ;
\coordinate (E) at ([shift=(110:0.8cm)] 0,0) ;
\coordinate (F) at ([shift=(-102:0.8cm)] 0,0) ;
\coordinate (G) at ([shift=(62:0.8cm)] 0,0) ;
\coordinate (H) at ([shift=(-92:0.8cm)] 0,0) ;
\end{scope}
\draw[very thin] (A)--(E) (B)--(F) (C)--(G) (D)--(H) ;
\end{tikzpicture}
\caption{The notion of a phase is local in the sense that it is defined on an open disk.}
\label{fig:phase_local_disk}
\end{figure}
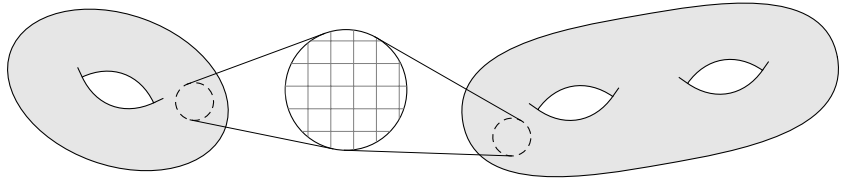

\medskip
One more thing can be said about the notion of a phase. It is often stated in literature to consider a phase defined on a closed manifold. There is nothing wrong about this statement. But it is wrong to regard the notion of a phase as a global notion (or an observable) defined on a closed manifold. 
\begin{itemize}

\item The only reason we can regard the same FQH state on topological surfaces with arbitrary genus as the same FQH phase is because they are the same phase on each open 2-disk covering the surfaces (see Figure \ref{fig:phase_local_disk}). This fact becomes even more obvious by considering a 2-sphere with a 2d quantum phase $\SA$ defined on the northern hemisphere, a different phase $\SB$ defined on the southern hemisphere and a 1d domain wall on the equator. Therefore, the notion of an $n$d phase is actually a `local' notion in the sense that it is defined on an open $n$-disk. 

\item But this term `local' is misleading because each $n$-disk must be in the thermodynamic limit thus `non-local' with respect to certain length scale. Therefore, the notion of a (quantum) phase is a notion defined on an open disk of infinite size, or on $\Rb^n$ for simplicity. To distinguish those `observables' defined on $\Rb^n$ (in the long wave length limit) from those global observables defined on a closed manifold, we refer to the former as $\Rb^n$-observables. In other words, a quantum phase should be characterized by all $\Rb^n$-observables in the long wave length limit.  

\end{itemize}
In Section \ref{sec:factorization_homology}, we explain briefly on how to obtain global observables (or invariants) defined on a closed manifold $\Sigma$ by integrating $\Rb^n$-observables on $\Sigma$ via the so-called factorization homology \cite{AKZ17}. 

\subsection{Anomaly-free/anomalous quantum phases} \label{sec:anomaly-free_anomalous}

The notion of an anomaly-free/anomalous topological order was first introduced in \cite{KW14}. We slightly generalize it to quantum phases. 

\begin{defn}
An $n$d quantum phase is called \emph{anomaly-free} if it can be realized by an $n$d lattice model with only local interactions. Otherwise it is called \emph{anomalous}.
\end{defn}

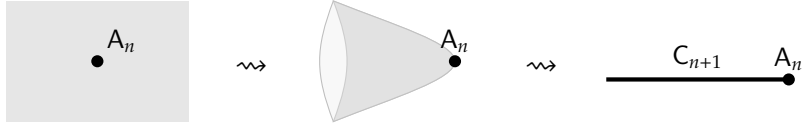
\begin{figure}[htbp]
\[
\begin{array}{c}
\begin{tikzpicture}[scale=0.8]
\fill[gray!20] (-1.5,-1) rectangle (1.5,1) ;
\fill (0,0) circle (0.1) node[above right] {$\SA_n$} ;
\end{tikzpicture}
\end{array}
\quad\rightsquigarrow\quad
\begin{array}{c}
\begin{tikzpicture}[scale=0.8]
\draw[gray,fill=gray!10,opacity=0.5] (-2,1) .. controls (-1,0.6) and (0,0.3) .. (0,0) .. controls (0,-0.3) and (-1,-0.6) .. (-2,-1) .. controls (-2.3,-0.2) and (-2.3,0.2) .. cycle ;
\draw[gray!50,fill=gray!30,opacity=0.7] (-2,1) .. controls (-1,0.6) and (0,0.3) .. (0,0) .. controls (0,-0.3) and (-1,-0.6) .. (-2,-1) .. controls (-1.7,-0.2) and (-1.7,0.2) .. cycle ;
\fill (0,0) circle (0.1) node[above] {$\SA_n$} ;
\end{tikzpicture}
\end{array}
\quad\rightsquigarrow\quad
\begin{array}{c}
\begin{tikzpicture}[scale=0.8]
\draw[ultra thick] (-3,0) rectangle (0,0) node[midway,above] {$\SC_{n+1}$} ;
\fill (0,0) circle (0.1) node[above] {$\SA_n$} ;
\end{tikzpicture}
\end{array}
\]
\caption{a dimensional reduction process}
\label{fig:dimensional_reduction}
\end{figure}

An $n$d anomalous quantum phase $\SA_n$, if exists, must be realized as a defect in a higher (but still finite)
dimensional lattice model. By a process of dimensional reduction illustrated in Figure \ref{fig:dimensional_reduction}, one can always realize the quantum phase $\SA_n$ as a boundary of an $\nao$d lattice model, which realizes an $\nao$d anomaly-free  quantum phase $\SC_{n+1}$ (see also \cite[Section VI.B]{KW14}). 

\begin{pthm}
An $n$d anomalous quantum phase can always be realized by a boundary of a 1-dimension-higher lattice model, or equivalently, a boundary of a 1-dimension-higher anomaly-free quantum phase. 
\end{pthm}

Although the realizations of $\SA_n$ as a higher codimensional defect are almost never unique, it was shown in \cite{KW14} that if $\SA_n$ is a topological order, after the dimensional reduction, the anomaly-free phase $\SC_{n+1}$ is unique (see \cite[Lemma 1]{KW14} for a physical proof). It was later proposed as a general principle for all quantum liquids $\SA_n$ \cite[Section 1]{KZ20a} (see Remark \ref{rem:liquid}). 
\begin{quote}
{\bf Unique bulk principle}: An $n$d quantum liquid $\SA_n$ can always be realized as a boundary of an anomaly-free $\nao$d quantum liquid $\SC_{n+1}$. Moreover, such $\SC_{n+1}$ is unique and is called the {\it bulk of $\SA_n$} and denoted by $\Bulk(\SA_n)$. 
\end{quote}

\begin{rem} \label{rem:liquid}
The bulk $\Bulk(\SA_n)$ of $\SA_n$ is also called the gravitational anomaly of $\SA_n$ \cite{KW14}. When quantum liquid $\SA_n$ is gapped (possibly with certain symmetries), the proof given in \cite[Lemma 1]{KW14} does apply to these cases (if its bulk $\SC_{n+1}$ have the same symmetry as that of $\SA_n$). When $\SA_n$ is gapless, the proof given in \cite[Lemma 1]{KW14} does not apply. Nevertheless, we propose the uniqueness of the bulk as a general principle for all quantum liquids. 

There is no precise definition of a gapless quantum liquid. We use the term in the sense of \cite[Section 1]{KZ20a}. Roughly speaking, a gapless quantum liquid is a gapless quantum phase that is `soft' enough so that it does not rigidly depend on the local spacetime geometry. Moreover, the unique bulk principle is proposed as a defining property of a quantum liquid. 
Gapless quantum liquids should include certain CFT-type gapless phases. 
\end{rem}

We depict the geometric relation between the boundary phase $\SA_n$ and its bulk as follows: 
\[
\begin{array}{c}
\begin{tikzpicture}[scale=0.8]
\draw[ultra thick] (-3,0) rectangle (0,0) node[midway,above] {$\hspace{-1cm}\Bulk(\SA_n)$} ;
\fill (0,0) circle (0.1) node[above] {$\SA_n$} ;
\end{tikzpicture}
\end{array}
\]
where the $\nao$d bulk phase $\Bulk(\SA_n)$ is depicted as a spatial open interval to emphasize that a quantum phase is defined on the spatial manifold $\Rb^n$. The time axis is not shown. It is clear that $\Bulk(\tTO_n)=\tTO_{n+1}$. Moreover, we have the following result. 
\begin{pthm}
An $n$d quantum liquid $\SA_n$ is anomaly-free if and only if $\Bulk(\SA_n)=\tTO_{n+1}$. 
\end{pthm}

%
%
%

\subsection{Topological defects and topological skeletons} \label{sec:top_skeleton}

For a gapped liquid, all its correlation functions decay exponentially thus do not define observables in the long wave length limit. In this case, there are other $\Rb^n$-observables given by topological defects.

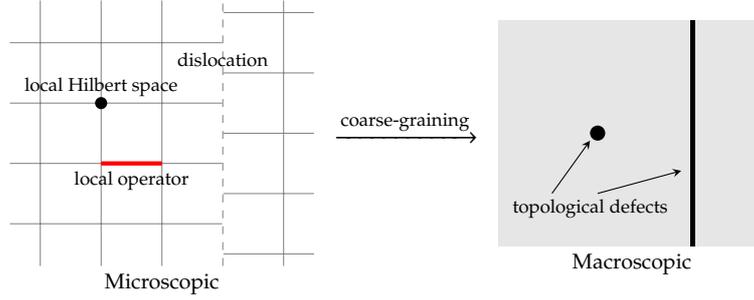
\begin{figure}[htbp]
\[
\begin{array}{c}
\begin{tikzpicture}[scale=0.8]
\useasboundingbox (-3.5,-1.7) rectangle (1.5,2.7) ;
\draw[help lines,step=1.0] (-3.5,-1.7) grid (-0.01,2.7)  ;
\draw[help lines,step=1.0,yshift=-0.5cm] (0.01,-1.2) grid (1.5,3.2) ;
\draw[help lines,dashed] (0,-1.7)--(0,2.7) ;

\fill (-2,1) circle (0.1) node[above,scale=0.7] {local Hilbert space} ;
\draw[ultra thick,red] (-2,0)--(-1,0) node[midway,below,black,scale=0.7] {local operator} ;
\node[above,scale=0.7] at (0,1.5) {dislocation} ;

\node[scale=0.8,below] at (-1,-1.7) {Microscopic} ;
\end{tikzpicture}
\end{array}
\xrightarrow{\text{coarse-graining}}
\begin{array}{c}
\begin{tikzpicture}[scale=1.0]
\useasboundingbox (-2.5,-1.5) rectangle (1,1.5) ;
\fill[color=gray!20] (-2.5,-1.5) rectangle (1,1.5) ;
\fill (-1.2,0) circle (0.1) ;
\fill (0.02,-1.5) rectangle (0.08,1.5) ;
\node[scale=0.7] at (-1.3,-1) {topological defects} ;
\draw[-stealth] (-1.8,-0.8)--(-1.3,-0.1) ;
\draw[-stealth] (-1.2,-0.8)--(-0.1,-0.5) ;

\node[scale=0.8,below] at (-0.75,-1.5) {Macroscopic} ;
\end{tikzpicture}
\end{array}
\]
\caption{A defect in a lattice model can be given by introducing dislocations, enlarging local Hilbert spaces, or adding local operators to the Hamiltonian. Topological defects are defects in the long wave length limit.}
\label{fig:topological_defect}
\end{figure}

\medskip
Microscopically, a \emph{defect} in a lattice model can be realized as a local modification of the lattice, local Hilbert spaces, or the Hamiltonian. We give explicit examples in Section\ \ref{sec:2d} (see also \cite[Section IV]{KW14}). A defect may break the uniformity of the phase, and determines an observable in the long wave length limit after coarse-graining. In a topological order, the equivalence class of such an observable with the equivalence relation defined by all admitted perturbations is called a \emph{topological defect} (see Figure \ref{fig:topological_defect}). 

\begin{rem} \label{rem:trivial_topological_defect}
It is possible that a defect does not break the uniformity of the phase and one can not see it in the long wave length limit. Such a topological defect is called the \emph{trivial topological defect}. In other words, the trivial topological defect is the equivalence class of `no defect'.
\end{rem}

A defect itself can be viewed as a lower-dimensional lattice model embedded into a higher-dimensional lattice model. We only consider gapped defects in this work. Hence in the long wave length limit, a topological defect in a topological order can be viewed as a lower-dimensional (potentially anomalous) topological order. 

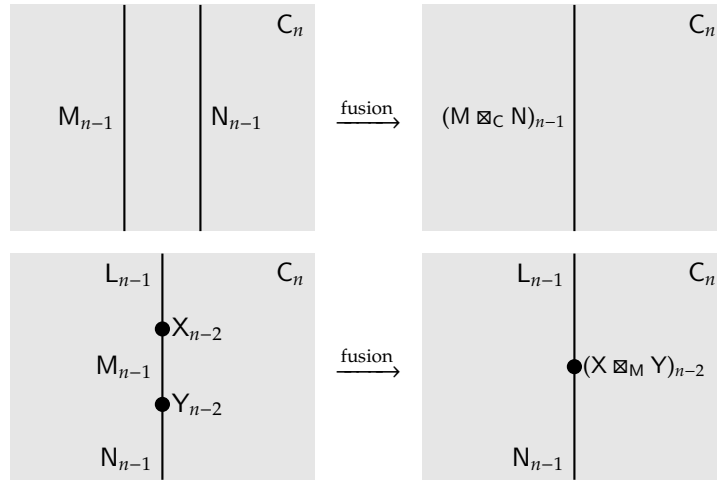
\begin{figure}[htbp]
\begin{gather*}
\begin{array}{c}
\begin{tikzpicture}[scale=1.0]
\fill[color=gray!20] (0,0) rectangle (4,3) node[below left,black] {$\SC_n$} ;
\draw[thick] (1.5,0)--(1.5,3) node[midway,left] {$\SM_{n-1}$} ;
\draw[thick] (2.5,0)--(2.5,3) node[midway,right] {$\SN_{n-1}$} ;
\end{tikzpicture}
\end{array}
\xrightarrow{\text{fusion}}
\begin{array}{c}
\begin{tikzpicture}[scale=1.0]
\fill[color=gray!20] (0,0) rectangle (4,3) node[below left,black] {$\SC_n$} ;
\draw[thick] (2,0)--(2,3) node[midway,left,scale=0.9] {$(\SM \boxtimes_\SC \SN)_{n-1}$} ;
\end{tikzpicture}
\end{array} \\
\begin{array}{c}
\begin{tikzpicture}[scale=1.0]
\fill[color=gray!20] (0,0) rectangle (4,3) node[below left,black] {$\SC_n$} ;
\draw[thick] (2,0) node[above left] {$\SN_{n-1}$}--(2,3) node[below left] {$\SL_{n-1}$} node[midway,left] {$\SM_{n-1}$} ;
\fill (2,1) circle (0.1) node[right] {$\SY_{n-2}$} ;
\fill (2,2) circle (0.1) node[right] {$\SX_{n-2}$} ;
\end{tikzpicture}
\end{array}
\xrightarrow{\text{fusion}}
\begin{array}{c}
\begin{tikzpicture}[scale=1.0]
\fill[color=gray!20] (0,0) rectangle (4,3) node[below left,black] {$\SC_n$} ;
\draw[thick] (2,0) node[above left] {$\SN_{n-1}$}--(2,3) node[below left] {$\SL_{n-1}$} ;
\fill (2,1.5) circle (0.1) node[right,scale=0.9] {$(\SX \boxtimes_\SM \SY)_{n-2}$};
\end{tikzpicture}
\end{array}
\end{gather*}
\caption{The topological defects in various dimensions can be fused together.}
\label{fig:fusion_topological_defect}
\end{figure}

\medskip
Given an $n$d topological order $\SC$ (also denoted by $\SC_n$), the collection of all topological defects (of all codimensions) in $\SC_n$ and their interrelation form a complicated structure called the \emph{topological skeleton} of $\SC_n$. A topological skeleton is not only a set but equipped with additional structures. For example, when two topological defects of the same codimension are closed to each other, they can be viewed as a single topological defect (using coarse-graining if necessary). This process is called the \emph{fusion} of two topological defects as depicted in Figure \ref{fig:fusion_topological_defect}. 

The main goal of this work is to explain that the topological skeleton of a topological order is naturally a (higher) category. 

\begin{rem}
In literature (see for example \cite{KW14,KWZ15,JF20}), the notion of a topological skeleton is often identified with that of a topological order (up to invertible topological orders). This point of view can sometimes cause confusion (see Remark \ref{rem:TO-neteq-TOSK}). In this work, we carefully distinguish them even in their notations. 
\end{rem}

%% file: 2d.tex

\section{Topological orders in 2d} \label{sec:2d}

In 1982, the fractional quantum Hall effect (FQHE) was experimentally discovered by Tsui, Stormer and Gossard \cite{TSG82}. Soon after, it was realized that there are two important features of the FQHE:
\bnu
\item Tao and Wu \cite{TW84} realized that a fractional quantum Hall state defined on a nontrivial surface has nontrivial ground state degeneracy (GSD). The source of this nontrivial GSD was unclear and caused a lot of confusion, which was often misguided by symmetry-breaking tradition (see for example \cite{And83,NTW85,Tho85}). It was finally clarified by Wen and Niu \cite{Wen90,WN90} that the GSD is robust against arbitrary weak perturbations. Hence the nontrivial GSD is not related to any symmetry and beyond Landau's paradigm.
\item The quasi-particles in a fractional quantum Hall states can carry fractional charges \cite{Lau83} and have fractional statistics \cite{ASW84,Hal84} or even non-abelian statistics \cite{Wu84,Wen91,RM92}. These quasi-particles are also called anyons \cite{Wil82} due to their anyonic statistics. Moreover, the nontrivial GSD is directly related to the fractional statistics of quasi-particles \cite{WN90}.
\enu
The emergence of anyons and the robust GSD indicate a new kind of order that is beyond Landau's paradigm. The proposed new kind of order was named ``topological order'' by Wen \cite{Wen90} because its low energy effective theory is a topological quantum field theory (see Remark \ref{rem:TO_EFT_TQFT}).

The anyons are special cases of topological defects introduced in Section \ref{sec:top_skeleton} because they are particle-like. In this section we explain in detail that particle-like topological defects in a 2d topological order form a mathematical structure called a unitary modular tensor category.

\subsection{Toric code model}

In this subsection we briefly review the 2d toric code model, which was introduced by Kitaev \cite{Kit03}, and explicitly compute the GSD.

\subsubsection{Definition of the 2d toric code model}

The 2d toric code model on a square lattice is defined as follows. There is a spin-$1/2$ on each edge (or link) of the lattice. In other words, the local degree of freedom $\mathcal H_i$ on each edge $i$ is a two-dimensional Hilbert space $\Cb^2$. The total Hilbert space is $\mathcal H_{\text{tot}} \coloneqq \medotimes_i \mathcal H_i = \medotimes_i \Cb^2$.

\begin{figure}[htbp]
\centering
\begin{tikzpicture}[scale=1.0]
\draw[step=1,help lines] (-1.5,-1.5) grid (2.5,2.5);

\draw[help lines,fill=m_ext] (0,0) rectangle (1,1) node[midway,black] {$p$} ;
\fill[e_ext] (0,1) node[above left,black] {$v$} circle (0.07) ;

\node[link_label] at (-0.5,1) {$1$} ;
\node[link_label] at (0,1.5) {$2$} ;
\node[link_label] at (0.5,1) {$3$} ;
\node[link_label] at (0,0.5) {$4$} ;
\node[link_label] at (0.5,0) {$5$} ;
\node[link_label] at (1,0.5) {$6$} ;
\end{tikzpicture}
\caption{the toric code model}
\label{fig:toric_code}
\end{figure}
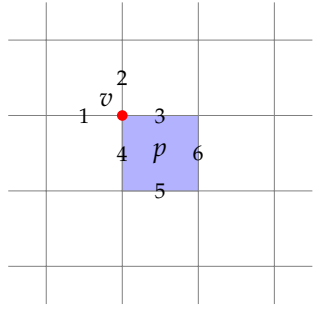

Let us recall the Pauli matrices acting on $\Cb^2$:
\[
\sigma_x = \begin{pmatrix} 0 & 1 \\ 1 & 0 \end{pmatrix} , \, \sigma_y = \begin{pmatrix} 0 & -\mathrm i \\ \mathrm i & 0 \end{pmatrix} , \, \sigma_z = \begin{pmatrix} 1 & 0 \\ 0 & -1 \end{pmatrix} .
\]
For each vertex $v$ and plaquette $p$ we define a vertex operator $A_v \coloneqq \prod_i \sigma_x^i$ and a plaquette $B_p \coloneqq \prod_j \sigma_z^j$ acting on adjacent edges. Here $\sigma_x^i = \cdots \otimes 1 \otimes \sigma_x \otimes 1 \otimes \cdots$ is the operator that acts on $\mathcal H_i$ as $\sigma_x$ and acts on other local Hilbert spaces as identities. For example, the operators in Figure \ref{fig:toric_code} are
\[
A_v = \sigma_x^1 \sigma_x^2 \sigma_x^3 \sigma_x^4 , \quad B_p = \sigma_z^3 \sigma_z^4 \sigma_z^5 \sigma_z^6 .
\]
The Hamiltonian of the toric code model is defined to be 
\be
H \coloneqq \sum_v (1-A_v) + \sum_p (1-B_p) ,
\ee
where the summation takes over all vertices $v$ and all plaquettes $p$.

\begin{rem}
The toric code model can be defined on an arbitrary lattice, and its behavior in the long wave length limit is independent of the choice of the lattice. This is due to the fact that the toric code model is at a fixed point of the renormalization flow. But for simplicity, we only consider square lattice here.
\end{rem}

A simple observation is that all $A_v$ and $B_p$ operators mutually commute. Indeed, any two plaquette operators commute because they consist of $\sigma_z$ operators. Similarly, any two vertex operators commute. So we only need to consider a plaquette operator $B_p$ and a vertex operator $A_v$. If $v$ and $p$ are not adjacent, they have no common edges and thus commute. If they are adjacent (for example see Figure \ref{fig:toric_code}), there are exactly two common edges labeled by $3$ and $4$. Since $\sigma_x^3$ anti-commutes with $\sigma_z^3$ and $\sigma_x^4$ anti-commutes with $\sigma_z^4$, we conclude that $A_v$ commutes with $B_p$.

%
%

It follows that the total Hilbert space can be decomposed as the direct sum of common eigenspaces of all $A_v$ and $B_p$ operators, whose eigenvalues are $\pm 1$. In particular, the ground state subspace has energy $0$ and is the common eigenspace of all $A_v$ and $B_p$ operators with eigenvalues $+1$. Moreover, the system is gapped, because the first excited state has energy at least $2$, no matter how large the system size is.

\subsubsection{The ground state degeneracy and ground state wave functions}

Let us compute the GSD of the toric code model defined on a genus $g$ closed surface. Suppose the lattice has $V$ vertices, $E$ edges and $F$ plaquettes. By Euler's formula, we have
\[
V-E+F=2-2g .
\]
The dimension of the total Hilbert space is $2^E$, and the ground state subspace is determined by the constraints that all $A_v = 1$ and $B_p$ = 1. Note that these constraints are not independent. We have
\[
\prod_v A_v = \prod_p B_p = 1
\]
because the model is defined on a closed surface. So there are only $(V+F-2)$ independent constraints, and each one halves the Hilbert space. Hence the GSD is $2^{E-(V+F-2)} = 2^{2g}$, which is a topological invariant.

\medskip
In the rest of this section we explicitly compute the ground state wave functions on a torus ($g = 1$).

\begin{figure}[htbp]
\centering
\subfigure[]{
\begin{tikzpicture}[scale=0.8]
\draw[step=1,help lines] (-2,-2) grid (2,2);

\draw[e_str] (0,2)|-(1,1)|-(0,0)--(0,-2) ;
\draw[e_str] (-1,0)|-(-2,-1) ;
\draw[e_str] (2,-1)--(1,-1) ;
\fill[e_ext] (-1,0) circle (0.07) ;
\fill[e_ext] (1,-1) circle (0.07) ;

\node at (0.5,0.5) {$p$} ;
\end{tikzpicture}
}
\hspace{5ex}
\subfigure[]{
\begin{tikzpicture}[scale=0.8]
\draw[step=1,help lines] (-2,-2) grid (2,2);

\draw[e_str] (0,2)--(0,-2) ;
\draw[e_str] (-1,0)|-(-2,-1) ;
\draw[e_str] (2,-1)--(1,-1) ;
\fill[e_ext] (-1,0) circle (0.07) ;
\fill[e_ext] (1,-1) circle (0.07) ;

\node at (0.5,0.5) {$p$} ;
\end{tikzpicture}
}
\caption{A string configuration corresponds to a wave function. We impose periodic boundary conditions in two directions so that the lattice model is defined on a torus. There is a long closed string and a short open string in these two figures. The action of the $B_p$ operator locally deforms the string from (a) to (b).}
\label{fig:string_wave_function}
\end{figure}
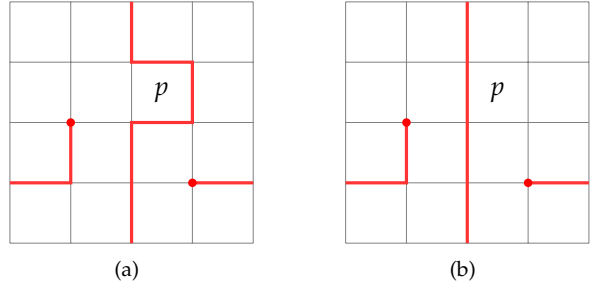

Choose eigenstates $\lvert \pm \rangle$ of $\sigma_x$ with eigenvalue $\pm 1$. We can decorate the lattice by red strings and construct a wave function from each string configuration (see Figure \ref{fig:string_wave_function}): if there is a red string on an edge, we put a state $\lvert - \rangle$ on this edge; if there is no red string on an edge, we put a $\lvert + \rangle$ on it. Thus each string configuration corresponds to a wave function is $\lvert + \rangle \otimes \lvert + \rangle \otimes \lvert - \rangle \otimes \cdots$, where $\lvert - \rangle$ only appears on red edges.

The condition that all $A_v$ operators acting on such a wave function as $+1$ means there is no open string. Indeed, if $v$ is an end point of an open string, then $A_v$ acts on the wave function as $-1$. It follows that a ground state wave function is the superposition of closed string wave functions. Moreover, the condition that all $B_p$ operators acting on the ground state wave function as $+1$ means that, if a closed string configuration can be obtained from another one by the action of several $B_p$ operators, their superposition coefficients must be equal. Clearly, the action of $B_p$ operators is locally creating, annihilating or deforming closed strings (see Figure \ref{fig:string_wave_function}). If there is a non-contractible closed string, such a configuration can not be obtained from the vacuum configuration by $B_p$ operators. Hence, by considering different non-contractible loops on a torus, we get four different (orthogonal) ground state wave functions:
\begin{gather*}
\vert \psi_{00} \rangle =
\biggl \lvert
\begin{tikzpicture}[scale=0.2,baseline={([yshift=-0.7ex]current bounding box.center)}]
\draw[step=1,help lines] (-2,-2) grid (2,2);
\draw[e_str] (0,0)|-(1,1)|-cycle ;
\end{tikzpicture}
\biggr \rangle +
\biggl \lvert
\begin{tikzpicture}[scale=0.2,baseline={([yshift=-0.7ex]current bounding box.center)}]
\draw[step=1,help lines] (-2,-2) grid (2,2);
\draw[e_str] (0,0)|-(-1,-1)|-(1,0)|-(0,1)--cycle ;
\end{tikzpicture}
\biggr \rangle +
\biggl \lvert
\begin{tikzpicture}[scale=0.2,baseline={([yshift=-0.7ex]current bounding box.center)}]
\draw[step=1,help lines] (-2,-2) grid (2,2);
\draw[e_str] (0,0)-|(1,1)-|(-1,-1)-|cycle ;
\end{tikzpicture}
\biggr \rangle + \cdots , \quad
\vert \psi_{01} \rangle =
\biggl \lvert
\begin{tikzpicture}[scale=0.2,baseline={([yshift=-0.7ex]current bounding box.center)}]
\draw[step=1,help lines] (-2,-2) grid (2,2);
\draw[e_str] (-2,0)-|(-1,-1)-|(0,1)-|(1,0)--(2,0) ;
\end{tikzpicture}
\biggr \rangle +
\biggl \lvert
\begin{tikzpicture}[scale=0.2,baseline={([yshift=-0.7ex]current bounding box.center)}]
\draw[step=1,help lines] (-2,-2) grid (2,2);
\draw[e_str] (-2,0)--(2,0) ;
\draw[e_str] (0,2)|-(-1,1)--(-1,2) ;
\draw[e_str] (0,-2)|-(-1,-1)--(-1,-2) ;
\end{tikzpicture}
\biggr \rangle +
\biggl \lvert
\begin{tikzpicture}[scale=0.2,baseline={([yshift=-0.7ex]current bounding box.center)}]
\draw[step=1,help lines] (-2,-2) grid (2,2);
\draw[e_str] (-2,-1)--(2,-1) ;
\draw[e_str] (0,0)|-(-1,1)|-cycle ;
\end{tikzpicture}
\biggr \rangle + \cdots , \\
\vert \psi_{10} \rangle =
\biggl \lvert
\begin{tikzpicture}[scale=0.2,baseline={([yshift=-0.7ex]current bounding box.center)}]
\draw[step=1,help lines] (-2,-2) grid (2,2);
\draw[e_str] (-1,2)|-(1,1)|-(-1,0)--(-1,-2) ;
\end{tikzpicture}
\biggr \rangle +
\biggl \lvert
\begin{tikzpicture}[scale=0.2,baseline={([yshift=-0.7ex]current bounding box.center)}]
\draw[step=1,help lines] (-2,-2) grid (2,2);
\draw[e_str] (-1,2)|-(0,1)|-(1,-1)|-(-1,0)--(-1,-2) ;
\end{tikzpicture}
\biggr \rangle +
\biggl \lvert
\begin{tikzpicture}[scale=0.2,baseline={([yshift=-0.7ex]current bounding box.center)}]
\draw[step=1,help lines] (-2,-2) grid (2,2);
\draw[e_str] (-1,2)|-(0,0)|-(1,1)|-(-1,-1)--(-1,-2) ;
\end{tikzpicture}
\biggr \rangle + \cdots , \quad
\vert \psi_{11} \rangle =
\biggl \lvert
\begin{tikzpicture}[scale=0.2,baseline={([yshift=-0.7ex]current bounding box.center)}]
\draw[step=1,help lines] (-2,-2) grid (2,2);
\draw[e_str] (-2,0)-|(0,-2) ;
\draw[e_str] (0,2)|-(1,1)|-(2,0) ;
\end{tikzpicture}
\biggr \rangle +
\biggl \lvert
\begin{tikzpicture}[scale=0.2,baseline={([yshift=-0.7ex]current bounding box.center)}]
\draw[step=1,help lines] (-2,-2) grid (2,2);
\draw[e_str] (1,2)|-(2,0) ;
\draw[e_str] (-2,0)-|(0,1)-|(-1,-1)-|(1,-2) ;
\end{tikzpicture}
\biggr \rangle +
\biggl \lvert
\begin{tikzpicture}[scale=0.2,baseline={([yshift=-0.7ex]current bounding box.center)}]
\draw[step=1,help lines] (-2,-2) grid (2,2);
\draw[e_str] (-1,2)--(-1,-2) ;
\draw[e_str] (2,0)-|(1,-1)-|(0,0)--(-2,0) ;
\end{tikzpicture}
\biggr \rangle + \cdots .
\end{gather*}
The closed strings in them belong to different $\Zb_2$-homology classes of a torus. Recall that $H^1(\mathrm{torus};\Zb_2) \simeq \Zb_2 \oplus \Zb_2$.

\begin{rem}
The ground state wave functions of the toric code model defined on an genus $g$ surface $\Sigma_g$ are similar. For each first homology class of $\Sigma_g$ (with coefficient $\Zb_2$) we can similarly construct a closed string wave function and hence there are $\lvert H^1(\Sigma_g;\Zb_2) \rvert = \lvert (\Zb_2)^{\oplus 2g} \rvert = 2^{2g}$ orthogonal ground state wave functions.
\end{rem}

This example also illustrates the following general principle \cite{LW05}:
\begin{quote}
The ground states are obtained from closed string condensation; the excitations are the end points of open strings.
\end{quote}

\subsection{The category of particle-like topological defects} \label{sec:category}

\subsubsection{Local operators and topological excitations}

Recall Section \ref{sec:top_skeleton} that a topological defect is an observable in the long wave length limit. How do we obtain a topological defect from microscopic data?

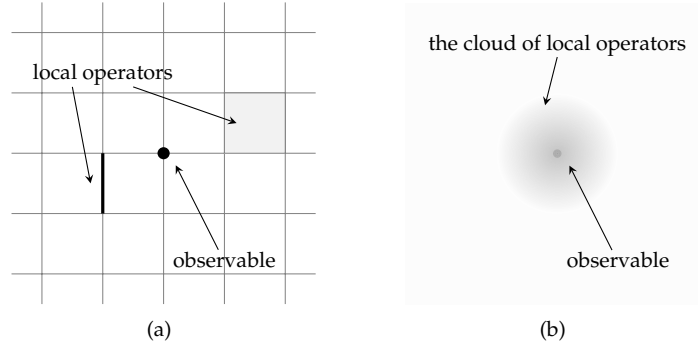
\begin{figure}[htbp]
\centering
\subfigure[]{
\begin{tikzpicture}[scale=0.8]
\draw[step=1,help lines] (-2.5,-2.5) grid (2.5,2.5);
\fill (0,0) circle (0.1) ;
\draw[very thick] (-1,0)--(-1,-1) ;
\draw[help lines,fill=gray!10] (1,0) rectangle (2,1) ;
\draw[-stealth] (-1.5,1.1)--(-1.2,-0.5) ;
\draw[-stealth] (-0.5,1.1)--(1.2,0.5) ;
\node[above,scale=0.8] at (-1,1) {local operators} ;
\draw[-stealth] (0.9,-1.6)--(0.2,-0.2) ;
\node[below,scale=0.8] at (1,-1.5) {observable} ;
\end{tikzpicture}
}
\hspace{5ex}
\subfigure[]{
\begin{tikzpicture}[scale=0.8]
\fill[gray!20,opacity=0.1] (-2.5,-2.5) rectangle (2.5,2.5) ;
\fill (0,0) circle (0.07) ;
\shade[inner color=gray!50,outer color=white,opacity=0.5] (0,0) circle (1) ;
\draw[-stealth] (0,1.6)--(-0.2,0.8) ;
\node[above,scale=0.8] at (0,1.5) {the cloud of local operators} ;
\draw[-stealth] (0.9,-1.6)--(0.2,-0.2) ;
\node[below,scale=0.8] at (1,-1.5) {observable} ;
\end{tikzpicture}
}
\caption{Observables are screened by the cloud of local operators. (a): an observable in the lattice model; (b): what we see in the long wave length limit.}
\label{fig:screening}
\end{figure}

The way to read macroscopic information from microscopic data is to do coarse-graining, which is a process of integrating and averaging out the microscopic degrees of freedom. For the study of topological defects, the effect of coarse-graining can be stated as the screening by local operators. A local operator is an operator defined in a bounded region. Local operators in a bounded region act on the local Hilbert space in the bounded region. 
Any macroscopic observables are those survived after the screening by the cloud of local operators (see Figure \ref{fig:screening}). 

%
%
%

The simplest way to obtain a defect is to add a local term in the Hamiltonian. Suppose $\delta H_\xi$ is an operator acting  around a site $\xi$. Then the ground state of the new Hamiltonian $(H + \delta H_\xi)$ is different from that of the original Hamiltonian $H$ in general. With respect to the original Hamiltonian $H$, the new ground state looks like an excitation around the site $\xi$, but coincides with the original ground state far from the site $\xi$. Intuitively, the local term $\delta H_\xi$ traps an excitation located at the site $\xi$. Conversely, any excitation located at a site $\xi$ can be trapped by some operators $\delta H_\xi$. Therefore, such a defect is characterized by an excitation of the Hamiltonian $H$.

What is the topological defect corresponding to a trap $\delta H_\xi$? Suppose $\lvert \psi \rangle$ is an excitation trapped by $\delta H_\xi$ (i.e., $\lvert \psi \rangle$ is a ground state of $(H + \delta H_\xi)$) and $A$ is a local operator, then $A \lvert \psi \rangle$ should also be trapped by $\delta H_\xi$ after coarse-graining. Thus all trapped states in the long wave length limit form a subspace which is invariant under the action of local operators. Such a subspace is called a \emph{topological excitation}. The topological defect corresponding to $\delta H_\xi$ is characterized by this subspace \cite{KW14}.

\begin{defn} \label{defn:topological_excitation}
A \emph{topological excitation} (or a particle-like topological defect) is a subspace of the total Hilbert space that is invariant under the action of local operators.
\end{defn}

Given a state $\vert \psi \rangle$, the minimal topological excitation containing $\lvert \psi \rangle$ is
\[
\{A \lvert \psi \rangle \mid A \text{ is a local operator}\} ,
\]
called the topological excitation \emph{generated} by $\lvert \psi \rangle$. In particular, the topological excitation generated by the ground state subspace is called the \emph{trivial topological excitation}, denoted by $\one$. The trivial topological excitation is exactly the trivial topological defect discussed in Remark \ref{rem:trivial_topological_defect}.


\begin{rem}
A \emph{nontrivial} topological excitation is usually defined to be an excitation which cannot be created from or annihilated to the ground state by local operators. This definition is equivalent to Definition \ref{defn:topological_excitation}. Indeed, if a state $\vert \psi \rangle$ can be created from the ground state by local operators, the topological excitation generated by $\vert \psi \rangle$ should be the same as the trivial topological excitation.

As shown in Remark \ref{rem:non-local_string_toric_code}, a nontrivial topological excitation can be created from or annihilated to the ground state by a non-local operator. Adding a local trap $\delta H_\xi$ to the original Hamiltonian $H$ is equivalent to introducing a non-local operator on the world line of $\xi$.
\end{rem}

\begin{rem}
The topological excitation given by a local trap $\delta H_\xi$ located at a site $\xi$ is particle-like. Particle-like topological excitations in 2d are also called \emph{anyons} due to their statistics (see Remark \ref{rem:braiding_statistic}). We can also consider a local trap acting around a higher-dimensional submanifold in the space, and such a local trap gives a higher-dimensional topological excitation \cite{KW14}.
\end{rem}

\begin{rem}
A particle-like topological defect is a special 0d topological defect, a name which is reserved for more general type of 0d defects, such as 0d domain walls between two 1d domain walls. 
\end{rem}

\begin{rem}
In Definition \ref{defn:topological_excitation}, we use the hand-waving term ``the action of local operators'' without defining it precisely. There are two different approaches to treat `local operators' precisely. In the first approach, the `local operators' were explicitly constructed in the quantum double models \cite{Kit03} and the Levin-Wen models \cite{LW05,KK12,Kon13} and were shown to form a Hopf-like algebra. Then a topological excitation can be defined to be a module over this algebra (see Section \ref{sec:local_operator_algebra}). In the second approach, the `local operator algebras' are replaced by the nets of `local operator algebras' \cite{Haa96,KZ22}. Then a topological excitation can be defined by a (superselection) sector of this net \cite{KZ22}. 
\end{rem}

\subsubsection{Topological excitations of the toric code model}

Let us find the topological excitations of the toric code model.

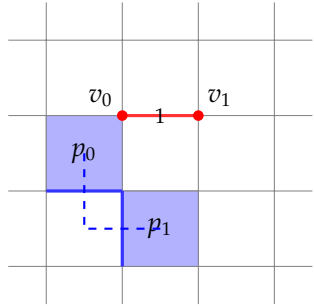
\begin{figure}[htbp]
\centering
\begin{tikzpicture}[scale=1.0]
\draw[step=1,help lines] (-1.5,-1.5) grid (2.5,2.5);

\draw[help lines,fill=m_ext] (-1,0) rectangle (0,1) node[midway,black] {$p_0$} ;
\draw[help lines,fill=m_ext] (0,-1) rectangle (1,0) node[midway,black] {$p_1$} ;
\draw[m_str] (-1,0)--(0,0) ;
\draw[m_str] (0,-1)--(0,0) ;
\draw[m_dual_str] (-0.5,0.5)--(-0.5,-0.5)--(0.5,-0.5) ;
\draw[e_str] (0,1)--(1,1) node[midway,link_label] {$1$} ;
\fill[e_ext] (0,1) circle (0.07) node[above left,black] {$v_0$} ;
\fill[e_ext] (1,1) circle (0.07) node[above right,black] {$v_1$} ;

\end{tikzpicture}
\caption{topological excitations of toric code}
\label{fig:ext_toric_code}
\end{figure}

The Hamiltonian of the toric code model is the sum of mutually commuting operators, so an excitation, as an energy eigenstate, is determined by its eigenvalue of all $A_v$ and $B_p$ operators. For example, given a vertex $v_0$, there is a state $\lvert \psi_{v_0} \rangle$ satisfying the following property:
\[
\begin{cases}
A_{v_0} \lvert \psi_{v_0} \rangle = -\lvert \psi_{v_0} \rangle , \\
A_v \lvert \psi_{v_0} \rangle = \lvert \psi_{v_0} \rangle \text{ for all vertices } v \neq v_0 , \\
B_p \lvert \psi_{v_0} \rangle = \lvert \psi_{v_0} \rangle \text{ for all plaquettes } p .
\end{cases}
\]
In other words, $\lvert \psi_{v_0} \rangle$ is an excitation located at $v_0$. Then we try to apply a local operator $\sigma_z^1$ on $\lvert \psi_{v_0} \rangle$ (see Figure \ref{fig:ext_toric_code}). Since $\sigma_z^1$ commutes with all plaquette operators and vertex operators except $A_{v_0}$ and $A_{v_1}$, we still have
\begin{gather*}
B_p \cdot \sigma_z^1 \lvert \psi_{v_0} \rangle = \sigma_z^1 \lvert \psi_{v_0} \rangle \\
A_v \cdot \sigma_z^1 \lvert \psi_{v_0} \rangle = \sigma_z^1 \lvert \psi_{v_0} \rangle
\end{gather*}
for all plaquettes $p$ and vertices $v \neq v_0,v_1$. But $\sigma_z^1$ anti-commutes with $A_{v_0}$ and $A_{v_1}$, so we have
\begin{gather*}
A_{v_0} \cdot \sigma_z^1 \lvert \psi_{v_0} \rangle = - \sigma_z^1 A_{v_0} \lvert \psi_{v_0} \rangle = \sigma_z^1 \lvert \psi_{v_0} \rangle , \\
A_{v_1} \cdot \sigma_z^1 \lvert \psi_{v_0} \rangle = - \sigma_z^1 A_{v_1} \lvert \psi_{v_0} \rangle = -\sigma_z^1 \lvert \psi_{v_0} \rangle .
\end{gather*}
Therefore, $\sigma_z^1 \lvert \psi_{v_0} \rangle$ is an excitation located at $v_1$.

As illustrated by the above discussion, local operators can not annihilate such an excitation $\lvert \psi_{v_0} \rangle$. So $\lvert \psi_{v_0} \rangle$ generates a nontrivial topological excitation, denoted by $e$. All $\lvert \psi_{v_0} \rangle$'s generate the same topological excitation $e$, because they can be connected by a string of $\sigma_z$ operators.

\begin{rem}
The state $\lvert \psi_{v_0} \rangle$ is the ground state of the Hamiltonian
\[
H' = H + 2 A_{v_0} = \sum_{v \neq v_0} (1 - A_v) + \sum_p (1 - B_p) + (1 + A_{v_0}) .
\]
Thus the corresponding defect is given by the local trap $\delta H_{v_0} = 2 A_{v_0}$. In the long wave length limit, all these defects give the same topological defect $e$.
\end{rem}

Similarly, for each plaquette $p_0$ there is a state $\lvert \psi_{p_0} \rangle$ satisfying the following property:
\[
\begin{cases}
B_{p_0} \lvert \psi_{p_0} \rangle = -\lvert \psi_{p_0} \rangle , \\
B_p \lvert \psi_{p_0} \rangle = \lvert \psi_{p_0} \rangle \text{ for all plaquettes } p \neq p_0 , \\
A_v \lvert \psi_{p_0} \rangle = \lvert \psi_{p_0} \rangle \text{ for all vertices } v .
\end{cases}
\]
Then all states $\lvert \psi_{p_0} \rangle$ generate the same topological excitation, denoted by $m$. Moreover, for every vertex $v_0$ and plaquette $p_0$ there is a state $\lvert \psi_{v_0,p_0} \rangle$ satisfying the following property:
\[
\begin{cases}
A_{v_0} \lvert \psi_{v_0,p_0} \rangle = -\lvert \psi_{v_0,p_0} \rangle , \\
A_v \lvert \psi_{v_0,p_0} \rangle = \lvert \psi_{v_0,p_0} \rangle \text{ for all vertices } v \neq v_0 , \\
B_{p_0} \lvert \psi_{v_0,p_0} \rangle = -\lvert \psi_{v_0,p_0} \rangle , \\
B_p \lvert \psi_{v_0,p_0} \rangle = \lvert \psi_{v_0,p_0} \rangle \text{ for all plaquettes } p \neq p_0 .
\end{cases}
\]
Then all states $\lvert \psi_{v_0,p_0} \rangle$ generate the same topological excitation, denoted by $f$. Intuitively, if an $e$ particle and an $m$ particle are adjacent, they can be viewed a single topological excitation $f$. Finally, there is always the trivial topological excitation $\one$, i.e., the topological excitation generated by the ground state.

Consequently, we have found 4 different topological excitations: $\one,e,m,f$.

\begin{rem} \label{rem:direct_sum_toric_code}
Indeed, there are more topological excitations. Recall that a topological excitation is a subspace of the total Hilbert space, thus we can talk about the direct sum of topological excitations. For example, $\one \oplus m$ can be generated by the ground state subspace of the following Hamiltonian:
\[
H' = H + B_{p_0} = \sum_v (1 - A_v) + \sum_{p \neq p_0} (1-B_p) + 1 .
\]
The details are discussed in Section \ref{sec:semisimple}.
\end{rem}

\begin{figure}[htbp]
\centering
\begin{tikzpicture}[scale=1.0]
\draw[step=1,help lines] (-1.5,-1.5) grid (3.5,2.5);
\draw[help lines,fill=m_ext] (0,0) rectangle (1,1) node[midway] (m) {} ;
\foreach \x in {1,2,3}
	\draw[m_str] (\x,0)--(\x,1) ;
\draw[m_dual_str] (0.5,0.5)--(3.5,0.5) ;
\draw[e_str] (0,1)--(4,1) ;
\fill[e_ext] (0,1) circle (0.07) node (e) {} ;
\draw[dashed,fill=gray!10,opacity=0.5] (0.2,0.8) circle (1.5) ;

\node[above left] at (e) {$e$} ;
\node at (m) {$m$} ;
\foreach \x/\xtext in {0.5/1,1.5/2,2.5/3,3.5/4}
	\node[link_label] at (\x,1) {$\xtext$} ;
\foreach \x/\xtext in {1/5,2/6,3/7}
	\node[link_label] at (\x,0.5) {$\xtext$} ;
\end{tikzpicture}
\caption{Topological excitations can be created or annihilated by non-local operators.}
\label{fig:ext_non_local}
\end{figure}
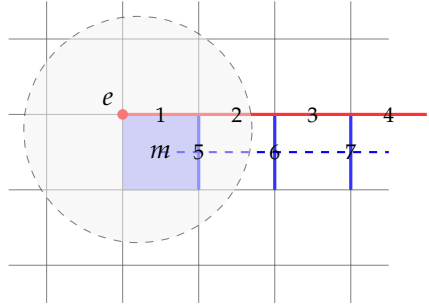

\begin{rem} \label{rem:non-local_string_toric_code}
The nontrivial topological excitations $e,m,f$ can be created from or annihilated to the ground state by non-local operators. As depicted in Figure \ref{fig:ext_non_local}, an $e$ particle can be created or annihilated by an infinitely long string operator $\cdots \sigma_z^4 \sigma_z^3 \sigma_z^2 \sigma_z^1$, and an $m$ particle can be created or annihilated by $\cdots \sigma_x^7 \sigma_x^6 \sigma_x^5$.
\end{rem}

\begin{rem} \label{rem:local_region_toric_code}
We do not really need an infinitely long string operators to create or annihilate topological excitations, because the notion of locality depends on the length scale. For example, if we treat operators in the shaded region in Figure \ref{fig:ext_non_local} as local, then $\sigma_z^3 \sigma_z^2 \sigma_z^1$ is already non-local and can create or annihilate an $e$ particle.

This example suggests that the notion of a local operator, as well as a topological excitation, depend on the length scale. We can choose an arbitrary region $R$, provided that it is particle-like when we look at it from far away, and define local operators and topological excitations in the region $R$. So a particle-like topological excitation does not necessarily located at a point. Its `shape' is the same as the chosen region $R$.
\end{rem}

\subsubsection{Instantons}

Topological excitations are subspaces of the total Hilbert space. A natural question is whether operators on these spaces are observables in the long wave length limit.

Suppose $x,y$ are two topological excitations located at a site $\xi$ and $f \colon x \to y$ is an operator between these two Hilbert spaces. The operator $f$ should be viewed as a 0D (spacetime dimension) defect between the world lines of $x$ and $y$. Thus it gives a 0D topological defect in the long wave length limit.

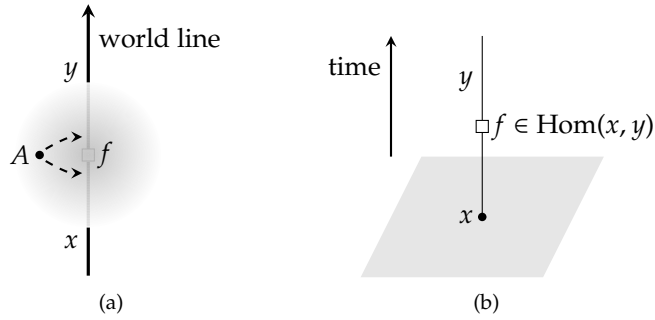
\begin{figure}[htbp]
\centering
\subfigure[]{
\begin{tikzpicture}[scale=0.8]
\draw[-stealth,very thick] (0,-2)--(0,2.5) node[very near end,right] {world line} node[near end,left] {$y$} node[very near start,left] {$x$} ;
\draw[fill=white] (-0.1,-0.1) rectangle (0.1,0.1) node[midway] (f) {} ;
\shade[inner color=gray!50,outer color=white,opacity=0.5] (0,0) circle (1.2) ;
\fill (-0.8,0) node[left] {$A$} circle (0.07) ;

\draw[-stealth,thick,dashed] (-0.7,0.1) to [out=30,in=180] (-0.1,0.3) ;
\draw[-stealth,thick,dashed] (-0.7,-0.1) to [out=-30,in=-180] (-0.1,-0.3) ;

\node[right] at (f) {$f$} ;
\end{tikzpicture}
}
\hspace{5ex}
\subfigure[]{
\begin{tikzpicture}[scale=0.8]
\fill[gray!20] (0,0)--(3,0)--(4,2)--(1,2)--cycle ;
\draw (2,1)--(2,4) node[near end,left] {$y$} ;
\draw[-stealth,thick] (0.5,2)--(0.5,4) node[near end,left] {time} ;

\fill (2,1) node[left] {$x$} circle (0.07) ;
\draw[fill=white] (1.9,2.4) rectangle (2.1,2.6) node[midway,right] {$f \in \Hom(x,y)$} ;
\end{tikzpicture}
}
\caption{(a): the screening on time axis; (b) an instanton can freely pass through the cloud of local operators.}
\label{fig:screening_time_instanton}
\end{figure}

As depicted in Figure \ref{fig:screening_time_instanton} (a), a local operator $A$ can act on the world line before or after $f$. In the long wave length limit, all these actions are averaged out, thus what we can see is not necessarily $f$, but an screened operator which commutes with all local operators. Hence, a 0D topological defect in the spacetime is an operator which commutes with all local operators. These operators are called \emph{instantons} because they are localized on time axis. Figure \ref{fig:screening_time_instanton} (b) is an intuitive picture of an instanton. We denote the space of all instantons from $x$ to $y$ by $\Hom(x,y)$.

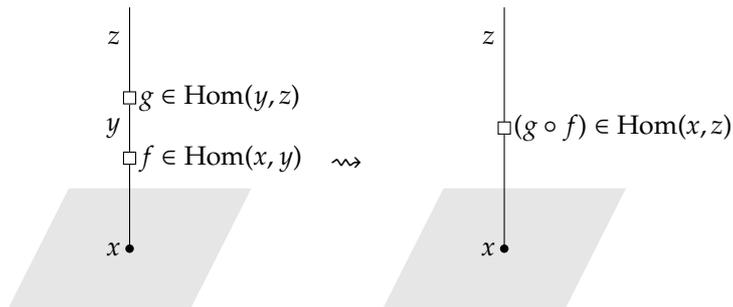
\begin{figure}[htbp]
\[
\begin{array}{c}
\begin{tikzpicture}[scale=0.8]
\fill[gray!20] (0,0)--(3,0)--(4,2)--(1,2)--cycle ;
\draw (2,1)--(2,5) node[midway,left] {$y$} node[very near end,left] {$z$} ;

\fill (2,1) circle (0.07) node[left] {$x$} ;
\draw[fill=white] (1.9,2.4) rectangle (2.1,2.6) node[midway,right] {$f \in \Hom(x,y)$} ;
\draw[fill=white] (1.9,3.4) rectangle (2.1,3.6) node[midway,right] {$g \in \Hom(y,z)$} ;
\end{tikzpicture}
\end{array}
\rightsquigarrow
\begin{array}{c}
\begin{tikzpicture}[scale=0.8]
\fill[gray!20] (0,0)--(3,0)--(4,2)--(1,2)--cycle ;
\draw (2,1)--(2,5) node[very near end,left] {$z$} ;

\fill (2,1) circle (0.07) node[left] {$x$} ;
\draw[fill=white] (1.9,2.9) rectangle (2.1,3.1) node[midway,right] {$(g \circ f) \in \Hom(x,z)$} ;
\end{tikzpicture}
\end{array}
\]
\caption{the fusion of instantons}
\label{fig:fusion_of_instantons}
\end{figure}

Instantons can be fused together. Intuitively, if two instantons $f \in \Hom(x,y)$ and $g \in \Hom(y,z)$ are very close to each other, they can be viewed as a single instanton from $x$ to $z$, denoted by $g \circ f$ (see Figure \ref{fig:fusion_of_instantons}). The fusion of instantons defines a map:
\begin{align*}
\circ \colon \Hom(y,z) \times \Hom(x,y) & \to \Hom(x,z) \\
(g,f) & \mapsto g \circ f .
\end{align*}
In the language of operators, the fusion of instantons is just the composition (or product) of operators.

We can also fuse more instantons. Suppose there are three instantons $f_i \colon x_{i-1} \to x_i$ for $i = 1,2,3$. By fusing them three together we get an instanton denoted by $f_3 \circ f_2 \circ f_1 \colon x_0 \to x_3$. On the other hand, fusing them two by two is also possible: we can first fuse $f_1$ and $f_2$ to get $f_2 \circ f_1$, then fuse it with $f_3$ to get $f_3 \circ (f_2 \circ f_1)$; or fuse $f_2$ and $f_3$ first then fuse $f_1$ with the result to get $(f_3 \circ f_2) \circ f_1$. It is physically obvious that these different ways of fusion should give the same result:
\[
f_3 \circ f_2 \circ f_1 = f_3 \circ (f_2 \circ f_1) = (f_3 \circ f_2) \circ f_1 .
\]
In other words, the fusion of instantons is associative. By considering instantons as operators on Hilbert spaces, the associativity is also obvious.

\begin{rem}
The associativity implies that the fusion of three instantons can be defined via the fusion of two instantons. Similarly, there are many different ways to fuse $n$ instantons two by two, and all the results are the same by the associativity. One can easily prove this statement by induction on $n$. Hence, it is enough to consider the fusion of two instantons and the associativity condition.
\end{rem}

\begin{figure}[htbp]
\[
\begin{array}{c}
\begin{tikzpicture}[scale=0.8]
\fill[gray!20] (0,0)--(3,0)--(4,2)--(1,2)--cycle ;
\draw (2,1)--(2,5) node[midway,left] {$y$} node[very near end,left] {$y$} ;

\fill (2,1) circle (0.07) node[left] {$x$} ;
\draw[fill=white] (1.9,2.4) rectangle (2.1,2.6) node[midway,right] {$f \in \Hom(x,y)$} ;
\draw[densely dotted,fill=white,opacity=0.5] (1.9,3.4) rectangle (2.1,3.6) node[midway,right,opacity=1] {$\id_y \in \Hom(y,y)$} ;
\end{tikzpicture}
\end{array}
\rightsquigarrow
\begin{array}{c}
\begin{tikzpicture}[scale=0.8]
\fill[gray!20] (0,0)--(3,0)--(4,2)--(1,2)--cycle ;
\draw (2,1)--(2,5) node[very near end,left] {$y$} ;

\fill (2,1) circle (0.07) node[left] {$x$} ;
\draw[fill=white] (1.9,2.9) rectangle (2.1,3.1) node[midway,right] {$f$} ;
\end{tikzpicture}
\end{array}
\leftsquigarrow
\begin{array}{c}
\begin{tikzpicture}[scale=0.8]
\fill[gray!20] (0,0)--(3,0)--(4,2)--(1,2)--cycle ;
\draw (2,1)--(2,5) node[midway,left] {$x$} node[very near end,left] {$y$} ;

\fill (2,1) circle (0.07) node[left] {$x$} ;
\draw[densely dotted,fill=white,opacity=0.5] (1.9,2.4) rectangle (2.1,2.6) node[midway,right,opacity=1] {$\id_x \in \Hom(x,x)$} ;
\draw[fill=white] (1.9,3.4) rectangle (2.1,3.6) node[midway,right] {$f \in \Hom(x,y)$} ;
\end{tikzpicture}
\end{array}
\]
\caption{The fusion of an instanton $f$ with a trivial instanton is still $f$.}
\label{fig:trivial_instanton}
\end{figure}
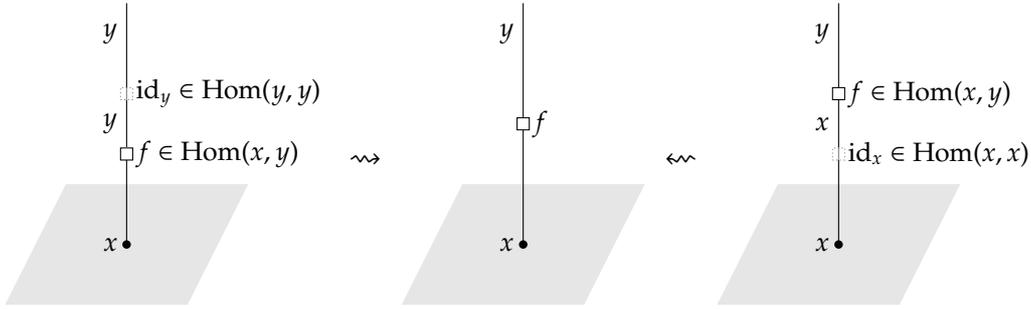

Moreover, for any topological defect $x$, clearly `nothing happening' is also an instanton. As an operator on the Hilbert space, it is just the identity operator. So we denote it by $\id_x$. It is trivial in the sense that any instanton $f \in \Hom(x,y)$ does not change by fusing with a trivial instanton (see Figure \ref{fig:trivial_instanton}):
\[
\id_y \circ f = f = f \circ \id_x .
\]

\subsubsection{The structure of a category}

We summarize the data and properties developed in previous subsections to a formal structure, called a category.

\begin{defn}
A \emph{category} $\CC$ consists of the following data:
\bit
\item a set $\ob(\CC) = \{x,y,\ldots\}$, whose elements are called \emph{objects} of $\CC$;
\item a set $\Hom_\CC(x,y)$ for every $x,y \in \ob(\CC)$, whose elements are called \emph{morphisms} from $x$ to $y$;
\item a map $\circ \colon \Hom_\CC(y,z) \times \Hom_\CC(x,y) \to \Hom_\CC(x,z) \colon (g,f) \mapsto g \circ f$ for any $x,y,z \in \ob(\CC)$, called the \emph{composition} of morphisms;
\item a distinguished morphism $\id_x \in \Hom_\CC(x,x)$ for each $x \in \ob(\CC)$, called the \emph{identity} morphism;
\eit
and these data satisfy the following conditions:
\bnu
\item (\textbf{associativity}) $(h \circ g) \circ f = h \circ (g \circ f)$ for any $f \in \Hom_\CC(x,y)$, $g \in \Hom_\CC(y,z)$, $h \in \Hom_\CC(z,w)$ and $x,y,z \in \ob(\CC)$;
\item (\textbf{unitality}) $\id_y \circ f = f = f \circ \id_x$ for any $f \in \Hom_\CC(x,y)$ and $x,y \in \ob(\CC)$.
\enu
\end{defn}

Then the discussion in previous subsections can be summarized to the following physical theorem.

\begin{pthm} \label{pthm:category_topological_defect}
Given a 2d topological order $\SC$, its particle-like (i.e., 0+1D) topological defects and (0D) instantons form a category $\CC$:
\bit
\item The set $\ob(\CC)$ of objects is the set of particle-like topological defects (topological excitations).
\item The set $\Hom_\CC(x,y)$ of morphisms is the space of instantons from $x$ to $y$.
\item The composition of morphisms is given by the fusion of instantons.
\item The identity morphisms are trivial instantons.
\eit
\end{pthm}

Indeed, the above discussions are independent of the dimension of the topological order, so this result holds for all $n$d topological orders where $n \geq 1$. When $\SC$ is a 1d topological order (either anomaly-free or anomalous), the category $\CC$ is the topological skeleton of $\SC$. When the dimension of $\SC$ is higher than one, there are also higher-dimensional topological defects, so the category $\CC$ is only a part of the topological skeleton. For 0d topological orders it is not reasonable to talk about particle-like topological defects, but such categorical structures still exist (see Section \ref{sec:0d_module_category}).

\begin{expl}
All particle-like topological defects and instantons of the toric code model form a category, denoted by $\Toric$. By definition, $\one,e,m,f$ are objects in $\Toric$, but there are also other objects in $\Toric$. The detail of the category $\Toric$ is discussed in Section \ref{sec:semisimple}.
\end{expl}

\begin{notation}
If $x$ is an object of a category $\CC$, we write $x \in \CC$ for simplicity instead of $x \in \ob(\CC)$.
\end{notation}

\begin{notation}
In this work, we use the same letter in different fonts to denote a topological order and its category of particle-like topological defects. Topological orders are labeled by \verb+\mathsf+ font: $\SC,\SD,\SE$\ldots, and categories are labeled by \verb+\EuScript+ font: $\CC,\CD,\CE$\ldots. However, 0d topological orders (for example, particle-like topological defects in another topological order) are also labeled by lower case letters: $a,b,c,x,y,z, \mathtt z$\ldots.
\end{notation}


\subsection{Basic category theory}

\subsubsection{Examples of categories} \label{sec:example_categories}


\begin{expl}
The category of sets, denoted by $\set$, is defined by the following data:
\bit
\item An object in $\set$ is a set.
\item Given two sets $X$ and $Y$, the set $\Hom_\set(X,Y)$ is the set of maps from $X$ to $Y$.
\item The composition of morphisms is the usual composition of maps.
\item The identity morphism $\id_X$ on a set $X$ is the usual identity map.
\eit
\end{expl}

\begin{expl}
The category of groups, denoted by $\grp$, is defined by the following data:
\bit
\item An object in $\grp$ is a group.
\item Given two groups $G$ and $H$, the set $\Hom_\grp(G,H)$ is the set of group homomorphisms from $G$ to $H$.
\item The composition of morphisms is the usual composition of maps.
\item The identity morphism $\id_G$ on a group $G$ is the usual identity map.
\eit
\end{expl}

\begin{expl}
The category of finite-dimensional vector spaces over $\Cb$ (the field of complex numbers), denoted by $\vect_\Cb$ or simply $\vect$, is defined by the following data:
\bit
\item An object in $\vect$ is a finite-dimensional vector space over $\Cb$.
\item Given two finite-dimensional vector spaces $V$ and $W$, the set $\Hom_\vect(V,W)$ is the set of $\Cb$-linear maps from $V$ to $W$.
\item The composition of morphisms is the usual composition of maps.
\item The identity morphism $\id_V$ on a finite-dimensional vector space $V$ is the usual identity map.
\eit
Similarly, all (not necessarily finite-dimensional) vector spaces over $\Cb$ and linear maps also form a category.
\end{expl}

\begin{expl}
Let $G$ be a group. The category of finite-dimensional $G$-representations (over $\Cb$), denoted by $\rep(G)$, is defined by the following data:
\bit
\item An object in $\rep(G)$ is a finite-dimensional $G$-representation (over $\Cb$), i.e., a finite-dimensional vector space $V$ equipped with a group homomorphism $G \to \mathrm{GL}(V)$. Here $\mathrm{GL}(V)$ denotes the group of linear automorphisms of $V$ (i.e., invertible linear maps from $V$ to itself).
\item Given two finite-dimensional $G$-representations $(V,\rho)$ and $(W,\sigma)$, a morphism $f \colon (V,\rho) \to (W,\sigma)$ is a $\Cb$-linear map $f \colon V \to W$ satisfying $f \circ \rho(g) = \sigma(g) \circ f$ for all $g \in G$.
\item The composition of morphisms is the usual composition of maps.
\item The identity morphism $\id_{(V,\rho)}$ on a finite-dimensional $G$-representation $V$ is the usual identity map.
\eit
\end{expl}

\begin{expl}
Let $G$ be a group. The category of locally finite-dimensional $G$-graded vector spaces (over $\Cb$), denoted by $\vect_G$, is defined by the following data:
\bit
\item An object in $\vect_G$ is a locally finite-dimensional $G$-graded vector space (over $\Cb$), i.e., a collection of finite-dimensional vector spaces $\{V_g \in \vect\}_{g \in G}$. The direct sum $V \coloneqq \medoplus_{g \in G} V_g$ is called the total space. By abuse of notation, we also use $V$ to denote this $G$-graded vector space.
\item Given two locally finite-dimensional $G$-graded vector spaces $V = \medoplus_{g \in G} V_g$ and $W = \medoplus_{g \in G} W_g$, a morphism from $V$ to $W$ is a collection of $\Cb$-linear maps $\{f_g \colon V_g \to W_g\}_{g \in G}$. Note that $f \coloneqq \medoplus_{g \in G} f_g \colon V \to W$ is a linear map between two total spaces.
\item The composition of morphisms is given by the usual composition of maps in each degree $g \in G$.
\item The identity morphism $\id_V$ on a locally finite-dimensional $G$-graded vector space $V = \medoplus_{g \in G} V_g$ is given by the usual identity map $\id_{V_g}$ on each degree $g \in G$.
\eit
\end{expl}

\begin{exercise} \label{exercise:vect_G_simple_object}
Let $G$ be a group. For every $g \in G$, we define a $G$-graded vector space $\Cb_{(g)}$ by the total space $\Cb$ equipped with the $G$-grading
\[
(\Cb_{(g)})_h \coloneqq \begin{cases} \Cb , & h = g , \\ 0 , & h \neq g .\end{cases}
\]
Find $\Hom_{\vect_G}(\Cb_{(g)},\Cb_{(h)})$ for $g,h \in G$.
\end{exercise}


\begin{expl} \label{expl:delooping_group}
Let $G$ be a group. There is a category $\mathrm B G$ defined by the following data:
\bit
\item There is only one object in $\mathrm B G$, i.e., $\ob(\mathrm B G) \coloneqq \{\ast\}$.
\item $\Hom_{\mathrm B G}(\ast,\ast) \coloneqq G$.
\item The composition is defined by the multiplication of $G$:
\[
\Hom_{\mathrm B G}(\ast,\ast) \times \Hom_{\mathrm B G}(\ast,\ast) = G \times G \xrightarrow{\text{multiplication}} G = \Hom_{\mathrm B G}(\ast,\ast) .
\]
\item The identity $\id_{\ast}$ is the unit of $G$.
\eit
The category $\mathrm B G$ is called the \emph{delooping} of $G$.
\end{expl}

%
%
%

\subsubsection{Construct new categories from old ones}

There are some method to construct new categories from old ones.

\begin{defn} \label{defn:opposite_category}
Let $\CC$ be a category. Its \emph{opposite category}, denoted by $\CC^\op$, is defined by the following data:
\bit
\item $\ob(\CC^\op) \coloneqq \ob(\CC)$.
\item $\Hom_{\CC^\op}(x,y) \coloneqq \Hom_\CC(y,x)$ for $x,y \in \CC^\op$.
\item The composition of morphisms in $\CC^\op$ is induced by that of $\CC$:
\begin{multline*}
\Hom_{\CC^\op}(y,z) \times \Hom_{\CC^\op}(x,y) = \Hom_\CC(z,y) \times \Hom_\CC(y,x) \\
\simeq \Hom_\CC(y,x) \times \Hom_\CC(z,y) \xrightarrow{\circ \text{ of } \CC} \Hom_\CC(z,x) = \Hom_{\CC^\op}(x,z) .
\end{multline*}
\item The identity morphisms in $\CC^\op$ are the same as those in $\CC$.
\eit
\end{defn}

Intuitively, the opposite category $\CC^\op$ is obtained by reversing the morphisms in $\CC$.

\begin{defn}
Let $\CC,\CD$ be categories. Their \emph{Cartesian product}, denoted by $\CC \times \CD$, is defined by the following data:
\bit
\item $\ob(\CC \times \CD) \coloneqq \ob(\CC) \times \ob(\CD)$.
\item $\Hom_{\CC \times \CD}((x,y),(x',y')) \coloneqq \Hom_\CC(x,x') \times \Hom_\CD(y,y')$ for $x,x' \in \CC$ and $y,y' \in \CD$.
\item The composition of morphisms in $\CC \times \CD$ is induced by those of $\CC$ and $\CD$:
\begin{multline*}
\Hom_{\CC \times \CD}((x',y'),(x'',y'')) \times \Hom_{\CC \times \CD}((x,y),(x',y')) \\
= \Hom_\CC(x',x'') \times \Hom_\CD(y',y'') \times \Hom_\CC(x,x') \times \Hom_\CD(y,y') \\
\simeq \Hom_\CC(x',x'') \times \Hom_\CC(x,x') \times \Hom_\CD(y',y'') \times \Hom_\CD(y,y') \\
\xrightarrow{(\circ \text{ of } \CC) \times (\circ \text{ of } \CD)} \Hom_\CC(x,x'') \times \Hom_\CD(y,y'') = \Hom_{\CC \times \CD}((x,y),(x'',y'')) .
\end{multline*}
\item $\id_{(x,y)} \coloneqq (\id_x,\id_y)$ for $x \in \CC,y \in \CD$.
\eit
\end{defn}

\begin{defn}
Let $\CC$ be a category. Suppose $\CD$ is a collection of the following data:
\bit
\item a subset $\ob(\CD) \subseteq \ob(\CC)$;
\item a subset $\Hom_\CD(x,y) \subseteq \Hom_\CC(x,y)$ for every $x,y \in \ob(\CD) \subseteq \ob(\CC)$;
\eit
such that $\CD$ is a category with the composition and identity morphisms induced from $\CC$ (i.e., the hom sets in $\CD$ is closed under composition and $\id_x \in \Hom_\CD(x,x)$ for each $x \in \CD$). Then $\CD$ is called a \emph{subcategory} of $\CC$. We say the subcategory $\CD \subseteq \CC$ is \emph{full} if $\Hom_\CD(x,y) = \Hom_\CC(x,y)$ for all $x,y \in \CD$.
\end{defn}

\subsubsection{Morphisms in a category}

The equations of morphisms in a category can be represented by \emph{commutative diagrams}. Here we do not give the definition of a commutative diagram, but intuitively explain this notion by some examples. If there are three morphisms $f \colon x \to y$, $g \colon y \to z$ and $h \colon x \to z$, then the equation $g \circ f = h$ is equivalent to say that the following diagram is commutative:
\[
\xymatrix{
x \ar[r]^{f} \ar[dr]_{h} & y \ar[d]^{g} \\
& z
}
\]
Also, the following commutative diagram
\[
\xymatrix{
x \ar[r]^{f} \ar[d]_{h} & y \ar[d]^{g} \\
z \ar[r]^{k} & w
}
\]
is equivalent to say that $g \circ f = k \circ h$.

\begin{defn}
Let $\CC$ be a category and $x,y \in \CC$. A morphism $f \colon x \to y$ is called an \emph{isomorphism} if there exists a morphism $g \colon y \to x$ such that $g \circ f = \id_x$ and $f \circ g = \id_y$. The morphism $g$, if exists, is called the \emph{inverse} of $f$ and denoted by $f^{-1}$. Two objects $x,y$ are said to be \emph{isomorphic} if there exists an isomorphism between them and we denote $x \simeq y$.
\end{defn}

\begin{exercise}
Prove that the inverse of an isomorphism is unique.
\end{exercise}

\begin{expl}
In the category $\set$ of sets, an isomorphism is a bijective map.
\end{expl}

\begin{defn}
Let $\CC$ be a category and $x \in \CC$. A morphism $e \colon x \to x$ is called an \emph{idempotent} if it satisfies $e = e^2 \coloneqq e \circ e$.
\end{defn}

\begin{expl}
Let $\CC$ be a category and $x,y \in \CC$. Suppose there are morphisms $r \colon x \to y$ and $s \colon y \to x$ such that $r \circ s = \id_y$. Then $s \circ r \colon x \to x$ is an idempotent because $(s \circ r) \circ (s \circ r) = s \circ (r \circ s) \circ r = s \circ r$.
\end{expl}

\begin{defn} \label{defn:idempotent_complete}
Let $\CC$ be a category and $x \in \CC$. We say that an idempotent $e \colon x \to x$ \emph{splits} if there exists an object $y \in \CC$ equipped with two morphisms $r \colon x \to y , s \colon y \to x$ such that $r \circ s = \id_y$ and $s \circ r = e$. The triple $(y,r,s)$ (or simply the object $y$) is called an \emph{image} of the idempotent $e$. A category $\CC$ is \emph{idempotent complete} if every idempotent in $\CC$ splits.
\end{defn}

\begin{rem}
An image of an idempotent, if exists, is unique up to a unique isomorphism. More precisely, suppose $(y,r,s)$ and $(y',r',s')$ are both images of an idempotent $e \colon x \to x$. Then there exists a unique isomorphism $f \colon y \to y'$ such that the following diagram commutes:
\[
\xymatrix{
x \ar[r]^{r} \ar[dr]_{r'} & y \ar@{-->}[d]^{f} \ar[dr]^{s} \\
 & y' \ar[r]^{s'} & x
}
\]
i.e., $f \circ r = r'$ and $s' \circ f = s$. Indeed, the morphism $f$ is equal to $r' \circ s$. Thus we can talk about \emph{the} image of an idempotent.
\end{rem}

\begin{expl}
All categories in Section \ref{sec:example_categories} are idempotent complete.
\end{expl}

%
%
%
%

\subsubsection{\texorpdfstring{$\Cb$}{C}-linear categories and direct sums}

The categories $\vect$, $\rep(G)$ and $\vect_G$ have more structures than an ordinary category. Recall for every vector spaces $V$ and $W$, the hom space $\Hom_\vect(V,W)$ is naturally a vector space:
\bit
\item The addition is defined by $(f + g)(v) \coloneqq f(v) + g(v)$ for $f,g \in \Hom_\vect(V,W)$.
\item The scalar product is defined by $(\lambda \cdot f)(v) \coloneqq \lambda \cdot f(v)$ for $\lambda \in \Cb$ and $f \in \Hom_\vect(V,W)$.
\eit
Moreover, the composition map $\Hom_\vect(V,W) \times \Hom_\vect(U,V) \to \Hom_\vect(U,W)$ is $\Cb$-bilinear.

\begin{defn}
A \emph{$\Cb$-linear category} is a category in which each hom set is equipped with a structure of a finite-dimensional vector space over $\Cb$, such that the composition of morphisms is $\Cb$-bilinear.
\end{defn}

\begin{expl}
The category $\vect$ is naturally a $\Cb$-linear category. For any group $G$, both $\rep(G)$ and $\vect_G$ are $\Cb$-linear categories.
\end{expl}

\begin{exercise} \label{exercise:zero_object}
Let $\CC$ be a $\Cb$-linear category and $x,y \in \CC$. Since the hom space $\Hom_\CC(x,y)$ is a vector space, there is a zero vector in it, denoted by $0 \colon x \to y$. This distinguished morphism is called the \emph{zero morphism}.
\bnu[(1)]
\item For any object $z \in \CC$ and morphisms $f \colon y \to z , g \colon z \to x$, prove that $f \circ 0 = 0 = 0 \circ g$. Note that these $0$'s are different morphisms in different hom spaces.
\item Prove that the following four conditions are equivalent for an object $a \in \CC$:
\bnu[(a)]
\item $\Hom_\CC(a,b) = 0$ for all $b \in \CC$.
\item $\Hom_\CC(b,a) = 0$ for all $b \in \CC$.
\item $\Hom_\CC(a,a) = 0$.
\item $\id_a = 0$.
\enu
An object satisfying one (hence all) of the above conditions is called a \emph{zero object}.
\item Prove that any two zero objects are isomorphic via a unique isomorphism. Thus we can talk about \emph{the} zero object, which is denoted by $0$.
\item Find the zero object in $\vect$, $\rep(G)$ and $\vect_G$ where $G$ is a group.
\enu
\end{exercise}

Given a topological order $\SC$, it is expected that the category $\CC$ of particle-like topological defects is $\Cb$-linear, because the instantons (morphisms) are microscopically realized by linear operators. By Remark \ref{rem:direct_sum_toric_code}, it is expected we can talk about direct sums in $\CC$.

What is the direct sum in a $\Cb$-linear category? Recall that the direct sum of two vector spaces $V$ and $W$ is
\[
V \oplus W \coloneqq \{(v,w) \mid v \in V , \, w \in W\}
\]
(which is just the Cartesian product $V \times W$ as a set) equipped with component-wise addition and scalar product. There are some obvious embedding and projection maps
\[
\iota_V \colon V \to V \oplus W , \, \pi_V \colon V \oplus W \to V , \, \iota_W \colon W \to V \oplus W , \, \pi_W \colon V \oplus W \to W ,
\]
and these linear maps satisfy the following equations:
\[
\pi_V \circ \iota_V = \id_V , \, \pi_W \circ \iota_W = \id_W , \, \pi_V \circ \iota_W = 0 , \, \pi_W \circ \iota_V = 0 , \, \iota_V \circ \pi_V + \iota_W \circ \pi_W = \id_{V \oplus W} .
\]
These maps and equations fully characterize the direct sum. Indeed, if there is a vector space $X$ equipped with such linear maps satisfying the above equations, one can prove that $X$ is canonically isomorphic to $V \oplus W$.

\begin{defn}
Let $\CC$ be a $\Cb$-linear category and $x_1,\ldots,x_n \in \CC$. A direct sum of $x_1,\ldots,x_n$ is an object $x \in \CC$ equipped with morphisms $\iota_i \colon x_i \to x$ and $\pi_i \colon x \to x_i$ for every $1 \leq i \leq n$, such that the following $(n^2+1)$ equations hold:
\[
\pi_i \circ \iota_j = \delta_{ij} \cdot \id_{x_j} \, (\text{for all } 1 \leq i,j \leq n) , \quad \sum_{j = 1}^n \iota_j \circ \pi_j = \id_x .
\]
\end{defn}

\begin{exercise}
Let $\CC$ be a $\Cb$-linear category such that the zero object exists. Prove that $x \oplus 0 \simeq x \simeq 0 \oplus x$ for every $x \in \CC$, i.e., $x$ is the direct sum of $x$ and $0$.
\end{exercise}

\begin{exercise} \label{exercise:direct_sum_universal_property}
Let $\CC$ be a $\Cb$-linear category and $x_1,\ldots,x_n \in \CC$. Suppose $(x,\{\iota_i\},\{\pi_i\})$ is a direct sum of $x_1,\ldots,x_n$.
\bnu[(1)]
\item Prove that $x$ satisfies the following universal property of a \emph{product}: for any object $y \in \CC$ and morphisms $\{f_i \colon y \to x_i\}_{i=1}^n$, there exists a unique morphisms $f \colon y \to x$ such that $\pi_i \circ f = f_i$ for every $1 \leq i \leq n$. Hint: $\pi_i \circ f = f_i$ implies that $\iota_i \circ \pi_i \circ f = \iota_i \circ f_i$.
\item Prove that $x$ satisfies the following universal property of a \emph{coproduct}: for any object $z \in \CC$ and morphisms $\{g_i \colon x_i \to z\}_{i=1}^n$, there exists a unique morphisms $g \colon x \to z$ such that $g \circ \iota_i = g_i$ for every $1 \leq i \leq n$.
\item Use the above universal properties to conclude that a morphism $f \colon x_1 \oplus \cdots \oplus x_n \to y_1 \oplus \cdots \oplus y_m$ can be equivalently represented as an $m$-by-$n$ matrix of morphisms $(f_{ji} \colon x_i \to y_j)$. Find the precise relation between the morphism $f$ and the matrix $(f_{ji})$.
\enu
\end{exercise}

\begin{rem}
A direct sum of some objects does not necessarily exist. But if it exists, it is unique up to a unique isomorphism. More precisely, suppose $(x,\{\iota_i\},\{\pi_i\})$ and $(x',\{\iota'_i\},\{\pi'_i\})$ are both direct sums of $x_1,\ldots,x_n$. Then there exists a unique isomorphism $f \colon x \to x'$ such that the following diagrams commute for all $1 \leq i \leq n$:
\[
\xymatrix{
 & x_i \ar[dr]^{\iota'_i} \ar[dl]_{\iota_i} & x \ar@{-->}[rr]_{f} \ar[dr]_{\pi_i} & & x' \ar[dl]^{\pi'_i} \\
x \ar@{-->}[rr]^{f} & & x' & x_i
}
\]
i.e., $f \circ \iota_i = \iota'_i$ and $\pi'_i \circ f = \pi_i$. Thus we can talk about \emph{the} direct sum of objects.
\end{rem}

\begin{expl}
The direct sum of $n$ objects is defined for every non-negative integer $n \in \Nb$. When $n = 0$, the direct sum of no object is an object $x$ equipped with no morphism, but satisfies a nontrivial equation (note that the sum of no morphism is $0$)
\[
0 = \id_x .
\]
By Exercise \ref{exercise:zero_object}, such an object $x$, i.e., the direct sum of no object, is the zero object.
\end{expl}

\begin{expl}
Let $\CC$ be a $\Cb$-linear category and $e \colon x \to x$ be an idempotent. Note that $e' \coloneqq (\id_x - e)$ is also an idempotent and $e' \circ e = 0 = e \circ e'$. Suppose $(y,r,s)$ and $(y',r',s')$ are the images of $e$ and $e'$, respectively. Then we have
\[
r' \circ s = r' \circ s' \circ r' \circ s \circ r \circ s = r' \circ e' \circ e \circ s = 0 .
\]
Similarly, $r \circ s' = 0$. Hence $(x,{s,s'},{r,r'})$ is the direct sum of $y$ and $y'$.
\end{expl}

\begin{defn} \label{defn:direct_sum_category}
Let $\CC,\CD$ be $\Cb$-linear categories. Their \emph{direct sum}, denoted by $\CC \oplus \CD$, is the $\Cb$-linear category defined by the following data:
\bit
\item The underlying category of $\CC \oplus \CD$ is the Cartesian product $\CC \times \CD$.
\item The $\Cb$-linear structure on a hom space $\Hom_{\CC \oplus \CD}((x,y),(x',y'))$ is given by the direct sum of vector spaces $\Hom_\CC(x,x') \oplus \Hom_\CD(y,y')$.
\eit
\end{defn}

\begin{exercise}
Let $\CC,\CD$ be $\Cb$-linear categories. Suppose $x \in \CC$ is the direct sum of $x_1,\ldots,x_n \in \CC$ and $y \in \CD$ is the direct sum of $y_1,\ldots,y_n \in \CD$. Prove that $(x,y) \in \CC \oplus \CD$ is the direct sum of $(x_1,y_1),\ldots,(x_n,y_n)$.
\end{exercise}

\subsubsection{Functors and natural transformations}

A structure-preserving map between categories is called a functor.

\begin{defn}
Let $\CC,\CD$ be categories. A \emph{functor} $F$ from $\CC$ to $\CD$, denoted by $F \colon \CC \to \CD$, consists of the following data:
\bit
\item a map $F \colon \ob(\CC) \to \ob(\CD)$;
\item a map $F_{x,y} \colon \Hom_\CC(x,y) \to \Hom_\CD(F(x),F(y))$ for each pair of objects $x,y \in \CC$ (for simplicity, we also denote $F_{x,y}(f)$ by $F(f)$ for a morphism $f \colon x \to y$);
\eit
and these data satisfy the following conditions:
\bnu
\item $F(g) \circ F(f) = F(g \circ f)$ for any $f \in \Hom_\CC(x,y) , g \in \Hom_\CC(y,z)$;
\item $F(\id_x) = \id_{F(x)}$.
\enu
\end{defn}

\begin{expl}
Let $\CC$ be a category. The \emph{identity functor} on $\CC$ is the functor $\id_\CC \colon \CC \to \CC$ defined by $\id_\CC(f \colon x \to y) \coloneqq (f \colon x \to y)$.
\end{expl}

\begin{expl}
Let $\CC,\CD,\CE$ be categories. The \emph{composition} of two functors $F \colon \CC \to \CD$ and $G \colon \CD \to \CE$ is the functor $G \circ F \colon \CC \to \CE$ defined by
\[
(G \circ F)(f \colon x \to y) \coloneqq (G(F(f)) \colon G(F(x)) \to G(F(y))) .
\]
\end{expl}

\begin{exercise}
Prove that all categories and functors also form a category, denoted by $\cat$.
\end{exercise}

\begin{exercise}
Prove that every functor preserves isomorphisms. More explicitly, let $F \colon \CC \to \CD$ be a functor and $f \colon x \to y$ be an isomorphism in $\CC$. Then $F(f) \colon F(x) \to F(y)$ is an isomorphism in $\CD$.
\end{exercise}

\begin{expl}
Suppose $\CC$ is a category and $\CD$ is a subcategory of $\CC$. Then the inclusion maps of object set and hom sets define a functor $\CD \to \CC$.
\end{expl}

\begin{expl}
Let $G$ be a group. A functor $F \colon \mathrm B G \to \vect$ from the delooping of $G$ (see Example \ref{expl:delooping_group}) to $\vect$ is equivalent to a finite-dimensional $G$-representation.
\end{expl}

\begin{expl} \label{expl:delooping_functor}
Let $G,H$ be groups. Given a group homomorphism $f \colon G \to H$, there is a functor $\mathrm B f \colon \mathrm B G \to \mathrm B H$ defined by $\mathrm B f(\ast) \coloneqq \ast$ and $\mathrm B f(g) \coloneqq f(g)$ for $g \in G = \Hom_{\mathrm B G}(\ast,\ast)$. Conversely, every functor $\mathrm B G \to \mathrm B H$ is equal to $\mathrm B f$ for some group homomorphism $f \colon G \to H$. It is not hard to see that the delooping construction defines a functor $\mathrm B \colon \grp \to \cat$.
\end{expl}

\begin{defn}
Let $\CC,\CD$ be $\Cb$-linear categories. A \emph{$\Cb$-linear functor} $F \colon \CC \to \CD$ is a functor $F \colon \CC \to \CD$ such that $F_{x,y} \colon \Hom_\CC(x,y) \to \Hom_\CD(F(x),F(y))$ is a $\Cb$-linear map for every $x,y \in \CC$.
\end{defn}

\begin{expl} \label{expl:forgetful_functor_vect}
Let $G$ be a finite group. There is a $\Cb$-linear functor $\rep(G) \to \vect$ defined by $(V,\rho) \mapsto V$. It is called the forgetful functor because it `forgets' the $G$-action. Similarly, there is a $\Cb$-linear forgetful functor $\vect_G \to \vect$ defined by $\{V_g\}_{g \in G} \mapsto V = \medoplus_{g \in G} V_g$.
\end{expl}

A structure-preserving map between functors is called a natural transformation.

\begin{defn}
Let $\CC,\CD$ be categories and $F,G \colon \CC \to \CD$ be functors from $\CC$ to $\CD$. A \emph{natural transformation} $\alpha \colon F \Rightarrow G$ from $F$ to $G$ is a family of morphisms $\{\alpha_x \colon F(x) \to G(x)\}_{x \in \CC}$ in $\CD$, such that the following diagram commutes for any morphism $f \colon x \to y$ in $\CC$:
\[
\xymatrix{
F(x) \ar[r]^{\alpha_x} \ar[d]^{F(f)} & G(x) \ar[d]^{G(f)} \\
F(y) \ar[r]^{\alpha_y} & G(y)
}
\]
A natural transformation $\alpha$ is called a \emph{natural isomorphism} if every morphism $\alpha_x$ is an isomorphism.
\end{defn}

\begin{expl}
Let $\CC,\CD$ be categories. The $\emph{identity natural transformation}$ on a functor $F \colon \CC \to \CD$ is the natural transformation $\id_F \colon F \Rightarrow F$ defined by $(\id_F)_x \coloneqq \id_{F(x)}$ for all $x \in \CC$.
\end{expl}

\begin{expl}
Let $\CC,\CD$ be categories and $F,G,H \colon \CC \to \CD$ be functors. The \emph{composition}, or \emph{vertical composition}, of two natural transformations $\alpha \colon F \Rightarrow G$ and $\beta \colon G \Rightarrow H$ is the natural transformation $\beta \cdot \alpha \colon F \Rightarrow H$ defined by $(\beta \cdot \alpha)_x \coloneqq \beta_x \circ \alpha_x$ for all $x \in \CC$.
\end{expl}

\begin{exercise} \label{exercise:functor_category}
Let $\CC,\CD$ be $\Cb$-linear categories. Prove that all $\Cb$-linear functors from $\CC$ to $\CD$ and natural transformations between them form a $\Cb$-linear category, denoted by $\fun(\CC,\CD)$.
\end{exercise}

\begin{expl} \label{expl:TK_duality_preparation}
Let $G$ be a finite group and $F \colon \rep(G) \to \vect$ be the forgetful functor defined in Example \ref{expl:forgetful_functor_vect}. For any $g \in G$ there is a natural isomorphism $\alpha^g \colon F \Rightarrow F$ defined by
\[
\alpha^g_{(V,\rho)} \coloneqq \rho(g) \colon V \to V , \quad (V,\rho) \in \rep(G) .
\]
Moreover, we have $\alpha^g \cdot \alpha^h = \alpha^{gh}$ for $g,h \in G$.
\end{expl}

\begin{exercise}
Let $\CC,\CD$ be categories and $F,G \colon \CC \to \CD$ be functors. Prove that a natural transformation $\alpha \colon F \Rightarrow G$ is a natural isomorphism if and only if there exists a natural transformation $\beta \colon G \Rightarrow F$ such that $\beta \cdot \alpha = \id_F$ and $\alpha \cdot \beta = \id_G$.
\end{exercise}

\begin{exercise} \label{exercise:linear_functor_preserve_direct_sum}
Let $\CC,\CD$ be $\Cb$-linear categories. Suppose $x_1,\ldots,x_n \in \CC$ and $(x,\{\iota_i\},\{\pi_i\})$ is a direct sum of $x_1,\ldots,x_n$.
\bnu[(1)]
\item Let $F \colon \CC \to \CD$ be a $\Cb$-linear functor. Prove that $(F(x),\{F(\iota_i)\},\{F(\pi_i)\})$ is a direct sum of $F(x_1),\ldots,F(x_n)$. In other words, $F$ preserves direct sums.
\item Let $F,G \colon \CC \to \CD$ be $\Cb$-linear functors and $\alpha \colon F \Rightarrow G$ be a natural transformations. Prove that $\alpha_x$ and $\{\alpha_{x_i}\}_{i=1}^n$ can determine each other. Hint: use Exercise \ref{exercise:direct_sum_universal_property} (3).
\enu
\end{exercise}

\subsubsection{Equivalence of categories}

\begin{defn}
Let $\CC,\CD$ be categories. A functor $F \colon \CC \to \CD$ is called an \emph{equivalence} if there exists a functor $G \colon \CD \to \CC$ such that $G \circ F$ is naturally isomorphic to $\id_\CC$ and $F \circ G$ is naturally isomorphic to $\id_\CD$. In this case $G$ is called the \emph{quasi-inverse} of $F$. We say $\CC$ and $\CD$ are \emph{equivalent} if there exists an equivalence between them.
\end{defn}

Equivalent categories can be viewed as having the same structure of a category. There is a useful theorem to prove whether a functor is an equivalence.

\begin{defn}
Let $\CC,\CD$ be categories and $F \colon \CC \to \CD$ be a functor. We say $F$ is \emph{faithful} if every map $F_{x,y} \colon \Hom_\CC(x,y) \to \Hom_\CD(F(x),F(y))$ is injective, and \emph{full} if every map $F_{x,y}$ is surjective. A \emph{fully faithful} functor is a functor that is both full and faithful.
\end{defn}

\begin{expl}
Let $\CC$ be a category and $\CD$ be a subcategory of $\CC$. Then the inclusion functor $\CD \to \CC$ is always faithful. It is full if and only if $\CD$ is a full subcategory.
\end{expl}

\begin{expl}
Let $G$ be a finite group. The forgetful functors $\rep(G) \to \vect$ and $\vect_G \to \vect$ (see Example \ref{expl:forgetful_functor_vect}) are faithful. They are full if and only if $G$ is the trivial group.
\end{expl}

\begin{expl}
The delooping functor $\mathrm B \colon \grp \to \cat$ defined in Example \ref{expl:delooping_functor} is fully faithful.
\end{expl}

\begin{thm} \label{thm:equivalence_fully_faithful_essentially_surjective}
A functor $F \colon \CC \to \CD$ between two categories $\CC,\CD$ is an equivalence if and only if it satisfies the following conditions:
\bnu
\item \textbf{(fully faithful)} For any $x,y \in \CC$, the map $F_{x,y} \colon \Hom_\CC(x,y) \to \Hom_\CD(F(x),F(y))$ is a bijection (i.e., $F$ is fully faithful).
\item \textbf{(essentially surjective)} For any $z \in \CD$, there exists an object $x \in \CC$ such that $F(x) \simeq z$.
\enu
\end{thm}

We do not give a proof here, but briefly explain why a functor $F$ satisfying these two conditions is an equivalence. Intuitively, an equivalence should be an identification of two categories, i.e., an identification on both morphisms and (isomorphism classes of) objects. The first condition that $F$ is fully faithful means that $F$ is an identification on morphisms. The second condition that $F$ is essentially surjective means that $F$ induces a surjective map from the set of isomorphism classes of objects of $\CC$ to that of $\CD$. This map is also injective by the first condition. Indeed, suppose $x,y \in \CC$ and $g \colon F(x) \to F(y)$ is an isomorphism. Since $F$ is fully faithful, there exists a unique morphism $f \colon x \to y$ such that $F(f) = g$. Consider the inverse of $g$, it is not hard to see that $f$ is also an isomorphism. Thus $x$ is isomorphic to $y$.

\subsubsection{Appendix: local operator algebras} \label{sec:local_operator_algebra}

\begin{defn}
An \emph{algebra} (over $\Cb$) is a vector space $A$ equipped with a $\Cb$-bilinear map (called the \emph{multiplication})
\begin{align*}
A \times A & \to A \\
(a,b) & \mapsto a \cdot b
\end{align*}
and a distinguished vector $1 \in A$ (called the \emph{identity}) such that $(a \cdot b) \cdot c = a \cdot (b \cdot c)$ and $1 \cdot a = a = a \cdot 1$ for every $a,b,c \in A$.
\end{defn}

\begin{expl}
The field $\Cb$ is an algebra with the usual multiplication of complex numbers.
\end{expl}

\begin{expl}
Let $G$ be a group. The \emph{group algebra} $\Cb[G]$ of $G$ is defined as follows. Its underlying vector space is freely generated by $G$, or equivalently, spanned by a family of symbols $\{v_g\}_{g \in G}$. The multiplication is defined by $v_g \cdot v_h \coloneqq v_{gh}$ for $g,h \in G$. The identity is $v_e$ where $e \in G$ is the unit.
\end{expl}

\begin{expl}
Let $V$ be a finite-dimensional vector space. Then $\End(V) \coloneqq \Hom_\vect(V,V)$ is an algebra with the multiplication given by the composition of linear maps and the identity given by the identity map. More generally, for any $\Cb$-linear category $\CC$ and an object $x \in \CC$, the space $\Hom_\CC(x,x)$ is an algebra.
\end{expl}

\begin{exercise}
Prove that the notion of a finite-dimensional algebra is equivalent to a $\Cb$-linear category with only one object. What is the precise meaning of `equivalent' here?
\end{exercise}

\begin{defn}
Let $A$ be an algebra. A \emph{left module} over $A$ or a \emph{left $A$-module} is a vector space $M$ equipped with a bilinear map (called the left $A$-action)
\begin{align*}
A \times M & \to M \\
(a,m) & \mapsto a \cdot m
\end{align*}
such that $(a \cdot b) \cdot m = a \cdot (b \cdot m)$ and $1 \cdot m = m$ for every $a,b \in A$ and $m \in M$.
\end{defn}

\begin{expl}
An algebra $A$ is a left module over itself.
\end{expl}

\begin{expl}
Let $G$ be a group. Suppose $M$ is a left module over the group algebra $\Cb[G]$. The left $\Cb[G]$ action is determined by $v_g \cdot m \eqqcolon \rho(g)(m)$ for all $g \in G$ and $m \in M$. Then we have
\[
\rho(gh)(m) = \rho(g)(\rho(h)(m)) = (\rho(g) \circ \rho(h))(m) , \quad \rho(e)(m) = m
\]
for all $g,h \in G$ and $m \in M$. In other words, $\rho \colon G \to \mathrm{GL}(M)$ is a group homomorphism. Thus the notion of a left $\Cb[G]$-module is equivalent to a $G$-representation.
\end{expl}

\begin{expl}
Let $V$ be a finite-dimensional vector space. Then $V$ is a left module over $\End(V)$. If $W$ is also a finite-dimensional vector space, then $\Hom_\vect(W,V)$ is a left module over $\End(V)$ with the left $\End(V)$-action given by the composition of linear maps. More generally, for any $\Cb$-linear category $\CC$ and objects $x,y \in \CC$, the space $\Hom_\CC(y,x)$ is a left $\Hom_\CC(x,x)$-module.
\end{expl}

\begin{rem}
Let $A$ be a finite-dimensional algebra. A \emph{homomorphism} or an \emph{$A$-module map} between two left $A$-modules $M,N$ is a linear map $f \colon M \to N$ satisfying $f(a \cdot m) = a \cdot f(m)$ for all $a \in A$ and $m \in M$. All finite-dimensional left $A$-modules and homomorphisms between them form a $\Cb$-linear category, denoted by $\LMod_A(\vect)$.
\end{rem}

Now we consider a lattice model that realizes a 2d topological order $\SC$. The local operators generates an algebra denoted by $\mathcal A_{\mathrm{loc}}$. By Definition \ref{defn:topological_excitation}, the local operator algebra acts on a topological excitation invariantly. In other words, a topological excitation is a left module over the local operator algebra $\mathcal A_{\mathrm{loc}}$. Thus Definition \ref{defn:topological_excitation} can be equivalently reformulated as follows.

\begin{defn}
A topological excitation is a module over the local operator algebra.
\end{defn}

\begin{rem}
An operator between topological excitations commuting with all local operators (i.e., an instanton) is a homomorphism of modules. Thus the category $\CC$ of particle-like topological excitations of $\SC$ is equivalent to the category $\LMod_{\mathcal A_{\mathrm{loc}}}(\vect)$.
\end{rem}

In the following we directly give the local operator algebra \cite{Kit03} of the toric code model and show its modules coincide with topological excitations $\one,e,m,f$. However, we do not know a general principle to derive the local operator algebra from the Hamiltonian.

\begin{figure}[htbp]
\centering
\begin{tikzpicture}[scale=1.0]
\draw[step=1,help lines] (-0.8,-0.5) grid (1.5,1.8);
\draw[help lines,fill=m_ext] (0,0) rectangle (1,1) node[midway] (m) {} ;
\fill[e_ext] (0,1) circle (0.07) node (e) {} ;
\draw[dashed,fill=gray!10,opacity=0.5] (0.2,0.8) circle (0.7) ;

\node at (m) {$p$} ;
\node[above left] at (e) {$v$} ;
\end{tikzpicture}
\caption{a site $\xi = (v,p)$}
\label{fig:site}
\end{figure}
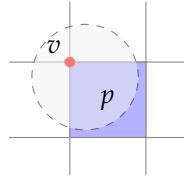

By a \emph{site} in the toric code model we mean a pair $(v,p)$ of adjacent vertex and plaquette (see Figure \ref{fig:site}). This is a minimal choice of a local region in the toric code model (see Remark \ref{rem:local_region_toric_code}). For a given site $\xi = (v,p)$, the local operator algebra $\mathcal A_\xi$ is generated by two operators $A_v$ and $B_p$, subject to the following relations:
\[
A_v^2 = B_p^2 = 1 , \quad A_v B_p = B_p A_v .
\]

Clearly $A_v$ generates a subalgebra which is isomorphic to $\Cb^2 = \Cb \oplus \Cb$ and so does $B_p$. Since $A_v$ commutes with $B_p$, the algebra $\mathcal A_\xi$ is isomorphic to tensor product of these two subalgebras, that is $\Cb^2 \otimes \Cb^2 \simeq \Cb^4$. It is a semisimple algebra with $4$ different simple modules (irreducible representations) up to isomorphism, which correspond to its $4$ primitive central idempotents:
\[
P_{\pm \pm} \coloneqq \frac{1 \pm A_v}{2} \frac{1 \pm B_p}{2} .
\]
These idempotents are projectors to the common eigenspaces of $A_v$ and $B_p$ and two $\pm$ signs are eigenvalues of $A_v$ and $B_p$ respectively. Then the module corresponding to $P_{-+}$ is an $e$ particle at $v$, the one corresponding to $P_{+-}$ is an $m$ particle at $p$, the one corresponding to $P_{--}$ is an $f$ particle at $\xi$, and the remaining one corresponding to $P_{++}$ is the trivial topological excitation $\one$.

\begin{rem}
A general module over $A_\xi$ may not be simple and isomorphic to a direct sum of simple modules. These non-simple modules are also topological excitations. So $\one,e,m,f$ are only simple topological excitations. We discuss the direct sum of topological excitations in Section \ref{sec:semisimple}.
\end{rem}

In Figure \ref{fig:ext_toric_code}, we say that the operator $\sigma_z^1$ is local, but why does it not appear in the local operator algebra $\mathcal A_\xi$? This is because a minimal local region (i.e., a site $\xi$) does not contain the edge $1$. If we choose a larger local region $R$, for example that in Figure \ref{fig:ext_non_local}, the local operator algebra $\mathcal A_R$ should be generated by all $A_v,B_p,\sigma_z,\sigma_x$ operators contained in $R$. For different regions, the local operator algebras may not be isomorphic, but one can prove that their modules are the same: there are always $4$ different simple modules denoted by $\one,e,m,f$ labeled by the eigenvalues of the product of all $A_v$ operators and the product of all $B_p$ operators. We do not show details here.

\begin{rem}
A $\Cb$-linear category is called a \emph{finite category} if it is equivalent to $\LMod_A(\vect)$ for some finite-dimensional $\Cb$-algebra $A$. If $\LMod_A(\vect)$ is equivalent to $\LMod_B(\vect)$ for two finite-dimensional $\Cb$-algebras $A$ and $B$, we say that $A$ and $B$ are \emph{Morita equivalent}.

So the above discussion can be reformulated as follows. The local operator algebras of the toric code model depend on the length scale, but the local operator algebras in different length scales are Morita equivalent (i.e., the categories of modules over different local operator algebras are equivalent). Hence the category $\CC$ of particle-like topological defects is a finite category and is an observable in the long wave length limit.
\end{rem}

\begin{rem}
In a general lattice model, the local operator algebras in different length scales may not be Morita equivalent. The toric code model is at a fixed point of the renormalization flow, and this is why we see the same topological excitations in all length scales. For an arbitrary lattice model, its local operator algebra and the modules should also flow under the renormalization flow.
\end{rem}

\subsection{More structures and properties} \label{sec:structure_2d}

We have known that particle-like topological defects in a topological order $\SC$ form a category $\CC$. When $\SC$ is a 2d topological order, $\CC$ is not only a category, but also equipped with more structures and has some nice properties.

In this subsection, we use $\SC$ to denote a 2d topological order. The category of particle-like topological defects of $\SC$ is denoted by $\CC$.

\subsubsection{Semisimplicity} \label{sec:semisimple}

As we have mentioned in Remark \ref{rem:direct_sum_toric_code}, it is reasonable to talk about the direct sum of topological excitations. First we consider an example in the toric code model. Given a plaquette $p_0$, we add a local trap $B_{p_0}$ to the Hamiltonian of the toric code model. So the new Hamiltonian is
\[
H + B_{p_0} = \sum_v (1 - A_v) + \sum_{p \neq p_0} (1-B_p) + 1 .
\]
Intuitively, this defect can be viewed as removing the plaquette $p_0$ in the lattice model. A simple calculation shows that the new ground state subspace is two-fold degenerate, and two ground states can be labeled by the eigenvalues $B_{p_0} = \pm 1$. The state with $B_{p_0} = +1$ is just the original ground state which generates the topological excitation $\one$, and the state with $B_{p_0} = -1$ generates an $m$ particle at $p$. In other words, the topological excitation given by the local trap $B_{p_0}$, as a subspace of the total Hilbert space, is the direct sum $\one \oplus m$ of $\one$ and $m$.

\begin{exercise}
Consider the $n$d lattice model ($n \geq 1$) with the local Hilbert space $\Cb^2$ and the Hamiltonian
\[
H = \sum_i (1 - \sigma_z^i) .
\]
The space dimension $n$ is not important in this exercise.
\bnu[(1)]
\item Prove that the ground state of $H$ is a product state. Thus $H$ realizes an $n$d trivial topological order (see Example \ref{expl:trivial_quantum_phase}). The trivial topological excitation (i.e., the topological excitation generated by the ground state), is denoted by $\one$.
\item Fix a site $j$ and add a local trap $2\sigma_z^j$ to the Hamiltonian $H$. Find the ground state of $H' = H + 2\sigma_z^j$. Show that the (particle-like) topological excitation realized by the local trap $2\sigma_z^j$ is $\one$. Hint: $\sigma_x^j$ is a local operator.
\item Fix a site $j$ and add a local trap $\sigma_z^j$ to the Hamiltonian $H$. Find the ground state subspace of $H' = H + \sigma_z^j$. Show that the (particle-like) topological excitation realized by the local trap $\sigma_z^j$ is $\one \oplus \one$.
\item Fix two sites $j,j'$. Show that the local trap $(\sigma_z^j + \sigma_z^{j'} \pm \sigma_z^j \sigma_z^{j'})$ realizes the topological excitation $\one \oplus \one$.
\item For each positive integer $k$, find a local trap that realizes the topological excitation $\one^{\oplus k}$. Hint: consider the term $-\prod_{\alpha=1}^k (1 \pm \sigma_z^{j_\alpha})$.
\enu
\end{exercise}

\begin{exercise}
Consider local traps in the toric code model.
\bnu[(1)]
\item Fix a vertex $v_0$ and a plaquette $p_0$. Show that the local trap $(A_{v_0} + B_{p_0} + A_{v_0} B_{p_0})$ realizes the topological excitation $e \oplus m$. What is the topological excitation realized by $(A_{v_0} + B_{p_0} - A_{v_0} B_{p_0})$?
\item For any non-negative integers $k_0,k_1,k_2,k_3$, find a local trap that realizes $\one^{\oplus k_0} \oplus e^{\oplus k_1} \oplus m^{\oplus k_2} \oplus f^{\oplus k_3}$.
\enu
\end{exercise}

\begin{rem} \label{rem:particle=0d-boundary-condition}
One can see that a local trap at a given site $i$ can also be viewed as defining a `0d boundary condition' of the 2d model at the site $i$. Therefore, the category $\CC$ of particle-like topological defects can also be viewed as that of `0d boundary conditions' at a given site $i$, which is denoted by $\CC_i$. This point of view is sometimes very useful. We can specify a given `0d boundary condition' $x$ at the site $i$ by a pair $(\CC_i,x)$.  
\end{rem}

The direct sum usually implies instability. In the above example, we can perturb the local trap $B_{p_0}$ by replacing it with $(1 + \varepsilon) B_{p_0}$. When $\varepsilon = 0$, the trapped topological excitation is $\one \oplus m$ as discussed above. When $\varepsilon < 0$, the new ground state is the same as the original ground state. When $\varepsilon > 0$, the new ground state is given by $B_{p_0} = -1$ and thus gives the topological excitation $m$. Therefore, the topological defect $\one \oplus m$ is unstable, and by applying a perturbation it collapse to $\one$ or $m$. Such an unstable topological defect is also called a \emph{composite topological defect} (or a \emph{composite anyon}), while a stable topological defect is called a \emph{simple topological defect} (or a \emph{simple anyon}). Each composite topological defect is a direct sum of simple topological defects. We say an $n$d topological order (where $n \geq 1$) is \emph{stable} if its ground state degeneracy on an open disk is $1$, or equivalently, the trivial topological defect $\one$ is stable.

We summarize the above discussions to the following definition.

\begin{defn}
Let $\CC$ be a $\Cb$-linear category. An object $x \in \CC$ is called \emph{simple} if $\Hom_\CC(x,x) \simeq \Cb$. Two objects $x,y \in \CC$ are called \emph{disjoint} if $\Hom_\CC(x,y) = 0 = \Hom_\CC(y,x)$. We say $\CC$ is \emph{semisimple} if it satisfies the following conditions:
\bnu
\item The direct sum of finitely many objects in $\CC$ exists.
\item There exists a collection of mutually disjoint simple objects $\{x_i\}_{i \in I}$ such that every object in $\CC$ is a finite direct sum of objects in $\{x_i\}_{i \in I}$.
\enu
If the index set $I$ is a finite set, then we say $\CC$ is a \emph{finite semisimple category}.
\end{defn}

\begin{rem}
There are many different definitions of simple objects and semisimple categories (for example, see \cite{Bai17,Pen21} and references therein).
\end{rem}

The category $\CC$ of particle-like topological defects of a $2$d topological order $\SC$ (where $n \geq 1$) is a finite semisimple category. The simple objects in $\CC$ are simple topological defects in $\SC$. The topological order $\SC$ is stable if and only if the trivial topological defect $\one \in \CC$ is simple.

\begin{rem}
Our analysis of particle-like topological defects works for any $n$d topological orders. In other words, the category of particle-like topological defects of an $n$d topological order is also a finite semisimple category. 
\end{rem}

\begin{rem}
Let $\CC$ be a $\Cb$-linear category. A nonzero object $x \in \CC$ is called \emph{indecomposable} if any decomposition $x = x_1 \oplus x_2$ is trivial, i.e., either $x_1 = 0$ or $x_2 = 0$. A simple object is indecomposable, but the converse is not true in general. When $\CC$ is a semisimple category, an object $x \in \CC$ is indecomposable if and only if it is simple.
\end{rem}

\begin{rem} \label{rem:structure_semisimple_category}
Let $\CC$ be a semisimple category. Suppose $\{x_i\}_{i \in I}$ is a collection of mutually disjoint simple objects such that every object in $\CC$ is a direct sum of objects in $\{x_i\}_{i \in I}$. By Exercise \ref{exercise:direct_sum_universal_property} (3), every morphism in $\CC$ can be written as a block-diagonal matrix with coefficients in $\Cb$. More precisely, we have
\[
\Hom_\CC \bigl( \bigoplus_{i \in I} x_i^{\oplus n_i} , \bigoplus_{j \in I} x_j^{\oplus m_j} \bigr) \simeq \bigoplus_{i \in I} \mathrm{M}_{m_i \times n_i}(\Cb) ,
\]
where $\mathrm{M}_{m \times n}(\Cb)$ denotes the space of $m$-by-$n$ matrices with coefficients in $\Cb$. Then we immediately see that every simple object in $\CC$ is isomorphic to exactly one object in $\{x_i\}_{i \in I}$. In other words, each isomorphism classes of simple objects in $\CC$ contains exactly one object in $\{x_i\}_{i \in I}$. Thus $\CC$ is finite semisimple if and only if there are finitely many simple objects in $\CC$ (up to isomorphism). We denote the set of isomorphism classes of simple objects in $\CC$ by $\Irr(\CC)$.
\end{rem}

\begin{expl}
Let $\CC,\CD$ be semisimple categories. Then their direct sum $\CC \oplus \CD$ is also semisimple. The simple objects in $\CC \oplus \CD$ are of the form $(x,0)$ for $x \in \CC$ simple or $(0,y)$ for $y \in \CD$ simple. Moreover, $\CC \oplus \CD$ is finite semisimple if and only if both $\CC$ and $\CD$ are finite semisimple.
\end{expl}

\begin{rem}
By Theorem \ref{thm:equivalence_fully_faithful_essentially_surjective} and Remark \ref{rem:structure_semisimple_category}, one can prove that a finite semisimple category $\CC$ is equivalent to the direct sum $\vect^{\oplus n}$ of $n$ copies of $\vect$, where $n$ is the number of isomorphism classes of simple objects of $\CC$. It follows that every finite semisimple category is idempotent complete. In practice, we can also define a finite semisimple category to be a category that is equivalent to the direct sum of several copies of $\vect$.
\end{rem}

\begin{expl}
The category $\Toric$ of particle-like topological defects of the toric code model is finite semisimple and has $4$ simple objects: $\one,e,m,f$. The other objects are direct sums of these simple objects. In other words, $\Toric$ is equivalent to $\vect^{\oplus 4}$.
\end{expl}

\begin{expl}
Let $G$ be a finite group. By Maschke's theorem \cite{Mas98,Mas99}, $\rep(G)$ is a finite semisimple category. A simple object in $\rep(G)$ is also called an \emph{irreducible $G$-representation}. The number of elements in $\Irr(\rep(G))$ is equal to the number of conjugacy classes of $G$. However, there is no canonical bijection between isomorphism classes of irreducible $G$-representations and conjugacy classes of $G$.
\end{expl}

\begin{exercise}
Let $G$ be a finite group. Prove that $\vect_G$ is a finite semisimple category. Hint: use Exercise \ref{exercise:vect_G_simple_object}.
\end{exercise}

\subsubsection{Monoidal structure}

\begin{figure}[htbp]
\[
\begin{array}{c}
\begin{tikzpicture}[scale=0.8]
\fill[gray!20] (0,0)--(3,0)--(4,2)--(1,2)--cycle ;
\draw (1.8,1)--(1.8,4) node[very near end,left] {$x'$} ;
\draw (2.2,1)--(2.2,4) node[very near end,right] {$y'$} ;

\fill (1.8,1) circle (0.07) node[left] {$x$} ;
\fill (2.2,1) circle (0.07) node[right] {$y$} ;
\draw[fill=white] (1.7,2.4) rectangle (1.9,2.6) node[midway,left] {$f$} ;
\draw[fill=white] (2.1,2.4) rectangle (2.3,2.6) node[midway,right] {$g$} ;
\end{tikzpicture}
\end{array}
\rightsquigarrow
\begin{array}{c}
\begin{tikzpicture}[scale=0.8]
\fill[gray!20] (0,0)--(3,0)--(4,2)--(1,2)--cycle ;
\draw (2,1)--(2,4) node[very near end,left] {$x' \otimes y'$} ;

\fill (2,1) circle (0.07) node[left] {$x \otimes y$} ;
\draw[fill=white] (1.9,2.4) rectangle (2.1,2.6) node[midway,right] {$f \otimes g$} ;
\end{tikzpicture}
\end{array}
\]
\caption{the fusion of particle-like topological defects}
\label{fig:fusion_intuition}
\end{figure}
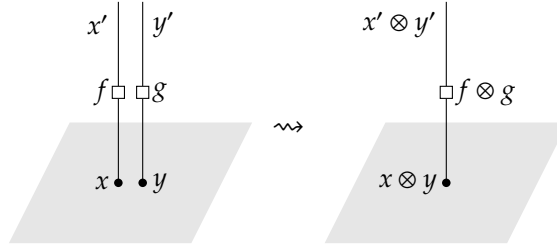

Roughly speaking, if two particle-like topological defects $x$ and $y$ are close to each other, one can view them as a single topological defect denoted by $x \otimes y$ (see Figure \ref{fig:fusion_intuition}).\footnote{This statement is a radical simplification of the real story. Since it has already been accepted by most working physicists in the field, we also adopt it and move on from here. However, the real story is much richer and physically more natural. We choose to discuss this technical but important issue in Remark \ref{rem:E1_algebra}.}  

\begin{expl}
In the toric code model, two $e$ particles can be annihilated to the ground state by a string of $\sigma_z$ operators which connects them (see Figure \ref{fig:ext_toric_code}). Thus we have $e \otimes e = \one$. Similarly $m \otimes m = f \otimes f = \one$. By the definition of $f$ we have $e \otimes m = f = m \otimes e$. Thus the fusion rule of simple topological excitations in the toric code model is the same as the multiplication of the group $\Zb_2 \times \Zb_2$.
\end{expl}

Moreover, the instantons attached on the world lines of topological defects are also fused together. Suppose $f \colon x \to x'$ and $g \colon y \to y'$ are two instantons. Then they fuse into an instanton from $x \otimes y$ to $x' \otimes y'$, denoted by $f \otimes g$. This defines a map
\begin{align*}
\Hom_\CC(x,x') \times \Hom_\CC(y,y') & \to \Hom_\CC(x \otimes y,x' \otimes y') \\
(f , g) & \mapsto f \otimes g .
\end{align*}
It follows that the fusion of particle-like topological defects defines a functor $\otimes \colon \CC \times \CC \to \CC$.

\begin{exercise}
Use physical intuitions to check the functoriality of $\otimes$.
\end{exercise}

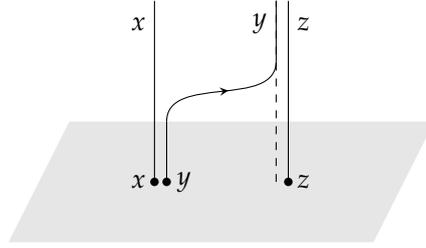
\begin{figure}[htbp]
\centering
\begin{tikzpicture}[scale=0.8]
\fill[gray!20] (-1,0)--(5,0)--(6,2)--(0,2)--cycle ;
\draw (1.4,1)--(1.4,4) node[very near end,left] {$x$} ;
\draw (1.6,1)--(1.6,2) ;
\draw[->-] (1.6,2) .. controls (1.6,2.8) and (3.4,2.2) .. (3.4,3) ;
\draw (3.4,3)--(3.4,4) ;
\draw[dashed] (3.4,1)--(3.4,4) node[very near end,left] {$y$} ;
\draw (3.6,1)--(3.6,4) node[very near end,right] {$z$} ;

\fill (1.4,1) circle (0.07) node[left] {$x$} ;
\fill (1.6,1) circle (0.07) node[right] {$y$} ;
\fill (3.6,1) circle (0.07) node[right] {$z$} ;

\end{tikzpicture}
\caption{the associator $\alpha_{x,y,z} \colon (x \otimes y) \otimes z \to x \otimes (y \otimes z)$}
\label{fig:associator}
\end{figure}

If there are three topological defects $x,y,z \in \CC$, there are different ways to fuse them together. We can first fuse $x$ and $y$ to get $x \otimes y$, then fuse it with $z$ to get $(x \otimes y) \otimes z$; we can also fuse $y$ and $z$ first then fuse the result with $x$ to get $x \otimes (y \otimes z)$. It is physically obvious that these two results should be the same. However, we do not require that $(x \otimes y) \otimes z$ and $x \otimes (y \otimes z)$ are equal on the nose, but require a distinguished isomorphism between them. We denote this isomorphism by $\alpha_{x,y,z} \colon (x \otimes y) \otimes z \to x \otimes (y \otimes z)$, called the associator. As an instanton, the associator $\alpha_{x,y,z}$ is depicted in Figure \ref{fig:associator}.

\begin{figure}[htbp]
\[
\begin{array}{c}
\begin{tikzpicture}[scale=0.8]
\fill[gray!20] (-1,0)--(5,0)--(6,2)--(0,2)--cycle ;
\draw (1.4,1)--(1.4,4) node[very near end,left] {$x'$} ;
\draw (1.6,1)--(1.6,2.5) ;
\draw[->-] (1.6,2.5) .. controls (1.6,3.3) and (3.4,2.7) .. (3.4,3.5) ;
\draw (3.4,3.5)--(3.4,4) ;
\draw[dashed] (3.4,1)--(3.4,4) node[very near end,left] {$y'$} ;
\draw (3.6,1)--(3.6,4) node[very near end,right] {$z'$} ;
\draw[fill=white] (1.3,1.65) rectangle (1.5,1.85) node[midway,left] {$f$} ;
\draw[fill=white] (1.5,1.65) rectangle (1.7,1.85) node[midway,right] {$g$} ;
\draw[fill=white] (3.5,1.65) rectangle (3.7,1.85) node[midway,right] {$h$} ;

\fill (1.4,1) circle (0.07) node[left] {$x$} ;
\fill (1.6,1) circle (0.07) node[right] {$y$} ;
\fill (3.6,1) circle (0.07) node[right] {$z$} ;

\end{tikzpicture}
\end{array}
=
\begin{array}{c}
\begin{tikzpicture}[scale=0.8]
\fill[gray!20] (-1,0)--(5,0)--(6,2)--(0,2)--cycle ;
\draw (1.4,1)--(1.4,4) node[very near end,left] {$x'$} ;
\draw (1.6,1)--(1.6,1.5) ;
\draw[->-] (1.6,1.5) .. controls (1.6,2.3) and (3.4,1.7) .. (3.4,2.5) ;
\draw (3.4,2.5)--(3.4,4) ;
\draw[dashed] (3.4,1)--(3.4,4) node[very near end,left] {$y'$} ;
\draw (3.6,1)--(3.6,4) node[very near end,right] {$z'$} ;
\draw[fill=white] (1.3,3.15) rectangle (1.5,3.35) node[midway,left] {$f$} ;
\draw[fill=white] (3.3,3.15) rectangle (3.5,3.35) node[midway,left] {$g$} ;
\draw[fill=white] (3.5,3.15) rectangle (3.7,3.35) node[midway,right] {$h$} ;

\fill (1.4,1) circle (0.07) node[left] {$x$} ;
\fill (1.6,1) circle (0.07) node[right] {$y$} ;
\fill (3.6,1) circle (0.07) node[right] {$z$} ;

\end{tikzpicture}
\end{array}
\]
\caption{the naturality of the associator $\alpha$}
\label{fig:naturality_associator}
\end{figure}
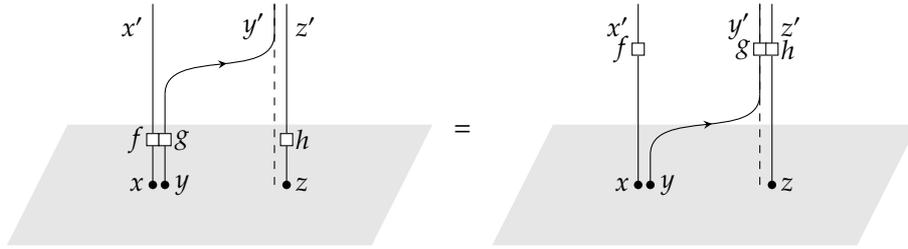

The family $\alpha = \{\alpha_{x,y,z}\}_{x,y,z \in \CC}$ is natural in all three variables because instantons can be freely moved along the world lines (see Figure \ref{fig:naturality_associator}). More precisely, for any morphisms $f \colon x \to x'$, $g \colon y \to y'$ and $h \colon z \to z'$ in $\CC$, we have 
\[
\alpha_{x',y',z'} \circ ((f \otimes g) \otimes h) = (f \otimes (g \otimes h)) \circ \alpha_{x,y,z} .
\]
Hence, the associator is a natural isomorphism $\alpha \colon \otimes \circ (\otimes \times \id_\CC) \Rightarrow \otimes \circ (\id_\CC \times \otimes)$, where these two functors are both from $\CC \times \CC \times \CC$ to $\CC$.

\begin{figure}[htbp]
\[
\begin{array}{c}
\begin{tikzpicture}[scale=0.8]
\fill[gray!20] (-0.5,0)--(5.5,0)--(6.5,2)--(0.5,2)--cycle ;
\draw (1.4,1)--(1.4,5) node[very near end,left] {$x$} ;
\draw (1.6,1)--(1.6,1.5) ;
\draw[->-] (1.6,1.5) .. controls (1.6,2) and (2.4,2) .. (2.4,2.5) ;
\draw (3.4,3.5)--(3.4,5) ;
\draw (2.6,1)--(2.6,2.5) ;
\draw[->-] (3.6,3.5) .. controls (3.6,4) and (4.4,4) .. (4.4,4.5) ;
\draw (4.4,4.5)--(4.4,5) ;
\draw (4.6,1)--(4.6,5) node[very near end,right] {$w$} ;
\draw[double=gray!5,double distance between line centers=0.16cm] (2.5,2.5) .. controls (2.5,3) and (3.5,3) .. (3.5,3.5) ;
\draw[dashed] (2.4,1)--(2.4,2.5) ;
\draw[dashed] (3.4,1)--(3.4,5) node[very near end,left] {$y$} ;
\draw[dashed] (3.6,1)--(3.6,3.5) ;
\draw[dashed] (4.4,1)--(4.4,5) node[very near end,left] {$z$} ;

\fill (1.4,1) circle (0.07) node[left] {$x$} ;
\fill (1.6,1) circle (0.07) node[right] {$y$} ;
\fill (2.6,1) circle (0.07) node[right] {$z$} ;
\fill (4.6,1) circle (0.07) node[right] {$w$} ;

\end{tikzpicture}
\end{array}
=
\begin{array}{c}
\begin{tikzpicture}[scale=0.8]
\fill[gray!20] (-0.5,0)--(5.5,0)--(6.5,2)--(0.5,2)--cycle ;
\draw (1.4,1)--(1.4,5) node[very near end,left] {$x$} ;
\draw (1.6,1)--(1.6,3) ;
\draw[->-] (1.6,3) .. controls (1.6,3.8) and (3.4,3.2) .. (3.4,4) ;
\draw (3.4,4)--(3.4,5) ;
\draw[dashed] (3.4,1)--(3.4,5) node[very near end,left] {$y$} ;
\draw (2.6,1)--(2.6,2) ;
\draw[->-] (2.6,2) .. controls (2.6,2.8) and (4.4,2.2) .. (4.4,3) ;
\draw (4.4,3)--(4.4,5) ;
\draw[dashed] (4.4,1)--(4.4,5) node[very near end,left] {$z$} ;
\draw (4.6,1)--(4.6,5) node[very near end,right] {$w$} ;

\fill (1.4,1) circle (0.07) node[left] {$x$} ;
\fill (1.6,1) circle (0.07) node[right] {$y$} ;
\fill (2.6,1) circle (0.07) node[right] {$z$} ;
\fill (4.6,1) circle (0.07) node[right] {$w$} ;

\end{tikzpicture}
\end{array}
\]
\caption{the pentagon equation}
\label{fig:pentagon}
\end{figure}
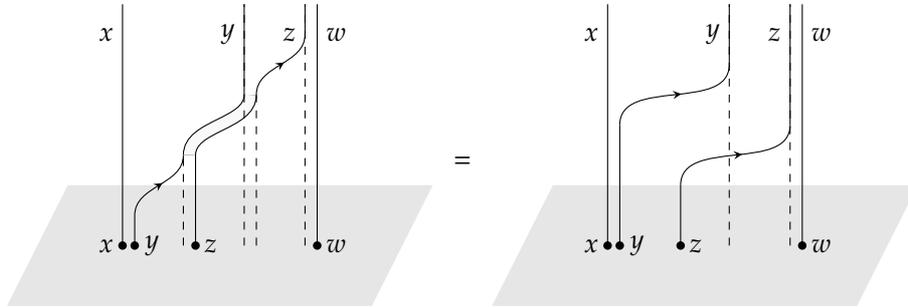

If there are four topological defects $x,y,z,w$, there are five different ways to fuse them two by two. The associators provide many isomorphisms between them. For example, Figure \ref{fig:pentagon} depicts two different isomorphisms (instantons) from $((x \otimes y) \otimes z) \otimes w$ to $x \otimes (y \otimes (z \otimes w))$. These two instantons should be equal because their world lines are homotopy:
\[
(\id_x \otimes \alpha_{y,z,w}) \circ \alpha_{x,y \otimes z,w} \circ (\alpha_{x,y,z} \otimes \id_w) = \alpha_{x,y,z \otimes w} \circ \alpha_{x \otimes y,z,w} .
\]
This is also called the pentagon equation because it can be interpreted as a pentagon-like commutative diagram \eqref{eq:pentagon}.

\begin{rem}
The associators and pentagon equation implies that all different ways to fuse $n \geq 4$ topological defects are isomorphic, and there is a unique isomorphism between two different fusion functors. This is called the coherence theorem \cite{Mac63}.
\end{rem}

Moreover, fusing a topological defect $x$ with the trivial topological defect $\one$ should not change it. So there are two isomorphisms $\lambda_x \colon \one \otimes x \to x$ and $\rho_x \colon x \otimes \one \to x$, called the left and right unitor. Also they are natural in $x$ and satisfy some coherence conditions.

We summarize the above data and properties to the following formal structure \cite{Ben63,Mac63}.

\begin{defn}
A \emph{monoidal category} consists of the following data:
\bit
\item a category $\CC$;
\item a functor $\otimes \colon \CC \times \CC \to \CC$, where $\otimes(x,y)$ is also denoted by $x \otimes y$;
\item a distinguished object $\one \in \CC$, called the \emph{tensor unit};
\item a natural isomorphism $\alpha_{x,y,z} \colon (x \otimes y) \otimes z \to x \otimes (y \otimes z)$, called the \emph{associator};
\item a natural isomorphism $\lambda_x \colon \one \otimes x \to x$, called the \emph{left unitor};
\item a natural isomorphism $\rho_x \colon x \otimes \one \to x$, called the \emph{right unitor};
\eit
and these data satisfy the following conditions:
\bnu
\item (\textbf{pentagon equation}) For any $x,y,z,w \in \CC$, the following diagram commutes:
\be \label{eq:pentagon}
\begin{array}{c}
\xymatrix{
 & ((x \otimes y) \otimes z) \otimes w \ar[dl]_{\alpha_{x,y,z} \otimes \id_w} \ar[dr]^{\alpha_{x \otimes y,z,w}} \\
(x \otimes (y \otimes z)) \otimes w \ar[d]_{\alpha_{x,y \otimes z,w}} & & (x \otimes y) \otimes (z \otimes w) \ar[d]^{\alpha_{x,y,z \otimes w}} \\
x \otimes ((y \otimes z) \otimes w) \ar[rr]^{\id_x \otimes \alpha_{y,z,w}} & & x \otimes (y \otimes (z \otimes w))
}
\end{array}
\ee
\item (\textbf{triangle equation}) For any $x,y \in \CC$, the following diagram commutes:
\[
\begin{array}{c}
\xymatrix{
(x \otimes \one) \otimes y \ar[rr]^{\alpha_{x,\one,y}} \ar[dr]_{\rho_x \otimes \id_y} & & x \otimes (\one \otimes y) \ar[dl]^{\id_x \otimes \lambda_y} \\
 & x \otimes y
}
\end{array}
\]
\enu
A \emph{$\Cb$-linear monoidal category} $\CC$ is both a $\Cb$-linear category and a monoidal category such that the tensor product functor $\otimes \colon \CC \times \CC \to \CC$ is $\Cb$-bilinear, i.e., both $x \otimes - \colon \CC \to \CC$ and $- \otimes x \colon \CC \to \CC$ are $\Cb$-linear functors for each $x \in \CC$.
\end{defn}

As a conclusion, the category $\CC$ of particle-like topological defects of a 2d topological order $\SC$ is a $\Cb$-linear monoidal category.

\begin{rem} \label{rem:E1_algebra}
We have derived the data and the axioms of a monoidal category from the physical intuitions behind the fusion product $\otimes$, which is a radical simplification of the real story. We briefly outline the real and much richer story in this remark. In a physical realization of a 2d topological order, suppose that two particle-like topological defects $x$ and $y$ are located at two different sites $\xi,\eta$, respectively. This physical configuration already defines a fusion product $x \otimes_{(\xi,\eta)} y$, which can be viewed a subspace of the total Hilbert space. In other words, there are infinitely many fusion products parameterized by the elements in the configuration space $\{ (\xi,\eta) \in \Rb^2 \mid \xi \neq \eta \}$. An adiabatic move of these two defects from $(\xi,\eta)$ to $(\xi',\eta')$ along a path $\gamma$ in the configuration space defines an isomorphism between two associated subspaces of the total Hilbert space, i.e., a linear and invertible map $T^\gamma_{x,y} \colon x\otimes_{(\xi,\eta)} y \Rightarrow x\otimes_{(\xi',\eta')} y$. Using Remark \ref{rem:particle=0d-boundary-condition}, $\otimes_{(\xi,\eta)}$ defines a tensor product functor $\otimes_{(\xi,\eta)} \colon \CC_\xi \times \CC_\eta \to \CC$, then $T^\gamma \coloneqq \{ T^\gamma_{x,y} \}_{x,y\in\CC} \colon \otimes_{(\xi,\eta)} \Rightarrow \otimes_{(\xi',\eta')}$ defines a natural isomorphism. 
The most of important physical properties of a topological order (almost a defining property) is stated as follows: 
\begin{equation*}
\text{if two paths $\gamma_1$ and $\gamma_2$ are homotopy equivalent in the configuration space, then $T^{\gamma_1}=T^{\gamma_2}$.} \tag{$\ast$} \label{cond:star}
\end{equation*}
This property holds for adiabatic moves or systems with an infinitely large gap\footnote{In reality, the gap of a physical system is always finite and an adiabatic move is practically impossible. In this case, there might be physically detectable difference between $T^{\gamma_1}$ and $T^{\gamma_2}$ due to the finiteness of the gap. In this case, it can be practically useful to replace the monoidal category by a monoidal $A_{\infty}$-category in order to encode the information of higher homotopies. For gapless phases, it is not clear how to define an adiabatic move. However, in 2D rational CFT's, the condition \eqref{cond:star} can still make sense \cite{HL94,Hua95,Hua08}.}, and allows us to radically reduce the number of the tensor products $\otimes_{(\xi,\eta)}$ that need to be considered. In 1d space, all fusion products can be reduced to a single one $\otimes \coloneqq \otimes_{(0,1)}$. A complete and rigorous proof of the monoidal structure on $\CC$ can be derived from here. It is equivalent to the proof of the well-known mathematical theorem that an $E_1$-algebra in the (2,1)-category of categories is a monoidal category \cite{SW03,Lur17,Fre17}. In 2d space, these fusion functors can be reduced to two compatible fusion functors, or equivalently, a single fusion functor $\otimes$ equipped with a braiding structure (see Remark \ref{rem:E2_algebra}). A physical work on adiabatic moves and braided monoidal structures can be found in \cite{KL20}. 
\end{rem}

\begin{rem} \label{rem:monoidal_gauge}
In other words, the monoidal structure on the category $\CC$ of particle-like topological defects of a 2d topological order $\SC$ depends on many artificial choices. For example, the monoidal structure can be obtained by microscopic realizations of adiabatic moves \cite{KL20}, and different realizations give rise to different (but equivalent) monoidal structures. Hence a topological order does not determine a single monoidal category $\CC$, but an equivalence class of monoidal categories. Physicists usually say that the monoidal structure (or simply the associator) is not `gauge invariant'.
\end{rem}

\begin{expl}
The tensor product $V \otimes_\Cb W$ of two (not necessarily finite-dimensional) vector spaces $V,W$ is defined by the following universal property: $V \otimes_\Cb W$ is equipped with a $\Cb$-bilinear map $\otimes_\Cb \colon V \times W \to V \otimes_\Cb W$, and for any vector space $X$ and $\Cb$-bilinear map $f \colon V \times W \to X$, there exists a unique $\Cb$-linear map $\underline f \colon V \otimes_\Cb W \to X$ such that $\underline f \circ \otimes_\Cb = f$, i.e., the following diagram commutes:
\[
\xymatrix{
V \times W \ar[r]^{\otimes_\Cb} \ar[dr]_{f} & V \otimes_\Cb W \ar[d]^{\underline f} \\
 & X
}
\]

This definition may be too abstract. In practice, it is enough to know the following facts:
\bnu
\item For every $v \in V$ and $w \in W$, there is a vector $v \otimes_\Cb w \in V \otimes_\Cb W$. In general, a vector in $V \otimes_\Cb W$ is the sum of vectors of this form.
\item Some obvious equations hold:
\begin{gather*}
(v + v') \otimes_\Cb w = v \otimes_\Cb w + v' \otimes_\Cb w , \\
v \otimes_\Cb (w + w') = v \otimes_\Cb w + v \otimes_\Cb w' , \\
(\lambda \cdot v) \otimes_\Cb w = \lambda \cdot (v \otimes_\Cb w) = v \otimes_\Cb (\lambda \cdot w) .
\end{gather*}
\item Suppose $\{v_i \mid 1 \leq i \leq n\}$ is a basis of $V$ and $\{w_j \mid 1 \leq j \leq m\}$ is a basis of $W$. Then $\{v_i \otimes_\Cb w_j \mid 1 \leq i \leq n , \, 1 \leq j \leq m\}$ is a basis of $V \otimes_\Cb W$. In particular, we have $\dim(V \otimes_\Cb W) = \dim(V) \dim(W)$.
\enu

The tensor product of vector spaces induces a $\Cb$-linear monoidal structure on $\vect$. Both the tensor product functor $\otimes_\Cb \colon \vect \times \vect \to \vect$, the associator and left/right unitor are induced by the universal property. For example, the associator $\alpha_{U,V,W} \colon (U \otimes_\Cb V) \otimes_\Cb W \to U \otimes_\Cb (V \otimes_\Cb W)$ is defined by $(u \otimes_\Cb v) \otimes_\Cb w \mapsto u \otimes_\Cb (v \otimes_\Cb w)$. The tensor unit is $\Cb$.
\end{expl}

\begin{expl}
Let $G$ be a finite group. For $(V,\rho) , (W,\sigma) \in \rep(G)$, their tensor product is the $G$-representation defined by the vector space $V \otimes_\Cb W$ equipped with the $G$-action
\[
g \mapsto \rho(g) \otimes_\Cb \sigma(g) , \quad g \in G .
\]
This tensor product induces a $\Cb$-linear monoidal structure on $\rep(G)$. The tensor unit is the trivial $G$-representation, i.e., the vector space $\Cb$ equipped with the $G$-action $g \mapsto \id_\Cb$.
\end{expl}

\begin{expl} \label{expl:vect_G_cocycle}
Let $G$ be a finite group. For $V,W \in \vect_G$, their tensor product is the $G$-graded vector space defined by the total space $V \otimes_\Cb W$ equipped with the $G$-grading
\[
(V \otimes_\Cb W)_g \coloneqq \bigoplus_{h \in G} V_h \otimes_\Cb W_{h^{-1}g} , \quad g \in G .
\]
This tensor product is well-define because the tensor product of vector spaces preserves direct sums (see Exercise \ref{exercise:linear_functor_preserve_direct_sum}). In particular, we have $\Cb_{(g)} \otimes_\Cb \Cb_{(h)} \simeq \Cb_{[gh]}$.

However, there are different $\Cb$-linear monoidal structures on $\vect_G$ with this tensor product functor. Suppose the associator $\alpha_{g,h,k} \colon (\Cb_{(g)} \otimes_\Cb \Cb_{(h)}) \otimes_\Cb \Cb_{(k)} \to \Cb_{(g)} \otimes_\Cb (\Cb_{(h)} \otimes_\Cb \Cb_{(k)})$ is defined by
\[
(1 \otimes_\Cb 1) \otimes_\Cb 1 \mapsto \omega(g,h,k) \cdot 1 \otimes_\Cb (1 \otimes_\Cb 1)
\]
for some nonzero complex number $\omega(g,h,k) \in \Cb^\times$ (by Exercise \ref{exercise:vect_G_simple_object}, we only need to consider these morphisms). Then the pentagon equation \eqref{eq:pentagon} translates to
\[
\omega(h,k,l) \omega(g,hk,l) \omega(g,h,k) = \omega(gh,k,l) \omega(g,h,kl) , \quad \forall g,h,k,l \in G .
\]
We say that $\omega \colon G \times G \times G \to \Cb^\times$ is a \emph{3-cocycle} valued in the abelian group $\Cb^\times$. The space of 3-cocycles valued in $\Cb^\times$ is denoted by $Z^3(G;\Cb^\times)$. Similarly, the triangle equation determines the left/right unitor up to a nonzero complex number $\varepsilon \in \Cb^\times$:
\[
\lambda_g = \varepsilon \cdot \omega(e,e,g)^{-1} , \quad \rho_g = \varepsilon \cdot \omega(g,e,e) .
\]
Usually we take $\varepsilon = 1$.

Hence, for each 3-cocycle $\omega \in Z^3(G;\Cb^\times)$ there is a monoidal category, denoted by $\vect_G^\omega$, whose underlying category is $\vect_G$ and associator is determined by $\omega$. Its tensor unit is $\Cb_{[e]}$ where $e \in G$ is the unit .
\end{expl}

\begin{expl} \label{expl:delooping_abelian_group}
Let $G$ be a group. Suppose there is a monoidal structure on the delooping category $\mathrm B G$ (see Example \ref{expl:delooping_group}). The tensor product functor $\otimes \colon \mathrm B G \times \mathrm B G \to \mathrm B G$ is determined by its action on morphisms $\otimes = \otimes_{(\ast,\ast),(\ast,\ast)} \colon G \times G \to G$. The functoriality implies that the map $\otimes \colon G \times G \to G$ satisfies $e \otimes e = e$ and the following equation for $g,h,k,l \in G$:
\[
(g \otimes h) \cdot (k \otimes l) = (g \cdot k) \otimes (h \cdot l) .
\]
The existence of left/right unitor implies that $e \otimes g = g = g \otimes e$ for every $g \in G$. Then
\begin{gather*}
g \cdot l = (g \otimes e) \cdot (e \otimes l) = (g \cdot e) \otimes (e \cdot l) = g \otimes l , \\
h \cdot k = (e \otimes h) \cdot (k \otimes e) = (e \cdot k) \otimes (h \cdot e) = k \otimes h .
\end{gather*}
Therefore, the map $\otimes$ is equal to the multiplication $\cdot$ of $G$ and is commutative. The pentagon equation implies that the associator is identity. The triangle equation implies that the left and right unitors are equal (and can be arbitrary). Hence we conclude that $\mathrm B G$ is a monoidal category if and only if $G$ is an abelian group.
\end{expl}

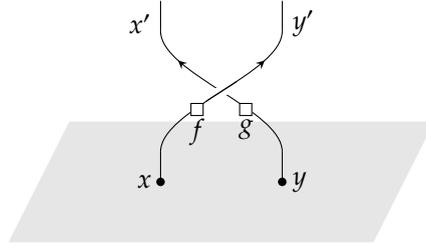
\begin{figure}[htbp]
\centering
\begin{tikzpicture}[scale=0.8]
\fill[gray!20] (-1,0)--(5,0)--(6,2)--(0,2)--cycle ;
\draw (1.5,1)--(1.5,1.5) ;
\draw (1.5,3.5)--(1.5,4) ;
\path (1.5,1)--(1.5,4) node[very near end,left] {$x'$} ;
\draw (3.5,1)--(3.5,1.5) ;
\draw (3.5,3.5)--(3.5,4) ;
\path (3.5,1)--(3.5,4) node[very near end,right] {$y'$} ;

\draw[->-=0.8] (3.5,1.5) .. controls (3.5,2.2) and (1.5,2.8) .. (1.5,3.5) ;
\begin{scope}
\clip (2.5,2.5) circle (0.3) ;
\draw[white,double=black,double distance=0.4pt,line width=2pt] (1.5,1.5) .. controls (1.5,2.2) and (3.5,2.8) .. (3.5,3.5) ;
\end{scope}
\draw[->-=0.8] (1.5,1.5) .. controls (1.5,2.2) and (3.5,2.8) .. (3.5,3.5) ;

\draw[fill=white] (2,2.1) rectangle (2.2,2.3) node[midway,below] {$f$} ;
\draw[fill=white] (3,2.1) rectangle (2.8,2.3) node[midway,below] {$g$} ;

\fill (1.5,1) circle (0.07) node[left] {$x$} ;
\fill (3.5,1) circle (0.07) node[right] {$y$} ;

\end{tikzpicture}
\caption{a `non-local' instanton}
\label{fig:spatial_fusion_anomaly}
\end{figure}

\begin{rem}
Let $\CC$ be a $\Cb$-linear monoidal category. Recall that the map
\[
\otimes \colon \Hom_\CC(x,x') \times \Hom_\CC(y,y') \to \Hom_\CC(x \otimes y,x' \otimes y')
\]
is $\Cb$-bilinear. By the universal property of the tensor product of vector spaces, it induces a $\Cb$-linear map
\[
\Hom_\CC(x,x') \otimes_\Cb \Hom_\CC(y,y') \to \Hom_\CC(x \otimes y,x' \otimes y') .
\]
In general, this linear map is not an isomorphism and we say there is a \emph{spatial fusion anomaly}. Physically, the presence of a spatial fusion anomaly means that the spatial fusion $\Hom_\CC(x,x') \otimes_\Cb \Hom_\CC(y,y')$ of two hom spaces is not at a fixed point of the renormalization flow, and it finally flows to the fixed point $\Hom_\CC(x \otimes y,x' \otimes y')$.

Microscopically, the presence of spatial fusion anomalies is due to the existence of `non-local' instantons in $\Hom_\CC(x \otimes y,x' \otimes y')$. For example, if $f \colon x \to y'$ and $g \colon y \to x'$ are two instantons, then the instanton depicted in Figure \ref{fig:spatial_fusion_anomaly} is an instanton in $\Hom_\CC(x \otimes y,x' \otimes y')$. However, it does not come from the spatial fusion $\Hom_\CC(x,x') \otimes_\Cb \Hom_\CC(y,y')$. When two sites are separated far enough, this instanton becomes non-local.
\end{rem}

\subsubsection{Unitarity}

Recall that an instanton can be realized as a linear map commuting with all local operators. Thus we can talk about the Hermitian conjugate of an instanton.

\begin{rem} \label{rem:Hilb_Hermitian}
We briefly review the Hermitian conjugate of a linear map between Hilbert spaces. Let $\mathcal H_1,\mathcal H_2$ be two Hilbert spaces and $f \colon \mathcal H_1 \to \mathcal H_2$ be a linear map. The Hermitian conjugate of $f$ is a linear map $f^\dagger \colon \mathcal H_2 \to \mathcal H_1$ defined by the following equation:
\[
\langle f(v),w \rangle_2 = \langle v,f^\dagger(w) \rangle_1 , \quad \forall v \in \mathcal H_1 , \, w \in \mathcal H_2 ,
\]
where $\langle -,- \rangle_i$ is the inner product of $\mathcal H_i$ for $i=1,2$. We list some important facts about the Hermitian conjugate:
\bnu
\item The map $\dagger \colon \Hom_\vect(\mathcal H_1,\mathcal H_2) \to \Hom_\vect(\mathcal H_2,\mathcal H_1)$ defined by $f \mapsto f^\dagger$ is anti-linear, i.e., $(f + g)^\dagger = f^\dagger + g^\dagger$ and $(\lambda \cdot f)^\dagger = \bar \lambda \cdot f^\dagger$ for any $f,g \in \Hom_\vect(\mathcal H_1,\mathcal H_2)$ and $\lambda \in \Cb$.
\item The map $\dagger$ is involutive, i.e., $(f^\dagger)^\dagger = f$.
\item The identity map $\id_\mathcal H$ on a Hilbert space $\mathcal H$ is invariant under the Hermitian conjugate, i.e., $\id_\mathcal H^\dagger = \id_\mathcal H$.
\item Suppose $\mathcal H_1,\mathcal H_2,\mathcal H_3$ are Hilbert space and $f \colon \mathcal H_1 \to \mathcal H_2$, $g \colon \mathcal H_2 \to \mathcal H_3$ are linear maps. Then $(g \circ f)^\dagger = f^\dagger \circ g^\dagger$.
\item Suppose $f \colon \mathcal H_1 \to \mathcal H_2$ is a linear map between two Hilbert spaces and satisfies $f^\dagger \circ f = 0$. Then $f = 0$.
\enu
\end{rem}

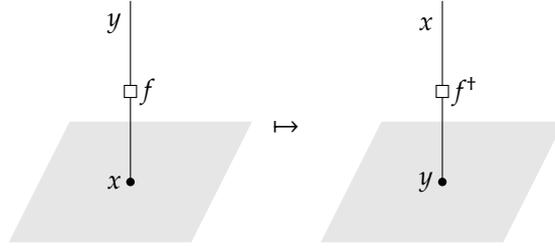
\begin{figure}[htbp]
\[
\begin{array}{c}
\begin{tikzpicture}[scale=0.8]
\fill[gray!20] (0,0)--(3,0)--(4,2)--(1,2)--cycle ;
\draw (2,1)--(2,4) node[very near end,left] {$y$} ;

\fill (2,1) circle (0.07) node[left] {$x$} ;
\draw[fill=white] (1.9,2.4) rectangle (2.1,2.6) node[midway,right] {$f$} ;
\end{tikzpicture}
\end{array}
\mapsto
\begin{array}{c}
\begin{tikzpicture}[scale=0.8]
\fill[gray!20] (0,0)--(3,0)--(4,2)--(1,2)--cycle ;
\draw (2,1)--(2,4) node[very near end,left] {$x$} ;

\fill (2,1) circle (0.07) node[left] {$y$} ;
\draw[fill=white] (1.9,2.4) rectangle (2.1,2.6) node[midway,right] {$f^\dagger$} ;
\end{tikzpicture}
\end{array}
\]
\caption{the dagger structure}
\label{fig:dagger}
\end{figure}

Since the Hermitian conjugate of a local operator is still a local operator, the Hermitian conjugate of an instanton is still an instanton. Thus taking the Hermitian conjugate of an instanton defines a map
\[
\dagger \colon \Hom_\CC(x,y) \to \Hom_\CC(y,x)
\]
for every $x,y \in \CC$. These maps form an additional structure on the category $\CC$, called a unitary structure. Intuitively, taking the Hermitian conjugate of an instanton is like a time-reversal operation (see Figure \ref{fig:dagger}).

The unitary structure on $\CC$ should satisfy the usual properties of the Hermitian conjugate. We summarize these data and properties to the following definition.

\begin{defn}
Let $\CC$ be a $\Cb$-linear category. A \emph{dagger structure} on $\CC$ is an involutive anti-linear functor $\dagger \colon \CC^\op \to \CC$ which acts on objects as identity. Equivalently, a dagger structure on $\CC$ is a collection of anti-linear maps $\dagger \colon \Hom_\CC(x,y) \to \Hom_\CC(y,x)$ for all $x,y \in \CC$, satisfying the following conditions:
\bnu
\item $(f^\dagger)^\dagger = f$ for every morphism $f$ in $\CC$.
\item $(g \circ f)^\dagger = f^\dagger \circ g^\dagger$ for all $f \in \Hom_\CC(x,y)$, $g \in \Hom_\CC(y,z)$ and $x,y,z \in \CC$.
\item $\id_x^\dagger = \id_x$ for all $x \in \CC$.
\enu
Moreover, a dagger structure is called a \emph{unitary structure} if it satisfies the following condition:
\bnu
\item[4.] A morphism $f$ in $\CC$ satisfies $f^\dagger \circ f = 0$ if and only if $f = 0$.
\enu
A \emph{unitary category} is a $\Cb$-linear category equipped with a unitary structure.
\end{defn}

\begin{rem}
A unitary category is automatically semisimple.
\end{rem}

\begin{rem}
The notion of a unitary category is equivalent to the notion of a $C^*$-category \cite[Proposition 2.1]{Mueg00}.
\end{rem}

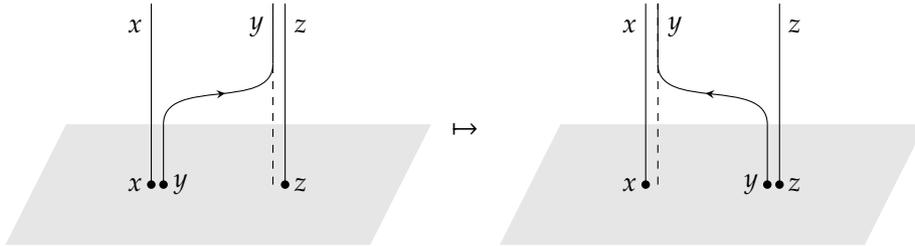
\begin{figure}[htbp]
\[
\begin{array}{c}
\begin{tikzpicture}[scale=0.8]
\fill[gray!20] (-1,0)--(5,0)--(6,2)--(0,2)--cycle ;
\draw (1.4,1)--(1.4,4) node[very near end,left] {$x$} ;
\draw (1.6,1)--(1.6,2) ;
\draw[->-] (1.6,2) .. controls (1.6,2.8) and (3.4,2.2) .. (3.4,3) ;
\draw (3.4,3)--(3.4,4) ;
\draw[dashed] (3.4,1)--(3.4,4) node[very near end,left] {$y$} ;
\draw (3.6,1)--(3.6,4) node[very near end,right] {$z$} ;

\fill (1.4,1) circle (0.07) node[left] {$x$} ;
\fill (1.6,1) circle (0.07) node[right] {$y$} ;
\fill (3.6,1) circle (0.07) node[right] {$z$} ;

\end{tikzpicture}
\end{array}
\mapsto
\begin{array}{c}
\begin{tikzpicture}[scale=0.8]
\fill[gray!20] (-1,0)--(5,0)--(6,2)--(0,2)--cycle ;
\draw (1.4,1)--(1.4,4) node[very near end,left] {$x$} ;
\draw (3.4,1)--(3.4,2) ;
\draw[->-] (3.4,2) .. controls (3.4,2.8) and (1.6,2.2) .. (1.6,3) ;
\draw (1.6,3)--(1.6,4) ;
\draw[dashed] (1.6,1)--(1.6,4) node[very near end,right] {$y$} ;
\draw (3.6,1)--(3.6,4) node[very near end,right] {$z$} ;

\fill (1.4,1) circle (0.07) node[left] {$x$} ;
\fill (3.4,1) circle (0.07) node[left] {$y$} ;
\fill (3.6,1) circle (0.07) node[right] {$z$} ;

\end{tikzpicture}
\end{array}
\]
\caption{the Hermitian conjugate of the associator: $\alpha_{x,y,z} \mapsto \alpha_{x,y,z}^\dagger$}
\label{fig:associator_dagger}
\end{figure}

For any topological order $\SC$, the above discussion implies that the category $\CC$ of particle-like topological defects is a unitary category. When $\SC$ is a 2d topological order, the monoidal structure and the unitary structure on $\CC$ should be compatible. For example, let us consider the Hermitian conjugate of the associator. The intuition of the time-reversal operation provides a picture of $\alpha_{x,y,z}^\dagger$, as depicted in Figure \ref{fig:associator_dagger}.

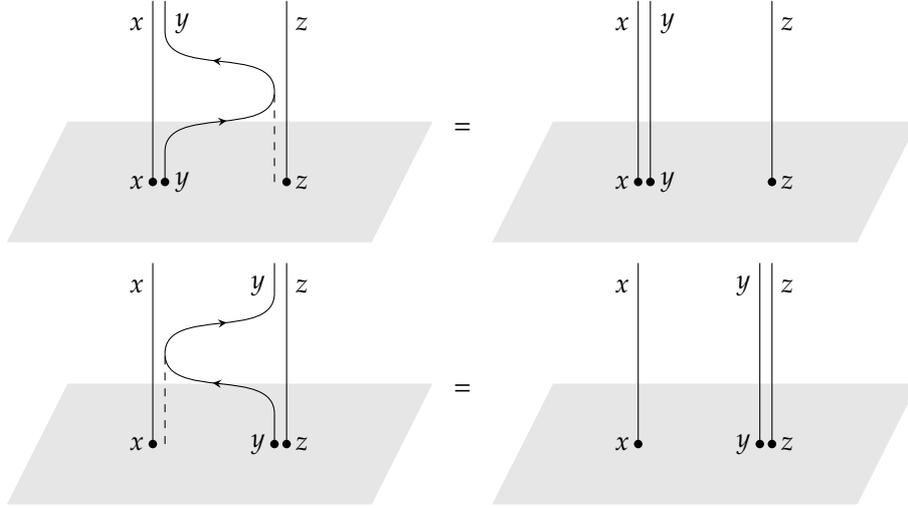
\begin{figure}[htbp]
\begin{gather*}
\begin{array}{c}
\begin{tikzpicture}[scale=0.8]
\fill[gray!20] (-1,0)--(5,0)--(6,2)--(0,2)--cycle ;
\draw (1.4,1)--(1.4,4) node[very near end,left] {$x$} ;
\draw (1.6,1)--(1.6,1.5) ;
\draw[->-] (1.6,1.5) .. controls (1.6,2.3) and (3.4,1.7) .. (3.4,2.5) ;
\draw[->-] (3.4,2.5) .. controls (3.4,3.3) and (1.6,2.7) .. (1.6,3.5) ;
\draw (1.6,3.5)--(1.6,4) ;
\draw[dashed] (3.4,1)--(3.4,2.5) ;
\path (1.6,1)--(1.6,4) node[very near end,right] {$y$} ;
\draw (3.6,1)--(3.6,4) node[very near end,right] {$z$} ;
\fill (1.4,1) circle (0.07) node[left] {$x$} ;
\fill (1.6,1) circle (0.07) node[right] {$y$} ;
\fill (3.6,1) circle (0.07) node[right] {$z$} ;
%
\end{tikzpicture}
\end{array}
=
\begin{array}{c}
\begin{tikzpicture}[scale=0.8]
\fill[gray!20] (-1,0)--(5,0)--(6,2)--(0,2)--cycle ;
\draw (1.4,1)--(1.4,4) node[very near end,left] {$x$} ;
\draw (1.6,1)--(1.6,4) node[very near end,right] {$y$} ;
\draw (3.6,1)--(3.6,4) node[very near end,right] {$z$} ;
\fill (1.4,1) circle (0.07) node[left] {$x$} ;
\fill (1.6,1) circle (0.07) node[right] {$y$} ;
\fill (3.6,1) circle (0.07) node[right] {$z$} ;
%
\end{tikzpicture}
\end{array} \\
\begin{array}{c}
\begin{tikzpicture}[scale=0.8]
\fill[gray!20] (-1,0)--(5,0)--(6,2)--(0,2)--cycle ;
\draw (1.4,1)--(1.4,4) node[very near end,left] {$x$} ;
\draw (3.4,1)--(3.4,1.5) ;
\draw[->-] (3.4,1.5) .. controls (3.4,2.3) and (1.6,1.7) .. (1.6,2.5) ;
\draw[->-] (1.6,2.5) .. controls (1.6,3.3) and (3.4,2.7) .. (3.4,3.5) ;
\draw (3.4,3.5)--(3.4,4) ;
\draw[dashed] (1.6,1)--(1.6,2.5) ;
\path (3.4,1)--(3.4,4) node[very near end,left] {$y$} ;
\draw (3.6,1)--(3.6,4) node[very near end,right] {$z$} ;
\fill (1.4,1) circle (0.07) node[left] {$x$} ;
\fill (3.4,1) circle (0.07) node[left] {$y$} ;
\fill (3.6,1) circle (0.07) node[right] {$z$} ;
%
\end{tikzpicture}
\end{array}
=
\begin{array}{c}
\begin{tikzpicture}[scale=0.8]
\fill[gray!20] (-1,0)--(5,0)--(6,2)--(0,2)--cycle ;
\draw (1.4,1)--(1.4,4) node[very near end,left] {$x$} ;
\draw (3.4,1)--(3.4,4) node[very near end,left] {$y$} ;
\draw (3.6,1)--(3.6,4) node[very near end,right] {$z$} ;
\fill (1.4,1) circle (0.07) node[left] {$x$} ;
\fill (3.4,1) circle (0.07) node[left] {$y$} ;
\fill (3.6,1) circle (0.07) node[right] {$z$} ;
%
\end{tikzpicture}
\end{array}
\end{gather*}
\caption{$\alpha_{x,y,z}^\dagger$ is the inverse of $\alpha_{x,y,z}$}
\label{fig:associator_unitary}
\end{figure}

Then the physical intuition implies that the following equations hold becuase their world lines are homotopy (see Figure \ref{fig:associator_unitary}):
\[
\alpha_{x,y,z}^\dagger \circ \alpha_{x,y,z} = \id_{(x \otimes y) \otimes z} , \quad \alpha_{x,y,z} \circ \alpha_{x,y,z}^\dagger = \id_{x \otimes (y \otimes z)} .
\]
In other words, $\alpha_{x,y,z}^\dagger$ is the inverse of $\alpha_{x,y,z}$.

\begin{defn}
A morphism $f$ in a unitary category is called \emph{unitary} if it is an isomorphism and satisfies $f^{-1} = f^\dagger$.
\end{defn}

\begin{defn}
A dagger structure on a $\Cb$-linear monoidal category $\CC = (\CC,\otimes,\one,\alpha,\lambda,\rho)$ is \emph{compatible} with the monoidal structure if
\bnu
\item $f^\dagger \otimes g^\dagger = (f \otimes g)^\dagger$ for all morphisms $f,g$ in $\CC$.
\item The morphisms $\alpha_{x,y,z},\lambda_x,\rho_x$ are unitary for all $x,y,z \in \CC$.
\enu
A \emph{unitary} monoidal category is a $\Cb$-linear monoidal category equipped with a compatible unitary structure.
\end{defn}

\begin{rem}
The condition that a dagger structure is compatible with the monoidal structure is equivalent to that the functor $\dagger \colon \CC^\op \to \CC$ is a monoidal functor. The notion of a monoidal functor is defined in Definition \ref{defn:monoidal_functor}.
\end{rem}

As a conclusion, the category $\CC$ of particle-like topological defects of a 2d topological order $\SC$ is a unitary monoidal category.

\begin{expl}
Let $\hilb$ be the category of finite-dimensional Hilbert spaces and linear maps. As explained in Remark \ref{rem:Hilb_Hermitian}, $\hilb$ is a unitary category with the unitary structure defined by the usual Hermitian conjugate. Moreover, $\hilb$ is a unitary monoidal category with the tensor product of two Hilbert spaces $\mathcal H_1,\mathcal H_2$ defined by the underlying vector space $\mathcal H_1 \otimes_\Cb \mathcal H_2$ and the inner product
\[
\langle v_1 \otimes_\Cb v_2,w_1 \otimes_\Cb w_2 \rangle \coloneqq \langle v_1,w_1 \rangle_1 \cdot \langle v_2,w_2 \rangle_2 .
\]
The category $\hilb$ can be viewed as a unitary version of $\vect$. Indeed, these two categories are equivalent.
\end{expl}

\begin{exercise}
Let $F \colon \hilb \to \vect$ be the functor that `forgets' the inner products of Hilbert spaces. Use Theorem \ref{thm:equivalence_fully_faithful_essentially_surjective} to prove that $F$ is an equivalence.
\end{exercise}

\begin{expl}
Let $G$ be a finite group. A \emph{unitary $G$-representation} is a Hilbert space $\mathcal H$ equipped with a group homomorphism $\rho \colon G \to \mathrm U(\mathcal H)$, where $\mathrm U(\mathcal H)$ is the group of unitary operators on $\mathcal H$. The category of finite-dimensional unitary $G$-representations is denoted by $\rep^{\mathrm u}(G)$. It is a unitary monoidal category with the unitary structure defined by the usual Hermitian conjugate. Also, the forgetful functor $\rep^{\mathrm u}(G) \to \rep(G)$ is an equivalence. Thus $\rep^{\mathrm u}(G)$ is a unitary version of $\rep(G)$.
\end{expl}

\begin{expl}
Let $G$ be a finite group. The unitary version of $\vect_G$ is the category $\hilb_G$ of finite-dimensional $G$-graded Hilbert spaces. Moreover, by considering functions $G \times G \times G \to \mathrm U(1)$ we can similarly define the group $Z^3(G;\mathrm U(1))$ of 3-cocycles valued in $\mathrm U(1)$. Then for every $\omega \in Z^3(G;\mathrm U(1))$, there is a unitary monoidal category $\hilb_G^\omega$ whose underlying category is $\hilb_G$ and the associator is determined by $\omega$.
\end{expl}

\subsubsection{Rigidity}

Recall that a nontrivial topological excitation can not be created alone by local operators from the ground state. Usually a non-local operator creates a pair of nontrivial topological excitations. Then one of them can be viewed as the `anti-particle' of another one.

\begin{figure}[htbp]
\centering
\subfigure[the creation instanton $b_x$]{
\begin{tikzpicture}[scale=0.8]
\fill[gray!20] (-1,0)--(5,0)--(6,2)--(0,2)--cycle ;
\draw (1.5,3.5)--(1.5,4) ;
\path (1.5,1)--(1.5,4) node[very near end,left] {$x$} ;
\draw (3.5,3.5)--(3.5,4) ;
\path (3.5,1)--(3.5,4) node[very near end,right] {$x^*$} ;

\draw[->-=0.8] (3.5,3.5) .. controls (3.5,1) and (1.5,1) .. (1.5,3.5) ;

\end{tikzpicture}
}
\hspace{5ex}
\subfigure[the annihilation instanton $d_x$]{
\begin{tikzpicture}[scale=0.8]
\fill[gray!20] (-1,0)--(5,0)--(6,2)--(0,2)--cycle ;
\draw (1.5,1)--(1.5,1.5) ;
\draw (3.5,1)--(3.5,1.5) ;

\draw[->-=0.2] (3.5,1.5) .. controls (3.5,4) and (1.5,4) .. (1.5,1.5) ;

\fill (1.5,1) circle (0.07) node[left] {$x^*$} ;
\fill (3.5,1) circle (0.07) node[right] {$x$} ;

\end{tikzpicture}
}
\caption{the creation and annihilation instantons associated to $x$}
\label{fig:creation_annihilation}
\end{figure}
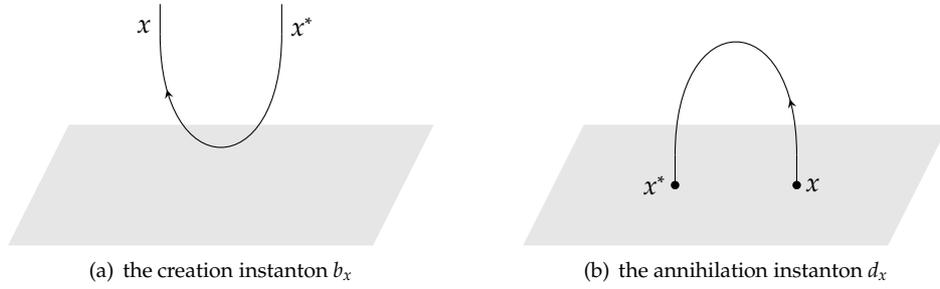

The simplest way to create a nontrivial topological defect $x$ is to `bend' the world line, or the path of an adiabatic move, of $x$ (see Figure \ref{fig:creation_annihilation} (a)). Microscopically, this instanton is realized by a string operator of $x$ acting on the ground state, which creates a topological excitation $x$ and another one denoted by $x^*$, called the dual (or anti-particle) of $x$. Similarly there is an annihilation instanton $d_x \colon x^* \otimes x \to \one$, as depicted in Figure \ref{fig:creation_annihilation} (b).

\begin{expl}
Microscopically, both the adiabatic move, the creation and the annihilation instantons are realized by the same string operator. The difference is how the operator act on different states. In the toric code model, for example, an $e$ string operator $\prod_i \sigma_z^i$ acts on an $e$ particle as adiabatically moving it to another site. However, the same $e$ string operator acting on the ground state creates an $e$ particle on each end of the string. This gives the creation instanton associated to $e$. The annihilation instanton can be obtained similarly by acting an $e$ string operator on two $e$ particles. It follows that $e^* = e$ in the toric code model. Similarly $m^* = m$ and $f^* = f$. The dual of the trivial topological excitation $\one^* = \one$ is itself in any topological order $\SC$.
\end{expl}

\begin{figure}[htbp]
\begin{gather*}
\begin{array}{c}
\begin{tikzpicture}[scale=0.8]
\fill[gray!20] (-1,0)--(5,0)--(6,2)--(0,2)--cycle ;
\draw[->-] (1.5,2)--(1.5,4) ;
\draw[dashed] (1.5,1)--(1.5,4) node[very near end,left] {$x$} ;
\draw (1.5,2) .. controls (1.5,1) and (2,1) .. (2.5,2.5) ;
\draw (2.5,2.5) .. controls (3,4) and (3.5,4) .. (3.5,3) ;
\draw[->-] (3.5,1)--(3.5,3) ;
\fill (3.5,1) circle (0.07) node[right] {$x$} ;
%
\end{tikzpicture}
\end{array}
=
\begin{array}{c}
\begin{tikzpicture}[scale=0.8]
\fill[gray!20] (-1,0)--(5,0)--(6,2)--(0,2)--cycle ;
\draw (1.5,3)--(1.5,4) ;
\draw[dashed] (1.5,1)--(1.5,4) node[very near end,left] {$x$} ;
\draw[->-] (3.5,2) .. controls (3.5,2.8) and (1.5,2.2) .. (1.5,3) ;
\draw (3.5,1)--(3.5,2) ;
\fill (3.5,1) circle (0.07) node[right] {$x$} ;
%
\end{tikzpicture}
\end{array} \\
\begin{array}{c}
\begin{tikzpicture}[scale=0.8]
\fill[gray!20] (-1,0)--(5,0)--(6,2)--(0,2)--cycle ;
\draw[->-] (3.5,4)--(3.5,2) ;
\draw[dashed] (3.5,1)--(3.5,4) node[very near end,right] {$x^*$} ;
\draw (3.5,2) .. controls (3.5,1) and (3,1) .. (2.5,2.5) ;
\draw (2.5,2.5) .. controls (2,4) and (1.5,4) .. (1.5,3) ;
\draw[->-] (1.5,3)--(1.5,1) ;
\fill (1.5,1) circle (0.07) node[left] {$x^*$} ;
%
\end{tikzpicture}
\end{array}
=
\begin{array}{c}
\begin{tikzpicture}[scale=0.8]
\fill[gray!20] (-1,0)--(5,0)--(6,2)--(0,2)--cycle ;
\draw (3.5,3)--(3.5,4) ;
\draw[dashed] (3.5,1)--(3.5,4) node[very near end,right] {$x^*$} ;
\draw[->-] (3.5,3) .. controls (3.5,2.2) and (1.5,2.8) .. (1.5,2) ;
\draw (1.5,1)--(1.5,2) ;
\fill (1.5,1) circle (0.07) node[left] {$x^*$} ;
%
\end{tikzpicture}
\end{array}
\end{gather*}
\caption{the zig-zag equations}
\label{fig:zig_zag}
\end{figure}
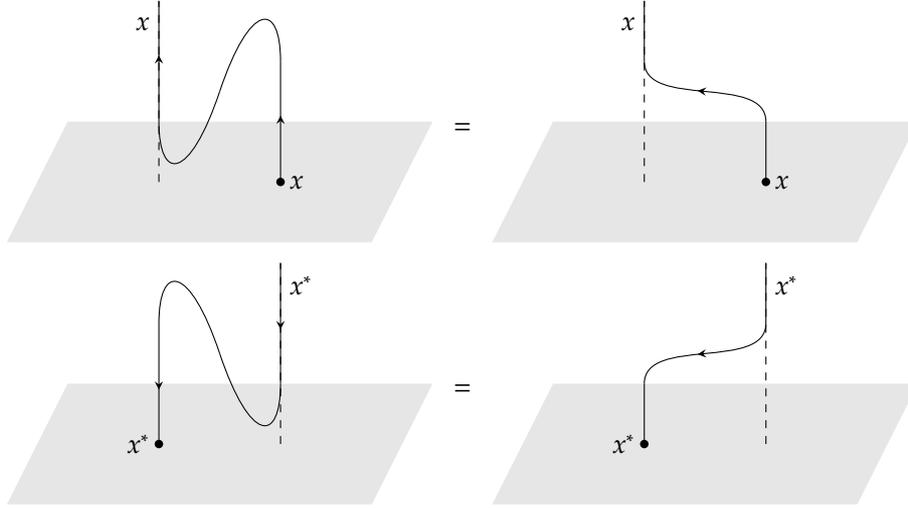

The physical intuition as depicted in Figure \ref{fig:zig_zag} implies that the creation and annihilation instantons should satisfy the following two zig-zag equations:
\begin{gather*}
\bigl( x \xrightarrow{\lambda_x^{-1}} \one \otimes x \xrightarrow{b_x \otimes \id_x} (x \otimes x^*) \otimes x \xrightarrow{\alpha_{x,x^*,x}} x \otimes (x^* \otimes x) \xrightarrow{\id_x \otimes d_x} x \otimes \one \xrightarrow{\rho_x} x \bigr) = \id_x , \\
\bigl( x^* \xrightarrow{\rho_{x^*}^{-1}} x^* \otimes \one \xrightarrow{\id_{x^*} \otimes b_x} x^* \otimes (x \otimes x^*) \xrightarrow{\alpha_{x^*,x,x^*}^{-1}} (x^* \otimes x) \otimes x^* \xrightarrow{d_x \otimes \id_{x^*}} \one \otimes x^* \xrightarrow{\lambda_{x^*}} x^* \bigr) = \id_{x^*} .
\end{gather*}

We summarize the properties of the dual of a topological defect in the following definition.

\begin{defn}
Let $\CC$ be a monoidal category and $x \in \CC$. A \emph{left dual} of $x$ is an object $x^L \in \CC$ equipped with two morphisms $b_x \colon \one \to x \otimes x^L$ and $d_x \colon x^L \otimes x \to \one$ satisfying the following two zig-zag equations:
\[
(\id_x \otimes d_x) \circ (b_x \otimes \id_x) = \id_x , \quad (d_x \otimes \id_{x^L}) \circ (\id_{x^L} \otimes b_x) = \id_{x^L} .
\]
Here we ignore associators and unitors for simplicity. Similarly, a right dual of $x$ is an object $x^R \in \CC$ equipped with two morphisms $b'_x \colon \one \to x^R \otimes x$ and $d'_x \colon x \otimes x^R \to \one$ satisfying the following two zig-zag equations:
\[
(d'_x \otimes \id_x) \circ (\id_x \otimes b'_x) = \id_x , \quad (\id_{x^R} \otimes d'_x) \circ (b'_x \circ \id_{x^R}) = \id_{x^R} .
\]
An object in $\CC$ is called \emph{dualizable} if it admits both left and right duals. If every object in $\CC$ is dualizable, we say that $\CC$ is a \emph{rigid monoidal category}.
\end{defn}

\begin{rem}
A left/right dual of an object, if exists, is unique up to a unique isomorphism. Thus we can talk about \emph{the} left/right dual of an object. It follows that there exists canonical isomorphisms $(x \otimes y)^L \simeq y^L \otimes x^L$ and $(x \otimes y)^R \simeq y^R \otimes x^R$ for every $x,y \in \CC$.
\end{rem}

\begin{expl}
Let $\CC$ be a monoidal category. The tensor unit $\one$ is both the left and right dual of itself.
\end{expl}

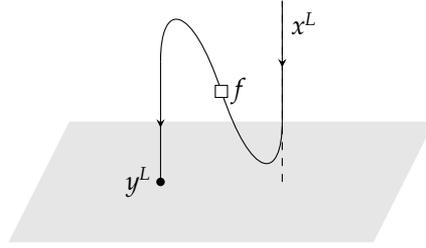
\begin{figure}[htbp]
\centering
\begin{tikzpicture}[scale=0.8]
\fill[gray!20] (-1,0)--(5,0)--(6,2)--(0,2)--cycle ;
\draw[->-] (3.5,4)--(3.5,2) ;
\draw[dashed] (3.5,1)--(3.5,4) node[very near end,right] {$x^L$} ;
\draw (3.5,2) .. controls (3.5,1) and (3,1) .. (2.5,2.5) ;
\draw (2.5,2.5) .. controls (2,4) and (1.5,4) .. (1.5,3) ;
\draw[->-] (1.5,3)--(1.5,1) ;
\draw[fill=white] (2.4,2.4) rectangle (2.6,2.6) node[midway,right] {$f$} ;
\fill (1.5,1) circle (0.07) node[left] {$y^L$} ;
%
\end{tikzpicture}
\caption{The left dual of a morphism $f \colon x \to y$ is a morphism $f^L \colon y^L \to x^L$.}
\label{fig:dual_instanton}
\end{figure}

\begin{rem}
In a rigid monoidal category $\CC$, the left dual $f^L \colon y^L \to x^L$ of a morphism $f \colon x \to y$ is defined by
\[
f^L \coloneqq (d_y \otimes \id_{x^L}) \circ (\id_{y^L} \otimes f \otimes \id_{x^L}) \circ (\id_{x^L} \otimes b_x) ,
\]
as depicted in Figure \ref{fig:dual_instanton}. Similarly we can define the right dual of $f$. The zig-zag equations imply that taking left duals defines a functor $\delta^L \colon \CC^\op \to \CC$. Note that this functor depends on many choices: we need to explicitly choose a left dual $(y^L,b_y,d_y)$ for every object $y \in \CC$. So indeed there are many different left dual functors, but these functors are canonically isomorphic. Similarly, we can define the right dual of a morphism and the right dual functor $\delta^R \colon \CC^\op \to \CC$.
\end{rem}

\begin{exercise} \label{exercise:dual_preserve_direct_sum}
Let $\CC$ be a $\Cb$-linear rigid monoidal category. Prove that taking left duals defines a $\Cb$-linear functor $\delta^L \colon \CC^\op \to \CC$. In particular, taking left duals preserves direct sums, i.e., there is a canonical isomorphism $(x_1 \oplus \cdots \oplus x_n)^L \simeq x_1^L \oplus \cdots \oplus x_n^L$ for $x_1,\ldots,x_n \in \CC$.
\end{exercise}

\begin{rem}
The instantons can be freely moved along the world line. However, the instantons $b_x$ and $d_x$ are not natural transformations, but satisfy the following property: for any morphism $f \colon x \to y$, we have
\[
(f \otimes \id_{x^L}) \circ b_x = (\id_y \otimes f^L) \circ b_y , \quad d_y \circ (\id_{y^L} \otimes f) = d_x \circ (f^L \otimes \id_x) .
\]
Also, there are two similar equations for right duals.
\end{rem}

\begin{expl}
Let $V \in \vect$. Its left dual is given by the usual dual space $V^* \coloneqq \Hom_\vect(V,\Cb)$. The annihilation map $d_V \colon V^* \otimes_\Cb V \to \Cb$ is defined by $\phi \otimes_\Cb v \mapsto \phi(v)$. Note that under the natural identification $V^* \otimes_\Cb V \simeq \Hom_\vect(V,V)$ the annihilation map $d_V$ is taking the trace. The creation map $b_V \colon \Cb \to V \otimes_\Cb V^*$ is defined by
\[
1 \mapsto \sum_{i=1}^n v_i \otimes_\Cb v^i ,
\]
where $\{v_i\}_{i=1}^n$ is a basis of $V$ and $\{v^i\}_{i=1}^n$ is the dual basis in $V^*$ defined by $v^i(v_j) = \delta^i_j$. It is straightforward to check that the creation map is independent of the choice of the basis.

Similarly, one can verify that $V^*$ is also the right dual of $V$. Therefore, $\vect$ is a rigid monoidal category. It is worth noting that in the monoidal category of all vector spaces, an object is dualizable if and only if it is finite-dimensional.
\end{expl}

\begin{exercise}
Let $V,W \in \vect$. Prove that the dual of a linear map $f \colon V \to W$ is the usual dual map $f^* \colon W^* \to V^*$ defined by $f^*(\phi) \coloneqq \phi \circ f$.
\end{exercise}

\begin{expl}
Let $G$ be a finite group. Both the left and right dual of a finite-dimensional $G$-representation $(V,\rho) \in \rep(G)$ is defined by the dual vector space $V^*$ equipped with the $G$-action
\[
g \mapsto \rho(g^{-1})^* , \quad g \in G .
\]
Therefore, $\rep(G)$ is a rigid monoidal category.
\end{expl}

\begin{expl}
Let $G$ be a finite group and $\omega \in Z^3(G;\Cb^\times)$. Both the left and right dual of a locally finite-dimensional $G$-graded vector space $V \in \vect_G^\omega$ is defined by the total space $V^*$ equipped with the $G$-grading
\[
(V^*)_g \coloneqq (V_{g^{-1}})^* , \quad g \in G .
\]
This dual space is well-defined because taking duals in $\vect$ preserves direct sums (see Exercise \ref{exercise:dual_preserve_direct_sum}). Equivalently, both the left and right dual of $\Cb_{(g)}$ is $\Cb_{(g^{-1})}$ for every $g \in G$. Therefore, $\vect_G^\omega$ is a rigid monoidal category.
\end{expl}

\begin{exercise}
Let $G$ be a finite group and $\omega \in Z^3(G;\Cb^\times)$. Find the creation and annihilation morphisms of $\Cb_{(g)}$ for every $g \in G$. The answer is unique up to a nonzero complex number.
\end{exercise}

In the above discussion we only consider the left dual $x^*$ of a topological defect $x$. What about the right dual? One may wish to rotate or braid the instantons in Figure \ref{fig:creation_annihilation} to get the creation and annihilation instantons for a right dual. However, these operations do not give the correct answer, and they can not be defined in 1d space.

Another solution which works in any dimension is to bend the world line in another direction. But then it is not clear whether the left dual and right dual of $x$ coincide. If not, the theory would be complicated: a topological defect could have different anti-particles, and each one could have different anti-anti-particles. Mathematically, we need a so-called pivotal structure or spherical structure to solve this problem. For unitary theories, the answer is simpler.

\begin{figure}[htbp]
\centering
\subfigure[annihilation $b_x^\dagger$]{
\begin{tikzpicture}[scale=0.8]
\fill[gray!20] (-1,0)--(5,0)--(6,2)--(0,2)--cycle ;
\draw (1.5,1)--(1.5,1.5) ;
\draw (3.5,1)--(3.5,1.5) ;

\draw[->-=0.2] (1.5,1.5) .. controls (1.5,4) and (3.5,4) .. (3.5,1.5) ;

\fill (1.5,1) circle (0.07) node[left] {$x$} ;
\fill (3.5,1) circle (0.07) node[right] {$x^*$} ;

\end{tikzpicture}
}
\hspace{5ex}
\subfigure[creation $d_x^\dagger$]{
\begin{tikzpicture}[scale=0.8]
\fill[gray!20] (-1,0)--(5,0)--(6,2)--(0,2)--cycle ;
\draw (1.5,3.5)--(1.5,4) ;
\path (1.5,1)--(1.5,4) node[very near end,left] {$x^*$} ;
\draw (3.5,3.5)--(3.5,4) ;
\path (3.5,1)--(3.5,4) node[very near end,right] {$x$} ;

\draw[->-=0.8] (1.5,3.5) .. controls (1.5,1) and (3.5,1) .. (3.5,3.5) ;

\end{tikzpicture}
}
\caption{the time-reversal of the creation and annihilation instantons}
\label{fig:creation_annihilation_dagger}
\end{figure}
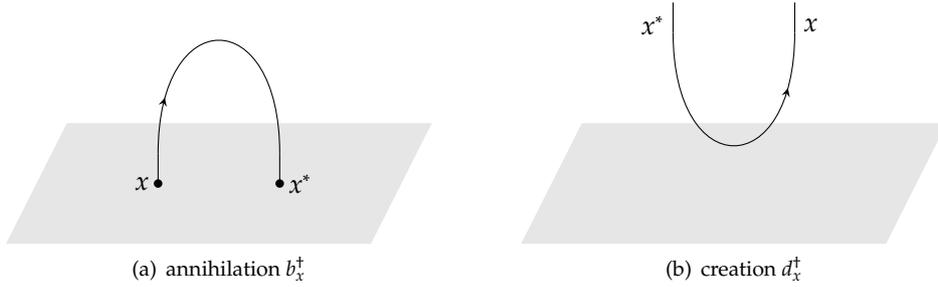

For every topological defect $x$, the unitary structure provides another two instantons $d_x^\dagger \colon \one \to x^* \otimes x$ and $b_x^\dagger \colon x \otimes x^* \to \one$, which are time-reversal of $d_x$ and $b_x$ (see Figure \ref{fig:creation_annihilation_dagger}). So they are nothing but the world line of $x$ bending in another direction. One may immediately realize that $(x^*,d_x^\dagger,b_x^\dagger)$ is a right dual of $x$, because it satisfies the time-reversal of the definition of a left dual. Mathematically, we have
\begin{gather*}
(b_x^\dagger \otimes \id_x) \circ (\id_x \otimes d_x^\dagger) = [(\id_x \otimes d_x) \circ (b_x \otimes \id_x)]^\dagger = \id_x^\dagger = \id_x , \\
(\id_{x^*} \otimes b_x^\dagger) \circ (d_x^\dagger \circ \id_{x^*}) = [(d_x \circ \id_{x^*}) \circ (\id_{x^*} \otimes b_x)]^\dagger = \id_{x^*}^\dagger = \id_{x^*} .
\end{gather*}
Therefore, $x^*$ is both the left and right dual of $x$. For simplicity, we say that $x^*$ is the dual of $x$.

\begin{figure}[htbp]
\[
\begin{array}{c}
\begin{tikzpicture}[scale=0.8]
\fill[gray!20] (-1,0)--(5,0)--(6,2)--(0,2)--cycle ;
\draw[->-] (3.5,4)--(3.5,2) ;
\draw[dashed] (3.5,1)--(3.5,4) node[very near end,right] {$x^*$} ;
\draw (3.5,2) .. controls (3.5,1) and (3,1) .. (2.5,2.5) ;
\draw (2.5,2.5) .. controls (2,4) and (1.5,4) .. (1.5,3) ;
\draw[->-] (1.5,3)--(1.5,1) ;
\draw[fill=white] (2.4,2.4) rectangle (2.6,2.6) node[midway,right] {$f$} ;
\fill (1.5,1) circle (0.07) node[left] {$y^*$} ;
\node at (1.8,4) {$d_y$} ;
\node at (3.2,1) {$b_x$} ;
%
\end{tikzpicture}
\end{array}
=
\begin{array}{c}
\begin{tikzpicture}[scale=0.8]
\fill[gray!20] (-1,0)--(5,0)--(6,2)--(0,2)--cycle ;
\draw[->-] (1.5,4)--(1.5,2) ;
\draw[dashed] (1.5,1)--(1.5,4) node[very near end,left] {$x^*$} ;
\draw (1.5,2) .. controls (1.5,1) and (2,1) .. (2.5,2.5) ;
\draw (2.5,2.5) .. controls (3,4) and (3.5,4) .. (3.5,3) ;
\draw[->-] (3.5,3)--(3.5,1) ;
\draw[fill=white] (2.4,2.4) rectangle (2.6,2.6) node[midway,left] {$f$} ;
\fill (3.5,1) circle (0.07) node[right] {$y^*$} ;
\node at (1.8,1) {$d_x^\dagger$} ;
\node at (3.2,4) {$b_y^\dagger$} ;
%
\end{tikzpicture}
\end{array}
\]
\caption{The left and right dual of a morphism $f \colon x \to y$ are equal by properly choosing the dual.}
\label{fig:left=right_dual}
\end{figure}
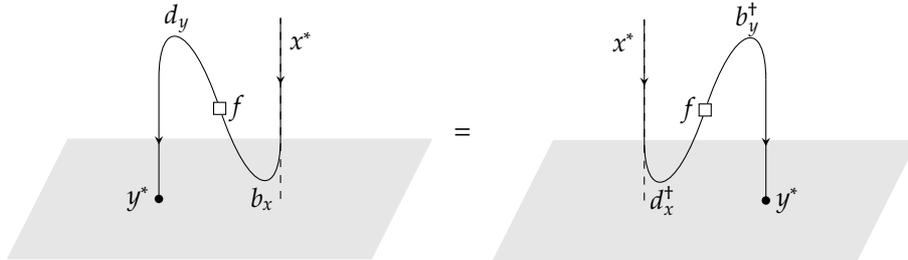

\begin{rem} \label{rem:unitary_rigid}
Moreover, by properly choosing the left dual $(x,b_x,d_x)$ for every $x \in \CC$ (see Section \ref{sec:fusion_quantum_dim}), the left and right dual of every morphism in $\CC$ are also the same, denoted by $f^*$ (see Figure \ref{fig:left=right_dual}):
\[
f^* \coloneqq (d_y \otimes \id_{x^*}) \circ (\id_{y^*} \otimes f \otimes \id_{x^*}) \circ (\id_{x^*} \otimes b_x) = (\id_{x^*} \otimes b_y^\dagger) \circ (\id_{x^*} \otimes f \otimes \id_{y^*}) \circ (d_x^\dagger \otimes \id_{x^*}) .
\]
Thus the left and right dual functors are equal, denoted by $\delta^* \colon \CC^\op \to \CC$. Moreover, in this case taking the dual commutes with taking the dagger, i.e., $(f^*)^\dagger = (f^\dagger)^*$ for every morphism $f$.
\end{rem}

\subsubsection{Fusion categories and dimensions} \label{sec:fusion_quantum_dim}

We summarize the structures developed in previous subsections to the following definition.

\begin{defn}
A \emph{multi-fusion category} is a $\Cb$-linear rigid monoidal category that is also finite semisimple. A \emph{fusion category} is a multi-fusion category such that the tensor unit $\one$ is a simple object. A \emph{unitary (multi-)fusion category} is a (multi-)fusion category equipped with a compatible unitary structure.
\end{defn}

Hence, the category $\CC$ of particle-like topological defects of a 2d topological order $\SC$ is a unitary multi-fusion category. If $\SC$ is stable, $\CC$ is a unitary fusion category.

\begin{exercise}
Prove that $\vect$ is a fusion category.
\end{exercise}

\begin{exercise}
Let $G$ be a finite group. Prove that $\rep(G)$ is a fusion category.
\end{exercise}

\begin{exercise}
Let $G$ be a finite group. Prove that $\vect_G^\omega$ is a fusion category for every $\omega \in Z^3(G;\Cb^\times)$.
\end{exercise}

\begin{rem} \label{rem:Grothendieck_ring}
Let $\CC$ be a multi-fusion category. Since $\CC$ is semisimple, the tensor product of two simple objects $x,y \in \CC$ is the direct sum of simple objects. Thus we have
\[
x \otimes y \simeq \bigoplus_{z \in \Irr(\CC)} N_{xy}^z \cdot z
\]
for some non-negative integers $N_{xy}^z \in \Nb$, where $n \cdot z$ denotes the direct sum of $n$ copies of $z$. These numbers $\{N_{xy}^z\}_{x,y,z \in \Irr(\CC)}$ are called the \emph{fusion rules} of $\CC$.

Equivalently, the isomorphism classes of simple objects of $\CC$ generate a ring, where the multiplication is given by the tensor product of $\CC$. Then its structure constants are the fusion rules of $\CC$. This ring is called the \emph{fusion ring} or \emph{Grothendieck ring} of $\CC$, denoted by $\Gr(\CC)$.

The Grothendieck ring does not determine the multi-fusion category. As an example, the fusion ring of $\vect_G^\omega$ for every $\omega \in Z^3(G;\Cb^\times)$ is isomorphic to the group ring $\Zb[G]$.
\end{rem}

\begin{figure}[htbp]
\[
\begin{array}{c}
\begin{tikzpicture}[scale=0.8]
\fill[gray!20] (-1,0)--(5,0)--(6,2)--(0,2)--cycle ;

\draw[->-=0.9] (3.5,2.5) .. controls (3.5,1) and (1.5,1) .. (1.5,2.5) ;
\draw[->-=0.9] (1.5,2.5) .. controls (1.5,4) and (3.5,4) .. (3.5,2.5) ;

\node[left] at (1.5,2.5) {$x$} ;
\node[right] at (3.5,2.5) {$x^*$} ;
\node at (2.5,1) {$b_x$} ;
\node at (2.5,4) {$b_x^\dagger$} ;
\end{tikzpicture}
\end{array}
=
\begin{array}{c}
\begin{tikzpicture}[scale=0.8]
\fill[gray!20] (-1,0)--(5,0)--(6,2)--(0,2)--cycle ;

\draw[->-=0.9] (1.5,2.5) .. controls (1.5,1) and (3.5,1) .. (3.5,2.5) ;
\draw[->-=0.9] (3.5,2.5) .. controls (3.5,4) and (1.5,4) .. (1.5,2.5) ;

\node[left] at (1.5,2.5) {$x^*$} ;
\node[right] at (3.5,2.5) {$x$} ;
\node at (2.5,1) {$d_x^\dagger$} ;
\node at (2.5,4) {$d_x$} ;
\end{tikzpicture}
\end{array}
\]
\caption{the quantum dimension of $x$ with a normalized choice of duals}
\label{fig:quantum_dimension}
\end{figure}
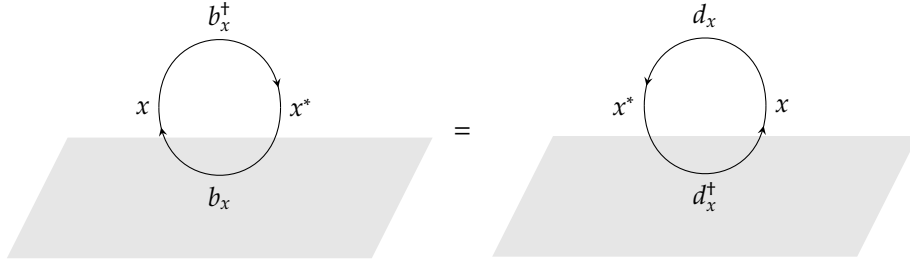

In the following we consider the case that $\SC$ is stable, i.e., $\CC$ is a unitary fusion category. With the help of four creation and annihilation instantons, we can close the world line of a particle-like topological defect $x \in \CC$ to a loop. As depicted in Figure \ref{fig:quantum_dimension}, for every simple object $x \in \CC$ there are two ways to close the world line of $x$. The results are morphisms $b_x^\dagger \circ b_x , d_x \circ d_x^\dagger \colon \one \to \one$. Since $\Hom_\CC(\one,\one) \simeq \Cb$, both $b_x^\dagger \circ b_x$ and $d_x \circ d_x^\dagger$ can be viewed as positive real numbers. We define the \emph{quantum dimension} of $x$ by
\[
\dim(x) \coloneqq \sqrt{(b_x^\dagger \circ b_x) \cdot (d_x \circ d_x^\dagger)} > 0 .
\]
The \emph{quantum dimension} of $\CC$ is defined to be
\[
\dim(\CC) \coloneqq \sum_{x \in \Irr(\CC)} \dim(x)^2 .
\]
Note that the morphisms $b_x$ and $d_x$ can be properly rescaled such that
\[
b_x^\dagger \circ b_x = d_x \circ d_x^\dagger = \dim(x) .
\]
We say that a choice of duals in $\CC$ are normalized if the above equation holds for every obejct in $\CC$. With a normalized choice of duals, the left and right dual functors are equal (see Remark \ref{rem:unitary_rigid}) \cite{Yam04}.

\begin{exercise}
Prove that the quantum dimension of a finite-dimensional Hilbert space $\mathcal H \in \hilb$ is equal to the usual dimension $\dim(\mathcal H)$.
\end{exercise}

\begin{rem}
Let $\CC$ be a unitary fusion category. For every $x,y \in \CC$ we have:
\[
\dim(x \oplus y) = \dim(x) + \dim(y) , \quad \dim(x \otimes y) = \dim(x) \cdot \dim(y) .
\]
In other words, $\dim$ defines a ring homomorphism $\Gr(\CC) \to \Rb$. In particular, we have
\[
\dim(x) \cdot \dim(y) = \sum_{z \in \Irr(\CC)} N_{xy}^z \cdot \dim(z)
\]
for every simple objects $x,y \in \CC$.
\end{rem}

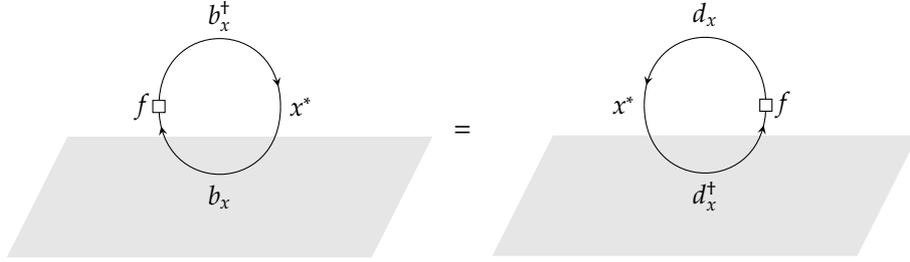
\begin{figure}[htbp]
\[
\begin{array}{c}
\begin{tikzpicture}[scale=0.8]
\fill[gray!20] (-1,0)--(5,0)--(6,2)--(0,2)--cycle ;

\draw[->-=0.9] (3.5,2.5) .. controls (3.5,1) and (1.5,1) .. (1.5,2.5) ;
\draw[->-=0.9] (1.5,2.5) .. controls (1.5,4) and (3.5,4) .. (3.5,2.5) ;
\draw[fill=white] (1.4,2.4) rectangle (1.6,2.6) node[midway,left] {$f$} ;

\node[right] at (3.5,2.5) {$x^*$} ;
\node at (2.5,1) {$b_x$} ;
\node at (2.5,4) {$b_x^\dagger$} ;
\end{tikzpicture}
\end{array}
=
\begin{array}{c}
\begin{tikzpicture}[scale=0.8]
\fill[gray!20] (-1,0)--(5,0)--(6,2)--(0,2)--cycle ;

\draw[->-=0.9] (1.5,2.5) .. controls (1.5,1) and (3.5,1) .. (3.5,2.5) ;
\draw[->-=0.9] (3.5,2.5) .. controls (3.5,4) and (1.5,4) .. (1.5,2.5) ;
\draw[fill=white] (3.4,2.4) rectangle (3.6,2.6) node[midway,right] {$f$} ;

\node[left] at (1.5,2.5) {$x^*$} ;
\node at (2.5,1) {$d_x^\dagger$} ;
\node at (2.5,4) {$d_x$} ;
\end{tikzpicture}
\end{array}
\]
\caption{the trace of an instanton $f \in \Hom_\CC(x,x)$ with a normalized choice of duals}
\label{fig:trace}
\end{figure}

More generally, given an instanton $f \in \Hom_\CC(x,x)$, there are two ways to close the world line of $x$ (see Figure \ref{fig:trace}). With a normalized choice of duals, these two instantons are equal and called the \emph{trace} of $f$:
\[
\tr(f) \coloneqq b_x^\dagger \circ (f \otimes \id_{x^*}) \circ b_x = d_x \circ ({\id_{x^*}} \otimes f) \circ d_x^\dagger \in \Hom_\CC(\one,\one) \simeq \Cb .
\]

\begin{rem} \label{rem:pivotal_spherical}
In a rigid monoidal category without unitary structure, given an object $x$, possibly there is no (canonical) isomorphism between $x^L$ and $x^R$, or equivalently, no isomorphism between $x$ and $x^{LL}$. A \emph{pivotal structure} on a rigid monoidal category $\CC$ is a natural isomorphism $a_x \colon x \to x^{LL}$ satisfying $a_x \otimes a_y = a_{x \otimes y}$ (this condition means that $a$ is a monoidal natural isomorphism; see Definition \ref{defn:monoidal_natural_transformation}).

In a pivotal fusion category, we can similarly close the world line of an object $x$. However, in this case there are two different ways to close the world line and they may not be equal, and these two results are called left and right quantum dimensions of $x$, respectively:
\[
\dim^L(x) \coloneqq d_{x^L} \circ (a_x \otimes \id_{x^L}) \circ b_x , \quad \dim^R(x) \coloneqq d_x \circ ({\id_{x^L}} \otimes a_x^{-1}) \circ b_{x^L} .
\]
A pivotal structure on a fusion category is called \emph{spherical} if the left and right quantum dimensions are equal \cite{BW99}. A unitary fusion category admits a unique pivotal structure defined by (with a normalized choice of duals)
\[
a_x \coloneqq (b_x^\dagger \otimes \id_{x^{LL}}) \circ ({\id_x} \otimes b_{x^L}) = ({\id_{x^{LL}}} \otimes d_x) \circ (d_{x^L}^\dagger \otimes \id_x) .
\]
This pivotal structure is automatically spherical and compatible with the unitary structure \cite{Yam04,Mueg03} (this fact is also mentioned in a footnote in \cite{Kit06}).
\end{rem}

\begin{rem}
The quantum dimension of a fusion category is independent of the choice of pivotal or spherical structures, but the quantum dimensions of objects depend on the choice of pivotal or spherical structures.
\end{rem}

\begin{expl} \label{expl:spherical_rep_G}
For every vector space $V$, there is a canonical $\Cb$-linear map $a_V \colon V \to V^{**}$ defined by $v \mapsto \ev_v$, where $\ev_v \colon V^* \to \Cb$ is defined by $\ev_v(\phi) \coloneqq \phi(v)$. Moreover, $a_V$ is an isomorphism if and only if $V$ is finite-dimensional. Hence each finite-dimensional vector space is canonically isomorphic to its double dual space.

The family $\{a_V\}_{V \in \vect}$ defines a spherical structure on $\vect$. With this spherical structure, the quantum dimension of $V \in \vect$ is equal to the usual dimension of $V$. Also, such canonical isomorphisms induce spherical structures on $\rep(G)$ and $\vect_G$ for every finite group $G$, and the quantum dimensions are equal to the usual dimensions.
\end{expl}

In a fusion category $\CC$, there is another version of dimension. Here is the physical intuition. Suppose $\CC$ is the fusion category of particle-like topological defects of a stable topological order. If we put $n$ copies of a topological defect $x \in \CC$ on a disk, the GSD grows exponentially on $n$:
\[
\mathrm{GSD} \sim \mathrm{constant} \cdot d^n .
\]
Mathematically, this number $d$ is equal to the largest non-negative eigenvalue of the matrix $N_x$ defined by $(N_x)_{zy} \coloneqq N_{xy}^z$, and the existence is provided by the Frobenius-Perron theorem \cite{Per07,Fro12}. Therefore, it is called the \emph{Frobenius-Perron dimension} of $x$ and denoted by $\fpdim(x)$ \cite{ENO05}. The Frobenius-Perron dimension of a fusion category $\CC$ is defined by
\[
\fpdim(\CC) \coloneqq \sum_{x \in \Irr(\CC)} \fpdim(x)^2 .
\]

\begin{rem} \label{rem:FPdim_unique}
Let $\CC$ be a fusion category. The Frobenius-Perron dimension also defines a ring homomorphism $\fpdim \colon \Gr(\CC) \to \Rb$. Moreover, it is the unique ring homomorphism $\Gr(\CC) \to \Cb$ that takes non-negative values (indeed, positive values) on simple objects. So the Frobenius-Perron dimensions can be directly obtained from the fusion rules. For example, the Frobenius-Perron dimensions in $\rep(G)$ or $\vect_G^\omega$ for a finite group $G$ and $\omega \in Z^3(G;\Cb^\times)$ have to be equal to the usual dimensions.
\end{rem}

\begin{rem} \label{rem:pseudo_unitary}
A fusion category $\CC$ is called \emph{pseudo-unitary} if $\dim(\CC) = \fpdim(\CC)$. A unitary fusion category is pseudo-unitary because the quantum dimensions of simple objects are positive and thus coincide with the Frobenius-Perron dimensions (see Remark \ref{rem:FPdim_unique}). Any pseudo-unitary fusion category admits a unique spherical structure such that the quantum dimensions of objects are equal to the Frobenius-Perron dimensions \cite{ENO05}. It is believed that for any pseudo-unitary fusion category there is an equivalent unitary fusion category. So in the following, a \emph{unitary fusion category} can be understood as a pseudo-unitary fusion category equipped with the unique spherical structure such that the quantum dimensions of objects are equal to the Frobenius-Perron dimensions.
\end{rem}

\subsubsection{Braiding structure}

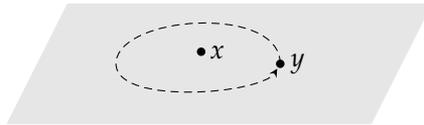
\begin{figure}[htbp]
\centering
\begin{tikzpicture}[scale=0.8]
\fill[gray!20] (-1,0)--(5,0)--(6,2)--(0,2)--cycle ;

\fill (3.5,1) circle (0.07) node[right] {$y$} ;
\fill (2.2,1.2) circle (0.07) node[right] {$x$} ;

\draw[-stealth,densely dashed] (3.5,1) .. controls (3.5,2) and (0.8,1.8) .. (0.8,1) .. controls (0.8,0.3) and (3.2,0.5) .. (3.45,0.92) ;

\end{tikzpicture}
\caption{a double braiding or a full braiding}
\label{fig:double_braiding}
\end{figure}

In a 2d topological order, a topological defect $x$ can be adiabatically moved around another $y$ (see Figure \ref{fig:double_braiding}). This process should not change the fusion of $x$ and $y$, and thus induces an isomorphism from $x \otimes y$ to itself.

\begin{figure}[htbp]
\centering
\subfigure[braiding]{
\begin{tikzpicture}[scale=0.8]
\fill[gray!20] (-1,0)--(5,0)--(6,2)--(0,2)--cycle ;
\draw (1.5,1)--(1.5,1.5) ;
\draw (1.5,3.5)--(1.5,4) ;
\path (1.5,1)--(1.5,4) node[very near end,left] {$y$} ;
\draw (3.5,1)--(3.5,1.5) ;
\draw (3.5,3.5)--(3.5,4) ;
\path (3.5,1)--(3.5,4) node[very near end,right] {$x$} ;
\draw[->-=0.8] (3.5,1.5) .. controls (3.5,2.2) and (1.5,2.8) .. (1.5,3.5) ;
\begin{scope}
\clip (2.5,2.5) circle (0.3) ;
\draw[white,double=black,double distance=0.4pt,line width=3pt] (1.5,1.5) .. controls (1.5,2.2) and (3.5,2.8) .. (3.5,3.5) ;
\end{scope}
\draw[->-=0.8] (1.5,1.5) .. controls (1.5,2.2) and (3.5,2.8) .. (3.5,3.5) ;

\fill (1.5,1) circle (0.07) node[left] {$x$} ;
\fill (3.5,1) circle (0.07) node[right] {$y$} ;

\node at (4.5,1.5) {$R$} ;
\end{tikzpicture}
}
\hspace{5ex}
\subfigure[anti-braiding]{
\begin{tikzpicture}[scale=0.8]
\fill[gray!20] (-1,0)--(5,0)--(6,2)--(0,2)--cycle ;
\draw (1.5,1)--(1.5,1.5) ;
\draw (1.5,3.5)--(1.5,4) ;
\path (1.5,1)--(1.5,4) node[very near end,left] {$y$} ;
\draw (3.5,1)--(3.5,1.5) ;
\draw (3.5,3.5)--(3.5,4) ;
\path (3.5,1)--(3.5,4) node[very near end,right] {$x$} ;
\draw[->-=0.8] (1.5,1.5) .. controls (1.5,2.2) and (3.5,2.8) .. (3.5,3.5) ;
\begin{scope}
\clip (2.5,2.5) circle (0.3) ;
\draw[white,double=black,double distance=0.4pt,line width=3pt] (3.5,1.5) .. controls (3.5,2.2) and (1.5,2.8) .. (1.5,3.5) ;
\end{scope}
\draw[->-=0.8] (3.5,1.5) .. controls (3.5,2.2) and (1.5,2.8) .. (1.5,3.5) ;

\fill (1.5,1) circle (0.07) node[left] {$x$} ;
\fill (3.5,1) circle (0.07) node[right] {$y$} ;

\end{tikzpicture}
}
\caption{the braiding and anti-braiding of $x$ and $y$}
\label{fig:braiding}
\end{figure}
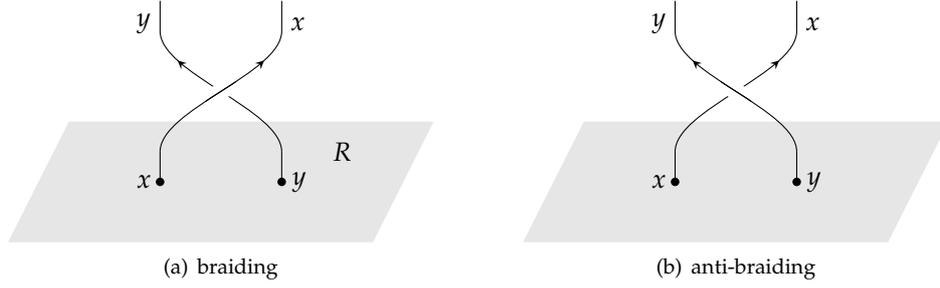

In particular, as depicted in Figure \ref{fig:braiding} (a), $x$ and $y$ can be adiabatically exchanged in the counterclockwise direction. This process gives an isomorphism (instanton) $c_{x,y} \colon x \otimes y \to y \otimes x$, called the braiding of $x$ and $y$. Also, $x$ and $y$ can be adiabatically exchanged in the clockwise direction. This process gives another isomorphism, called the anti-braiding of $x$ and $y$. Clearly the anti-braiding is the inverse of the braiding.

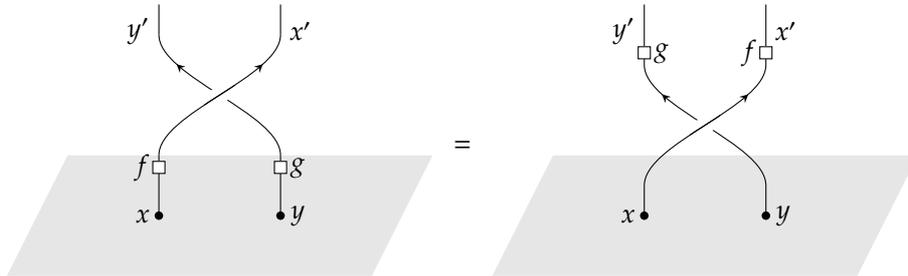
\begin{figure}[htbp]
\centering
\[
\begin{array}{c}
\begin{tikzpicture}[scale=0.8]
\fill[gray!20] (-1,0)--(5,0)--(6,2)--(0,2)--cycle ;
\draw (1.5,1)--(1.5,2) ;
\draw (1.5,4)--(1.5,4.5) ;
\path (1.5,1)--(1.5,4.5) node[very near end,left] {$y'$} ;
\draw (3.5,1)--(3.5,2) ;
\draw (3.5,4)--(3.5,4.5) ;
\path (3.5,1)--(3.5,4.5) node[very near end,right] {$x'$} ;
\draw[->-=0.8] (3.5,2) .. controls (3.5,2.7) and (1.5,3.3) .. (1.5,4) ;
\begin{scope}
\clip (2.5,3) circle (0.3) ;
\draw[white,double=black,double distance=0.4pt,line width=3pt] (1.5,2) .. controls (1.5,2.7) and (3.5,3.3) .. (3.5,4) ;
\end{scope}
\draw[->-=0.8] (1.5,2) .. controls (1.5,2.7) and (3.5,3.3) .. (3.5,4) ;
\draw[fill=white] (1.4,1.7) rectangle (1.6,1.9) node[midway,left] {$f$} ;
\draw[fill=white] (3.4,1.7) rectangle (3.6,1.9) node[midway,right] {$g$} ;

\fill (1.5,1) circle (0.07) node[left] {$x$} ;
\fill (3.5,1) circle (0.07) node[right] {$y$} ;

\end{tikzpicture}
\end{array}
=
\begin{array}{c}
\begin{tikzpicture}[scale=0.8]
\fill[gray!20] (-1,0)--(5,0)--(6,2)--(0,2)--cycle ;
\draw (1.5,1)--(1.5,1.5) ;
\draw (1.5,3.5)--(1.5,4.5) ;
\path (1.5,1)--(1.5,4.5) node[very near end,left] {$y'$} ;
\draw (3.5,1)--(3.5,1.5) ;
\draw (3.5,3.5)--(3.5,4.5) ;
\path (3.5,1)--(3.5,4.5) node[very near end,right] {$x'$} ;
\draw[->-=0.8] (3.5,1.5) .. controls (3.5,2.2) and (1.5,2.8) .. (1.5,3.5) ;
\begin{scope}
\clip (2.5,2.5) circle (0.3) ;
\draw[white,double=black,double distance=0.4pt,line width=3pt] (1.5,1.5) .. controls (1.5,2.2) and (3.5,2.8) .. (3.5,3.5) ;
\end{scope}
\draw[->-=0.8] (1.5,1.5) .. controls (1.5,2.2) and (3.5,2.8) .. (3.5,3.5) ;
\draw[fill=white] (1.4,3.6) rectangle (1.6,3.8) node[midway,right] {$g$} ;
\draw[fill=white] (3.4,3.6) rectangle (3.6,3.8) node[midway,left] {$f$} ;

\fill (1.5,1) circle (0.07) node[left] {$x$} ;
\fill (3.5,1) circle (0.07) node[right] {$y$} ;

\end{tikzpicture}
\end{array}
\]
\caption{the naturality of the braiding}
\label{fig:naturality_braiding}
\end{figure}

The family $\{c_{x,y}\}_{x,y \in \CC}$ is natural in two variables because instantons can be freely moved along the world lines. More precisely, for any morphisms $f \colon x \to x'$ and $g \colon y \to y'$, the physical intuition implies the following equation (see Figure \ref{fig:naturality_braiding}):
\[
c_{x',y'} \circ (f \otimes g) = (g \otimes f) \circ c_{x,y} .
\]
Therefore, the braiding is a natural isomorphism $c \colon \otimes \Rightarrow \otimes \circ \tau$, where $\tau \colon \CC \times \CC \to \CC \times \CC$ is the permutation of two components defined by $\tau(x,y) \coloneqq (y,x)$. Intuitively, the existence of the braiding means that the tensor product is `commutative', although there are different commutativities given by the braiding or anti-braiding.

%

\begin{figure}[htbp]
\[
\begin{array}{c}
\begin{tikzpicture}[scale=0.8]
\draw[step=1,help lines] (-2.5,-2.5) grid (3.5,3.5);
\draw[help lines,fill=m_ext] (0,0) rectangle (1,1) node[midway,black] {$m$} ;
\draw[-stealth,e_str] (-1,2)--(-1,-1)--(2,-1)--(2,2)--(-0.93,2) ;
\fill[e_ext] (-1,2) circle (0.07) node[above left,black] {$e$} ;

\foreach \x in {-0.5,0.5,1.5} {
	\node[above,link_label] at (\x,2) {$\sigma_z$} ;
	\node[right,link_label] at (2,\x) {$\sigma_z$} ;
	\node[below,link_label] at (\x,-1) {$\sigma_z$} ;
	\node[left,link_label] at (-1,\x) {$\sigma_z$} ;
}
\end{tikzpicture}
\end{array}
=
\begin{array}{c}
\begin{tikzpicture}[scale=0.8]
\draw[step=1,help lines] (-2.5,-2.5) grid (3.5,3.5);
\draw[help lines,fill=m_ext] (0,0) rectangle (1,1) ;
\draw[-stealth,e_str] (-1,2)--(-1,-1)--(2,-1)--(2,2)--(-0.93,2) ;
\fill[e_ext] (-1,2) circle (0.07) ;
\foreach \x in {-0.5,0.5,1.5}
	\foreach \y in {-0.5,0.5,1.5}
		\node[link_label] at (\x,\y) {$B_p$} ;
\end{tikzpicture}
\end{array}
\]
\caption{the double braiding of $e$ and $m$ in the toric code model}
\label{fig:braiding_toric_code}
\end{figure}
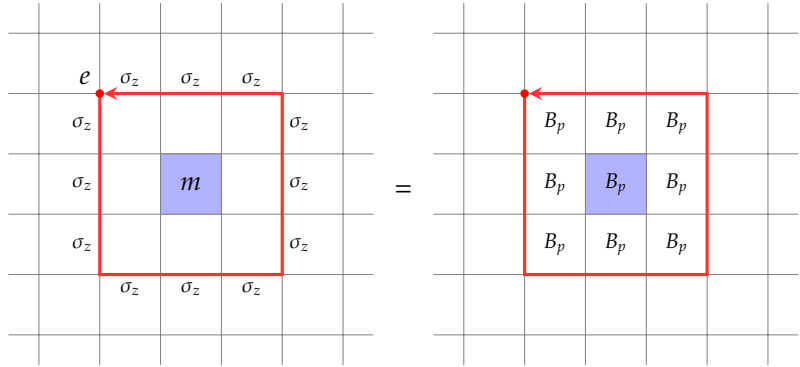

\begin{expl} \label{expl:double_braiding_toric_code}
Let us compute the double braiding of $e$ and $m$ in the toric code model. As depicted in Figure \ref{fig:braiding_toric_code}, an $e$ particle can be moved around an $m$ particle by an $e$ string operator $\prod_j \sigma_z^j$. This string operator is equal to the product of all $B_p$ operators on the plaquettes which are encircled by the path of $e$, because all $\sigma_z$ on inner links appear twice in the product. Since there is only one $m$ particle, the action of this string operator is equal to $-1$. Consequently, the double braiding $c_{m,e} \circ c_{e,m} \colon e \otimes m \to e \otimes m$ is equal to $-\id_{e \otimes m}$.
\end{expl}

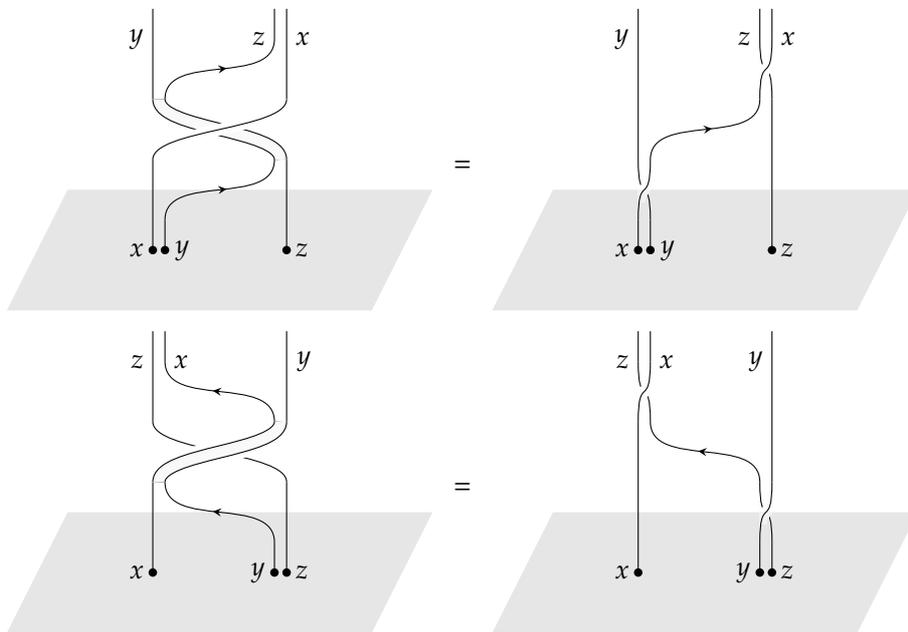
\begin{figure}[htbp]
\begin{gather*}
\begin{array}{c}
\begin{tikzpicture}[scale=0.8]
\fill[gray!20] (-1,0)--(5,0)--(6,2)--(0,2)--cycle ;
\draw (1.4,1)--(1.4,2.5) ;
\draw (3.6,3.5)--(3.6,5) ;
\path (3.6,1)--(3.6,5) node[very near end,right] {$x$} ;
\draw (1.6,1)--(1.6,1.5) ;
\draw[->-] (1.6,1.5) .. controls (1.6,2.3) and (3.4,1.7) .. (3.4,2.5) ;
\draw (1.4,3.5)--(1.4,5) ;
\path (1.4,1)--(1.4,5) node[very near end,left] {$y$} ;
\draw (3.6,1)--(3.6,2.5) ;
\draw[->-] (1.6,3.5) .. controls (1.6,4.3) and (3.4,3.7) .. (3.4,4.5) ;
\draw (3.4,4.5)--(3.4,5) ;
\path (3.4,1)--(3.4,5) node[very near end,left] {$z$} ;
\draw[double=gray!5,double distance between line centers=0.16cm] (3.5,2.5) .. controls (3.5,3) and (1.5,3) .. (1.5,3.5) ;
\draw[white,double=black,double distance=0.4pt,line width=2pt] (1.4,2.5) .. controls (1.4,3) and (3.6,3) .. (3.6,3.5) ;
\fill (1.4,1) circle (0.07) node[left] {$x$} ;
\fill (1.6,1) circle (0.07) node[right] {$y$} ;
\fill (3.6,1) circle (0.07) node[right] {$z$} ;
\end{tikzpicture}
\end{array}
=
\begin{array}{c}
\begin{tikzpicture}[scale=0.8]
\fill[gray!20] (-1,0)--(5,0)--(6,2)--(0,2)--cycle ;
\draw (1.4,1)--(1.4,1.5) ;
\draw (1.4,1.5) .. controls (1.4,2.3) and (1.6,1.7) .. (1.6,2.5) ; 
\draw[->-] (1.6,2.5) .. controls (1.6,3.3) and (3.4,2.7) .. (3.4,3.5) ;
\draw (3.4,3.5) .. controls (3.4,4.3) and (3.6,3.7) .. (3.6,4.5) ; 
\draw (3.6,4.5)--(3.6,5) ;
\path (3.6,1)--(3.6,5) node[very near end,right] {$x$} ;
\draw (1.6,1)--(1.6,1.5) ;
\draw (1.6,1.5) .. controls (1.6,2.3) and (1.4,1.7) .. (1.4,2.5) ;
\draw (1.4,2.5)--(1.4,5) ;
\path (1.4,1)--(1.4,5) node[very near end,left] {$y$} ;
\draw (3.6,1)--(3.6,3.5) ;
\draw (3.6,3.5) .. controls (3.6,4.3) and (3.4,3.7) .. (3.4,4.5) ;
\draw (3.4,4.5)--(3.4,5) ;
\path (3.4,1)--(3.4,5) node[very near end,left] {$z$} ;
\begin{scope}
\clip (1.4,2) rectangle (1.6,2.2) ;
\draw[white,double=black,double distance=0.4pt,line width=2pt] (1.4,1.5) .. controls (1.4,2.3) and (1.6,1.7) .. (1.6,2.5) ;
\end{scope}
\begin{scope}
\clip (1.4,2) rectangle (1.6,1.8) ;
\draw[gray!10,double=black,double distance=0.4pt,line width=2pt] (1.4,1.5) .. controls (1.4,2.3) and (1.6,1.7) .. (1.6,2.5) ;
\end{scope}
\draw[white,double=black,double distance=0.4pt,line width=2pt] (3.4,3.5) .. controls (3.4,4.3) and (3.6,3.7) .. (3.6,4.5) ;
\fill (1.4,1) circle (0.07) node[left] {$x$} ;
\fill (1.6,1) circle (0.07) node[right] {$y$} ;
\fill (3.6,1) circle (0.07) node[right] {$z$} ;
\end{tikzpicture}
\end{array} \\
\begin{array}{c}
\begin{tikzpicture}[scale=0.8]
\fill[gray!20] (-1,0)--(5,0)--(6,2)--(0,2)--cycle ;
\draw (1.4,1)--(1.4,2.5) ;
\draw[->-] (3.4,3.5) .. controls (3.4,4.3) and (1.6,3.7) .. (1.6,4.5) ;
\draw (1.6,4.5)--(1.6,5) ;
\path (1.6,1)--(1.6,5) node[very near end,right] {$x$} ;
\draw (3.4,1)--(3.4,1.5) ;
\draw[->-] (3.4,1.5) .. controls (3.4,2.3) and (1.6,1.7) .. (1.6,2.5) ;
\draw (3.6,3.5)--(3.6,5) ;
\path (3.6,1)--(3.6,5) node[very near end,right] {$y$} ;
\draw (3.6,1)--(3.6,2.5) ;
\draw (3.6,2.5) .. controls (3.6,3) and (1.4,3) .. (1.4,3.5) ;
\draw (1.4,3.5)--(1.4,5) ;
\path (1.4,1)--(1.4,5) node[very near end,left] {$z$} ;
\draw[white,line width=0.16cm+4.4pt] (1.5,2.5) .. controls (1.5,3) and (3.5,3) .. (3.5,3.5) ;
\draw[double=gray!5,double distance between line centers=0.16cm] (1.5,2.5) .. controls (1.5,3) and (3.5,3) .. (3.5,3.5) ;
\fill (1.4,1) circle (0.07) node[left] {$x$} ;
\fill (3.4,1) circle (0.07) node[left] {$y$} ;
\fill (3.6,1) circle (0.07) node[right] {$z$} ;
\end{tikzpicture}
\end{array}
=
\begin{array}{c}
\begin{tikzpicture}[scale=0.8]
\fill[gray!20] (-1,0)--(5,0)--(6,2)--(0,2)--cycle ;
\draw (1.4,1)--(1.4,3.5) ;
\draw (1.4,3.5) .. controls (1.4,4.3) and (1.6,3.7) .. (1.6,4.5) ; 
\draw (1.6,4.5)--(1.6,5) ;
\path (1.6,1)--(1.6,5) node[very near end,right] {$x$} ;
\draw (3.4,1)--(3.4,1.5) ;
\draw (3.4,1.5) .. controls (3.4,2.3) and (3.6,1.7) .. (3.6,2.5) ; 
\draw (3.6,2.5)--(3.6,5) ;
\path (3.6,1)--(3.6,5) node[very near end,left] {$y$} ;
\draw (3.6,1)--(3.6,1.5) ;
\draw (3.6,1.5) .. controls (3.6,2.3) and (3.4,1.7) .. (3.4,2.5) ;
\draw[->-] (3.4,2.5) .. controls (3.4,3.3) and (1.6,2.7) .. (1.6,3.5) ;
\draw (1.6,3.5) .. controls (1.6,4.3) and (1.4,3.7) .. (1.4,4.5) ;
\draw (1.4,4.5)--(1.4,5) ;
\path (1.4,1)--(1.4,5) node[very near end,left] {$z$} ;
\begin{scope}
\clip (3.4,2) rectangle (3.6,2.2) ;
\draw[white,double=black,double distance=0.4pt,line width=2pt] (3.4,1.5) .. controls (3.4,2.3) and (3.6,1.7) .. (3.6,2.5) ;
\end{scope}
\begin{scope}
\clip (3.4,2) rectangle (3.6,1.8) ;
\draw[gray!10,double=black,double distance=0.4pt,line width=2pt] (3.4,1.5) .. controls (3.4,2.3) and (3.6,1.7) .. (3.6,2.5) ;
\end{scope}
\draw[white,double=black,double distance=0.4pt,line width=2pt] (1.4,3.5) .. controls (1.4,4.3) and (1.6,3.7) .. (1.6,4.5) ;
\fill (1.4,1) circle (0.07) node[left] {$x$} ;
\fill (3.4,1) circle (0.07) node[left] {$y$} ;
\fill (3.6,1) circle (0.07) node[right] {$z$} ;
\end{tikzpicture}
\end{array}
\end{gather*}
\caption{the hexagon equations}
\label{fig:hexagon}
\end{figure}

Now suppose there are three topological defects $x,y,z \in \CC$. There are two ways to braid $x$ with $y$ and $z$: we can either braid $x$ with $y$ and $z$ one by one, or braid $x$ with the fusion $y \otimes z$. These two instantons should be equal because their world lines are homotopy (see Figure \ref{fig:hexagon}). Thus we get the following two equations, called the hexagon equations:
\begin{gather*}
\alpha_{y,z,x} \circ c_{x,y \otimes z} \circ \alpha_{x,y,z} = ({\id_y} \otimes c_{x,z}) \circ \alpha_{y,x,z} \circ (c_{x,y} \otimes \id_z) , \\
\alpha_{z,x,y}^{-1} \circ c_{x \otimes y,z} \circ \alpha_{x,y,z}^{-1} = (c_{x,z} \otimes \id_y) \circ \alpha_{x,z,y}^{-1} \circ ({\id_x} \otimes c_{y,z}) .
\end{gather*}

We summarize the properties of the braiding to the following formal structure.

\begin{defn}
A \emph{braided monoidal category} consists of the following data:
\bit
\item a monoidal category $\CC = (\CC,\otimes,\one,\alpha,\lambda,\rho)$;
\item a natural isomorphism $c_{x,y} \colon x \otimes y \to y \otimes x$;
\eit
and these data satisfy the following conditions:
\bnu
\item (\textbf{hexagon equation 1}) For any $x,y,z \in \CC$, the following diagram commutes:
\[
\begin{array}{c}
\xymatrix{
 & x \otimes (y \otimes z) \ar[r]^{c_{x,y \otimes z}} & (y \otimes z) \otimes x \ar[dr]^{\alpha_{y,z,x}} \\
(x \otimes y) \otimes z \ar[ur]^{\alpha_{x,y,z}} \ar[dr]_{c_{x,y} \otimes \id_z} & & & y \otimes (z \otimes x) \\
 & (y \otimes x) \otimes z \ar[r]^{\alpha_{y,x,z}} & y \otimes (x \otimes z) \ar[ur]_{\id_y \otimes c_{x,z}}
}
\end{array}
\]
\item (\textbf{hexagon equation 2}) For any $x,y,z \in \CC$, the following diagram commutes:
\[
\begin{array}{c}
\xymatrix{
 & (x \otimes y) \otimes z \ar[r]^{c_{x \otimes y,z}} & z \otimes (x \otimes y) \ar[dr]^{\alpha_{z,x,y}^{-1}} \\
x \otimes (y \otimes z) \ar[ur]^{\alpha_{x,y,z}^{-1}} \ar[dr]_{\id_x \otimes c_{y,z}} & & & (z \otimes x) \otimes y \\
 & x \otimes (z \otimes y) \ar[r]^{\alpha_{x,z,y}^{-1}} & (x \otimes z) \otimes y \ar[ur]_{c_{x,z} \otimes \id_y}
}
\end{array}
\]
\enu
Moreover, we say $\CC$ is a \emph{symmetric monoidal category} if $c_{y,x} \circ c_{x,y} = \id_{x \otimes y}$ for all $x,y \in \CC$.

If $\CC$ is a $\Cb$-linear braided monoidal category equipped with a dagger structure, we say that the dagger structure is \emph{compatible} with the braiding if $c_{x,y}$ is unitary for every $x,y \in \CC$. A \emph{unitary braided monoidal category} is a $\Cb$-linear braided monoidal category equipped with a compatible unitary structure.
\end{defn}

\begin{rem}
The condition that a dagger structure is compatible with the braiding structure is equivalent to that the functor $\dagger \colon \CC^\op \to \CC$ is a braided monoidal functor. The notion of a braided monoidal functor is defined in Definition \ref{defn:braided_monoidal_functor}.
\end{rem}

As a conclusion, the category $\CC$ of particle-like topological defects of a stable 2d topological order $\SC$ is a unitary braided fusion category.

\begin{rem} \label{rem:E2_algebra}
As we have mentioned in Remark \ref{rem:E1_algebra}, the above discussion of braidings is a radical simplification of the real story. Using adiabatic moves and the property \eqref{cond:star} in Remark \ref{rem:E1_algebra}, we can reduce all different but isomorphic fusion functors $\otimes_{(\xi,\eta)}$ to a single one $\otimes \coloneqq \otimes_{(0,1)}$. The braiding structure or the double braidings naturally appear because moving one anyon $x$ around another $y$ defines a homotopy non-trivial path in the configuration space. The complete and rigorous proof of the natural emergence of a braided monoidal structure on $\CC$ is equivalent to the proof of the well-known mathematical theorem that an $E_2$-algebra in the (2,1)-category of categories is a braided monoidal category \cite{SW03,Lur17,Fre17}. A physical work on adiabatic moves and braided monoidal structures can be found in \cite{KL20}. 
\end{rem}

\begin{rem} \label{rem:braiding_gauge}
In other words, the braiding structure on the category $\CC$ of particle-like topological defects of a 2d topological order $\SC$ depends on many artificial choices (see Remark \ref{rem:monoidal_gauge}).  Hence a topological order does not determine a single braided fusion category $\CC$, but an equivalence class of braided fusion categories. Physicists usually say that the braiding is not `gauge invariant'. Only the double braiding $c_{y,x} \circ c_{x,y}$ is gauge invariant and this is because that the initial and final states are exactly the same $x \otimes y$.

In the toric code model, we have known that the double braiding of $e$ and $m$ is $-1$ (see Example \ref{expl:double_braiding_toric_code}), but there are (at least) two different braiding structures on $\toric$: one satisfies $c_{e,m} = +1 , c_{m,e} = -1$, and the other one satisfies $c_{e,m} = -1 , c_{m,e} = +1$.

To determine the braiding structure explicitly, we need to `fix the gauge'. One way to fix the gauge is to choose a boundary and use the half-braiding to compute the braiding (see Remark \ref{rem:braiding_gauge_toric_code} and Definition \ref{defn:Drinfeld_center}).
\end{rem}

\begin{rem} \label{rem:braiding_statistic}
Physically, the braiding is related to the statistics of particle-like topological defects. It is well-known that there are two types of particles: bosons and fermions. They are defined by the phase factor $\pm 1$ of exchanging two particles. Such a phase factor must be $\pm 1$ because exchanging twice returns to the original state. However, this is the story in the space with dimension higher than two. In 2d, the double braiding $c_{y,x} \circ c_{x,y}$ of $x$ and $y$ is not necessarily equal to the identity because the loop of $x$ around $y$ can not contract to a trivial loop (see Figure \ref{fig:double_braiding}). So in 2d the phase factor of exchanging two particles can be \emph{any} complex number (or even a matrix), which gives a representation of the braid group \cite{Wu84}. This is why a particle-like topological defect in a 2d topological order is called an \emph{anyon} \cite{Wil82}.
\end{rem}

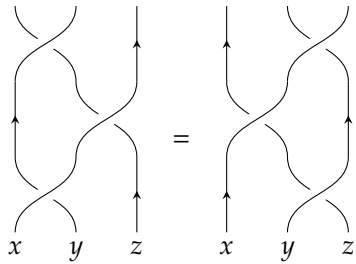
\begin{figure}[htbp]
\[
\begin{array}{c}
\begin{tikzpicture}[scale=1.0]
\draw[->-] (0.8,0)--(0.8,1) ;
\draw (0,0) .. controls (0,0.5) and (-0.8,0.5) .. (-0.8,1) ;
\draw[white,double=black,double distance=0.4pt,line width=3pt] (-0.8,0) .. controls (-0.8,0.5) and (0,0.5) .. (0,1) ;
\draw (0.8,1) .. controls (0.8,1.5) and (0,1.5) .. (0,2) ;
\draw[white,double=black,double distance=0.4pt,line width=3pt] (0,1) .. controls (0,1.5) and (0.8,1.5) .. (0.8,2) ;
\draw[->-] (-0.8,1)--(-0.8,2) ;
\draw[->-] (0.8,2)--(0.8,3) ;
\draw (0,2) .. controls (0,2.5) and (-0.8,2.5) .. (-0.8,3) ;
\draw[white,double=black,double distance=0.4pt,line width=3pt] (-0.8,2) .. controls (-0.8,2.5) and (0,2.5) .. (0,3) ;
\node[below] at (-0.8,0) {$x$} ;
\node[below] at (0,0) {$y$} ;
\node[below] at (0.8,0) {$z$} ;
\end{tikzpicture}
\end{array}
=
\begin{array}{c}
\begin{tikzpicture}[scale=1.0]
\draw (0.8,0) .. controls (0.8,0.5) and (0,0.5) .. (0,1) ;
\draw[white,double=black,double distance=0.4pt,line width=3pt] (0,0) .. controls (0,0.5) and (0.8,0.5) .. (0.8,1) ;
\draw[->-] (-0.8,0)--(-0.8,1) ;
\draw[->-] (0.8,1)--(0.8,2) ;
\draw (0,1) .. controls (0,1.5) and (-0.8,1.5) .. (-0.8,2) ;
\draw[white,double=black,double distance=0.4pt,line width=3pt] (-0.8,1) .. controls (-0.8,1.5) and (0,1.5) .. (0,2) ;
\draw[->-] (-0.8,2)--(-0.8,3) ;
\draw (0.8,2) .. controls (0.8,2.5) and (0,2.5) .. (0,3) ;
\draw[white,double=black,double distance=0.4pt,line width=3pt] (0,2) .. controls (0,2.5) and (0.8,2.5) .. (0.8,3) ;
\node[below] at (-0.8,0) {$x$} ;
\node[below] at (0,0) {$y$} ;
\node[below] at (0.8,0) {$z$} ;
\end{tikzpicture}
\end{array}
\]
\caption{The Yang-Baxter equation}
\label{fig:Yang_Baxter}
\end{figure}

\begin{rem}
Suppose $\CC$ is a braided monoidal category. For any $x,y,z \in \CC$, we have the following Yang-Baxter equation:
\[
(c_{y,z} \otimes \id_x) \circ ({\id_y} \otimes c_{x,z}) \circ (c_{x,y} \otimes \id_z) = ({\id_z} \otimes c_{x,y}) \circ (c_{x,z} \otimes \id_y) \circ ({\id_x} \otimes c_{y,z}) .
\]
Here we ignore associators. This equation is a direct corollary of the hexagon equations and the naturality of the braiding (for example, see \cite[Proposition 8.1.10]{EGNO15}). The physical intuition of the Yang-Baxter equation is depicted in Figure \ref{fig:Yang_Baxter}.
\end{rem}

\begin{expl}
There is a natural braiding on $\vect$ defined by
\begin{align*}
c_{V,W} \colon V \otimes W & \to W \otimes V \\
v \otimes w & \mapsto w \otimes v .
\end{align*}
Moreover, this braiding is symmetric.
\end{expl}

\begin{expl} \label{expl:symmetric_classification}
Let $G$ be a finite group $G$. There is a natural braiding on $\rep(G)$ defined by
\begin{align}
c_{V,W} \colon V \otimes W & \to W \otimes V \nonumber \\
v \otimes w & \mapsto w \otimes v . \label{eq:braiding_Rep_G}
\end{align}
Then $\rep(G)$ equipped with this braiding structure is a symmetric fusion category.

In general, there are other braiding structures on $\rep(G)$. Suppose $z \in G$ is a central element (i.e., $z$ commutes with all elements in $G$) and satisfies $z^2 = e$. For any $(V,\rho) \in \rep(G)$ we define
\[
V_0 \coloneqq \{v \in V \mid \rho(z)(v) = v\} , \quad V_1 \coloneqq \{v \in V \mid \rho(z)(v) = -v\} .
\]
We denote $\vert v \vert = i$ if $v \in V_i$ for $i = 0,1$. Then both $V_0$ and $V_1$ are $G$-representations and $V = V_0 \oplus V_1$. In other words, any $G$-representation can be decomposed with respect to the eigenvalues of $z$, and thus admits a canonical $\Zb_2$-grading. Then the following morphisms define a braiding structure on $\rep(G)$:
\begin{align}
c_{V,W} \colon V \otimes W & \to W \otimes V \nonumber \\
v \otimes w & \mapsto (-1)^{\vert v \vert \vert w \vert} \cdot w \otimes v , \label{eq:braiding_Rep_G_z}
\end{align}
The fusion category $\rep(G)$ equipped with this braiding structure is denoted by $\rep(G,z)$. It is clear that $\rep(G,z)$ is a symmetric fusion category. When $z = e$, the braiding \eqref{eq:braiding_Rep_G_z} is the obvious one \eqref{eq:braiding_Rep_G}, and we denote $\rep(G,e)$ by $\rep(G)$ for simplicity. Also, $\rep(\Zb_2,z)$ where $z = 1 \in \Zb_2$ is the nontrivial element is usually denoted by $\svect$, because a $\Zb_2$-graded vector space is also called a \emph{super vector space}. Physically, the symmetric fusion category $\svect$ describes the fermion parity symmetry. 

There is a mathematical theorem \cite{Del07,Del02} stated that every symmetric fusion category is equivalent to $\rep(G,z)$ for some finite group $G$ and central element $z \in G$ satisfying $z^2 = e$.
\end{expl}

\begin{expl}
Let $G$ be a finite group and $\omega \in Z^3(G;\Cb^\times)$. When $G$ is non-abelian, there is no braiding structure on $\vect_G^\omega$. When $G$ is abelian, the braiding structures on $\vect_G^\omega$ can be equivalently described by quadratic forms on $G$ (see Section \ref{sec:pointed_category}).
\end{expl}

\begin{figure}[htbp]
\[
\begin{array}{c}
\begin{tikzpicture}[scale=0.8]
\fill[gray!20] (-1,0)--(5,0)--(6,2)--(0,2)--cycle ;
%
\draw[->-=0.9] (3,2.5) .. controls (3,1) and (1,1) .. (1,2.5) ;
\draw[->-=0.9] (1,2.5) .. controls (1,4) and (3,4) .. (3,2.5) ;
\draw[->-=0.9] (4,2.5) .. controls (4,1) and (2,1) .. (2,2.5) ;
\draw[->-=0.9] (2,2.5) .. controls (2,4) and (4,4) .. (4,2.5) ;
\begin{scope}
\clip (2.5,1.5) circle (0.3) ;
\draw[gray!20,double=black,double distance=0.4pt,line width=3pt] (3,2.5) .. controls (3,1) and (1,1) .. (1,2.5) ;
\end{scope}
\begin{scope}
\clip (2.5,3.5) circle (0.3) ;
\draw[white,double=black,double distance=0.4pt,line width=3pt] (2,2.5) .. controls (2,4) and (4,4) .. (4,2.5) ;
\end{scope}
\node[left] at (1,2.5) {$x$} ;
\node[left] at (2,2.5) {$y$} ;
\node[right] at (3,2.5) {$x^*$} ;
\node[right] at (4,2.5) {$y^*$} ;
\node at (2,1) {$b_x$} ;
\node at (2,4) {$b_x^\dagger$} ;
\node at (3,1) {$b_y$} ;
\node at (3,4) {$b_y^\dagger$} ;
\end{tikzpicture}
\end{array}
=
\begin{array}{c}
\begin{tikzpicture}[scale=0.8]
\fill[gray!20] (-1,0)--(5,0)--(6,2)--(0,2)--cycle ;
%
\draw[->-=0.9] (1,2.5) .. controls (1,1) and (3,1) .. (3,2.5) ;
\draw[->-=0.9] (3,2.5) .. controls (3,4) and (1,4) .. (1,2.5) ;
\draw[->-=0.9] (2,2.5) .. controls (2,1) and (4,1) .. (4,2.5) ;
\draw[->-=0.9] (4,2.5) .. controls (4,4) and (2,4) .. (2,2.5) ;
\begin{scope}
\clip (2.5,1.5) circle (0.3) ;
\draw[gray!20,double=black,double distance=0.4pt,line width=3pt] (1,2.5) .. controls (1,1) and (3,1) .. (3,2.5) ;
\end{scope}
\begin{scope}
\clip (2.5,3.5) circle (0.3) ;
\draw[white,double=black,double distance=0.4pt,line width=3pt] (4,2.5) .. controls (4,4) and (2,4) .. (2,2.5) ;
\end{scope}
\node[left] at (1,2.5) {$x^*$} ;
\node[left] at (2,2.5) {$y^*$} ;
\node[right] at (3,2.5) {$x$} ;
\node[right] at (4,2.5) {$y$} ;
\node at (2,1) {$d_x^\dagger$} ;
\node at (2,4) {$d_x$} ;
\node at (3,1) {$d_y^\dagger$} ;
\node at (3,4) {$d_y$} ;
\end{tikzpicture}
\end{array}
\]
\caption{the $S$ matrix $S_{xy}$}
\label{fig:S_matrix}
\end{figure}
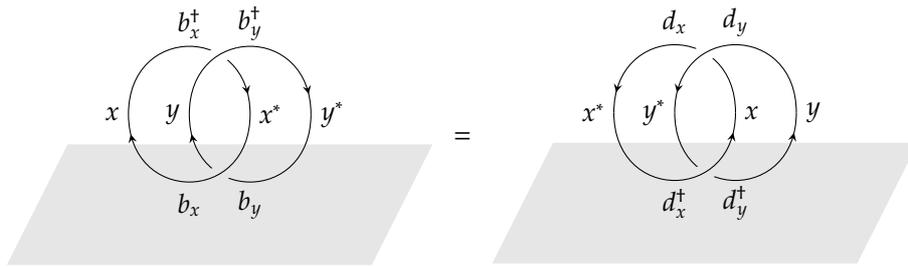

\begin{defn}
Let $\CC$ be a unitary braided fusion category. For two simple objects $x,y \in \CC$, define (see Figure \ref{fig:S_matrix})
\[
S_{xy} \coloneqq \tr(c_{y,x^*} \circ c_{x^*,y}) = \tr(c_{y^*,x} \circ c_{x,y^*}) .
\]
The matrix $S \coloneqq (S_{xy})_{x,y \in \Irr(\CC)}$ is called the $S$ matrix of $\CC$.
\end{defn}

The $S$ matrix encodes the information of mutual-statistics.

\begin{expl}
In the basis $(\one,e,m,f)$, the $S$ matrix of the toric code model is
\[
S =
\begin{pmatrix*}[r]
1 & 1 & 1 & 1 \\
1 & 1 & -1 & -1 \\
1 & -1 & 1 & -1 \\
1 & -1 & -1 & 1
\end{pmatrix*}.
\]
\end{expl}

\begin{rem}
The $S$ matrix can also be defined for a braided fusion category equipped with a spherical structure.
\end{rem}

\subsubsection{Ribbon structure}

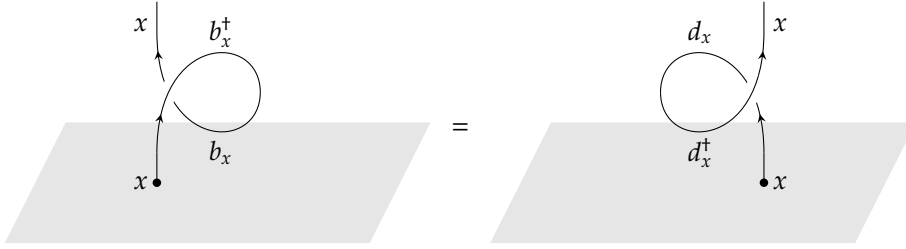
\begin{figure}[htbp]
\[
\begin{array}{c}
\begin{tikzpicture}[scale=0.8]
\fill[gray!20] (-1,0)--(5,0)--(6,2)--(0,2)--cycle ;
\draw (1.5,1)--(1.5,1.5) ;
\draw (1.5,3.5)--(1.5,4) ;
\path (1.5,1)--(1.5,4) node[very near end,left] {$x$} ;

\draw[->-=0.2] (1.5,1.5) .. controls (1.5,3.5) and (3.2,3.5) .. (3.2,2.5) ;
\node at (2.55,3.5) {$b_x^\dagger$} ;
\draw[->-=0.9] (3.2,2.5) .. controls (3.2,1.5) and (1.5,1.5) .. (1.5,3.5) ;
\node at (2.55,1.5) {$b_x$} ;
\begin{scope}
\clip (1.7,2.5) circle (0.3) ;
\draw[white,double=black,double distance=0.4pt,line width=3pt] (1.5,1.5) .. controls (1.5,3.5) and (3.2,3.5) .. (3.2,2.5) ;
\end{scope}

\fill (1.5,1) circle (0.07) node[left] {$x$} ;

\end{tikzpicture}
\end{array}
=
\begin{array}{c}
\begin{tikzpicture}[scale=0.8]
\fill[gray!20] (-1,0)--(5,0)--(6,2)--(0,2)--cycle ;
\draw (3.5,1)--(3.5,1.5) ;
\draw (3.5,3.5)--(3.5,4) ;
\path (3.5,1)--(3.5,4) node[very near end,right] {$x$} ;

\draw[->-=0.2] (3.5,1.5) .. controls (3.5,3.5) and (1.8,3.5) .. (1.8,2.5) ;
\node at (2.45,3.5) {$d_x$} ;
\draw[->-=0.9] (1.8,2.5) .. controls (1.8,1.5) and (3.5,1.5) .. (3.5,3.5) ;
\node at (2.45,1.5) {$d_x^\dagger$} ;
\begin{scope}
\clip (3.3,2.5) circle (0.3) ;
\draw[white,double=black,double distance=0.4pt,line width=3pt] (1.8,2.5) .. controls (1.8,1.5) and (3.5,1.5) .. (3.5,3.5) ;
\end{scope}

\fill (3.5,1) circle (0.07) node[right] {$x$} ;

\end{tikzpicture}
\end{array}
\]
\caption{the topological spin $\theta_x$ of $x$}
\label{fig:T_matrix}
\end{figure}

Let $\CC$ be a unitary braided fusion category and $x \in \CC$. Consider the following two morphisms as depicted in Figure \ref{fig:T_matrix} (with a normalized choice of duals):
\begin{gather}
x \xrightarrow{\id_x \otimes b_x} x \otimes x \otimes x^* \xrightarrow{c_{x,x} \otimes \id_{x^*}} x \otimes x \otimes x^* \xrightarrow{\id_x \otimes b_x^\dagger} x , \label{eq:ribbon_1} \\
x \xrightarrow{d_x^\dagger \otimes \id_x} x^* \otimes x \otimes x \xrightarrow{\id_{x^*} \otimes c_{x,x}} x^* \otimes x \otimes x \xrightarrow{d_x \otimes \id_x} x . \label{eq:ribbon_2}
\end{gather}
By taking traces one can easily show that these two morphisms are equal. This morphism is denoted by $\theta_x \colon x \to x$ and called the twist (or topological spin) of $x$. The family $\{\theta_x\}_{x \in \CC}$ is natural.

\begin{exercise}
Use the physical intuition in Figure \ref{fig:T_matrix} to prove the following statements:
\bnu[(1)]
\item Each $\theta_x$ is an isomorphism.
\item $\theta_x^* = \theta_{x^*}$.
\item $\theta_{x \otimes y} = (\theta_x \otimes \theta_y) \circ c_{y,x} \circ c_{x,y}$.
\enu
\end{exercise}

\begin{figure}[htbp]
\[
\begin{array}{c}
\begin{tikzpicture}
\draw[double distance between line centers=0.3cm] (1.5,4)--(1.5,3.5) .. controls (1.5,1.5) and (3.2,1.5) .. (3.2,2.5) ;
\draw[double distance between line centers=0.3cm] (3.2,2.5) .. controls (3.2,3.5) and (1.5,3.5) .. (1.5,1.5)--(1.5,1) ;


\draw (1.35,1)--(1.65,1) (1.35,4)--(1.65,4) ;
\end{tikzpicture}
\end{array}
=
\begin{array}{c}
\begin{tikzpicture}
\draw (0.3,1) .. controls (0.3,1.5) and (0,1.5) .. (0,2) ;
\draw (0.3,0) .. controls (0.3,0.5) and (0,0.5) .. (0,1) ;
\draw[white,double=black,double distance=0.4pt,line width=3pt] (0,0) .. controls (0,0.5) and (0.3,0.5) .. (0.3,1) (0,1) .. controls (0,1.5) and (0.3,1.5) .. (0.3,2) ;

\draw (0,0)--(0,-0.5)--(0.3,-0.5)--(0.3,0) (0,2)--(0,2.5)--(0.3,2.5)--(0.3,2) ;
\end{tikzpicture}
\end{array}
=
\begin{array}{c}
\begin{tikzpicture}
\draw[double distance between line centers=0.3cm] (3.5,1)--(3.5,1.5) .. controls (3.5,3.5) and (1.8,3.5) .. (1.8,2.5) ;
\draw[double distance between line centers=0.3cm] (1.8,2.5) .. controls (1.8,1.5) and (3.5,1.5) .. (3.5,3.5)--(3.5,4) ;

\draw (3.35,1)--(3.65,1) (3.35,4)--(3.65,4) ;
\end{tikzpicture}
\end{array}
\]
\caption{the twist of a ribbon}
\label{fig:twist}
\end{figure}
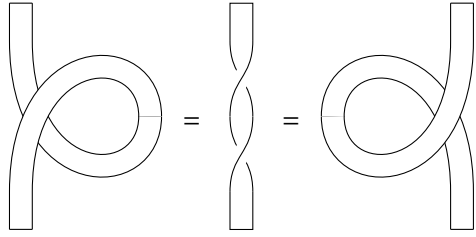

\begin{rem}
One may find that two world lines in Figure \ref{fig:T_matrix} are homotopy to the trivial paths and conclude that $\theta_x$ is just $\id_x$. This is not true. The world line should be viewed as a ribbon (indeed, a line equipped with a framing), not just a line. As depicted in Figure \ref{fig:twist}, these two process are both equal to twisting the ribbon counterclockwisely. Intuitively, this twisting process is rotating a topological defect by 360 degrees, and that is why we call $\theta_x$ the topological spin of $x$.
\end{rem}

For each simple object $x \in \CC$, since $\Hom_\CC(x,x) \simeq \Cb$ we know that $\theta_x = T_x \cdot \id_x$ for some scalar $T_x \in \Cb$. The diagonal matrix $T \coloneqq (T_x \delta_{xy})_{x,y \in \Irr(\CC)}$ is called the $T$ matrix of the unitary braided fusion category $\CC$. It encodes the information of self-statistics.

\begin{figure}[htbp]
\centering
\begin{tikzpicture}[scale=1.0]
\draw[step=1,help lines] (-1.5,-1.5) grid (3.5,2.5) ;
\draw[help lines,fill=m_ext] (0,1) rectangle (-1,2) node[near start,black] {$f$} ;
\foreach \y in {0,1} {
	\draw[m_str] (-1,\y)--(0,\y) ;
	\draw[m_str] (2,\y)--(3,\y) ;
}
\foreach \x in {0,1,2} {
	\draw[m_str] (\x,1)--(\x,2) ;
	\draw[m_str] (\x,-1)--(\x,0) ;
}
\draw[-stealth,m_dual_str] (-0.5,1.5)--(-0.5,-0.5)--(2.5,-0.5)--(2.5,1.5)--(-0.5,1.5) ;
\draw[-stealth,e_str] (0,1)--(0,0)--(2,0)--(2,1)--(0.07,1) ;
\fill[e_ext] (0,1) circle (0.07) ;

\draw[->-=0.6,densely dotted,opacity=0.5] (0,1)--(-0.5,1.5) ;
\draw[->-=0.6,densely dotted,opacity=0.5] (0,1)--(-0.5,0.5) ;
\draw[->-=0.6,densely dotted,opacity=0.5] (0,0)--(-0.5,0.5) ;
\draw[->-=0.6,densely dotted,opacity=0.5] (0,0)--(-0.5,-0.5) ;
\draw[->-=0.6,densely dotted,opacity=0.5] (0,0)--(0.5,-0.5) ;
\draw[->-=0.6,densely dotted,opacity=0.5] (1,0)--(0.5,-0.5) ;
\draw[->-=0.6,densely dotted,opacity=0.5] (1,0)--(1.5,-0.5) ;
\draw[->-=0.6,densely dotted,opacity=0.5] (2,0)--(1.5,-0.5) ;
\draw[->-=0.6,densely dotted,opacity=0.5] (2,0)--(2.5,-0.5) ;
\draw[->-=0.6,densely dotted,opacity=0.5] (2,0)--(2.5,0.5) ;
\draw[->-=0.6,densely dotted,opacity=0.5] (2,1)--(2.5,0.5) ;
\draw[->-=0.6,densely dotted,opacity=0.5] (2,1)--(2.5,1.5) ;
\draw[->-=0.6,densely dotted,opacity=0.5] (2,1)--(1.5,1.5) ;
\draw[->-=0.6,densely dotted,opacity=0.5] (1,1)--(1.5,1.5) ;
\draw[->-=0.6,densely dotted,opacity=0.5] (1,1)--(0.5,1.5) ;
\draw[->-=0.6,densely dotted,opacity=0.5] (0,1)--(0.5,1.5) ;

\end{tikzpicture}
\caption{the topological spin of $f$ in the toric code model}
\label{fig:spin_f}
\end{figure}
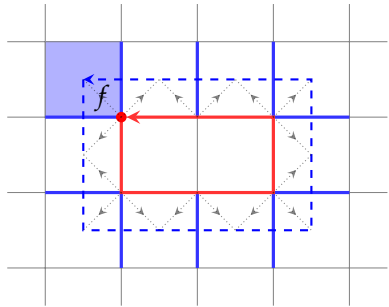

\begin{expl}
Let us compute the topological spin of $f$ in the toric code model. Consider an $f$ ribbon which consists of an $e$ string and an $m$ string (see Figure \ref{fig:spin_f}). The $f$ ribbon operator is the product of the $e$ string operator and $m$ string operator. Intuitively, the action of this ribbon operator is rotating the $f$ particle by 360 degrees.

Note that a closed $e$ string (or $m$ string) operator is equal to the product of $B_p$ (or $A_v$) operators encircled by the closed loop. Therefore, this ribbon operator is equal to the product of two $B_p$ operators and six $A_v$ operators. Only the upper left $A_v$ operator acts as $-1$ because there is an $f$ particle, and all other operators act as $+1$. Hence the topological spin of $f$ is $-1$.

Similarly we can compute the $T$ matrix of the toric code model:
\[
T =
\begin{pmatrix}
1 & & & \\
 & 1 & & \\
 & & 1 & \\
 & & & -1
\end{pmatrix}.
\]
\end{expl}

\begin{rem}
Let $\CC$ be a braided fusion category. A \emph{ribbon structure} on $\CC$ is a natural isomorphism $\theta_x \colon x \to x$ such that $\theta_{x \otimes y} = (\theta_x \otimes \theta_y) \circ c_{y,x} \circ c_{x,y}$ and $\theta_x^* = \theta_{x^*}$. There is a one-to-one correspondence between the ribbon structures and spherical structures on a braided fusion category \cite{Yet92}. By Remark \ref{rem:pivotal_spherical}, a unitary braided fusion category is equipped with a canonical ribbon structure \cite{LR97,Mueg00} (i.e., the one defined by \eqref{eq:ribbon_1} or \eqref{eq:ribbon_2}), and this ribbon structure is compatible with the unitary structure, in the sense that $\theta_x$ is unitary.
\end{rem}

\begin{defn}[\cite{Bru00}]
A \emph{pre-modular tensor category} (or simply a \emph{pre-modular category}) is a braided fusion category equipped with a ribbon structure, or equivalently, a braided fusion category equipped with a spherical structure.
\end{defn}

In particular, a unitary braided fusion category is automatically a pre-modular category.

\begin{expl} \label{expl:symmetric_ribbon}
Let $G$ be a finite group. Since $\rep(G)$ is symmetric, $\theta_{(V,\rho)} \coloneqq \id_V$ defines a ribbon structure on $\rep(G)$. The corresponding spherical structure on $\rep(G)$ is the one defined in Example \ref{expl:spherical_rep_G}.

More generally, suppose $z \in G$ is a central element satisfying $z^2 = e$. Then the family $\theta_{(V,\rho)} \coloneqq \rho(z)$ also defines a ribbon structure on $\rep(G,z)$. The corresponding spherical structure on $\rep(G,z)$, which is the same as $\rep(G)$ as a fusion category, is also the one defined in Example \ref{expl:spherical_rep_G}.
\end{expl}

\subsection{Unitary modular tensor categories}

\subsubsection{Nondegeneracy}

We have known that the category $\CC$ of particle-like topological defects of a stable 2d topological order $\SC$ is a unitary braided fusion category. If we require that $\SC$ is anomaly-free (see Section \ref{sec:anomaly-free_anomalous}), the unitary braided fusion category $\CC$ should satisfy a nondegeneracy condition. The following principle first appeared in \cite{Lev13,KW14}.

\begin{principle}[Remote-detectable principle] \label{principle:remote-detectable}
A topological order is anomaly-free if and only if all topological defects of codimension 2 and higher are able to detect themselves via braidings.
\end{principle}

Note that the braiding can only be defined between topological defects of codimension 2 or higher. We focus on 2d topological orders. For a stable 2d topological order $\SC$, the braiding can be used to remotely detect topological defects. As depicted in Figure \ref{fig:double_braiding}, if we want to detect whether there is a nontrivial topological defect $x \in \CC$ in a region, we can move a topological defect $y$ around this region and do a double braiding $c_{y,x} \circ c_{x,y}$. If there is at least one $y \in \CC$ such that the double braiding is nontrivial, the topological defect $x$ is detectable by this double braiding.

The trivial topological defect $x = \one$ is not detectable in this sense because for all $y \in \CC$ the double braiding $c_{y,\one} \circ c_{\one,y} = \id_y$ is trivial. However, if all nontrivial simple topological defect $x \neq \one$ are detectable, then $\one$ can also be detected in the following sense: $\one$ is the only simple topological defect which has trivial double braiding with all $y \in \CC$.

Hence, the condition that all topological defects are able to detect themselves via braidings can be reformulated as follows: if a simple topological defect $x \in \CC$ has trivial mutual statistics with all $y \in \CC$, i.e., $c_{y,x} \circ c_{x,y} = \id_{x \otimes y}$ for all $y \in \CC$, then $x = \one$ is trivial.

We propose the following definitions.

\begin{defn}
Let $\CC$ be a braided fusion category. The \emph{M\"{u}ger center} or \emph{symmetric center} $\FZ_2(\CC)$ of $\CC$ is the full subcategory consisting of the objects which have trivial double braiding with all objects in $\CC$, i.e.,
\[
\FZ_2(\CC) \coloneqq \{x \in \CC \mid c_{y,x} \circ c_{x,y} = \id_{x \otimes y} , \, \forall y \in \CC\} .
\]
\end{defn}

It is not hard to see that the M\"{u}ger center $\FZ_2(\CC)$ is a symmetric fusion category.

\begin{defn}
A braided fusion category $\CC$ is called \emph{nondegenerate} if its M\"{u}ger center $\FZ_2(\CC)$ contains only one simple object $\one$, or equivalently, $\FZ_2(\CC) \simeq \vect$.
\end{defn}

Hence the remote-detectable principle \ref{principle:remote-detectable} implies the following physical theorem.

\begin{pthm} \label{pthm:2d_anomaly-free_nondegenerate}
A stable 2d topological order $\SC$ is anomaly-free if and only if the unitary braided fusion category $\CC$ of particle-like topological defects of $\SC$ is nondegenerate.
\end{pthm}

\begin{defn}
A \emph{modular tensor category} (or simply a \emph{modular category}) is a nondegenerate pre-modular category. A \emph{unitary modular tensor category} (UMTC) is a nondegenerate unitary braided fusion category.
\end{defn}

\begin{rem}
A modular tensor category was defined by Turaev \cite{Tur20} to be a pre-modular category whose $S$ matrix is nondegenerate. It is clear that a pre-modular category with the nondegenerate $S$-matrix must have the trivial M\"{u}ger center. Indeed, the converse is also true \cite{Reh90,Bru00,Mueg00,BB01,Mueg03a,Mueg03b}. Hence these two definitions are equivalent.
\end{rem}

\begin{expl}
Let us consider a symmetric fusion category equipped with a ribbons structure. Its $S$ matrix satisfies $S_{x,y} = \dim(x) \dim(y)$ and has rank $1$. By contrast, the $S$ matrix of a modular tensor category has the maximal rank. Thus symmetric fusion categories and modular tensor categories are two extreme cases of pre-modular categories.
\end{expl}

\begin{rem}
The complete classification of unitary modular tensor categories with the number of isomorphism classes of simple objects less than 5 can be found in \cite{RSW09}.
\end{rem}

\begin{rem}
In the following, a \emph{unitary modular tensor category} can be understood as a nondegenerate pseudo-unitary braided fusion category equipped with the unique spherical structure such that the quantum dimensions of objects are equal to the Frobenius-Perron dimensions (see Remark \ref{rem:pseudo_unitary}).
\end{rem}

Finally, we get the following physical theorem \cite{MS89,FRS89,FG90,Reh90} which describes the structure of particle-like topological defects of an anomaly-free stable 2d topological order (see also \cite{Kit06} for a review).

\begin{pthm}
The particle-like topological defects of an anomaly-free stable 2d topological order $\SC$ form a unitary modular tensor category $\CC$.
\end{pthm}

\begin{rem} \label{rem:categorical_description_UMTC}
Given an anomaly-free stable 2d topological order $\SC$, the unitary modular tensor category $\CC$ of particle-like topological defects can be viewed as a categorical description of $\SC$. If $\SC$ is not stable, this is not an appropriate description.
\end{rem}

\begin{figure}[htbp]
\centering
\begin{tikzpicture}[scale=0.8]
\fill[gray!30,opacity=0.7] (2,0) circle (2) ;
\draw[dashed,-stealth] (30:2.5cm)++(2cm,0) arc [start angle=30,end angle=-45,radius=2.5cm] ;
\draw[densely dashed,rotate=30] (0,0) arc [start angle=180,end angle=0,x radius=1.732cm,y radius=0.5cm] ;
\draw[densely dashed] (0,0) arc [start angle=180,end angle=0,x radius=2cm,y radius=0.7cm] ;
\draw[densely dashed,rotate=-30] (0,0) arc [start angle=180,end angle=0,x radius=1.732cm,y radius=0.5cm] ;
\draw[densely dashed,rotate=-60] (0,0) arc [start angle=180,end angle=0,x radius=1cm,y radius=0.2cm] ;
\draw[fill=gray!30,opacity=0.7] (2,0) circle (2) ;
\draw[->-,densely dashed,rotate=30] (0,0) arc [start angle=180,end angle=360,x radius=1.732cm,y radius=0.5cm] ;
\draw[->-,densely dashed] (0,0) arc [start angle=180,end angle=360,x radius=2cm,y radius=0.8cm] ;
\draw[->-,densely dashed,rotate=-30] (0,0) arc [start angle=180,end angle=360,x radius=1.732cm,y radius=0.5cm] ;
\draw[->-,densely dashed,rotate=-60] (0,0) arc [start angle=180,end angle=360,x radius=1cm,y radius=0.2cm] ;
\fill (0,0) circle (0.07) node[left] {$x$} ;
\fill (120:2cm)++(2cm,0) circle (0.07) node[above left] {$y$} ;
\end{tikzpicture}
\caption{The braiding is not well-defined on a sphere.}
\label{fig:braiding_sphere}
\end{figure}
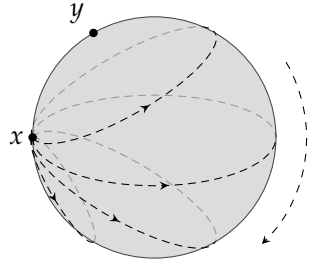

\begin{rem} \label{rem:double_braiding_not_global}
This unitary braided fusion category is an observable defined only on a disk. For example, the braiding of topological defects are not well-defined on a sphere or other closed surfaces. Figure \ref{fig:braiding_sphere} depicts a paradox that every double braiding is trivial because the loop of a double braiding on a sphere is contractible.
\end{rem}

\begin{rem} \label{rem:TO-TQFT}
The anomaly-free condition for a 2d topological order $\SX$ is different from that of the TQFT associated to $\SX$. More precisely, a UMTC is anomaly-free from the perspective of a 2d topological order; but is anomalous from the perspective of an Atiyah's TQFT \cite{Ati88} due to the so-called framing anomaly \cite{Tur20}. This subtle difference is due to the fact that the spacetime manifolds considered in a TQFT is much richer than those considered in a topological order. More precisely, the spacetime 3-manifolds in the later case are restricted to the type $\Sigma^2 \times S^1$ (or $\Sigma^2 \times \Rb^1$), where $\Sigma^2$ is an arbitrary spatial 2-manifold and $S^1$ (or $\Rb^1$) is the temporal 1-manifold. 
\end{rem}

\begin{rem}
There is also a 3+1D generalization of the toric code model \cite{HZW05}. For intrigued readers, it is beneficial to read \cite{KTZ20a} for the `modular 2-category' description of the 3+1D toric code model.
\end{rem}

\subsubsection{Basic properties of modular tensor categories}

In the following we list some important properties of a (not necessarily unitary) modular tensor category $\CC$ (for example, see \cite{Tur20,BK01,EGNO15}):
\bnu
\item The $S$ matrix satisfies $S_{x y} = S_{y x} = S_{x^* y^*} = \overline{S_{x y^*}}$ and $S_{\one x} = S_{x \one} = d_x \coloneqq \dim(x)$.
\item For each simple object $x \in \CC$, the assignment $h_x \colon y \mapsto S_{xy}/S_{x \one}$ defines a ring homomorphism $\Gr(\CC) \to \Cb$, i.e.,
\[
S_{xy} S_{xz} = d_x \sum_{w \in \Irr(\CC)} N_{yz}^w S_{xw} .
\]
Note that $h_\one(y) = \dim(y)$ is the quantum dimension.
\item We have $S^2 = \dim(\CC) \cdot C$, where $C$ is the charge-conjugate matrix defined by $C_{x y} \coloneqq \delta_{x^* y}$. Thus, $(S^{-1})_{x y} = \dim(\CC)^{-1} \cdot S_{x y^*}$.
\item The $S$ matrix determines the fusion rule:
\[
\sum_{x \in \Irr(\CC)} \frac{S_{xy} S_{xz} S_{xw^*}}{S_{x \one}} = \dim(\CC) \cdot N_{yz}^w .
\]
This equation is called the \emph{Verlinde formula} \cite{Ver88,MS89,MS90,Hua08a}.\footnote{A generalization of the Verlinde formula can be found in \cite{SH19}.}
\item The \emph{Gauss sums} of $\CC$ are defined by
\[
\tau^\pm(\CC) \coloneqq \sum_{x \in \Irr(\CC)} T_x^{\pm 1} d_x^2 .
\]
Then $\tau^+(\CC) \tau^-(\CC) = \dim(\CC)$. The \emph{multiplicative central charge} of $\CC$ is defined by
\[
\xi(\CC) \coloneqq \frac{\tau^+(\CC)}{\sqrt{\dim(\CC)}} = \frac{\sqrt{\dim(\CC)}}{\tau^-(\CC)} .
\]
Usually we also write $\xi(\CC) = \exp(2 \pi \mathrm i c(\CC) / 8)$, where $c(\CC)$ is called the \emph{additive central charge} of $\CC$. It can be proved that $\xi(\CC)$ is a root of unity \cite{AM88,Vaf88}, so $c(\CC) \in \Qb / 8 \Zb$.
\item We have $(S T)^3 = \tau^+(\CC) \cdot S^2$. Recall that the group $\mathrm{SL}(2,\Zb)$, which consists of $2$-by-$2$ integral matrices of determinant $1$, is generated by
\[
\mathfrak s \coloneqq \begin{pmatrix}0 & -1 \\ 1 & 0\end{pmatrix} , \, \mathfrak t \coloneqq \begin{pmatrix}1 & 1 \\ 0 & 1\end{pmatrix}
\]
subject to the relations $(\mathfrak s \mathfrak t)^3 = \mathfrak s^2$ and $\mathfrak s^4 = 1$. Hence, the assignment
\be \label{eq:modular_rep}
\mathfrak s \mapsto \frac{S}{\sqrt{\dim(\CC)}} , \, \mathfrak t \mapsto T
\ee
defines a projective $\mathrm{SL}(2,\Zb)$-representation. The obstruction of this projective representation being a linear representation is the multiplicative central charge $\xi(\CC)$.
\item The matrix $\sqrt{\dim(\CC)}^{-1} \cdot S$ is unitary \cite{ENO05}. Also each $T_x$ is a root of unity \cite{AM88,Vaf88}, thus the $T$-matrix is also unitary. So for any (not necessarily unitary) modular tensor category $\CC$ the above projective $\mathrm{SL}(2,\Zb)$-representation is unitary.
\enu

\begin{rem}
If we put an anomaly-free 2d topological order $\SC$ on a torus, the automorphims of the torus induce transformations of the ground state subspace. This defines a projective representation of the mapping class group of the torus, which is isomorphic to the modular group $\mathrm{SL}(2,\Zb)$. This representation is isomorphic to the representation \eqref{eq:modular_rep} associated to $\CC$ and was used to characterize 2d topological orders \cite{Wen90}. These modular reprensentations can also be generalized to higher dimensions \cite{WL14,JMR14,WW15,BD21a}. 
\end{rem}

\begin{rem}
Modular tensor categories are not determined by $S$ and $T$ matrices \cite{MS21}.
\end{rem}

\subsubsection{Construct new modular tensor categories from old ones}

There are some method to construct new modular tensor categories from old ones.

\begin{defn}[\cite{Del07}] \label{defn:Deligne_tensor_product}
Let $\CC,\CD$ be two finite semisimple categories. Their \emph{Deligne tensor product}, denoted by $\CC \boxtimes \CD$, is a finite semisimple category satisfying the following universal property: $\CC \boxtimes \CD$ is equipped with a $\Cb$-bilinear functor $\boxtimes \colon \CC \times \CD \to \CC \boxtimes \CD$, and for any finite semisimple category $\CX$ and $\Cb$-bilinear functor $F \colon \CC \times \CD \to \CX$, there exists a unique $\Cb$-linear functor $\underline F \colon \CC \boxtimes \CD \to \CX$ (up to isomorphism) such that the following diagram commutes:
\[
\xymatrix{
\CC \times \CD \ar[r]^{\boxtimes} \ar[dr]_{F} & \CC \boxtimes \CD \ar[d]^{\underline F} \\
 & \CX
}
\]
or more precisely, the functor $- \circ \boxtimes$ from $\fun(\CC \boxtimes \CD,\CX)$ to the category of $\Cb$-bilinear functors from $\CC \times \CD$ to $\CX$.
\end{defn}

This definition may be too abstract. In practice, it is enough to know the following facts:
\bnu
\item For every $x \in \CC$ and $y \in \CD$, there is an object $x \boxtimes y \in \CC \boxtimes \CD$. In general, an object in $\CC \boxtimes \CD$ is the direct sum of objects of this form.
\item $\Hom_{\CC \boxtimes \CD}(x \boxtimes y,x' \boxtimes y') \simeq \Hom_\CC(x,x') \otimes_\Cb \Hom_\CD(y,y')$ for $x,x' \in \CC$ and $y,y' \in \CD$.
\item An object $x \boxtimes y \in \CC \boxtimes \CD$ is simple if and only if $x \in \CC$ and $y \in \CD$ are both simple. Thus $\Irr(\CC \boxtimes \CD) \simeq \Irr(\CC) \times \Irr(\CD)$. In particular, $\vect \boxtimes \CC \simeq \CC \simeq \CC \boxtimes \vect$.
\enu

When $\CC$ and $\CD$ are both pre-modular categories, their Deligne tensor product $\CC \boxtimes \CD$ is also a pre-modular category:
\bit
\item The tensor product of $\CC \boxtimes \CD$ is defined by
\[
(x \boxtimes y) \otimes (x' \boxtimes y') \coloneqq (x \otimes x') \boxtimes (y \otimes y') , \quad x,x' \in \CC , \, y,y' \in \CD .
\]
The tensor unit of $\CC \boxtimes \CD$ is $\one \boxtimes \one$, which is a simple object. The associator and left/right unitor are induced by those of $\CC$ and $\CD$.
\item The left/right dual of an object $x \boxtimes y \in \CC \boxtimes \CD$ is $x^{L/R} \boxtimes y^{L/R}$. Thus $\CC \boxtimes \CD$ is a fusion category. Also we have $\fpdim(x \boxtimes y) = \fpdim(x) \cdot \fpdim(y)$ and $\fpdim(\CC \boxtimes \CD) = \fpdim(\CC) \cdot \fpdim(y)$.
\item The spherical structure on $\CC \boxtimes \CD$ is defined by $a^{\CC \boxtimes \CD}_{x \boxtimes y} \coloneqq a^\CC_x \boxtimes a^\CD_y$. It follows that $\dim(x \boxtimes y) = \dim(x) \cdot \dim(y)$ and $\dim(\CC \boxtimes \CD) = \dim(\CC) \cdot \dim(\CD)$.
\item The braiding structure on $\CC \boxtimes \CD$ is defined by
\[
c^{\CC \boxtimes \CD}_{x \boxtimes y,x' \boxtimes y'} = \bigl( (x \boxtimes y) \otimes (x' \boxtimes y') \coloneqq (x \otimes x') \boxtimes (y \otimes y') \xrightarrow{c^\CC_{x,x'} \boxtimes c^\CD_{y,y'}} (x' \otimes x) \boxtimes (y' \otimes y) = (x' \boxtimes y') \otimes (x \boxtimes y) \bigr) .
\]
So the $S$ matrix of $\CC \boxtimes \CD$ is
\be \label{eq:S_matrix_Deligne_tensor_product}
S^{\CC \boxtimes \CD}_{x \boxtimes y,x' \boxtimes y'} = S^\CC_{x,x'} \cdot S^\CD_{y,y'} .
\ee
\item The ribbon structure on $\CC \boxtimes \CD$ is defined by $\theta^{\CC \boxtimes \CD}_{x \boxtimes y} \coloneqq \theta^\CC_x \boxtimes \theta^\CD_y$. So the $T$ matrix of $\CC \boxtimes \CD$ is
\[
T^{\CC \boxtimes \CD}_{x \boxtimes y} = T^\CC_x \cdot T^\CD_y .
\]
\eit
It is not hard to see that $\FZ_2(\CC \boxtimes \CD) \simeq \FZ_2(\CC) \boxtimes \FZ_2(\CD)$. Thus $\CC \boxtimes \CD$ is a modular tensor category if $\CC$ and $\CD$ are modular. Equivalently, the $S$ matrix of $\CC \boxtimes \CD$ \eqref{eq:S_matrix_Deligne_tensor_product} is nondegenerate if the $S$ matrices of $\CC$ and $\CD$ are both nondegenerate.

\begin{exercise}
Suppose $\CC$ and $\CD$ are modular tensor categories. Prove that the Gauss sums satisfy $\tau^\pm(\CC \boxtimes \CD) = \tau^\pm(\CC) \cdot \tau^\pm(\CD)$, the multiplicative central charge satisfies $\xi(\CC \boxtimes \CD) = \xi(\CC) \cdot \xi(\CD)$, and the additive central charge satisfies $c(\CC \boxtimes \CD) = c(\CC) \cdot c(\CD)$.
\end{exercise}

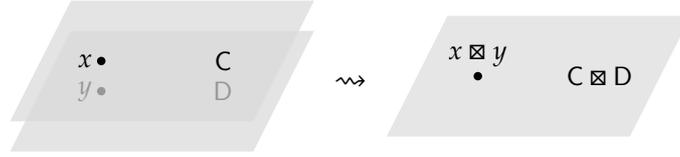
\begin{figure}[htbp]
\[
\begin{array}{c}
\begin{tikzpicture}[scale=0.8]
\fill[gray!30,opacity=0.7] (0,0)--(4,0)--(5,2)--(1,2)--cycle ;
\node at (3.5,1) {$\SD$} ;
\fill (1.5,1) circle (0.07) node[left] {$y$} ;
\fill[gray!30,opacity=0.7,yshift=0.5cm] (0,0)--(4,0)--(5,2)--(1,2)--cycle ;
\node at (3.5,1.5) {$\SC$} ;
\fill (1.5,1.5) circle (0.07) node[left] {$x$} ;
\end{tikzpicture}
\end{array}
\rightsquigarrow
\begin{array}{c}
\begin{tikzpicture}[scale=0.8]
\fill[gray!20] (0,0)--(4,0)--(5,2)--(1,2)--cycle ;
\node at (3.5,1) {$\SC \boxtimes \SD$} ;
\fill (1.5,1) circle (0.07) node[above] {$x \boxtimes y$} ;
\end{tikzpicture}
\end{array}
\]
\caption{stacking of 2d topological orders}
\label{fig:stacking_2d}
\end{figure}

Physically, the Deligne tensor product corresponds to the stacking of 2d topological orders (see Example \ref{expl:stacking_quantum_phase}). As depicted in Figure \ref{fig:stacking_2d}, suppose $x$ and $y$ are particle-like topological defects in anomaly-free stable 2d topological orders $\SC$ and $\SD$, respectively. Then their stacking is a particle-like topological defect, denoted by $x \boxtimes y$, in the stacking topological order $\SC \boxtimes \SD$. It is believed that the modular tensor category of particle-like topological defects of $\SC \boxtimes \SD$ is the Deligne tensor product $\CC \boxtimes \CD$.

\begin{defn}
Let $\CC$ be a braided monoidal category. Define $\overline{\CC}$ to be the same underlying monoidal category of $\CC$ equipped with the braiding structure
\[
c^{\overline{\CC}}_{x,y} \coloneqq (c^\CC_{y,x})^{-1} , \quad x,y \in \CC .
\]
\end{defn}

\begin{rem}
By definition a braided monoidal category $\CC$ is symmetric if and only if $\CC = \overline{\CC}$. However, it is possible that $\CC \simeq \overline{\CC}$ as braided monoidal categories but $\CC$ is not symmetric. The modular tensor category $\Toric$ is such an example.
\end{rem}

When $\CC$ is a (pre-)modular tensor category, $\overline{\CC}$ is also a (pre-)modular tensor category:
\bit
\item By definition $\overline{\CC}$ is a braided fusion category and $\FZ_2(\overline{\CC}) = \FZ_2(\CC)$.
\item The spherical structure on $\overline{\CC}$ is the same as $\CC$. In particular, the quantum dimension of objects in $\overline{\CC}$ are the same as $\CC$. The $S$ matrix of $\overline{\CC}$ is
\[
S^{\overline{\CC}}_{xy} = S^\CC_{xy^*} .
\]
\item Equivalently, the ribbon structure on $\overline{\CC}$ is defined by $\theta^{\overline{\CC}}_x \coloneqq (\theta^\CC_x)^{-1}$. The $T$ matrix of $\overline{\CC}$ is
\[
T^{\overline{\CC}}_x = (T^\CC_x)^{-1} .
\]
\eit

\begin{exercise}
Suppose $\CC$ is a modular tensor category. Prove that the Gauss sums satisfy $\tau^\pm(\overline{\CC}) = \tau^\mp(\CC)$, the multiplicative central charge satisfies $\xi(\overline{\CC}) = \xi(\CC)^{-1}$, and the additive central charge satisfies $c(\overline{\CC}) = -c(\CC)$.
\end{exercise}

Physically, $\overline{\CC}$ corresponds to the time-reversal of a 2d topological order. Here time-reversal means reversing the orientation of the spacetime. In 0+1D, there is only one way to reverse the the orientation of the spacetime, i.e., time-reversal. If $\CC$ is the category of particle-like topological defects of a topological order $\SC$, then the opposite category $\CC^\op$ (see Definition \ref{defn:opposite_category}) is the category of particle-like topological defects of the time-reversal $\overline{\SC}$ of $\SC$.

In 1+1D, there are two ways to reverse the orientation of the spacetime. If we choose to reverse the orientation of the 1d space, the fusion of topological defects is also reversed. The following definition is the precise meaning of reversing the fusion.

\begin{defn}
Let $\CC$ be a monoidal category. Its \emph{reversed category} $\CC^\rev$ is the same underlying category as $\CC$ equipped with the tensor product $\otimes^\rev$ defined by
\[
x \otimes^\rev y \coloneqq y \otimes x , \quad x,y \in \CC .
\]
\end{defn}

Therefore, if $\CC$ is the fusion category of particle-like topological defects of a 1d topological order $\SC$, then the reversed category $\CC^\rev$ is the fusion category of particle-like topological defects of the time-reversal $\overline{\SC}$ of $\SC$. When $\CC$ is a fusion category, $\CC^\op$ is also a fusion category and $\CC^\op \simeq \CC^\rev$ as fusion categories.

\begin{figure}[htbp]
\[
\begin{array}{c}
\begin{tikzpicture}
\fill[gray!20] (0,0) rectangle (3,2) node[below left,black] {$\overline{\SC}$} ;
\fill (1.5,1) circle (0.07) node[above] {$x$} ;
\fill (0.8,1) circle (0.07) node[left] {$y$} ;
\draw[densely dashed,-stealth] (0.8,1) arc [radius=0.7,start angle=180,end angle=-172] ;
\end{tikzpicture}
\end{array}
\xleftarrow[\text{flip}]{\text{horizontal}}
\begin{array}{c}
\begin{tikzpicture}
\fill[gray!20] (0,0) rectangle (3,2) node[below left,black] {$\SC$} ;
\fill (1.5,1) circle (0.07) node[above] {$x$} ;
\fill (2.2,1) circle (0.07) node[right] {$y$} ;
\draw[densely dashed,-stealth] (2.2,1) arc [radius=0.7,start angle=0,end angle=352] ;
\end{tikzpicture}
\end{array}
\xrightarrow[\text{flip}]{\text{vertical}}
\begin{array}{c}
\begin{tikzpicture}
\fill[gray!20] (0,0) rectangle (3,2) node[below left,black] {$\overline{\SC}$} ;
\fill (1.5,1) circle (0.07) node[below] {$x$} ;
\fill (2.2,1) circle (0.07) node[right] {$y$} ;
\draw[densely dashed,-stealth] (2.2,1) arc [radius=0.7,start angle=0,end angle=-352] ;
\end{tikzpicture}
\end{array}
\]
\caption{two ways to reverse the orientation of the 2d space}
\label{fig:reverse_2d}
\end{figure}
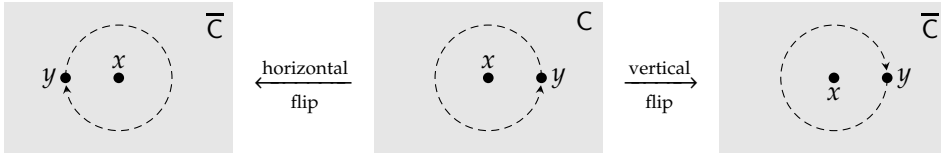

In 2+1D, there are two ways to reverse the orientation of the 2d space, as depicted in Figure \ref{fig:reverse_2d}. The first way reverses the fusion of topological defects. Another way does not reverse the fusion of topological defects, but reverses the braiding to the anti-braiding. Therefore, if $\CC$ is the braided fusion category of particle-like topological defects of a stable 2d topological order $\SC$, then $\overline{\CC}$ is the braided fusion category of particle-like topological defects in the time-reversal $\overline{\SC}$ of $\SC$. When $\CC$ is a modular tensor category, both $\CC^\op$ and $\CC^\rev$ are also modular tensor categories and $\CC^\op \simeq \CC^\rev \simeq \overline{\CC}$ as modular tensor categories.

\subsection{Examples of (unitary) modular tensor categories}

In this subsection we give some examples of (unitary) modular tensor categories.

\subsubsection{The categories \texorpdfstring{$\vect$}{Vec} and \texorpdfstring{$\hilb$}{Hilb}} \label{sec:example_hilb}

The simplest modular tensor category is the category $\vect$ of finite-dimensional vector spaces. Its structure of a modular tensor category has been defined in Section \ref{sec:structure_2d}. We list the structure in the following:
\bit
\item The objects of $\vect$ are finite-dimensional vector spaces (over $\Cb$), and the morphisms are linear maps. It is a finite semisimple category, and the only simple object (up to isomorphism) is $\Cb$.
\item The tensor product of $\vect$ is the usual tensor product $\otimes_\Cb$ of vector spaces. The associator and left/right unitor are induced by the universal property of $\otimes_\Cb$.
\item Both the left and right dual of a finite-dimensional vector space $V$ are given by the usual dual space $V^* \coloneqq \Hom(V,\Cb)$.
\item The spherical structure is
\begin{align*}
a_V \colon V & \to V^{**} \\
v & \mapsto \ev_v
\end{align*}
where $\ev_v \colon V^* \to \Cb$ is defined by $\ev_v(\phi) \coloneqq \phi(v)$.
\item Both the quantum dimension and Frobenius-Perron dimension are equal to the usual dimension of vector spaces.
\item The braiding structure is given by
\begin{align*}
c_{V,W} \colon V \otimes W & \to W \otimes V \\
v \otimes w & \mapsto w \otimes v .
\end{align*}
Note that $\vect$ is also a symmetric fusion category.
\item The ribbon structure is given by $\theta_V = \id_V$.
\eit

The unitary version of $\vect$ is the $\hilb$, which describes the particle-like topological defects of the trivial 2d topological order.

\subsubsection{The category \texorpdfstring{$\Toric$}{Toric}}

In Section \ref{sec:structure_2d} we have computed the unitary modular tensor category $\Toric$ of particle-like topological defects of the toric code model. We list the structure of $\Toric$ in the following:
\bit
\item There are only four simple objects $\one,e,m,f$.
\item The fusion rules are given by $e \otimes m = f = m \otimes e$ and $e \otimes e = m \otimes m = f \otimes f = \one$. This implies that each simple objects is self-dual and has Frobenius-Perron dimension $1$. The associator and left/right unitor are identities.
\item The $S$ matrix is
\[
S =
\begin{pmatrix*}[r]
1 & 1 & 1 & 1 \\
1 & 1 & -1 & -1 \\
1 & -1 & 1 & -1 \\
1 & -1 & -1 & 1
\end{pmatrix*}.
\]
\item The $T$ matrix is
\[
T =
\begin{pmatrix}
1 & & & \\
 & 1 & & \\
 & & 1 & \\
 & & & -1
\end{pmatrix}.
\]
\eit

As explained in Remark \ref{rem:monoidal_gauge} and \ref{rem:braiding_gauge}, the lattice model only determines $\Toric$ up to equivalence.

\subsubsection{Quantum double category of a finite group} \label{sec:quantum_double}

For any finite group $G$ there is a (unitary) modular tensor category $\CD_G$, called the quantum double category of $G$:
\bit
\item An object in $\CD_G$ is a vector space $V$ equipped with a $G$-action $\rho \colon G \to \mathrm{GL}(V)$ and a $G$-grading $V = \medoplus_{g \in G} V_g$, such that $\rho(g)$ maps $V_h$ to $V_{ghg^{-1}}$ for every $g,h \in G$. A morphism in $\CD_G$ is a $\Cb$-linear map that is both a morphism in $\rep(G)$ and a morphism in $\vect_G$.
\item Given an element $g \in G$, we denote the conjugacy class of $g$ by $[g]$ and the centralizer of $g$ by $Z(g)$. If $\pi$ is an irreducible $Z(g)$-representation, the induced representation $\Ind^G_{Z(g)}(\pi) \eqqcolon X_{(g,\pi)}$ is a $G$-representation and admits a canonical $G/Z(g) \simeq [g]$-grading. This $[g]$-grading on $X_{(g,\pi)}$ can be viewed as a $G$-grading with $h$-components being $0$ for all $h \notin [g]$. Then one can verify that $X_{(g,\pi)}$ is a simple object in $\CD_G$. Conversely, every simple object in $\CD_G$ is isomorphic to $X_{(g,\pi)}$ for some $g \in G$ and $\pi \in \Irr(\rep(Z(g)))$. The isomorphism classes of simple objects of $\CD_G$ are labeled by the pairs $([g],[\rho])$ where $[g]$ is a conjugacy class in $G$ and $[\rho]$ is an isomorphism class of irreducible representation of $Z(g)$. Note that $Z(g)$ only depends on the conjugacy class of $g$.
\item The tensor product of two objects in $\CD_G$ is the usual tensor product as a $G$-representation, equipped with the usual tensor product $G$-grading. In particular, the tensor unit of $\CD_G$ is $\Cb_{(e)}$ equipped with the trivial $G$-action.
\item The dual of an object in $\CD_G$ is the usual dual representation equipped with the usual dual $G$-grading. In particular, the dual of $X_{(g,\pi)}$ is $X_{(g^{-1},\pi^*)}$, where $\pi^*$ is the dual representation of $\pi$. Note that $Z(g) = Z(g^{-1})$.
\item The spherical structure on $\CD_G$ is given by that on $\vect$. Thus the Frobenius-Perron dimension and quantum dimension are both equal to the usual dimension of a vector space. In particular, both two dimension of $X_{(g,\pi)}$ are equal to $\lvert [g] \rvert \cdot \dim(\pi)$. The Frobenius-Perron dimension and quantum dimension of $\CD_G$ are both equal to $\lvert G \rvert^2$.
\item The braiding $c_{(V,\rho),(W,\sigma)} \colon V \otimes_\Cb W \to W \otimes_\Cb V$ is defined by $v \otimes_\Cb w \mapsto \sigma(g)(y) \otimes_\Cb x$ for $v \in V_g$ and $w \in W$. The $S$ matrix is given by
\[
S_{([g],[\pi]),([g'],[\pi'])} = \frac{\lvert G \rvert}{\lvert Z(g) \rvert \lvert Z(g') \rvert} \sum_{\{h \in G \mid hg'h^{-1} \in Z(g)} \tr_\pi(h(g')^{-1}h^{-1}) \tr_{\pi'}(h^{-1}g^{-1}h) .
\]
\item The ribbon structure on $\CD_G$ is defined by $\theta_{(V,\rho)}(v) \coloneqq \rho(g)(v)$ for $v \in V_g$. The $T$ matrix is given by
\[
T_{([g],[\pi])} = \frac{\tr_\pi(g)}{\tr_\pi(e)} .
\]
\item The multiplicative central charge of $\CD_G$ is $1$ and the additive central charge is $0 \pmod 8$.
\eit

\begin{rem}
Let $G$ be a finite group. There is an exactly solvable 2d lattice model associated to $G$, called Kitaev's quantum double model \cite{Kit03}. The unitary modular tensor category of particle-like topological defects of Kitaev's quantum double model is equivalent to $\CD_G$. When $G = \Zb_2$, it is the toric code model. In particular, $\Toric$ is equivalent to $\CD_{\Zb_2}$ as a modular tensor category.
\end{rem}

\begin{rem}
Let $G$ be a finite group. There is a quasi-triangular Hopf algebra $D(G)$ associated to $G$, called the quantum double algebra of $G$. The category of finite-dimensional representations of $D(G)$ is equivalent to $\CD_G$ as modular tensor categories (see for example \cite{Kas95,EGNO15}). Moreover, $D(G)$ is the minimal local operator algebra of Kitaev's quantum double model (see Section \ref{sec:local_operator_algebra}).
\end{rem}

\begin{rem}
The low energy effective theory of the quantum double model associated to a finite group $G$ is the 2+1D Dijkgraaf-Witten theory \cite{DW90} of $G$ with the trivial 3-cocycle.
\end{rem}

\begin{rem}
Kitaev's quantum double lattice model associated to a group can be generalized to a lattice model associated to a (weak) Hopf algebra \cite{BMCA13,BK12,Cha14}.
\end{rem}

\subsubsection{Ising type categories}

There is an important family of (unitary) modular tensor categories called Ising type (unitary) modular tensor categories.

An Ising type fusion category is a fusion category with three simple objects $\one,\sigma,\psi$ and fusion rules
\[
\psi \otimes \psi = \one , \, \psi \otimes \sigma = \sigma = \sigma \otimes \psi , \, \sigma \otimes \sigma = \one \oplus \psi .
\]
An Ising type modular tensor category  is a modular tensor category whose underlying fusion category is of Ising type. We list the classification of Ising type modular tensor categories as follows \cite{DGNO10}:
\bit
\item The fusion rules imply that each simple object in an Ising type fusion category is self-dual, and $\fpdim(\one) = \fpdim(\psi) = 1$, $\fpdim(\sigma) = \sqrt 2$.
\item There are 2 Ising type fusion categories (up to equivalence) labeled by a sign $\pm$. Their nonzero associators are given by
\begin{gather*}
\alpha_{\psi,\sigma,\psi} \colon (\psi \otimes \sigma) \otimes \sigma = \sigma \xrightarrow{-1} \sigma = \psi \otimes (\sigma \otimes \psi) , \\
\alpha_{\sigma,\psi,\sigma} \colon (\sigma \otimes \psi) \otimes \sigma = \one \oplus \psi \xrightarrow{1 \oplus -1} \one \oplus \psi = \sigma \otimes (\psi \otimes \sigma) , \\
\alpha_{\sigma,\sigma,\sigma} \colon (\sigma \otimes \sigma) \otimes \sigma = (\one \otimes \sigma) \oplus (\psi \otimes \sigma) \xrightarrow{A_\pm} (\sigma \otimes \one) \oplus (\sigma \otimes \psi) = \sigma \otimes (\sigma \otimes \sigma) ,
\end{gather*}
where 
\[
A_\pm = \frac{\pm 1}{\sqrt 2} \begin{pmatrix*}[r] 1 & 1 \\ 1 & -1\end{pmatrix*}
\]
is the only difference of these two monoidal structures.
\item Each Ising type fusion category has 4 different braiding structures (up to equivalence), so there are 8 different Ising type braided fusion categories (up to equivalence). These braided fusion categories are labeled by a complex number $\zeta$ satisfying $\zeta^8 = -1$, and the underlying Ising type fusion categories are determined by the sign $\pm$ in the following equation:
\[
\zeta^2 + \zeta^{-2} = \pm \sqrt 2 .
\]
Explicitly, the braiding structures of Ising type braided fusion categories are given by
\begin{gather*}
c_{\psi,\psi} \colon \psi \otimes \psi = \one \xrightarrow{-1} \one = \psi \otimes \psi , \\
c_{\sigma,\sigma} \colon \sigma \otimes \sigma = \one \oplus \psi \xrightarrow{\zeta \oplus \zeta^{-3}} \one \oplus \psi = \sigma \otimes \sigma , \\
c_{\psi,\sigma} \colon \psi \otimes \sigma = \sigma \xrightarrow{\zeta^4} \sigma = \sigma \otimes \psi , \\
c_{\sigma,\psi} \colon \sigma \otimes \psi = \sigma \xrightarrow{\zeta^4} \sigma = \psi \otimes \sigma .
\end{gather*}
The other braidings are identities. Since $c_{\psi,\sigma} \circ c_{\sigma,\psi} = \zeta^8 = -1$, all Ising type braided fusion categories have the trivial M\"{u}ger center and thus are nondegenerate.
\item For each Ising type braided fusion category, there are 2 different spherical structures, labeled by a number $\epsilon = \pm 1$. Their difference is the quantum dimension
\[
\dim(\one) = \dim(\psi) = 1 , \, \dim(\sigma) = \epsilon(\zeta^2 + \zeta^{-2}) .
\]
So there are 16 different Ising type modular tensor categories (up to equivalence) labeled by a pair $(\zeta,\epsilon)$, and only 8 of them with $\dim(\sigma) = +\sqrt 2$ are unitary. The $S$ matrix of an Ising type modular category is
\[
S =
\begin{pmatrix}
1 & 1 & \phantom{-}\epsilon(\zeta^2 + \zeta^{-2})  \\
1 & 1 & -\epsilon(\zeta^2 + \zeta^{-2}) \\
\epsilon(\zeta^2 + \zeta^{-2}) & -\epsilon(\zeta^2 + \zeta^{-2}) & 0
\end{pmatrix}.
\]
If we only consider Ising type unitary modular tensor categories, the $S$ matrix is
\[
S =
\begin{pmatrix}
1 & 1 & \phantom{-}\sqrt 2  \\
1 & 1 & -\sqrt 2 \\
\sqrt 2 & -\sqrt 2 & 0
\end{pmatrix}.
\]
\item The $T$ matrix of an Ising type modular category is
\[
T =
\begin{pmatrix}
1 & & \\
 & -1 & \\
 & & \epsilon \zeta^{-1}
\end{pmatrix}.
\]
\item The multiplicative central charge of an Ising type modular category is equal to $T_\sigma = \epsilon \zeta^{-1}$. All 8 Ising type unitary modular tensor categories have different additive central charges, so we can label them by their additive central charge $c = 1/2 , \ldots , 15/2 \pmod 8$.
\eit

\begin{rem}
The well-known Ising vertex operator algebra, which appears in the minimal model $\mathcal M(4,3)$, has central charge $1/2$. Its module category is the Ising type unitary modular tensor category with additive central charge $1/2 \pmod 8$, i.e., the one corresponding to $\zeta = \exp(-2 \pi \mathrm i / 16)$ and $\epsilon = +1$.
\end{rem}

\begin{rem} \label{rem:Ising_16-fold_way}
In a 2+1D topological superconductor (which can be viewed as an SPT order with only the fermion parity symmetry), the particle-like topological defects form a symmetric fusion category equivalent to $\svect$ (see Example \ref{expl:symmetric_classification}). These particle-like topological defects, together with vortices form a unitary modular tensor category $\CM$ which contains $\svect$.
The well-known Kitaev's 16-fold way states that $\CM$ depends on the Chern number $\nu$ modulo $16$ \cite{Kit06}. When $\nu$ is odd, $\CM$ is the Ising type unitary modular tensor category with additive central charge $\nu/2 \pmod 8$; when $\nu$ is even, $\CM$ is a pointed unitary modular tensor category (see Section \ref{sec:pointed_category}). All 16 different unitary modular tensor categories are minimal modular extensions of $\svect$ \cite{Kit06,DGNO10}.
\end{rem}

\subsubsection{Pointed modular tensor categories and metric groups} \label{sec:pointed_category}

\begin{thm} \label{thm:pointed}
Let $\CC$ be a fusion category and $x \in \CC$. Then the following statements are equivalent:
\bnu[(a)]
\item The creation morphism $b_x \colon \one \to x \otimes x^L$ and annihilation morphism $d_x \colon x^L \otimes x \to \one$ are both isomorphisms.
\item There exists an object $y \in \CC$ such that $x \otimes y \simeq \one \simeq y \otimes x$.
\item $\fpdim(x) = 1$.
\enu
\end{thm}

\pf
Clearly (a) implies (b). Suppose there exists an object $y \in \CC$ such that $x \otimes y \simeq \one \simeq y \otimes x$. Then the matrix $N_x$ of left multiplication by $x$ is a permutation matrix, whose largest non-negative eigenvalue must be $1$. Thus (b) implies (c). To prove that (c) implies (a), note that $\fpdim(x^L) = \fpdim(x)$ (because the matrix $N_{x^L}$ is the transpose of $N_x$) and thus $\fpdim(x \otimes x^L) = \fpdim(x) \fpdim(x^L) = 1$. Then the nonzero morphisms $b_x$ and $d_x$ have to be isomorphisms.
\epf

\begin{defn}
Let $\CC$ be a fusion category. An object in $\CC$ is called \emph{invertible} if it satisfies one (and hence all) of the conditions in Theorem \ref{thm:pointed}. We say $\CC$ is \emph{pointed} if every simple object in $\CC$ is invertible.
\end{defn}

Given a pointed fusion category $\CC$, its isomorphism classes of simple object form a finite group $\Irr(\CC) \eqqcolon G$ under the tensor product. Thus $\CC \simeq \vect_G$ as categories and the tensor product functor of $\CC$ is the same as $\vect_G$. Moreover, the pentagon equation implies that the associator is determined by a 3-cocycle $\omega \in Z^3(G;\Cb^\times)$ (see Example \ref{expl:vect_G_cocycle}). Thus $\CC \simeq \vect_G^\omega$ as fusion categories.

Now suppose $\CC$ is a pointed braided fusion category. As a fusion category we can assume that $\CC = \vect_G^\omega$ for some finite group $G$ and $\omega \in Z^3(G;\Cb^\times)$. The existence of braiding implies that $G$ is an abelian group. Similar to the associator, the braiding structure is determined by a function $c \colon G \times G \to \Cb^\times$ that satisfies the following two hexagon equations for any $g,h,k \in G$:
\begin{gather*}
\omega(g,h,k) c(g,hk) \omega(h,k,g) = c(g,h) \omega(h,g,k) c(g,k) , \\
\omega(g,h,k)^{-1} c(gh,k) \omega(k,g,h)^{-1} = c(h,k) \omega(g,k,h)^{-1} c(g,k) .
\end{gather*}
Such a pair of functions $(\omega,c)$ is called an \emph{abelian 3-cocycle}. This braided fusion category is also denoted by $\vect_G^{(\omega,c)}$.

\medskip
In the following we give another description of pointed braided fusion categories.

\begin{defn}
Let $G$ be an abelian group. A \emph{quadratic form} on $G$ is a map $q \colon G \to \Cb^\times$ such that $q(g) = q(g^{-1})$ and the symmetric function $b \colon G \times G \to \Cb^\times$ defined by
\[
b(g,h) \coloneqq \frac{q(gh)}{q(g)q(h)}
\]
is a \emph{bicharacter}, i.e., $b(g_1 g_2,h) = b(g_1,h) b(g_2,h)$ for all $g_1,g_2,h \in G$. The symmetric bicharacter $b$ is called the \emph{associated bicharacter} of $q$. We say $q$ is \emph{nondegenerate} if $b$ is, i.e., $b(g,h) = 1$ for all $h \in G$ if and only if $g = e$ is the unit.

A \emph{pre-metric group} is a finite abelian group $G$ equipped with a quadratic form $q$. If $q$ is nondegenerate, then $(G,q)$ is called a \emph{metric group}.
\end{defn}

Consider a pointed braided fusion category $\CC$. By the above discussion, we can assume that $\CC = \vect_G^{(\omega,c)}$. Define
\[
q(g) \coloneqq c(g,g) , \quad g \in G .
\]
It follows from the pentagon and hexagon equations that $q$ is a quadratic form and the associated bicharacter is
\[
b(g,h) \coloneqq c(g,h) c(h,g) , \quad g,h \in G .
\]
Then $(G,q)$ is called the \emph{associated pre-metric group} of $\CC$.

Conversely, two pointed braided fusion categories are equivalent if and only if their associated pre-metric groups are isomorphic \cite{JS93,Qui99,DGNO10}. In other words, for any pre-metric group $(G,q)$ there exists a unique (up to equivalence) pointed braided fusion category, denoted by $\CC(G,q)$, such that its associated pre-metric group is isomorphic to $(G,q)$.

\begin{exercise}
Let $(G,q)$ be a pre-metric group. Prove that $\CC(G,q)$ is symmetric if and only if $q \colon G \to \Cb^\times$ is a group homomorphisms satisfying $q(g) = \pm 1$ for all $g \in G$. Furthermore, in this case we have $\CC(G,q) \simeq \rep(\hat G,q)$ as symmetric fusion categories, where $\hat G \coloneqq \Hom(G,\Cb^\times)$ is the dual group of $G$.
\end{exercise}

There is a unique spherical structure on $\CC(G,q)$ such that the quantum dimension of every simple object is $1$ (see Remark \ref{rem:pseudo_unitary}). Thus $\CC(G,q)$ is a pre-modular category. Its $S$ matrix is give by
\[
S_{g,h} = b(g,h^{-1}) = b(g,h^{-1}) ,
\]
and the $T$ matrix is given by
\[
T_g = q(g) .
\]
Hence $\CC(G,q)$ is a modular tensor category if and only if $q$ is nondegenerate, i.e., $(G,q)$ is a metric group.

\begin{rem}
An anomaly-free 2d topological order $\SC$ is called an \emph{abelian topological order} if the modular tensor category $\CC$ of particle-like topological defects is pointed. This name is because the mutual-statistics, i.e., the braiding $c_{x,y}$ for simple topological defects $x,y$, are complex numbers (abelian phase), not matrices (non-abelian phase).
\end{rem}

\begin{rem}
The toric code model realizes an abelian topological order. Mathematically, $\Toric$ is a pointed modular tensor category and the corresponding metric group is $(\Zb_2 \times \Zb_2,q)$ where $q(a,b) \coloneqq (-1)^{ab} , \, a,b \in \{0,1\} = \Zb_2$. More generally, the quantum double category $\CD_A$ of a finite abelian group $A$ is a pointed modular tensor category and the corresponding metric group is $(A \times \hat A , q)$, where $\hat A \coloneqq \Hom(A,\Cb^\times)$ is the dual group of $A$ and the quadratic form $q$ is defined by $q(a,\chi) \coloneqq \chi(a)$.
\end{rem}

\begin{rem}
Physically, every anomaly-free 2d abelian topological order can be realized by a 2+1D $\mathrm U(1)$ Chern-Simons theory \cite{BW90,BW90a,BW91,Rea90}:
\be \label{eq:CS}
S = \frac{1}{4 \pi} \int K_{IJ} \, a^I \wedge \mathrm d a^J ,
\ee
where $a^I$ for $I = 1,\ldots,n$ are $\mathrm U(1)$ gauge fields ($1$-forms), and the $n$-by-$n$ matrix $K = (K_{IJ})$ satisfies the following conditions:
\bnu
\item All components of $K$ are integers, and the diagonal components of $K$ are even.
\item $K$ is symmetric.
\item The determinant of $K$ is non-zero.
\enu
Mathematically, $K$ can be equivalently replaced by a nondegenerate even integral lattice $L$. There is a metric group $(G,q)$ associated to such a matrix $K$. First, the matrix $K$ can be viewed as a homomorphism $K \colon \Zb^n \to \Zb^n$, and we define $G \coloneqq \coker(K) = \Zb^n / \im(K)$, called the \emph{discriminant group} of $K$. In other words, $G$ is the group of equivalence classes of integral column vectors where the equivalence relation is generated by $v \sim v + K l$. In the language of lattices, $G$ is the quotient $L^\vee/L$, where $L^\vee \coloneqq \Hom_\Zb(L,\Zb)$ is the dual of $L$. The quadratic form $q$ is defined by
\[
q([v]) \coloneqq \exp(\pi \mathrm i \cdot v^{\mathrm T} K^{-1} v), \quad v \in \Zb^n .
\]
The associated bicharacter is
\[
b([v],[w]) = \exp(2 \pi \mathrm i \cdot v^{\mathrm T} K^{-1} w) , \quad v,w \in \Zb^n .
\]
It is not hard to verify that $q$ is nondegenerate. The category of particle-like topological defects of the topological order realized by \eqref{eq:CS} is equivalent to the pointed modular tensor category $\CC(G,q)$ \cite{WZ92,BW90a}.

Some properties of $\CC(G,q)$ can be obtained directly from the matrix $K$. The number of isomorphism classes of simple objects is $\lvert G \rvert = \lvert \det K \rvert$. The additive central charge of $\CC(G,q)$ is equal to the number of positive eigenvalues of $K$ minus the number of negative eigenvalues modulo $8$ \cite{Mil74} (see also \cite[Appendix 4]{MH73} for a proof).

Different $K$ matrices are said to be equivalent if they realize physically equivalent theories. The equivalence relation is generated by:
\bnu[(a)]
\item $K \sim PKP^{-1}$ where $P \in \mathrm{GL}(n,\Zb)$. This relation means the basis change of gauge fields is irrelevant.
\item $K \sim K \oplus \sigma_x$. The matrix $\sigma_x$ realizes the trivial 2d topological order. So this relation means the trivial 2d topological order is the unit under stacking.
\enu

Every metric group can be realized as the discriminant group of a $K$ matrix \cite{Wal63}. As an example, the modular tensor category $\Toric$ can be constructed from
\[
K = \begin{pmatrix}0 & 2 \\ 2 & 0\end{pmatrix} .
\]
In general, different $K$ matrices may have isomorphic discriminant groups.
\end{rem}

\subsubsection{Categories associated to the Lie algebra \texorpdfstring{$\mathfrak{sl}_2$}{sl2}}

Let us recall the basic representation theory of the Lie algebra $\mathfrak{sl}_2 \coloneqq \mathfrak{sl}(2,\Cb)$. The category $\rep(\mathfrak{sl}_2)$ of finite-dimensional representations of $\mathfrak{sl}_2$ is almost a fusion category, except that there are infinite simple objects up to isomorphism. Indeed, for any $n \in \Nb$, there exists a unique (up to isomorphism) irreducible representation of dimension $(n+1)$, denoted by $V_n$. Usually $V_n$ is called the spin-$(n/2)$ representation. The tensor product of two irreducible representations is decomposed as the direct sum of irreducible representations by the Clebsch-Gordan rule:
\[
V_n \otimes V_m = \bigoplus_{i=0}^{\min\{n,m\}} V_{n+m-2i} = V_{\vert n - m \vert} \oplus V_{\vert n - m \vert + 2} \oplus \cdots \oplus V_{n+m} .
\]

We may expect to `truncate' $\rep(\mathfrak{sl}_2)$ to get a fusion category. One way is to `deform' the enveloping algebra of $\mathfrak{sl}_2$ and consider certain representations of the deformed algebra.

Let $k \in \Nb$ be a non-negative integer (called the level). For every integer $a \in \Zb$ such that the greatest common divisor of $a$ and $(k+2)$ is $1$, let $q = \mathrm e^{a \pi \mathrm i / (k+2)}$ (i.e., $q^2$ is a primitive $(k+2)$-th root of unity) and $t$ be a square root of $q$. Then there is an associated pre-modular category denoted by $\CC(\mathfrak{sl}_2,q,t)$ \cite{RT91}. We list some data in the following:
\bit
\item There are $(k+1)$ simple objects (up to isomorphism) denoted by $V_0,\ldots,V_k$.
\item The fusion rule is given by the truncated Clebsch-Gordan rule:
\[
V_n \otimes V_m = \bigoplus_{i=\max\{n+m-k,0\}}^{\min\{n,m\}} V_{n+m-2i} = \begin{cases} V_{\vert n - m \vert} \oplus V_{\vert n - m \vert + 2} \oplus \cdots \oplus V_{n+m} , & n+m \leq k ; \\ V_{\vert n - m \vert} \oplus V_{\vert n - m \vert + 2} \oplus \cdots \oplus V_{2k-(n+m)} , & n+m>k . \end{cases}
\]
In other words, the fusion rule is the same as the usual Clebsch-Gordan rule if $n+m \leq k$; when $n+m>k$, the undefined terms ($V_i$ with $i>k$) are deleted and also their mirror images with respect to the point $(k+1)$. The tensor unit is $V_0$ and each $V_n$ is self-dual.
\item Define 
\[
[n]_q \coloneqq \frac{q^n - q^{-n}}{q - q^{-1}} = q^{n-1} + q^{n-3} + \cdots + q^{-n+1} .
\]
The Frobenius-Perron dimension of $V_n$ is $\fpdim(V_n) = [n+1]_{q_1}$, where $q_1 \coloneqq \mathrm e^{\pi \mathrm i / (k+2)}$. The Frobenius-Perron dimension of the fusion category $\CC(\mathfrak{sl}_2,q,t)$ is
\[
\fpdim(\CC(\mathfrak{sl}_2,q,t)) = \frac{k+2}{2 \sin^2(\frac{\pi}{k+2})} .
\]
The quantum dimension of $V_n$ is $\dim(V_n) = [n+1]_q$, and the quantum dimension of $\CC(\mathfrak{sl}_2,q)$ is
\[
\dim(\CC(\mathfrak{sl}_2,q,t)) = \frac{k+2}{2 \sin^2(\frac{a \pi}{k+2})} .
\]
\item The $S$ matrix is
\[
S_{nm} \coloneqq S_{V_n,V_m} = [(n+1)(m+1)]_q = \frac{\sin(\frac{a (n+1)(m+1) \pi}{k+2})}{\sin(\frac{a \pi}{k+2})} .
\]
\item The $T$ matrix is
\[
T_n \coloneqq T_{V_n} = t^{n(n+2)} .
\]
\eit
The underlying fusion category of $\CC(\mathfrak{sl}_2,q,t)$ only depends on $q$ but not $t$. The pre-modular category $\CC(\mathfrak{sl}_2,q,t)$ is modular if and only if $a$ is odd \cite{Bru00}. In particular, when $k$ is even, $a$ must be odd and hence $\CC(\mathfrak{sl}_2,q,t)$ is modular.

\begin{rem}
Let us prove that $\CC(\mathfrak{sl}_2,q,t)$ is modular if and only if $a$ is odd by direct computations. We have
\begin{align*}
\sin \bigl( \frac{a\pi}{k+2} \bigr)^2 \sum_{l=0}^k S_{nl} S_{ml} & = \sum_{l=1}^{k+1} \sin \bigl( \frac{a(n+1)l\pi}{k+2} \bigr) \sin \bigl( \frac{a(m+1)l\pi}{k+2} \bigr) \\
& = -\frac12 \sum_{l=0}^{k+1} \biggl( \cos \bigl( \frac{a(n+m+2)l\pi}{k+2} \bigr) - \cos \bigl( \frac{a(n-m)l\pi}{k+2} \bigr) \biggr) .
\end{align*}
Note that for $\theta = a r \pi / (k+2)$ where $r \in \Zb$, we have
\begin{align*}
\sum_{l=0}^{k+1} \cos(l \theta) & = \mathrm{Re} \bigl( \sum_{l=0}^{k+1} \mathrm e^{\mathrm i l \theta} \bigr) \\
& =
\begin{cases}
k+2 , & \mathrm e^{\mathrm i \theta} = 1 , \\
\mathrm{Re} \bigl( \frac{1-\mathrm e^{\mathrm i (k+2) \theta}}{1-\mathrm e^{\mathrm i \theta}} \bigr) , & \mathrm e^{\mathrm i \theta} \neq 1 ,
\end{cases} \\
& =
\begin{cases}
k+1 , & \mathrm e^{\mathrm i \theta} = 1 , \\
0 , & \mathrm e^{\mathrm i \theta} \neq 1 , \, ar \text{ even} , \\
\mathrm{Re} \bigl( \frac{2}{1-\mathrm e^{\mathrm i \theta}} \bigr) = 1 , & \mathrm e^{\mathrm i \theta} \neq 1 , \, ar \text{ odd} .
\end{cases}
\end{align*}
Thus when $a$ is odd we have
\[
\sum_{l=0}^k S_{nl} S_{ml} = \sin \bigl( \frac{a\pi}{k+2} \bigr)^{-2} \cdot \frac{k+2}{2} \delta_{n,m} = \dim(\CC(\mathfrak{sl}_2,q,t)) \delta_{n,m} .
\]
This orthogonality relation implies the nondegeneracy of the $S$ matrix. When $a$ is even we have $S_{n,m} = -S_{n,k-m}$, thus the $S$ matrix is degenerate.
\end{rem}


\begin{rem} \label{rem:pointed_sl2}
Note that $V_k \otimes V_k = V_0$ and the only invertible objects in $\CC(\mathfrak{sl}_2,q)$ are $V_0$ and $V_k$ (up to isomorphism). The fusion subcategory generated by $V_0$ and $V_k$ is denoted by $\CC(\mathfrak{sl}_2,q,t)^\times$. We have
\[
\dim(V_k) = [k+1]_q = (-1)^{a+1} ,
\]
and
\[
T_k = t^{k(k+2)} .
\]
Hence, when $k$ is even, the pre-modular category $\CC(\mathfrak{sl}_2,q,t)^\times$ only depends on $q$ but not $t$. It is symmetric and equipped with the canonical ribbon structure (see Example \ref{expl:symmetric_ribbon}). When $k$ is odd, the pre-modular category $\CC(\mathfrak{sl}_2,q,t)^\times$ is modular if and only if a is odd.
\end{rem}

Let us consider the fusion subcategory of $\CC(\mathfrak{sl}_2,q,t)$ generated by integer-spin representations $\{V_n \mid n \text{ even}\}$. We denote this fusion subcategory by $\CC(\mathfrak{sl}_2,q,t)_+$. It only depends on $q$ but not $t$. When $k$ is odd, $\CC(\mathfrak{sl}_2,q,t)_+$ is modular. When $k$ is even, the pre-modular category $\CC(\mathfrak{sl}_2,q,t)_+$ is not modular because $S_{n,m} = S_{n,k-m}$ for even $n,m$.

\begin{rem}
Let us prove that $\CC(\mathfrak{sl}_2,q,t)_+$ is modular when $k$ is odd by direct computations. For $n,m$ even, we have
\begin{align*}
\sin \bigl( \frac{a\pi}{k+2} \bigr)^2 \sum_{l=0}^{(k-1)/2} S_{n,2l} S_{m,2l} & = \sum_{l=0}^{(k-1)/2} \sin \bigl( \frac{a(n+1)(2l+1)\pi}{k+2} \bigr) \sin \bigl( \frac{a(m+1)(2l+1)\pi}{k+2} \bigr) \\
& = -\frac12 \sum_{l=0}^{(k-1)/2} \biggl( \cos \bigl( \frac{a(n+m+2)(2l+1)\pi}{k+2} \bigr) - \cos \bigl( \frac{a(n-m)(2l+1)\pi}{k+2} \bigr) \biggr) .
\end{align*}
Note that for $\theta = a r \pi / (k+2)$ where $r \in 2\Zb$, we have
\begin{align*}
\sum_{l=0}^{(k-1)/2} \cos((2l+1) \theta) & = \mathrm{Re} \bigl( \sum_{l=0}^{(k-1)/2} \mathrm e^{\mathrm i (2l+1) \theta} \bigr) \\
& =
\begin{cases}
\mathrm{Re} \bigl( \mathrm e^{\mathrm i \theta} \cdot (k+1)/2  \bigr) , & \mathrm e^{\mathrm i 2\theta} = 1 , \\
\mathrm{Re} \bigl( \frac{\mathrm e^{\mathrm i \theta}-\mathrm e^{\mathrm i (k+2) \theta}}{1-\mathrm e^{\mathrm i 2 \theta}} \bigr) , & \mathrm e^{\mathrm i 2 \theta} \neq 1 ,
\end{cases} \\
& =
\begin{cases}
(k+1)/2 , & r/(k+2) \in \Zb , \\
\mathrm{Re} \bigl( \frac{-1}{1+\mathrm e^{\mathrm i \theta}} \bigr) = -1/2 , & r/(k+2) \notin \Zb .
\end{cases}
\end{align*}
Thus we have
\[
\sum_{l=0}^{(k-1)/2} S_{n,2l} S_{m,2l} = \sin \bigl( \frac{a\pi}{k+2} \bigr)^{-2} \cdot \frac{k+2}{4} \delta_{n,m} = \frac12 \dim(\CC(\mathfrak{sl}_2,q,t)) \delta_{n,m} .
\]
Also we know that $2 \dim(\CC(\mathfrak{sl}_2,q,t))_+ = \dim(\CC(\mathfrak{sl}_2,q,t))$.
\end{rem}

\begin{rem}
When $k$ is odd, we also have $\CC(\mathfrak{sl}_2,q,t) \simeq \CC(\mathfrak{sl}_2,q,t)_+ \boxtimes \CC(\mathfrak{sl}_2,q,t)^\times$.
\end{rem}

\begin{expl}
Let us consider the case $k = 3$. In this case the fusion subcategory $\CC(\mathfrak{sl}_2,q,t)_+$ has two simple objects (up to isomorphism) $\one \coloneqq V_0 , X \coloneqq V_2$. The fusion rule is given by $X \otimes X = \one \oplus X$. It follows that the Frobenius-Perron dimension of $X$ is $\fpdim(X) = (1 + \sqrt 5) / 2$.

When $q = \mathrm e^{\pi \mathrm i / 5}$, this fusion subcategory is unitary and the quantum dimension of $X$ is $1 + 2 \cos(\frac{2\pi}{5}) = (1 + \sqrt 5) / 2$. When $q = \mathrm e^{2 \pi \mathrm i / 5}$, this fusion subcategory is not unitary and the quantum dimension of $X$ is $1 + 2 \cos(\frac{4 \pi}{5}) = (1 - \sqrt 5) / 2$. Both of them are modular categories. 
\end{expl}

We are mainly interested in the case that $q = \mathrm e^{\pi \mathrm i / (k+2)}$ and $t = \mathrm e^{\pi \mathrm i / 2(k+2)}$. In this case $\CC(\mathfrak{sl}_2,q,t)$ is a unitary modular tensor category and also denoted by $\CC(\mathfrak{sl}_2,k)$. It is also equivalent to the category of integrable modules of level $k$ over the affine Lie algebra $\widehat{\mathfrak{sl}_2}$ \cite{BK01}. Its additive central charge is $3 k / (k+2) \pmod 8$. Some examples are listed below:
\bit
\item For $k = 0$, $\CC(\mathfrak{sl}_2,0)$ is the trivial unitary modular tensor category $\hilb$.
\item For $k = 1$, $\CC(\mathfrak{sl}_2,1)$ is a pointed modular category, and the associated metric group is $(\Zb_2,\phi)$ where $\phi(n) = \mathrm i^n$ for $n \in \Zb_2$.
\item For $k = 2$, $\CC(\mathfrak{sl}_2,2)$ is an Ising type unitary modular tensor category with $\one \coloneqq V_0 , \sigma \coloneqq V_1 , \psi \coloneqq V_2$. The additive central charge is $3/2 \pmod 8$. This Ising type unitary modular tensor category corresponds to $\zeta = \exp(10 \pi \mathrm i/16)$ and $\epsilon = -1$.
\eit

\begin{rem}
The above constructions can also be generalized to any finite-dimensional simple Lie algebras. See the appendix of \cite{TW97} for references of constructions.
\end{rem}

\begin{rem}
The $\mathrm{SU}(2)$ Chern-Simons theory
\[
S = \frac{k}{4 \pi} \int \tr \bigl(A \wedge \mathrm d A + \frac23 \, A \wedge A \wedge A \bigr)
\]
realizes a 2d topological order (2+1D topological quantum field theory \cite{Wit89}) and the corresponding modular tensor category is $\CC(\mathfrak{sl}_2,k)$ \cite{RT91}. 
\end{rem}

\begin{rem}
The modular tensor category $\CC(\mathfrak{sl}_2,k)$ can also be obtained from the vertex operator algebra (VOA) associated to the affine Lie algebra $\hat{\mathfrak{sl}}_2$ at level $k$ \cite{Hua08,Hua08a}. More generally, it was proved by Huang that the category of modules over a VOA, which satisfies certain properties (i.e. a rational VOA), is a modular tensor category \cite{Hua08,Hua08a}. 
\end{rem}

\subsection{Chiral central charges}

Besides the topological defects, there is another observable of an anomaly-free stable 2d topological order, called the chiral central charge.

An anomaly-free stable 2d topological order $\SC$ can have gapless 1d boundaries. The gapless modes on a boundary can carry energy and leads to thermal transport. The energy current $I$ along the boundary is proportional to the square of the temperature $T$ (which is assumed to be much smaller than the energe gap) \cite{KF97,CHZ02}:
\[
I = \frac{\pi}{12} c T^2 .
\]
This is called the thermal Hall effect. The coefficient $c$ is called the \emph{chiral central charge}. Intuitively, it counts the number of chiral modes minus the number of anti-chiral modes. Surprisingly, it only depends on the bulk topological order $\SC$.

There is another interpretation of the chiral central charge. Usually the low energy effective theory of a gapless 1d boundary of $\SC$ is a conformal field theory (CFT). A CFT gives a representation of the Virasoro algebra and thus determines a chiral central charge $c$:
\[
[L_m,L_n] = (m+n) L_{m-n} + \frac{c}{12}(m^3-m) \delta_{m+n,0} .
\]
All gapless 1d boundaries of $\SC$ have the same chiral central charge \cite{KZ20,KZ21}. Therefore, the chiral central charge is an observable of the bulk topological order $\SC$ and we denote it by $c(\SC)$. Moreover, the additive central charge $c(\CC)$ of the modular tensor category $\CC$ of particle-like topological defects of $\CC$ is equal to $c(\SC)$ modulo $8$ \cite{FG90,Reh90,KZ20,KZ21}.

It is believed that an anomaly-free 2d topological order $\SC$ is determined by the pair $(\CC,c(\SC))$, where $\CC$ is the unitary modular tensor category of particle-like topological defects of $\SC$ and $c(\SC)$ is the chiral central charge of $\SC$.

\begin{expl}
Consider an abelian topological order realized by a $\mathrm U(1)$ Chern-Simons theory \eqref{eq:CS}. Its chiral central charge is equal to the number of positive eigenvalues of the $K$ matrix minus the number of negative eigenvalues.
\end{expl}

\begin{rem}
An $n$d anomaly-free topological order $\SC_n$ is called \emph{invertible} if it is invertible under the stacking operation \cite{KW14}, i.e., there exists an $n$d topological $\SD_n$ such that the stacking $\SC_n \boxtimes \SD_n = \tTO_n$ is the trivial $n$d topological order. The invertible topological orders (or invertible topological field theories), as well as symmetry protected topological (SPT) orders, can be classified by using methods in algebraic topology (for example, see \cite{FT14,Fre14,FH21,Kap14b,Kap14a,KTTW15,Kit13,Kit15,GJF19a}).
\end{rem}

\begin{rem}
In 2d, it is believed that a topological order $\SC$ is invertible if and only if there is no nontrivial particle-like topological defect, i.e., $\CC \simeq \vect$. There is a nontrivial invertible 2d topological order called the $E_8$ topological order \cite{Kit06,Kit11}. It has a gapless boundary described by the so-called $E_8$ CFT, whose chiral central charge is $8$.

If two topological orders $\SC,\SD$ have the same unitary modular tensor category of particle-like topological defects $\CC = \CD$, then it is believed that they are only differed by several copies of the $E_8$ topological order. More precisely, the stacking of $\SC$ and $n$ copies of the $E_8$ topological order is $\SD$, where $n = (c(\SD) - c(\SC))/8 \in \Zb$.
\end{rem}

\begin{rem}
The $E_8$ topological order can be realized by a 2+1D $\mathrm U(1)$ Chern-Simons theory \eqref{eq:CS} with the $K$ matrix give by the Cartan matrix of the $E_8$ Lie algebra (or equivalently, the $E_8$ root lattice). Conversely, an invertible 2d topological order is an abelian topological order, thus can be realized by a $\mathrm U(1)$ Chern-Simons theory with $\lvert \det(K) \rvert = 1$. Such $K$ matrices (or equivalently, unimodular even lattices), up to equivalence, is the direct sum of several copies of the $E_8$ Cartan matrix (or equivalently, the $E_8$ root lattice) \cite{Ser73}.
\end{rem}

%% file: 1d.tex

\section{Topological orders in 1d} \label{sec:1d}

\subsection{Boundary theories of the toric code model} \label{sec:toric_code_boundary}

\begin{figure}[htbp]
\centering
\begin{tikzpicture}[scale=1.0]
\draw[help lines,step=1.0] (-2.5,-1.99) grid (2.5,2);
\draw[help lines,dashed] (-2.5,-2)--(2.5,-2) ;

\fill[e_ext] (1,2) circle (0.07) node[above,black] {$v$} ;
\begin{scope}
\clip (-1,-1.98) rectangle (0,-1) ;
\draw[help lines,fill=m_ext] (-1,-2) rectangle (0,-1) node[midway,black] {$p$} ;
\end{scope}

\node[scale=0.8] at (0.5,2) {$1$} ;
\node[scale=0.8] at (1.5,2) {$2$} ;
\node[scale=0.8] at (1,1.5) {$3$} ;
\node[scale=0.8] at (-0.5,-1) {$5$} ;
\node[scale=0.8] at (-1,-1.5) {$6$} ;
\node[scale=0.8] at (0,-1.5) {$7$} ;
\end{tikzpicture}
\caption{the smooth and rough boundary of the toric code model}
\label{fig:toric_code_boundary}
\end{figure}
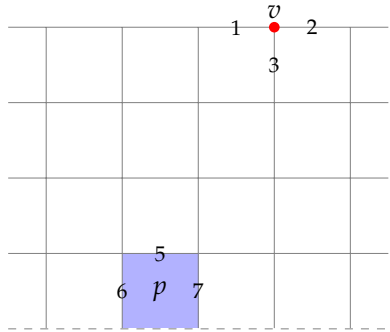

There are two different gapped boundaries of the toric code model, which was proposed by Bravyi and Kitaev \cite{BK98}. As depicted in Figure \ref{fig:toric_code_boundary}, the top one is called the smooth boundary, while the bottom one is called the rough boundary. The Hilbert space of the model is the same as the original toric code model, i.e., there is a spin-$1/2$ on each edge. But there is no spin on the dashed edges on the rough boundary. The $A_v$ and $B_p$ operators in the bulk are the same as in the original toric code model, and the $A_v$ and $B_p$ operators on the boundaries are defined to be, for example, $A_v = \sigma_x^1 \sigma_x^2 \sigma_x^3$ and $B_p = \sigma_z^5 \sigma_z^6 \sigma_z^7$ in Figure \ref{fig:toric_code_boundary}.

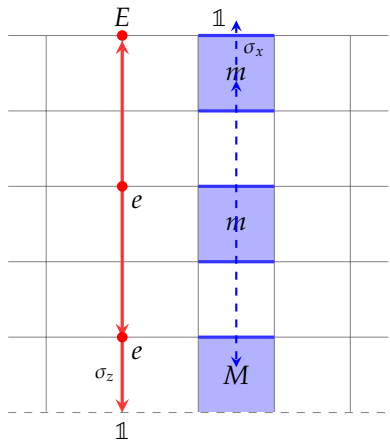
\begin{figure}[htbp]
\centering
\begin{tikzpicture}[scale=1.0]
\draw[help lines,step=1.0] (-2.5,-1.99) grid (2.5,3);
\draw[help lines,dashed] (-2.5,-2)--(2.5,-2) ;
\draw[-stealth,e_str] (-1,1)--(-1,2.93) ;
\draw[-stealth,e_str] (-1,1)--(-1,-1) ;
\draw[-stealth,e_str] (-1,-1)--(-1,-2) ;
\fill[e_ext] (-1,1) circle (0.07) node[below right,black] {$e$} ;
\fill[e_ext] (-1,-1) circle (0.07) node[below right,black] {$e$} ;
\fill[e_ext] (-1,3) circle (0.07) node[above,black] {$E$} ;
\node[below] at (-1,-2) {$\one$} ;
\node[left,link_label] at (-1,-1.5) {$\sigma_z$} ;
\draw[help lines,fill=m_ext] (0,0) rectangle (1,1) node[midway,black] {$m$} ;
\draw[help lines,fill=m_ext] (0,2) rectangle (1,3) node[midway,black] {$m$} ;
\begin{scope}
\clip (0,-1.98) rectangle (1,-1) ;
\draw[help lines,fill=m_ext] (0,-2) rectangle (1,-1) node[midway,black] {$M$} ;
\end{scope}
\foreach \y in {-1,...,3}
	\draw[m_str] (0,\y)--(1,\y) ;
\draw[-stealth,m_dual_str] (0.5,0.5)--(0.5,2.4) ;
\draw[-stealth,m_dual_str] (0.5,2.4)--(0.5,3.2) ;
\draw[-stealth,m_dual_str] (0.5,0.5)--(0.5,-1.4) ;
\node[above left] at (0.5,3) {$\one$} ;
\node[below right,link_label] at (0.5,3) {$\sigma_x$} ;
\end{tikzpicture}
\caption{topological excitations on two boundaries}
\label{fig:ext_boundary}
\end{figure}

In the presence of two types of boundaries, all $A_v$ and $B_p$ operators are still mutually commutative. Thus we can easily find the topological excitations as in the original toric code model. If there is an $m$ particle near the smooth boundary, it can be annihilated to the ground state by a $\sigma_x$ operator, which is a local operator on the smooth boundary. But if there is an $e$ particle near the smooth boundary, it cannot be annihilated by local operators on the boundary (see Figure \ref{fig:ext_boundary}). It follows that there are two simple topological excitations on the smooth boundary: the trivial one $\one$ and the nontrivial one, which is denoted by $E$.

Similarly, on the rough boundary one can use a $\sigma_z$ to annihilate an $e$ particle; an $m$ particle stays at the boundary and is denoted by $M$ (see Figure \ref{fig:ext_boundary}).

\begin{rem}
The smooth boundary is also called the $m$-condensed boundary while the rough boundary is called the $e$-condensed boundary, because they can be obtained by condensing $m$ particles or $e$ particles respectively. Mathematically, these condensation processes are described by Lagrangian algebras $\one \oplus m$ and $\one \oplus e$ in the modular tensor category $\toric$, respectively. More generally, all gapped boundaries of a 2d topological order can be obtained by anyon condensation \cite{Kon14}.
\end{rem}


To compare the topological defects in the bulk and on the boundaries, we can move bulk topological defects to boundaries by using string operators. Such operations define two maps: the bulk-to-smooth-boundary map is
\be \label{eq:bulk-to-smooth-boundary_map}
\one,m \mapsto \one , \quad e,f \mapsto E ,
\ee
and the bulk-to-rough-boundary map is
\be \label{eq:bulk-to-rough-boundary_map}
\one,e \mapsto \one , \quad m,f \mapsto M .
\ee 

It seems that moving bulk topological defects to boundaries loses information. For example, both $\one$ and $e$ become trivial after moving to the rough boundary. Is it possible to distinguish $e$ from $\one$ by using the data on the rough boundary?

\begin{figure}[htbp]
\[
\begin{array}{c}
\begin{tikzpicture}[scale=1.0]
\draw[help lines,step=1.0] (-2.5,0.01) grid (3.5,2.5) ;
\draw[help lines,dashed] (-2.5,0)--(3.5,0) ;

\begin{scope}
\clip (0,0.02) rectangle (1,1) ;
\draw[help lines,fill=m_ext] (0,0) rectangle (1,1) node[midway,black] {$M$} ;
\end{scope}
\draw[->-=0.6,e_str] (2,0)--(2,1) ;
\draw[->-=0.6,e_str] (-1,1)--(-1,0) ;
\draw[-stealth,e_str] (2,1)--(2,2)--(-1,2)--(-1,1.07) ;
\fill[e_ext] (2,1) circle (0.07) node[below left,black] {$e$} ;
\fill[e_ext] (-1,1) circle (0.07) node[below right,black] {$e$} ;

\foreach \x in {-0.5,0.5,1.5}
	\node[above,link_label] at (\x,2) {$\sigma_z$} ;
\foreach \y in {0.5,1.5} {
	\node[right,link_label] at (2,\y) {$\sigma_z$} ;
	\node[left,link_label] at (-1,\y) {$\sigma_z$} ;
}
\end{tikzpicture}
\end{array}
=
\begin{array}{c}
\begin{tikzpicture}[scale=1.0]
\draw[help lines,step=1.0] (-2.5,0.01) grid (3.5,2.5) ;
\draw[help lines,dashed] (-2.5,0)--(3.5,0) ;

\begin{scope}
\clip (0,0.02) rectangle (1,1) ;
\draw[help lines,fill=m_ext] (0,0) rectangle (1,1) ;
\end{scope}
\draw[->-=0.6,e_str] (2,0)--(2,1) ;
\draw[->-=0.6,e_str] (-1,1)--(-1,0) ;
\draw[-stealth,e_str] (2,1)--(2,2)--(-1,2)--(-1,1.07) ;
\fill[e_ext] (2,1) circle (0.07) ;
\fill[e_ext] (-1,1) circle (0.07) ;

\foreach \x in {-0.5,0.5,1.5}
	\foreach \y in {0.5,1.5}
		\node[link_label] at (\x,\y) {$B_p$} ;
\end{tikzpicture}
\end{array}
\]
\caption{the half-braiding of $e$ and $M$}
\label{fig:half_braiding_e_M}
\end{figure}
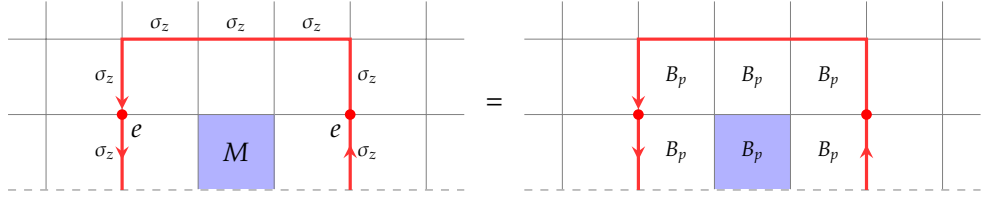

Consider the following configuration as depicted in Figure \ref{fig:half_braiding_e_M}. There is an $M$ particle on the rough boundary and an $e$ particle on the right hand side of $M$, which is created by a local operator $\sigma_z$ near the rough boundary. Then we use an $e$ string operator to move it from the right hand side of $M$ to the left. Finally we annihilate it by a $\sigma_z$ operator near the rough boundary.

This process can be viewed as happening on the rough boundary, thus defines an instanton from $M = M \otimes \one$ to $M = \one \otimes M$. Let us compute this instanton explicitly. As an operator, this process is an $e$ string operator, which is precisely equal to the product of all $B_p$ operators encircled by the path of $e$. So it acts on the $M$ particle as $-1$. In other words, this instanton is
\[
M \otimes \one = M \xrightarrow{-1} M = \one \otimes M .
\]
We say the half-braiding of $e$ with $M$ is $-1$. Similarly, we can consider the case that there is no $M$ particle in the region enclosed by the path of $e$. This gives the half-braiding of $e$ with $\one$,  which should be $+1$.

We list these properties of $e$ particles in the following:
\[
e = (\one, \one \otimes \one \xrightarrow{+1} \one \otimes \one , M \otimes \one \xrightarrow{-1} \one \otimes M) .
\]
\bnu[(a)]
\item The first component $\one$ means an $e$ particle becomes $\one$ after moving to the rough boundary.
\item The second component means the half-braiding of $e$ with $\one$ is $+1$.
\item The third component means the half-braiding of $e$ with $M$ is $-1$.
\enu
Here the equal sign means we label $e$ by these data. Note that this triple only involves the data on the rough boundary.

Similarly, all particle-like topological defects in the bulk can be labeled in this way \cite{HZZKWY20}:
\be \label{eq:toric_code_half_braiding_M}
\begin{cases}
\one = (\one, \one \otimes \one \xrightarrow{+1} \one \otimes \one , M \otimes \one \xrightarrow{+1} \one \otimes M) , \\
e = (\one, \one \otimes \one \xrightarrow{+1} \one \otimes \one , M \otimes \one \xrightarrow{-1} \one \otimes M) , \\
m = (M, \one \otimes M \xrightarrow{+1} M \otimes \one , M \otimes M \xrightarrow{+1} M \otimes M) , \\
f = (M, \one \otimes M \xrightarrow{+1} M \otimes \one , M \otimes M \xrightarrow{-1} M \otimes M) .
\end{cases}
\ee
Although different bulk topological defects may become the same after moving to the boundary, we can still distinguish them by their half-braidings. In this sense, we recover the bulk topological order from the rough boundary!

We can also use the smooth boundary to recover the bulk topological order:
\be \label{eq:toric_code_half_braiding_E}
\begin{cases}
\one = (\one, \one \otimes \one \xrightarrow{+1} \one \otimes \one , E \otimes \one \xrightarrow{+1} \one \otimes E) , \\
m = (\one, \one \otimes \one \xrightarrow{+1} \one \otimes \one , E \otimes \one \xrightarrow{-1} \one \otimes E) , \\
e = (E, \one \otimes E \xrightarrow{+1} E \otimes \one , E \otimes E \xrightarrow{+1} E \otimes E) , \\
f = (E, \one \otimes E \xrightarrow{+1} E \otimes \one , E \otimes E \xrightarrow{-1} E \otimes E) .
\end{cases}
\ee

These examples suggest that an anomaly-free stable 2d topological order can be recovered from a gapped boundary.

\begin{rem} \label{rem:braiding_gauge_toric_code}
With respect to the rough boundary, we have $c_{e,m} = +1$ and $c_{m,e} = -1$ because the half-braiding of $e$ with $M$ is $-1$ and the half-braiding of $m$ with $\one$ is $+1$. Similarly, with respect to the smooth boundary, we have $c_{e,m} = -1$ and $c_{m,e} = +1$.
\end{rem}

\begin{rem}
A topological phase transition between two boundaries of the toric code model was studied in \cite{CJKYZ20}.
\end{rem}

\subsection{Gapped boundaries of 2d topological orders}

Most structures and properties developed in Section \ref{sec:structure_2d} are still valid in 1d, except the braiding and ribbon structures. So we conclude that the topological defects of a stable 1d topological order $\SA$ form a (unitary) fusion category $\CA$. By definition, $\CA$ is the topological skeleton of $\SA$.

\begin{rem}
In 1d we are also interested in unstable topological orders. Without the stability, the topological skeleton of a 1d topological order is a (unitary) multi-fusion category, whose tensor unit $\one$ is potentially non-simple: $\one = \medoplus_i \one_i$. Physically it means that the ground state subspace is degenerate, and each simple direct summand $\one_i$ is generated by a ground state. Unstable topological orders and multi-fusion categories usually appear in dimensional reduction processes, which are discussed later.
\end{rem}

\subsubsection{Bulk-to-boundary map as a monoidal functor}

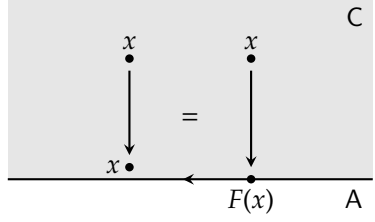
\begin{figure}[htbp]
\centering
\begin{tikzpicture}[scale=0.8]
\fill[gray!20] (-3,0) rectangle (3,3) node[at end,below left,black] {$\SC$} ;
\draw[->-=0.52,thick] (3,0)--(-3,0) node[at start,below left] {$\SA$} ;

\fill (-1,0.2) circle (0.07) node[left] {$x$} ;
\fill (-1,2) circle (0.07) node[above] {$x$} ;
\fill (1,0) circle (0.07) node[below] {$F(x)$} ;
\fill (1,2) circle (0.07) node[above] {$x$} ;

\node at (0,1) {$=$} ;

\draw[-stealth,thick] (-1,1.8) to (-1,0.4) ;
\draw[-stealth,thick] (1,1.8) to (1,0.2) ;
\end{tikzpicture}
\caption{the bulk-to-boundary map $F$}
\label{fig:central_functor}
\end{figure}

Consider an anomaly-free stable 2d topological order $\SC$ and a stable gapped 1d boundary $\SA$, whose particle-like topological defects form a (unitary) modular tensor category $\CC$ and a (unitary) fusion category $\CA$, respectively. Let $x \in \CC$ be a bulk topological defect. If we move it to the boundary, it becomes a boundary topological defect, denoted by $F(x) \in \CA$ (see Figure \ref{fig:central_functor}). This defines a map $F$ from the set of bulk topological defects (i.e., objects of $\CC$) to the set of boundary topological defects (i.e., objects of $\CA$).

\begin{rem}
The boundary of $\SC$ is not a geometric boundary (the bottom straight line in Figure \ref{fig:central_functor}), but the neighborhood of this line. The `width' of the boundary depends on the length scale and is not important at a fixed point of the renormalization flow. When we say ``moving $x$ to the boundary'', what we mean is ``moving $x$ very close to the boundary'', and how $x$ is close to the boundary depends on the length scale. So the boundary topological defect $F(x)$ is indeed a topological defect $x$ near the boundary.
\end{rem}

We do not only move topological defects, but also the world lines and instantons attached to them. Hence $F$ is indeed a functor $F \colon \CC \to \CA$. Moreover, $F$ should preserve the fusion of topological defects, i.e., two different ways of moving two topological defects to the boundary should give the same result. In other words, $F(x) \otimes F(y)$ should be the same as $F(x \otimes y)$. We do not require that they are precisely equal, but expect an isomorphism $F^2_{x,y} \colon F(x) \otimes F(y) \to F(x \otimes y)$, which is a part of the structure of the bulk-to-boundary-map. Moreover, $F^2_{x,y}$ should satisfy some obvious coherence conditions. These data and properties of the bulk-to-boundary map can be summarized to the following definition.

%
%
%
%

\begin{defn} \label{defn:monoidal_functor}
Let $\CC,\CD$ be monoidal categories. A \emph{monoidal functor} $F \colon \CC \to \CD$ consists of the following data:
\bit
\item a functor $F \colon \CC \to \CD$;
\item a natural isomorphism $F^2_{x,y} \colon F(x) \otimes F(y) \to F(x \otimes y)$;
\item an isomorphism $F^0 \colon \one_\CD \to F(\one_\CC)$;
\eit
and these data satisfy the following conditions:
\bnu
\item For any $x,y,z \in \CC$, the following diagram commutes:
\be \label{diag:monoidal_functor}
\begin{array}{c}
\xymatrix@C=4em{
(F(x) \otimes F(y)) \otimes F(z) \ar[r]^{\alpha_{F(x),F(y),F(z)}^\CD} \ar[d]_{F^2_{x,y} \otimes \id_{F(z)}} & F(x) \otimes (F(y) \otimes F(z)) \ar[d]^{\id_{F(x)} \otimes F^2_{y,z}} \\
F(x \otimes y) \otimes F(z) \ar[d]_{F^2_{x \otimes y,z}} & F(x) \otimes F(y \otimes z) \ar[d]^{F^2_{x,y \otimes z}} \\
F((x \otimes y) \otimes z) \ar[r]^{F(\alpha_{x,y,z}^\CC)} & F(x \otimes (y \otimes z))
}
\end{array}
\ee
\item For any $x \in \CC$, the following two diagrams commute:
\[
\begin{array}{c}
\xymatrix{
F(\one_\CC) \otimes F(x) \ar[r]^-{F^2_{\one,x}} & F(\one_\CC \otimes x) \ar[d]^{F(\lambda_x^\CC)} \\
\one_\CD \otimes F(x) \ar[r]^-{\lambda_{F(x)}^\CD} \ar[u]^{F^0 \otimes \id_{F(x)}} & F(x)
}
\end{array} \quad
\begin{array}{c}
\xymatrix{
F(x) \otimes F(\one_\CC) \ar[r]^{F^2_{x,\one}} & F(x \otimes \one_\CC) \ar[d]^{F(\rho_x^\CC)} \\
F(x) \otimes \one_\CD \ar[r]^{\rho_{F(x)}^\CD} \ar[u]^{\id_{F(x)} \otimes F^0} & F(x)
}
\end{array}
\]
\enu
A monoidal functor that is also an equivalence is called a \emph{monoidal equivalence}.
\end{defn}

Hence, the bulk-to-boundary map $F$ is a monoidal functor.

\begin{expl} \label{expl:rep_G_forget_H}
Let $G$ be a finite group and $H$ be a subgroup of $G$. Each $G$-representation is naturally an $H$-representation by `forgetting' the action of the group elements that are not in $H$. This defines a forgetful functor
\begin{align*}
F \colon \rep(G) & \to \rep(H) \\
(V,\rho) & \mapsto (V,\rho \vert_H) .
\end{align*}
Then $F$ is a monoidal functor with $F^2$ and $F^0$ are both identities. Note that when $H$ is the trivial group this recovers the forgetful functor $\rep(G) \to \vect$ defined in Example \ref{expl:forgetful_functor_vect}.
\end{expl}

\begin{expl} \label{expl:vect_G_omega_monoidal_functor}
Let $G$ be a finite group and $\omega_1,\omega_2 \in Z^3(G;\Cb^\times)$. Suppose $F \colon \vect_G^{\omega_1} \to \vect_G^{\omega_2}$ is a monoidal functor whose underlying functor is identity. The natural isomorphism $F^2$ is defined by
\begin{align*}
F^2_{g,h} \colon \Cb_{(g)} \otimes_\Cb \Cb_{(h)} & \to \Cb_{(g)} \otimes_\Cb \Cb_{(h)} \\
1 \otimes_\Cb 1 & \mapsto \beta(g,h) \cdot 1 \otimes_\Cb 1
\end{align*}
for some nonzero complex number $\beta(g,h) \in \Cb^\times$. Then the commutative diagram \eqref{diag:monoidal_functor} translates to
\be \label{eq:monoidal_functor_pointed}
\omega_1(g,h,k) \beta(gh,k) \beta(g,h) = \beta(g,hk) \beta(h,k) \omega_2(g,h,k) , \quad \forall g,h,k \in G .
\ee
Define
\[
(\mathrm d \beta)(g,h,k) \coloneqq \frac{\beta(h,k) \beta(g,hk)}{\beta(gh,k) \beta(g,h)} .
\]
Then \eqref{eq:monoidal_functor_pointed} can be rewritten as $\omega_1 = \omega_2 \cdot \mathrm d \beta$. The operator $\mathrm d$ is usually called the \emph{differential}. It is straightforward to check that $\mathrm d \beta \in Z^3(G;\Cb^\times)$ is a 3-cocycle. A 3-cocycle obtained by the differential is called a \emph{3-coboundary}. The space of 3-coboundaries (valued in $\Cb^\times$) is denoted by $B^3(G;\Cb^\times)$. The quotient group $H^3(G;\Cb^\times) \coloneqq Z^3(G;\Cb^\times) / B^3(G;\Cb^\times)$ is called \emph{the third group cohomology group} of $G$ (with coefficient $\Cb^\times$).

Similarly, the other two commutative diagrams imply that the morphism $F^0$ is equal to $\beta(e,e)^{-1} \cdot \id_{\Cb_{[e]}}$. Hence we conclude that such a monoidal equivalence $F$ exists if and only if $\omega_1$ and $\omega_2$ are contained in the same cohomology class. In other words, the monoidal structures on the category $\vect_G$ are classified by $H^3(G;\Cb^\times)$ up to equivalence.
\end{expl}

\begin{rem}
For each integer $n>1$ we have $H^3(\Zb_n;\Cb^\times) \simeq \Zb_n$. For $n=2$, consider the 3-cocycle $\omega \in Z^3(\Zb_2;\Cb^\times)$ defined as follows:
\[
\omega(a,b,c) = \begin{cases} -1 , & a=b=c=1 , \\ +1 , & \text{otherwise} .\end{cases}
\]
This 3-cocyle is not a 3-coboundary. In other words, it represents the nontrivial cohomology class in $H^3(\Zb_2;\Cb^\times) \simeq \Zb_2$.
\end{rem}

\begin{rem}
For every finite group $G$, the inclusion $\mathrm U(1) \hookrightarrow \Cb^\times$ induces a group isomorphism $H^3(G;\mathrm U(1)) \to H^3(G;\Cb^\times)$.
\end{rem}

\begin{defn} \label{defn:braided_monoidal_functor}
Let $\CC,\CD$ be braided monoidal categories. A \emph{braided monoidal functor} (or simply a \emph{braided functor}) $F \colon \CC \to \CD$ is a monoidal functor $F \colon \CC \to \CD$ such that the following diagram commutes for any $x,y \in \CC$:
\[
\xymatrix{
F(x) \otimes F(y) \ar[r]^-{F^2_{x,y}} \ar[d]_{c_{F(x),F(y)}^\CD} & F(x \otimes y) \ar[d]^{F(c_{x,y}^\CC)} \\
F(y) \otimes F(x) \ar[r]^-{F^2_{y,x}} & F(y \otimes x)
}
\]
A braided monoidal functor that is also an equivalence is called a \emph{braided monoidal equivalence}.
\end{defn}

\begin{defn} \label{defn:monoidal_natural_transformation}
Let $\CC,\CD$ be monoidal categories and $F,G \colon \CC \to \CD$ be monoidal functors. A \emph{monoidal natural transformation} $\alpha \colon F \Rightarrow G$ is a natural transformation $\alpha \colon F \Rightarrow G$ satisfying the following conditions:
\bnu
\item For any $x,y \in \CC$, the following diagram commutes:
\[
\xymatrix{
F(x) \otimes F(y) \ar[r]^-{F^2_{x,y}} \ar[d]_{\alpha_x \otimes \alpha_y} & F(x \otimes y) \ar[d]^{\alpha_{x \otimes y}} \\
G(x) \otimes G(y) \ar[r]^-{G^2_{x,y}} & G(x \otimes y)
}
\]
\item The following diagram commutes:
\[
\xymatrix{
\one_\CD \ar[r]^{F^0} \ar[dr]_{G^0} & F(\one_\CC) \ar[d]^{\alpha_{\one}} \\
 & G(\one_\CC)
}
\]
\enu
A monoidal natural transformation that is also a natural isomorphism is called a \emph{monoidal natural isomorphism}.
\end{defn}

\begin{expl}
Let $G$ be a finite group and $F \colon \rep(G) \to \vect$ be the forgetful functor. Let us find the group of monoidal isomorphisms from $F$ to itself (also called a \emph{monoidal natural automorphism} on $F$).

In Example \ref{expl:TK_duality_preparation} we have constructed a natural isomorphism $\alpha^g \colon F \Rightarrow F$ for each $g \in G$ defined by
\[
\alpha^g_{(V,\rho)} \coloneqq \rho(g) \colon V \to V , \quad (V,\rho) \in \rep(G) .
\]
It is easy to see that every $\alpha^g$ is a monoidal natural isomorphism. Conversely, every monoidal natural automorphism on $F$ has to be of the form $\alpha^g$ for some $g \in G$. The proof is not very hard but we do not give it here. In other words, the assignment $g \mapsto \alpha^g$ defines a group isomorphism from $G$ to the group $\Aut_\otimes(F)$ of monoidal natural automorphisms on $F$. This is a special case of the Tannaka-Krein duality \cite{Tan39,Kre49}.
\end{expl}

\subsubsection{Half-braidings and central functors} \label{sec:central_structure}

\begin{figure}[htbp]
\centering
\subfigure[in 2d space]{
\begin{tikzpicture}[scale=0.8]
\fill[gray!20] (-3,0) rectangle (3,3) node[at end,below left,black] {$\CC$} ;
\draw[thick] (-3,0)--(3,0) node[at end,below left] {$\CA$} ;

\fill (0,0) circle (0.07) node[below] {$a$} ;
\fill (0,2) circle (0.07) node[above] {$x$} ;
\fill (-1,0) circle (0.07) node[below] {$F(x)$} ;
\fill (1,0) circle (0.07) node[below] {$F(x)$} ;

\draw[-stealth,dashed,thick] (0.9,0.15) to [out=135,in=45] (-0.9,0.15) ;
\draw[-stealth,thick] (-0.15,1.9) to [out=225,in=75] (-1,0.15) ;
\draw[-stealth,thick] (0.15,1.9) to [out=-45,in=105] (1,0.15) ;
\end{tikzpicture}
}
\hspace{5ex}
\subfigure[in 2+1D spacetime]{
\begin{tikzpicture}[scale=0.8]
\fill[gray!10] (-1,0)--(5,0)--(6,2)--(0,2)--cycle ;

\draw (3,0)--(3,0.3) .. controls (3,0.4) and (3.1,0.7) .. (3.2,0.9) ; 
\draw[->-=0.4] (3.2,0.9) .. controls (3.5,1.5) and (1.5,1.5) .. (1.2,2.1) ;
\draw (1.2,2.1) .. controls (1.1,2.3) and (1,2.6) .. (1,2.7)--(1,3) ;
\begin{scope}
\clip (2,1.6) circle (0.3) ;
\draw[gray!10,double=black,double distance=0.4pt,line width=3pt] (2,0)--(2,3) ;
\end{scope}
\draw[->-=0.4] (2,0)--(2,3) ;

\node[below left] at (6,2) {$\CC$} ;
\draw[thick] (-1,0)--(5,0) node[at end,below left] {$\CA$} ;
\fill (2,0) circle (0.07) node[below] {$a$} ;
\fill (3,0) circle (0.07) node[below] {$F(x)$} ;

\fill[blue!10,opacity=0.3] (-1,0) rectangle (5,3) ;
\end{tikzpicture}
}
\caption{the half-braiding of $F(x)$ with $a$}
\label{fig:central_structure}
\end{figure}
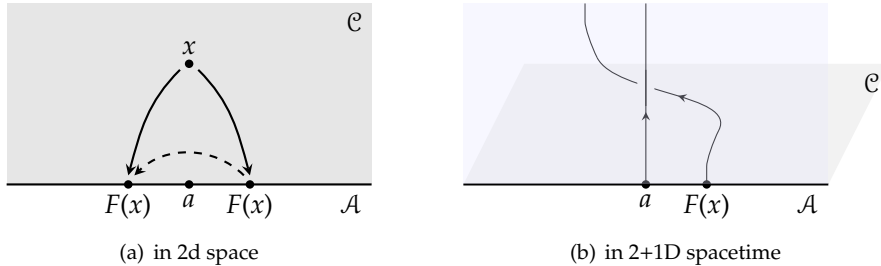

Now suppose there is a bulk topological defect $x \in \CC$ and a boundary topological defect $a \in \CA$. Recall that the boundary topological defect $F(x)$ is just $x$ near the boundary, so we can freely move $F(x)$ along the boundary, regardless of the topological defects `on' the boundary. In other words, moving $x$ to the left or right side of $a$ and then fusing them together should give the same result (see Figure \ref{fig:central_structure}). Thus we get an isomorphism (instanton) $\sigma_{a,x} \colon a \otimes F(x) \to F(x) \otimes a$. The family $\{\sigma_{a,x} \colon a \otimes F(x) \to F(x) \otimes a \}$ should be natural in both $a \in \CA$ and $x \in \CC$ because instantons can be freely moved along the worlds lines. 

\begin{figure}[htbp]
\[
\begin{array}{c}
\begin{tikzpicture}[scale=0.8]
\useasboundingbox[fill,gray!20] (-3,0) rectangle (3,3) node[at end,below left,black] {$\CC$} ;
\draw[thick] (-3,0)--(3,0) node[at end,below left] {$\CA$} ;
\fill (-1,0) circle (0.07) node[below] {$a$} ;
\fill (1,0) circle (0.07) node[below] {$b$} ;
\fill (0,2) circle (0.07) node[above] {$x$} ;
\fill (-2,0) circle (0.07) node[below] {$F(x)$} ;
\fill (0,0) circle (0.07) node[below] {$F(x)$} ;
\fill (2,0) circle (0.07) node[below] {$F(x)$} ;
\draw[-stealth,dashed,thick] (-0.1,0.1) to [out=150,in=30] (-1.9,0.1) ;
\draw[-stealth,dashed,thick] (1.9,0.1) to [out=150,in=30] (0.1,0.1) ;
\draw[-stealth,thick] (-0.15,2) to [out=180,in=90] (-2,0.15) ;
\draw[-stealth,thick] (0.15,2) to [out=0,in=90] (2,0.15) ;
\draw[-stealth,dashed,thick] (1.9,0.2) to [out=120,in=60] (-1.9,0.2) ;
\end{tikzpicture}
\end{array}
\implies
\begin{array}{c}
\xymatrix@C=1em{
a \otimes b \otimes F(x) \ar[rr]^{\sigma_{a \otimes b,x}} \ar[dr]_{\id_a \otimes \sigma_{b,x}} & & F(x) \otimes a \otimes b \\
 & a \otimes F(x) \otimes b \ar[ur]_{\sigma_{a,x} \otimes \id_b} 
}
\end{array}
\]
\caption{the first property of the half-braiding}
\label{fig:central_prop_1}
\end{figure}
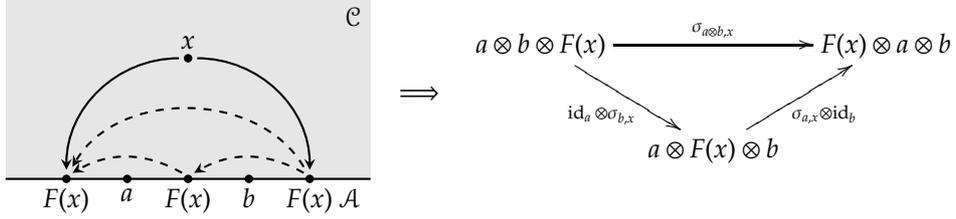

The following figures show that the family $\{\sigma_{a,x}\}$ should satisfy some properties. Figure \ref{fig:central_prop_1} shows that if there are two boundary topological defects and one bulk topological defects, two different ways of half-braidings should be the same. 


\begin{figure}[htbp]
\[
\begin{array}{c}
\begin{tikzpicture}[scale=0.8]
\useasboundingbox[fill,gray!20] (-3,0) rectangle (3,3) node[at end,below left,black] {$\CC$} ;
\draw[thick] (-3,0)--(3,0) node[at end,below left] {$\CA$} ;
\fill (0,0) circle (0.07) node[below] {$a$} ;
\fill (-0.2,2) circle (0.07) node[above] {$x$} ; 
\fill (0.2,2) circle (0.07) node[above] at (0.2,2) {$y$} ;
\fill (-1.2,0) circle (0.07) node[below left] {$F(x)$} ;
\fill (0.8,0) circle (0.07) node[below] {$F(x)$} ;
\fill (-0.8,0) circle (0.07) node[below] {$F(y)$} ;
\fill (1.2,0) circle (0.07) node[below right] {$F(y)$} ;
\draw[-stealth,dashed,thick] (0.7,0.15) to [out=135,in=45] (-1.1,0.15) ;
\draw[-stealth,thick] (-0.35,1.9) to [out=225,in=75] (-1.2,0.15) ;
\draw[-stealth,thick] (-0.05,1.9) to [out=-45,in=105] (0.8,0.15) ;
\draw[-stealth,dashed,thick] (1.1,0.15) to [out=135,in=45] (-0.7,0.15) ;
\draw[-stealth,thick] (0.05,1.9) to [out=225,in=75] (-0.8,0.15) ;
\draw[-stealth,thick] (0.35,1.9) to [out=-45,in=105] (1.2,0.15) ;
\end{tikzpicture}
\end{array}
\implies
\begin{array}{c}
\xymatrix@R=1.5em{
a \otimes F(x) \otimes F(y) \ar[rr]^-{\id_a \otimes F^2_{x,y}} \ar[d]^{\sigma_{a,x} \otimes \id_{F(y)}} & & a \otimes F(x \otimes y) \ar[dd]^{\sigma_{a,x \otimes y}} \\
F(x) \otimes a \otimes F(y) \ar[d]^{\id_{F(x)} \otimes \sigma_{a,y}} \\
F(x) \otimes F(y) \otimes a \ar[rr]^-{F^2_{x,y} \otimes \id_a} & & F(x \otimes y) \otimes a
}
\end{array}
\]
\caption{the second property of the half-braiding}
\label{fig:central_prop_2}
\end{figure}
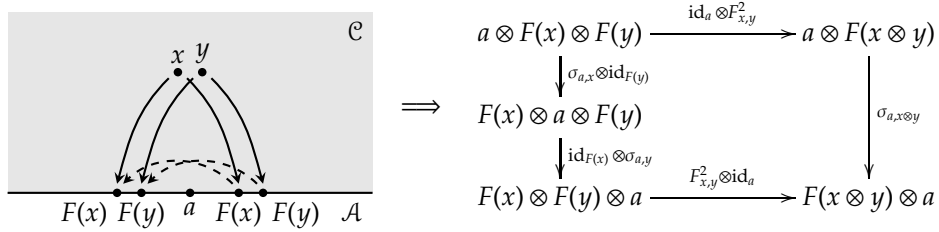

Figure \ref{fig:central_prop_2} shows that if there are two bulk topological defects and one boundary topological defects, two different ways of half-braidings should be the same.


\begin{figure}[htbp]
\[
\begin{array}{c}
\begin{tikzpicture}[scale=0.8]
\useasboundingbox[fill,gray!20] (-3,0) rectangle (3,3) node[at end,below left,black] {$\CC$} ;
\draw[thick] (-3,0)--(3,0) node[at end,below left] {$\CA$} ;
\fill (0,0) circle (0.07) node[below] {$F(x)$} ;
\fill (-1,0) circle (0.07) node[below] {$F(y)$} ;
\fill (1,0) circle (0.07) node[below] {$F(y)$} ;
\fill (0,2) circle (0.07) node[above] {$x$} ;
\fill (-1,2) circle (0.07) node[left] {$y$} ;
\fill (1,2) circle (0.07) node[right] {$y$} ;
\draw[-stealth,dashed,thick] (0.9,0.15) to [out=135,in=45] (-0.9,0.15) ;
\draw[-stealth,thick] (0.9,2.15) to [out=135,in=45] (-0.9,2.15) ;
\draw[-stealth,thick] (0,1.9) to [out=-90,in=90] (0,0.15) ;
\draw[-stealth,thick] (-1,1.9) to [out=-90,in=90] (-1,0.15) ;
\draw[-stealth,thick] (1,1.9) to [out=-90,in=90] (1,0.15) ;
\end{tikzpicture}
\end{array}
\implies
\begin{array}{c}
\xymatrix{
F(x) \otimes F(y) \ar[r]^-{F^2_{x,y}} \ar[d]^{\sigma_{F(x),y}} & F(x \otimes y) \ar[d]^{F(c_{x,y})} \\
F(y) \otimes F(x) \ar[r]^-{F^2_{y,x}} & F(y \otimes x)
}
\end{array}
\]
\caption{the third property of the half-braiding}
\label{fig:central_prop_3}
\end{figure}
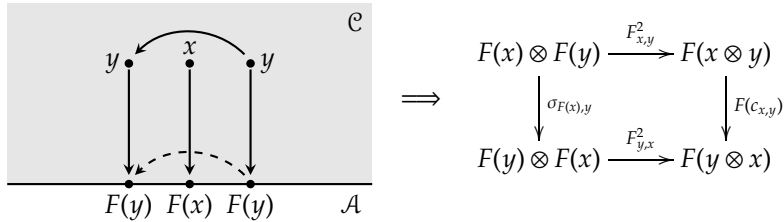

Figure \ref{fig:central_prop_3} shows that by moving bulk topological defects to the boundary, the braiding of two bulk topological defects becomes the half-braiding of two boundary topological defects.


In the above three commutative diagrams, we ignore the monoidal structures (associators) of $\CC$ and $\CA$. The half-braidings and their properties are summarized to the following definition \cite{Bez04}.

\begin{defn}
Let $\CC$ be a braided monoidal category and $\CA$ be a monoidal category. A \emph{central structure} of a monoidal functor $F \colon \CC \to \CA$ is a natural isomorphism $\sigma_{a,x} \colon a \otimes F(x) \to F(x) \otimes a$, where $x \in \CC$ and $a \in \CA$, such that the following conditions hold:
\bnu
\item For any $x \in \CC$ and $a,b \in \CA$, the following diagram commutes:
\be \label{eq:central_1}
\begin{array}{c}
\xymatrix@C=1em{
a \otimes b \otimes F(x) \ar[rr]^{\sigma_{a \otimes b,x}} \ar[dr]_{\id_a \otimes \sigma_{x,b}} & & F(x) \otimes a \otimes b \\
 & a \otimes F(x) \otimes b \ar[ur]_{\sigma_{a,x} \otimes \id_b} 
}
\end{array}
\ee
\item For any $x,y \in \CC$ and $a \in \CA$, the following diagram commutes:
\be \label{eq:central_2}
\begin{array}{c}
\xymatrix@R=1.5em{
a \otimes F(x) \otimes F(y) \ar[rr]^-{\id_a \otimes F^2_{x,y}} \ar[d]^{\sigma_{a,x} \otimes \id_{F(y)}} & & a \otimes F(x \otimes y) \ar[dd]^{\sigma_{a,x \otimes y}} \\
F(x) \otimes a \otimes F(y) \ar[d]^{\id_{F(x)} \otimes \sigma_{a,y}} \\
F(x) \otimes F(y) \otimes a \ar[rr]^-{F^2_{x,y} \otimes \id_a} & & F(x \otimes y) \otimes a
}
\end{array}
\ee
\item For any $x,y \in \CC$, the following diagram commutes:
\be \label{eq:central_3}
\begin{array}{c}
\xymatrix{
F(x) \otimes F(y) \ar[r]^-{F^2_{x,y}} \ar[d]^{\sigma_{F(x),y}} & F(x \otimes y) \ar[d]^{F(c_{x,y})} \\
F(y) \otimes F(x) \ar[r]^-{F^2_{y,x}} & F(y \otimes x)
}
\end{array}
\ee
\enu
A \emph{central functor} is a monoidal functor equipped with a central structure.
\end{defn}

Hence the bulk-to-boundary map $F \colon \CC \to \CA$ is a central functor.

\subsubsection{Drinfeld center} \label{sec:Drinfeld-center}

We have known that a bulk topological defect $x \in \CC$ determines the following data:
\bit
\item a boundary topological defect $F(x) \in \CA$;
\item a natural isomorphism $\sigma_{-,x} \coloneqq \{\sigma_{a,x} : a \otimes F(x) \to F(x) \otimes a\}_{a \in \CA}$, called the half-braidings of $F(x)$, satisfying the property given by Figure \ref{fig:central_prop_1}, i.e., the following diagram commutes for all $a,b \in \CA$:
\[
\xymatrix@C=1em{
a \otimes b \otimes F(x) \ar[rr]^{\sigma_{x,a \otimes b}} \ar[dr]_{\id_a \otimes \sigma_{x,b}} & & F(x) \otimes a \otimes b \\
 & a \otimes F(x) \otimes b \ar[ur]_{\sigma_{x,a} \otimes \id_{b}} 
}
\]
\eit
Motivated by this construction and the example in the toric code model \eqref{eq:toric_code_half_braiding_M} \eqref{eq:toric_code_half_braiding_E}, we propose the following definition \cite{Maj91,JS91}.

\begin{defn} \label{defn:Drinfeld_center}
Let $\CA$ be a monoidal category. The \emph{Drinfeld center} or \emph{monoidal center} of $\CA$ is the braided monoidal category $\FZ_1(\CA)$ defined as follows:
\bit
\item The objects are all pairs $(a,\gamma_{-,a})$, where $a \in \CA$ and $\gamma_{-,a} \colon {- \otimes a} \Rightarrow {a \otimes -} \colon \CA \to \CA$ is a natural isomorphism (called a half-braiding) such that the following diagram commutes for all $b,c \in \CA$,
\[
\xymatrix{
 & b \otimes (a \otimes c) \ar[r]^{\alpha_{b,a,c}^{-1}} & (b \otimes a) \otimes c \ar[dr]^{\gamma_{b,a} \otimes \id_c} \\
b \otimes (c \otimes a) \ar[ur]^{\id_b \otimes \gamma_{c,a}} \ar[dr]^{\alpha_{b,c,a}^{-1}} & & & (a \otimes b) \otimes c \\
 & (b \otimes c) \otimes a \ar[r]^{\gamma_{b \otimes c,a}} & a \otimes (b \otimes c) \ar[ur]^{\alpha_{a,b,c}^{-1}}
}
\]
where $\alpha$ is the associator of $\CA$.
\item The morphisms from $(a,\gamma_{-,a})$ to $(b,\gamma_{-,b})$ are morphisms $f \in \Hom_{\CA}(a,b)$ such that the following diagram commutes for all $c \in \CA$.
\[
\xymatrix{
c \otimes a \ar[r]^{\gamma_{c,a}} \ar[d]_{\id_c \otimes f} & a \otimes c \ar[d]^{f \otimes \id_c} \\
c \otimes b \ar[r]^{\gamma_{c,b}} & b \otimes c
}
\]
\item The tensor product of two objects $(a,\gamma_{-,a})$ and $(b,\gamma_{-,b})$ is $(a \otimes b,\gamma_{-,a \otimes b})$, where $a \otimes b$ is the tensor product in $\CA$ and the half-braiding $\gamma_{c,a \otimes b} \colon c \otimes a \otimes b \to a \otimes b \otimes c$ is the composition
\[
c \otimes a \otimes b \xrightarrow{\gamma_{c,a} \otimes \id_b} a \otimes c \otimes b \xrightarrow{\id_a \otimes \gamma_{c,b}} a \otimes b \otimes c .
\]
Here we omit the associators of $\CA$.
\item The tensor unit is $(\one,\gamma_{-,\one})$, where $\one \in \CA$ is the tensor unit of $\CA$ and the half-braiding $\gamma_{a,\one} \colon a \otimes \one \to \one \otimes a$ is
\[
a \otimes \one \xrightarrow{r_a} a \xrightarrow{l_a^{-1}} \one \otimes a ,
\]
or simply $\gamma_{a,\one} = \id_a$ if we ignore the unitors.
\item The associator and left/right unitor of $\FZ_1(\CA)$ are induced by those of $\CA$.
\item The braiding of $(a,\gamma_{-,a})$ and $(b,\gamma_{-,b})$ is given by $\gamma_{a,b} \colon a \otimes b \to b \otimes a$. Note that it is independent of the half-braiding of $a$.
\eit
\end{defn}

\begin{rem}
The Drinfeld center is a kind of $E_1$-center \cite{Lur17}.
\end{rem}

\begin{rem}
Given an object $(a,\gamma_{-,a}) \in \FZ_1(\CA)$, the following diagram commutes:
\[
\xymatrix{
\one \otimes a \ar[rr]^{\gamma_{\one,a}} \ar[dr]_{\lambda_a} & & a \otimes \one \ar[dl]^{\rho_a} \\
 & a
}
\]
where $\lambda,\rho$ are the left/right unitor of $\CA$. In other words, $\gamma_{\one,a} = \id_a$ if we ignore the unitors.
\end{rem}

\begin{rem} \label{rem:braiding_half_braiding}
Note that the definition of a half-braiding is similar to the hexagon equations in the definition of a braiding structure. It follows that for any braided monoidal category $\CC$ and $x \in \CC$, the braiding structure induces a half-braiding $c_{-,x} \colon {-} \otimes x \to x \otimes -$ on $x$. Moreover, there is a canonical braided functor $\CC \to \FZ_1(\CC)$ defined by $x \mapsto (x , c_{-,x})$. Similarly, using the anti-braiding there is a canonical braided functor $\overline{\CC} \to \FZ_1(\CC)$ defined by $x \mapsto (x , c_{x,-}^{-1})$. When $\CC$ is symmetric, these two functors are equal.
\end{rem}

\begin{expl}
The Drinfeld center $\FZ_1(\vect)$ is equivalent to $\vect$ as braided monoidal categories. Indeed, suppose $V \in \vect$ and $\gamma_{-,V} \colon {-} \otimes V \to V \otimes -$ is a half-braiding. Then we have the following commutative diagram for every vector space $W$ and linear map $f \colon \Cb \to W$:
\[
\xymatrix@R=1ex{
 & \Cb \otimes V \ar[r]^{f \otimes 1} \ar[dd]^{\gamma_{\Cb,V}} & W \otimes V \ar[dd]^{\gamma_{W,V}} \\
V \ar[ur]^{\simeq} \ar[dr]_{\simeq} \\
 & V \otimes \Cb \ar[r]^{1 \otimes f} & V \otimes W
}
\]
Therefore, we have $\gamma_{W,V}(w \otimes v) = v \otimes w$ for every $W \in \vect$ and $w \in W$. Hence the canonical inclusion $\vect \to \FZ_1(\vect)$ (see Remark \ref{rem:braiding_half_braiding}) is an equivalence of braided monoidal categories.
\end{expl}

\begin{expl} \label{expl:toric_code_boundary_bulk}
Recall there are two simple topological defects $\one$ and $E$ on the smooth boundary of the toric code model. Clearly the fusion rule is given by $E \otimes E = \one$. Thus the topological skeleton of the smooth boundary is equivalent to $\rep(\Zb_2)$ as fusion categories, where $\one$ corresponds to the trivial representation of $\Zb_2$ and $E$ corresponds to the nontrivial irreducible representation of $\Zb_2$. Then \eqref{eq:toric_code_half_braiding_E} means $\Toric$ is equivalent to $\FZ_1(\rep(\Zb_2))$ as a braided fusion category.

Similarly, the topological skeleton of the rough boundary is equivalent to $\vect_{\Zb_2}$ as fusion categories, where $\one$ corresponds to $\Cb$ with the trivial grading and $M$ corresponds to $\Cb$ with the nontrivial grading, and \eqref{eq:toric_code_half_braiding_M} is precisely a braided equivalence $\Toric \simeq \FZ_1(\vect_{\Zb_2})$.
\end{expl}

\begin{rem}
For any monoidal category $\CA$, there is an obvious functor
\begin{align*}
I \colon \FZ_1(\CA) & \to \CA \\
(a,\gamma_{-,a}) & \mapsto a ,
\end{align*}
called the forgetful functor because it `forgets' the half-braiding. Clearly it is a faithful monoidal functor. Moreover, $I$ is a central functor with the central structure defined by
\[
\sigma_{a,(b,\gamma_{-,b})} \coloneqq \gamma_{a,b} .
\]
This central structure satisfies \eqref{eq:central_1} by the definition of $\gamma_{-,b}$; \eqref{eq:central_2} and \eqref{eq:central_3} follow from the definition of tensor product and braiding of $\FZ_1(\CA)$, respectively.

Furthermore, suppose $\CC$ is a braided monoidal category, $\CA$ is a monoidal category and $F \colon \CC \to \CA$ is a monoidal functor. Then a central structure on $F$ is equivalent to a braided monoidal functor $F' \colon \CC \to \FZ_1(\CA)$ such that $I \circ F' = F$. Indeed, given a braided monoidal functor $F' \colon \CC \to \FZ_1(\CA)$ such that $I \circ F' = F$, the central structure of $I$ induces that of $F$, i.e.,
\[
\sigma_{a,x} \coloneqq \gamma_{a,F'(x)} .
\]
Conversely, given a central structure $\sigma_{a,x} \colon a \otimes F(x) \to F(x) \otimes a$, it is not hard to see that
\begin{align*}
F' \colon \CC & \to \FZ_1(\CA) \\
x & \mapsto (F(x) , \sigma_{-,x})
\end{align*}
is a braided monoidal functor and $I \circ F' = F$. Hence a central structure of a monoidal functor $F \colon \CC \to \CA$ can be equivalently defined as a braided monoidal functor $F' \colon \CC \to \FZ_1(\CA)$ such that $I \circ F' = F$. For example, \eqref{eq:toric_code_half_braiding_E} and \eqref{eq:toric_code_half_braiding_M} are central structures of the bulk-to-smooth-boundary map \eqref{eq:bulk-to-smooth-boundary_map} and the bulk-to-rough-boundary map \eqref{eq:bulk-to-rough-boundary_map}, respectively.
\end{rem}

\begin{rem}
It is possible that different central functors have the same underlying monoidal functor.
\end{rem}

\subsubsection{Boundary-bulk relation} \label{sec:boundary_bulk}

We have also shown model-independently in Section \ref{sec:central_structure} and \ref{sec:Drinfeld-center} that the bulk modular tensor category necessarily factors through the Drinfeld center of the boundary fusion category. It turns out that a stronger result holds for all 2d topological orders and their gapped boundaries. 

\begin{pthm}[Boundary-bulk relation in 2d, 1st version] \label{pthm:boundary_bulk}
Suppose $\SC$ is an anomaly-free stable 2d topological order with a stable gapped 1d boundary $\SA$, and their particle-like topological defects form a (unitary) modular tensor category $\CC$ and a (unitary) fusion category $\CA$, respectively.
\bnu[(a)]
\item The bulk-to-boundary map $F \colon \CC \to \CA$ is a central functor.
\item The central structure $F' \colon \CC \to \FZ_1(\CA)$ is a braided equivalence.
\enu
\end{pthm}

\begin{rem}
In the 2d toric code model, the braided equivalences \eqref{eq:toric_code_half_braiding_E} and \eqref{eq:toric_code_half_braiding_M} imply that the bulk modular tensor category $\Toric$ is precisely the Drinfeld center of the boundary fusion categories (see Example \ref{expl:toric_code_boundary_bulk}).
\end{rem}

\begin{rem}
The boundary-bulk relation in 2d was first discovered in the Levin-Wen model \cite{KK12} (see also \cite{Kon13,LW14,HWW17,CCW17} for further developments). Model independently, it was shown that the bulk modular tensor category factors through the Drinfeld center of a boundary fusion category \cite{FSV13}. Since a gapped boundary of a 2d topological order is necessarily the result of a 2d anyon condensation to the trivial phase, we obtain a complete model-independent proof of the boundary-bulk relation for 2d topological orders and their gapped boundaries based on the anyon condensation theory \cite{Kon14} (see also Theorem$^{\mathrm{ph}}$ \ref{pthm:anyon-condensation}). More general boundary-bulk relation for potentially gapless higher dimensional quantum liquid phases and their boundaries was proposed and proved in \cite{KWZ15,KWZ17} (see also Theorem$ {\mathrm{ph}}$ \ref{pthm:bbr}). 
\end{rem}

\begin{rem}
The smooth and rough boundaries of the toric code model are the same as (anomalous) 1d topological orders, but they are different as 1d boundaries. Mathematically, there is a monoidal equivalence (by identifying $E$ and $M$) of $\rep(\Zb_2)$ and $\vect_{\Zb_2}$, but this equivalence does not preserve the bulk-to-boundary maps.

In Kitaev's quantum double model associated to a finite group $G$, there are also (at least) two types of boundaries \cite{BSW11,CCW17}, whose topological skeletons are $\rep(G)$ and $\vect_G$ respectively. If $G$ is non-abelian, $\rep(G)$ and $\vect_G$ are not equivalent as categories.
\end{rem}

\begin{rem}
For any unitary (or spherical) fusion category $\CA$, there is a 2d fixed point lattice model associated to $\CA$ , called the Levin-Wen model or the string-net model \cite{LW05}. The (unitary) modular tensor category particle-like topological defects of the Levin-Wen model is equivalent to $\FZ_1(\CA)$. When $\CA = \vect_G$ or $\rep(G)$ where $G$ is a finite group, the associated Levin-Wen model realizes the same topological order as the quantum double model associated to $G$. Mathematically, the Drinfeld center of $\rep(G)$ and $\vect_G$ are equivalent as modular tensor categories (see Example \ref{expl:quantum_double_as_Drinfeld_center}).
\end{rem}

\begin{rem} \label{rem:indecomposable_multi-fusion_category}
Our analyses do not depend on the stability of the boundary. So the boundary-bulk relation still holds when $\SA$ is unstable, i.e., $\CA$ is a multi-fusion category. We say a multi-fusion category $\CA$ is \emph{indecomposable} if it can not be written as a direct sum $\CA \simeq \CA_1 \oplus \CA_2$ for nonzero multi-fusion categories $\CA_1,\CA_2$. The Drinfeld center $\FZ_1(\CA)$ of a multi-fusion category $\CA$ is a fusion category (indeed, a nondegenerate braided fusion category) if and only if $\CA$ is indecomposable. A fusion category is an indecomposable multi-fusion category because the tensor unit $\one$ is an indecomposable object. Moreover, we have $\FZ_1(\CA_1 \oplus \CA_2) \simeq \FZ_1(\CA_1) \oplus \FZ_1(\CA_2)$.
\end{rem}

\begin{rem}
As a corollary of the boundary-bulk relation in 2d (Theorem$^{\text{ph}}$ \ref{pthm:boundary_bulk}), the topological skeleton of an anomaly-free 1d topological order $\SA$ is a (unitary) multi-fusion category $\CA$ satisfying $\FZ_1(\CA) \simeq \vect$ as braided fusion categories.
\end{rem}

\begin{rem}
Suppose $\SC$ is an anomaly-free stable 2d topological order that admits a gapped boundary. Then the chiral central charge of $\SC$ is $0$.
\end{rem}

%
%

\subsubsection{Properties of the Drinfeld center}

Since the particle-like topological defects of an anomaly-free stable 2d topological order form a (unitary) modular tensor category, the boundary-bulk relation (Theorem$^{\text{ph}}$ \ref{pthm:boundary_bulk}) immediately implies that the Drinfeld center of a (unitary) fusion category should be a (unitary) modular tensor category. This consequence can be made into a mathematically rigorous theorem.

\begin{thm}[\cite{Mueg03a}]
Let $\CA$ be a spherical fusion category. Then its Drinfeld center $\FZ_1(\CA)$ is a modular tensor category.
\end{thm}

\begin{rem}
There are several variants of the above theorem. If $\CA$ is only a fusion category, $\FZ_1(\CA)$ is a nondegenerate braided fusion category \cite{DGNO10}. If $\CA$ is a unitary fusion category, $\FZ_1(\CA)$ is a unitary modular tensor category \cite{Mueg03a,Gal14}.
\end{rem}

\begin{rem}
We do not go through the details of the above theorem, but list some data and properties of the Drinfeld center $\FZ_1(\CA)$ of a spherical fusion category $\CA$:
\bnu
\item The left dual of $(x,\gamma_{-,x}) \in \FZ_1(\CA)$ is $(x^L,\tilde \gamma_{-,x^L})$, where the half-braiding $\tilde \gamma_{y,x^L} \colon y \otimes x^L \to x^L \otimes y$ is the left dual of $\gamma_{y^R,x} \colon y^R \otimes x \to x \otimes y^R$.
\item The pivotal structure $a_{(x,\gamma_{-,x})} \colon (x,\gamma_{-,x}) \to (x,\gamma_{-,x})^{LL}$ is given by the pivotal structure $a_x$ of $\CA$. It is spherical and $\dim(x,\gamma_{-,x}) = \dim(x)$. Therefore, the forgetful functor $I \colon \FZ_1(\CA) \to \CA$ preserves the quantum dimension of objects.
\item The Gauss sums $\tau^{\pm}(\FZ_1(\CC))$ are both equal to $\dim(\CC)$ \cite{Mueg03a}. Thus $\dim(\FZ_1(\CC)) = \dim(\CC)^2$ \cite{Mueg03a,ENO05} and the additive central charge of $\FZ_1(\CC)$ is $c = 0 \pmod 8$.
\item The Frobenius-Perron dimension of $\FZ_1(\CA)$ is $\fpdim(\FZ_1(\CA)) = \fpdim(\CA)^2$ \cite{EO04}.
\enu
\end{rem}

\begin{expl} \label{expl:quantum_double_as_Drinfeld_center}
Let $G$ be a finite group. The Drinfeld center of both $\rep(G)$ and $\vect_G$ are equivalent to the quantum double category $\CD_G$ (see Section \ref{sec:quantum_double}) as modular tensor categories.
\end{expl}

\begin{exercise}
Prove that $\FZ_1(\vect_G)$ is equivalent to $\CD_G$ as categories. Hint: use Exercise \ref{exercise:linear_functor_preserve_direct_sum} (2) to determine a half-braiding in $\FZ_1(\vect_G)$.
\end{exercise}

\begin{expl} \label{expl:Drinfeld_center_vect_G_omega}
Let $G$ be a finite group and $\omega \in Z^3(G;\Cb^\times)$. The Drinfeld center $\FZ_1(\vect_G^\omega)$ is not equivalent to $\CD_G$ in general (see \cite{DS17} for an equivalent description of $\FZ_1(\vect_G^\omega)$ that is similar to $\CD_G$).

Suppose $(V,\rho) \in \rep(G)$ is a finite-dimensional $G$-representation. The vector space $V$ can be viewed as a $G$-graded vector space with the trivial $G$-grading so that $V \in \vect_G^\omega$, i.e.,
\[
V_g = \begin{cases} V , & g = e , \\ 0 , & g \neq e . \end{cases}
\]
For every $g \in G$, define a morphism in $\vect_G^\omega$ by
\begin{align*}
\gamma_{g,V} \colon \Cb_{(g)} \otimes_\Cb V & \to V \otimes_\Cb \Cb_{(g)} \\
1 \otimes_\Cb v & \mapsto \rho(g)(v) \otimes_\Cb \omega(g,e,g) .
\end{align*}
Then $\{\gamma_{g,V}\}_{g \in G}$ defines a half-braiding on $V \in \vect_G^\omega$. In other words, $(V,\{\gamma_{g,V}\}_{g \in G}) \in \FZ_1(\vect_G^\omega)$.

It is straightforward to verify that the above construction defines a fully faithful $\Cb$-linear braided monoidal functor $\rep(G) \to \FZ_1(\vect_G^\omega)$. We say that $\FZ_1(\vect_G^\omega)$ is a \emph{minimal modular extension} \cite{Mueg00} of $\rep(G)$. Moreover, all minimal modular extensions of $\rep(G)$ are of this form \cite{DGNO10}.
\end{expl}

\begin{rem}
If $\CA$ is a fusion category, the forgetful functor $I \colon \FZ_1(\CA) \to \CA$ is \emph{surjective}, in the sense that any simple object $a \in \CA$ appears as a direct summand of $I(x)$ for some object $x \in \FZ_1(\CA)$. However, this is not true if $\CA$ is a multi-fusion category.
\end{rem}

\begin{rem}
A simple topological defect in the bulk may not be simple after moving to the boundary. Mathematically, if $\CA$ is a fusion category, the forgetful functor $I \colon \FZ_1(\CA) \to \CA$ may carry a simple object to a non-simple object.
\end{rem}

\subsection{Gapped 1d domain walls} \label{sec:1d_domain_wall}

A gapped boundary of an anomaly-free stable 2d topological order $\SC$ can be viewed as a gapped domain wall between $\SC$ and the trivial 2d topological order $\tTO_2$. In this subsection we study general gapped 1d domain walls between two anomaly-free stable 2d topological orders by the boundary-bulk relation in 2d (Theorem$^{\text{ph}}$ \ref{pthm:boundary_bulk}).

\subsubsection{General theory of 1d domain walls}

\begin{figure}[htbp]
\[
\begin{array}{c}
\begin{tikzpicture}[scale=1.0]
\fill[gray!15] (-1.5,-1) rectangle (1.5,1) ;
\draw[very thick] (0,-1)--(0,1) node[very near end,left] {$\SM$} node[very near start,left] {$\SN$} ;
\fill (0,0) circle (0.1) node[right] {$x$} ;

\node at (-0.8,0) {$\SC$} ;
\node at (0.8,0) {$\SD$} ;
\end{tikzpicture}
\end{array}
\rightsquigarrow
\begin{array}{c}
\begin{tikzpicture}[scale=0.8]
\draw[color=gray,fill=gray!30,opacity=0.7] (-1.5,0)--(0,0)--(1,2)--(-0.5,2)--cycle ;
\draw[color=gray,fill=gray!30,opacity=0.7] (-1.3,0.7)--(0,0)--(1,2)--(-0.3,2.7)--cycle ;
\draw[very thick] (0,0)--(1,2) node[very near end,right] {$\SM$} node[very near start,right] {$\SN$} ;
\fill (0.5,1) circle (0.1) node[right] {$x$} ;

\node[above] at (-1,0) {$\SC$} ;
\node at (-0.2,1.2) {$\overline{\SD}$} ;
\end{tikzpicture}
\end{array}
\rightsquigarrow
\begin{array}{c}
\begin{tikzpicture}[scale=1.0]
\fill[gray!20] (-1.5,-1) rectangle (0,1) ;
\draw[very thick] (0,-1)--(0,1) node[very near end,right] {$\SM$} node[very near start,right] {$\SN$} ;
\fill (0,0) circle (0.1) node[right] {$x$} ;

\node at (-0.8,0) {$\SC \boxtimes \overline{\SD}$} ;
\end{tikzpicture}
\end{array}
\]
\caption{the folding trick}
\label{fig:folding}
\end{figure}
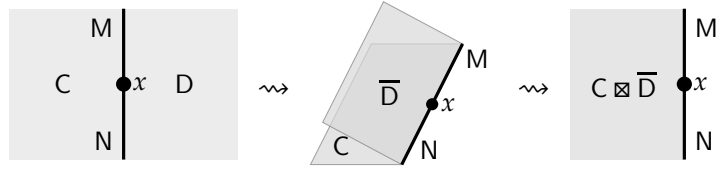

The study of general 1d domain walls can be reduced to the study of boundaries by the folding trick (see Figure \ref{fig:folding}).

Suppose $\SM$ is a gapped 1d domain wall between two anomaly-free stable 2d topological orders $\SC$ and $\SD$.
\bnu[(a)]
\item The topological skeleton $\CM$ of the 1d topological order $\SM$ is a (unitary) multi-fusion category. The particle-like topological defects in $\SC$ and $\SD$ form unitary modular tensor categories $\CC$ and $\CD$, respectively.
\item The left-bulk-to-domain-wall map $F \colon \CC \to \CM$ is a central functor. Similarly the right-bulk-to-domain-wall map is a central functor $G \colon \overline{\CD} \to \CM$. Note that the orientation is reversed.
\item The folding trick shows that $\SM$ is a boundary of $\SC \boxtimes \overline{\SD}$, and the bulk-to-boundary map is given by $F \boxtimes G \colon \CC \boxtimes \overline{\CD} \to \CM$. By the boundary-bulk relation in 2d, $F \boxtimes G$ is a central functor whose central structure is a braided equivalence $\CC \boxtimes \overline{\CD} \to \FZ_1(\CM)$.
\enu
These discussions directly give rise to the following definition \cite{Lur17,AFT16,KZ18,BZBJ18a}.

\begin{defn}
Let $\CC,\CD$ be braided fusion categories.
\bnu[(1)]
\item A \emph{(multi-)fusion left $\CC$-module} is a (multi-)fusion category $\CM$ equipped with a braided functor $F \colon \CC \to \FZ_1(\CM)$. It is called \emph{closed} if $F$ is an equivalence.
\item A \emph{(multi-)fusion right $\CD$-module} is a (multi-)fusion category $\CM$ equipped with a braided functor $G \colon \overline{\CD} \to \FZ_1(\CM)$. It is called \emph{closed} if $G$ is an equivalence.
\item A \emph{(multi-)fusion $(\CC,\CD)$-bimodule} is a (multi-)fusion category $\CM$ equipped with braided functors $F \colon \CC \to \FZ_1(\CM)$ and $G \colon \overline{\CD} \to \FZ_1(\CM)$. It is called \emph{closed} if $F \boxtimes G \colon \CC \boxtimes \overline{\CD} \to \FZ_1(\CM)$ is an equivalence.
\enu
\end{defn}

Using the new language of multi-fusion modules, we conclude that $\CM$ is a closed multi-fusion $(\CC,\CD)$-bimodule. In particular, the boundary-bulk relation in 2d (Theorem$^{\text{ph}}$ \ref{pthm:boundary_bulk}) simply says that $\CA$ is a closed multi-fusion left $\CC$-module.

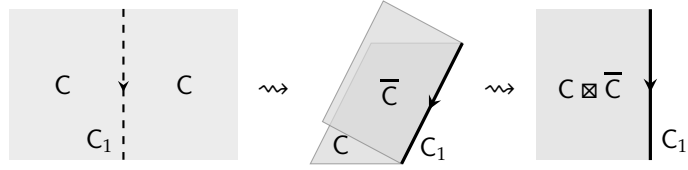
\begin{figure}[htbp]
\[
\begin{array}{c}
\begin{tikzpicture}[scale=1.0]
\fill[gray!15] (-1.5,-1) rectangle (1.5,1) ;
\draw[->-,thick,dashed] (0,1)--(0,-1) node[very near end,left] {$\SC_1$} ;

\node at (-0.8,0) {$\SC$} ;
\node at (0.8,0) {$\SC$} ;
\end{tikzpicture}
\end{array}
\rightsquigarrow
\begin{array}{c}
\begin{tikzpicture}[scale=0.8]
\draw[color=gray,fill=gray!30,opacity=0.7] (-1.5,0)--(0,0)--(1,2)--(-0.5,2)--cycle ;
\draw[color=gray,fill=gray!30,opacity=0.7] (-1.3,0.7)--(0,0)--(1,2)--(-0.3,2.7)--cycle ;
\draw[->-,very thick] (1,2)--(0,0) node[very near end,right] {$\SC_1$} ;

\node[above] at (-1,0) {$\SC$} ;
\node at (-0.2,1.2) {$\overline{\SC}$} ;
\end{tikzpicture}
\end{array}
\rightsquigarrow
\begin{array}{c}
\begin{tikzpicture}[scale=1.0]
\fill[gray!20] (-1.5,-1) rectangle (0,1) ;
\draw[->-,very thick] (0,1)--(0,-1) node[very near end,right] {$\SC_1$} ;

\node at (-0.8,0) {$\SC \boxtimes \overline{\SC}$} ;
\end{tikzpicture}
\end{array}
\]
\caption{By restricting $\SC$ to a line we get the the trivial domain wall $\SC_1$ in $\SC$. The folding trick shows that $\SC \boxtimes \overline{\SC}$ is the bulk of $\SC$ if $\SC$ is anomaly-free.}
\label{fig:trivial_domain_wall}
\end{figure}

\begin{expl} \label{expl:trivial_domain_wall}
Given a stable 2d topological order $\SC$, by restricting it to a line we get the trivial 1d domain wall in $\SC$, denoted by $\SC_1$ (see Figure \ref{fig:trivial_domain_wall}). Clearly the particle-like topological defects on the trivial domain wall are the same as those in $\SC$. In other words, the topological skeleton of the trivial domain wall is the fusion category $\CC$, which is obtained by forgetting the braiding of the braided fusion category $\CC$ of particle-like topological defects in $\SC$. Both the left-bulk-to-domain-wall-map and right-bulk-to-domain-wall-map are the identity functor $\id_\CC \colon \CC \to \CC$. Their central structures, i.e., the half-braiding of bulk topological defects with domain wall topological defects, are given by the braiding of $\CC$, because there is essentially no domain wall.

Mathematically, given a braided fusion category $\CC$, two canonical braided functors defined in Remark \ref{rem:braiding_half_braiding} give rise to a single braided functor $\CC \boxtimes \overline{\CC} \to \FZ_1(\CC)$, which equips $\CC$ with a canonical fusion $(\CC,\CC)$-bimodule structure. Moreover, the braided fusion category $\CC$ is nondegenerate (i.e., $\FZ_2(\CC) = \vect$ is trivial) if and only if this canonical functor $\CC \boxtimes \overline{\CC} \to \FZ_1(\CC)$ is an equivalence \cite{Mueg03a,DGNO10}, i.e., this fusion $(\CC,\CC)$-bimodule $\CC$ is closed.
\end{expl}

\begin{expl} \label{expl:domain_wall_automorphism}
Let us give a more general construction. Given a braided fusion category $\CC$ and a braided autoequivalence $\phi \colon \CC \to \CC$, there is a fusion $(\CC,\CC)$-bimodule structure on $\CC$ defined by
\[
\CC \boxtimes \overline{\CC} \xrightarrow{\id_\CC \boxtimes \phi} \CC \boxtimes \overline{\CC} \to \FZ_1(\CC) ,
\]
where the second functor is the canonical one defined in Example \ref{expl:trivial_domain_wall}. So the left-bulk-to-boundary-map is $\id_\CC$ and the right-bulk-to-boundary-map is $\phi$. We denote this fusion $(\CC,\CC)$-bimodule by $\CC_\phi$. Similarly, $\CC_\phi$ is closed if and only if $\CC$ is nondegenerate.
\end{expl}

\begin{figure}[htbp]
\[
\begin{array}{c}
\begin{tikzpicture}[scale=0.8]
\fill[gray!20] (-2.5,-1.5) rectangle (2.5,1.5) ;
\draw[->-,very thick] (-0.7,1.5)--(-0.7,-1.5) node[at start,above] {$\SM$} ;
\draw[->-,very thick] (0.7,1.5)--(0.7,-1.5) node[at start,above] {$\SN$} ;
\draw[thick,decorate,decoration=brace] (-0.7,1.55)--(0.7,1.55) ;

\node at (-1.6,0) {$\SC$} ;
\node at (0,0) {$\SD$} ;
\node at (1.6,0) {$\SE$} ;
\end{tikzpicture}
\end{array}
\rightsquigarrow
\begin{array}{c}
\begin{tikzpicture}[scale=0.8]
\fill[gray!20] (-2,-1.5) rectangle (2,1.5) ;
\draw[->-,ultra thick] (0,1.5)--(0,-1.5) node[at start,above] {$\SM \boxtimes_\SD \SN$} ;

\node at (-1,0) {$\SC$} ;
\node at (1,0) {$\SE$} ;
\end{tikzpicture}
\end{array}
\]
\caption{the fusion of two 1d domain walls}
\label{fig:relative_fusion_product}
\end{figure}
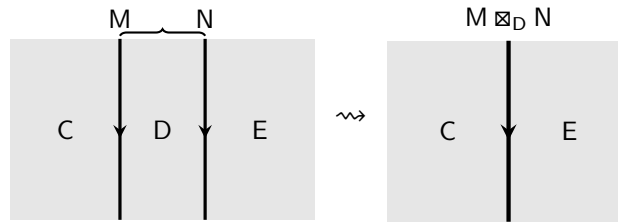

As depicted in Figure \ref{fig:relative_fusion_product}, if two domain walls $\SM$ and $\SN$ are very close to each other, they can be viewed as a single domain wall between $\SC$ and $\SE$, denoted by $\SM \boxtimes_\SD \SN$. This is an example of dimensional reduction. Clearly, if $\SD$ is the trivial 2d topological order, such a fusion process coincides with the stacking operation. The topological skeleton of $\SM \boxtimes_\SD \SN$ is denoted by $\CM \boxtimes_\CD \CN$, called the relative tensor product of $\CM$ and $\CN$ over $\CD$.

\begin{rem}
We only consider the case that $\SD$ is stable. Surprisingly, the relative tensor product $\CM \boxtimes_\CD \CN$ only depends on the multi-fusion categories $\CM,\CN$, the braided fusion category $\CD$ and the multi-fusion module structures of $\CM,\CN$ over $\CD$. Similar to the Deligne tensor product (see Definition \ref{defn:Deligne_tensor_product}), the relative tensor product $\CM \boxtimes_\CD \CN$ is defined by a universal property (for example, see \cite{Tam01,Mueg03,JL09,ENO10,Gre10,GS12,DNO12,DN13,GS15,FSV13,DSPS19,KZ18}). In particular, $\CM \boxtimes_\vect \CN$ agrees with the Deligne tensor product $\CM \boxtimes \CN$.
\end{rem}

In particular, when $\SC = \SD = \SE$, the fusion of 1d domain walls defines a multiplication on the set of gapped 1d domain walls in the 2d topological order $\SC$ (see Figure \ref{fig:fusion_topological_defect}).

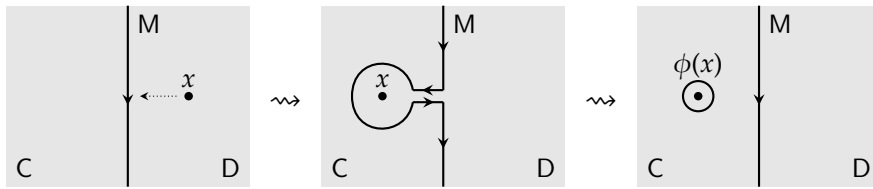
\begin{figure}[htbp]
\[
\begin{array}{c}
\begin{tikzpicture}[scale=0.8]
\fill[gray!20] (-2,0) rectangle (2,3) ;
\draw[->-,thick] (0,3)--(0,0) ;
\fill (1,1.5) circle (0.07) node[above] {$x$} ;
\draw[-stealth,densely dotted] (0.8,1.5)--(0.2,1.5) ;

\node[below right] at (0,3) {$\SM$} ;
\node[above right] at (-2,0) {$\SC$} ;
\node[above left] at (2,0) {$\SD$} ;
\end{tikzpicture}
\end{array}
\rightsquigarrow
\begin{array}{c}
\begin{tikzpicture}[scale=0.8]
\fill[gray!20] (-2,0) rectangle (2,3) ;
\draw[->-,thick] (0,3)--(0,1.6) ;
\draw[->-=0.7,thick] (0,1.6)--(-0.5,1.6) ;
\draw[thick] (-0.5,1.6) .. controls (-0.6,2.2) and (-1.5,2.2) .. (-1.5,1.5) .. controls (-1.5,0.8) and (-0.6,0.8) .. (-0.5,1.4) ;
\draw[->-=0.7,thick] (-0.5,1.4)--(0,1.4) ;
\draw[->-,thick] (0,1.4)--(0,0) ;
\fill (-1,1.5) circle (0.07) node[above] {$x$} ;

\node[below right] at (0,3) {$\SM$} ;
\node[above right] at (-2,0) {$\SC$} ;
\node[above left] at (2,0) {$\SD$} ;
\end{tikzpicture}
\end{array}
\rightsquigarrow
\begin{array}{c}
\begin{tikzpicture}[scale=0.8]
\fill[gray!20] (-2,0) rectangle (2,3) ;
\draw[->-,thick] (0,3)--(0,0) ;
\draw[thick] (-1,1.5) circle (0.25) ;
\fill (-1,1.5) circle (0.07) ;

\node at (-1,2) {$\phi(x)$} ;
\node[below right] at (0,3) {$\SM$} ;
\node[above right] at (-2,0) {$\SC$} ;
\node[above left] at (2,0) {$\SD$} ;
\end{tikzpicture}
\end{array}
\]
\caption{A topological defect $x \in \CD$ can be moved across an invertible domain $\SM$ and becomes a topological defect $\phi(x) \in \CC$. This process defines a braided equivalence $\phi \colon \CD \to \CC$.}
\label{fig:invertible_domain_wall}
\end{figure}

\begin{rem} \label{rem:invertible_domain_wall}
We say a 1d domain wall $\SM$ between two 2d topological orders $\SC$ and $\SD$ is \emph{invertible} if there exists another domain wall $\SN$ between $\SD$ and $\SC$ such that their fusions are trivial domain walls:
\[
\SM \boxtimes_\SD \SN = \SC_1 , \quad \SN \boxtimes_\SC \SM = \SD_1 .
\]
Also $\SN$ is called a inverse of $\CM$. Indeed, the inverse $\SN$, if exists, must be the time-reversal $\overline{\SM}$ by a general argument.

The particle-like topological defects can be moved across the invertible domain wall. Suppose $\SM$ is an invertible domain wall between two anomaly-free stable 2d topological orders $\SC$ and $\SD$. As depicted in Figure \ref{fig:invertible_domain_wall}, if we move a topological defect $x \in \CD$ across $\SM$ to $\SC$, this process produces a `bubble' in $\SC$. Two horizontal domain walls can be annihilated because $\SM \boxtimes_\SD \overline{\SM} = \SC_1$. We denote bubble containing the topological defect $x$ by $\phi(x)$. Then $x \mapsto \phi(x)$ defines a braided equivalence $\phi \colon \CD \to \CC$. Similarly, moving topological defects from $\SC$ to $\SD$ defines a braided equivalence $\psi \colon \CC \to \CD$, which is clearly a quasi-inverse of $\phi$. This argument also show that the topological skeleton of an invertible domain wall $\SM$ in a 2d topological order $\SC$ must have the form $\CC_\phi$ for some braided autoequivalence $\phi$ of $\CC$ (see Example \ref{expl:domain_wall_automorphism}).


Conversely, let $\CC$ be a nondegenerate braided fusion category. Suppose $\phi,\psi$ are two braided autoequivalence of $\CC$, then we have an equivalence
\[
\CC_\phi \boxtimes_\CC \CC_\psi \simeq \CC_{\phi \circ \psi}
\]
as fusion $(\CC,\CC)$-bimodules. Hence, the assignment $\phi \mapsto \CC_\phi$ is a group homomorphism from the group of isomorphism classes of braided autoequivalences of $\CC$ to the group of equivalence classes of invertible fusion $(\CC,\CC)$-bimodules. Indeed, this is a group isomorphism and both two groups are isomorphic to the Picard group of $\CC$ \cite{ENO10,KZ18}.
\end{rem}

\subsubsection{Gapped 1d domain walls in the toric code model} \label{sec:1d_domain_wall_toric_code}

\begin{figure}[htbp]
\centering
\subfigure[ss, $\rep(\Zb_2) \boxtimes \rep(\Zb_2)$]{
\begin{tikzpicture}[scale=0.8]
\draw[help lines,step=1.0] (-1.5,-1.7) grid (0,2.7) ;
\draw[help lines,step=1.0,xshift=-0.5cm] (1,-1.7) grid (2.5,2.7) ;
\end{tikzpicture}
}
\hspace{1ex}
\subfigure[sr, $\rep(\Zb_2) \boxtimes \vect_{\Zb_2}$]{
\begin{tikzpicture}[scale=0.8]
\draw[help lines,step=1.0] (-1.5,-1.7) grid (0,2.7) ;
\draw[help lines,dashed] (0.5,-1.7)--(0.5,2.7) ;
\draw[help lines,step=1.0,xshift=-0.5cm] (1.01,-1.7) grid (2.5,2.7) ;
\end{tikzpicture}
}
\hspace{1ex}
\subfigure[rs, $\vect_{\Zb_2} \boxtimes \rep(\Zb_2)$]{
\begin{tikzpicture}[scale=0.8]
\draw[help lines,step=1.0] (-1.5,-1.7) grid (-0.01,2.7) ;
\draw[help lines,dashed] (0,-1.7)--(0,2.7) ;
\draw[help lines,step=1.0,xshift=-0.5cm] (1,-1.7) grid (2.5,2.7) ;
\end{tikzpicture}
}
\hspace{1ex}
\subfigure[rr, $\vect_{\Zb_2} \boxtimes \vect_{\Zb_2}$]{
\begin{tikzpicture}[scale=0.8]
\draw[help lines,step=1.0] (-1.5,-1.7) grid (-0.01,2.7) ;
\draw[help lines,dashed] (0,-1.7)--(0,2.7) ;
\draw[help lines,dashed] (0.5,-1.7)--(0.5,2.7) ;
\draw[help lines,step=1.0,xshift=-0.5cm] (1.01,-1.7) grid (2.5,2.7) ;
\end{tikzpicture}
}
\caption{four non-invertible gapped 1d domain walls in the toric code model and their topological skeletons}
\label{fig:toric_code_domain_wall}
\end{figure}
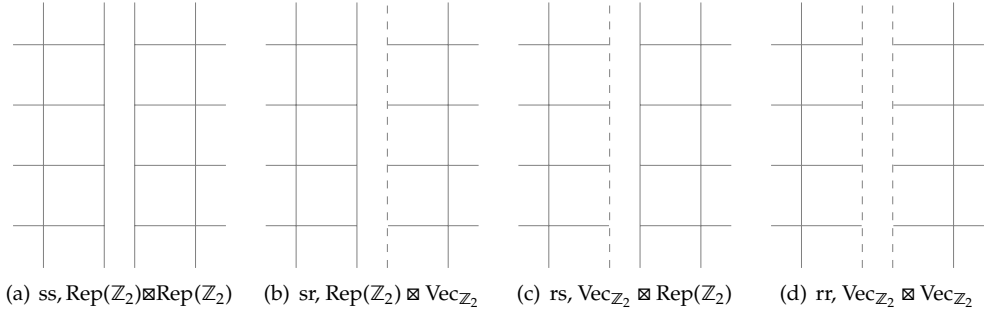

There are 6 different simple (i.e., stable) gapped 1d domain walls in the toric code model \cite{LWW15}. Four of them as depicted in Figure \ref{fig:toric_code_domain_wall} are non-invertible with respect to the fusion of 1d domain walls.

Both four non-invertible domain walls are obtained by fusing two boundaries of the toric code model. We denote them by `ss', `sr', `rs' and `rr', where `s' stands for the smooth boundary and `r' stands for the rough boundary. For the ss domain wall, it is clear that there are 4 simple topological excitations on this domain wall: the trivial one (denoted by $\one \boxtimes \one$), an $E$ particle on the left boundary (denoted by $E \boxtimes \one$), an $E$ particle on the right boundary (denoted by $\one \boxtimes E$), and two $E$ particles on both sides (denoted by $E \boxtimes E$). Thus the topological skeleton of this domain wall is $\rep(\Zb_2) \boxtimes \rep(\Zb_2)$. The left-bulk-to-domain-wall map is
\[
\one , m \mapsto \one \boxtimes \one , \quad e , f \mapsto E \boxtimes \one ,
\]
and the right-bulk-to-domain wall map is
\[
\one , m \mapsto \one \boxtimes \one , \quad e , f \mapsto \one \boxtimes E .
\]
Similarly one can find the topological skeletons of the other three domain walls. It is easy to show that these four domain walls are different by checking their bulk-to-domain-wall maps.

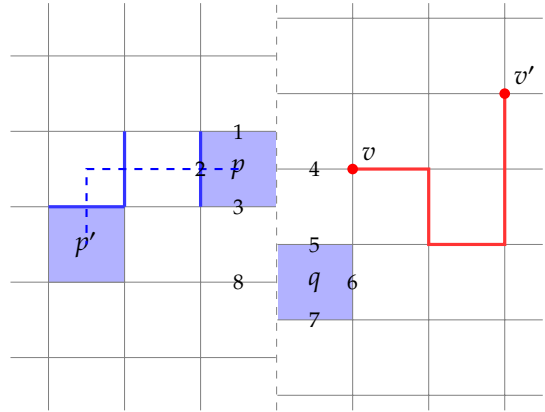
\begin{figure}[htbp]
\centering
\begin{tikzpicture}[scale=1.0]
\draw[help lines,step=1.0] (-3.5,-2.7) grid (-0.01,2.7)  ;
\draw[help lines,step=1.0,yshift=-0.5cm] (0.01,-2.2) grid (3.5,3.2) ;
\draw[help lines,dashed] (0,-2.7)--(0,2.7) ;

\draw[help lines,fill=m_ext] (-2,0) rectangle (-3,-1) node[midway,black] {$p'$} ;
\begin{scope}
\clip (-1,0) rectangle (-0.02,1) ;
\draw[help lines,fill=m_ext] (-1,0) rectangle (0,1) node[midway,black] {$p$} ;
\end{scope}
\begin{scope}
\clip (0.02,-0.5) rectangle (1,-1.5) ;
\draw[help lines,fill=m_ext] (0,-0.5) rectangle (1,-1.5) node[midway,black] {$q$} ;
\end{scope}
\draw[m_str] (-1,0)--(-1,1) ;
\draw[m_str] (-2,0)--(-2,1) ;
\draw[m_str] (-2,0)--(-3,0) ;
\draw[m_dual_str] (-0.5,0.5)-|(-2.5,-0.5) ;
\draw[e_str] (1,0.5)-|(2,-0.5)-|(3,1.5) ;
\fill[e_ext] (1,0.5) circle (0.07) node[above right,black] {$v$} ;
\fill[e_ext] (3,1.5) circle (0.07) node[above right,black] {$v'$} ;

\node[link_label] at (-0.5,1) {$1$} ;
\node[link_label] at (-1,0.5) {$2$} ;
\node[link_label] at (-0.5,0) {$3$} ;
\node[link_label] at (0.5,0.5) {$4$} ;
\node[link_label] at (0.5,-0.5) {$5$} ;
\node[link_label] at (1,-1) {$6$} ;
\node[link_label] at (0.5,-1.5) {$7$} ;
\node[link_label] at (-0.5,-1) {$8$} ;
\end{tikzpicture}
\caption{the $e$-$m$ exchange domain wall in toric code}
\label{fig:domain_wall_e_m}
\end{figure}

The other two simple 1d domain walls in the toric code model are invertible. One is the trivial domain wall (see Example \ref{expl:trivial_domain_wall}), another one is the dislocation as depicted in Figure \ref{fig:domain_wall_e_m}. There is still a spin-$1/2$ on each edge, except those on the dashed line. The $B_p$ operators on the domain wall are defined to be, for example, $B_p = \sigma_z^1 \sigma_z^2 \sigma_z^3 \sigma_x^4$ and $B_q = \sigma_z^5 \sigma_z^6 \sigma_z^7 \sigma_x^8$. The $A_v$ and $B_p$ operators in the bulk are the same as before.

There are $4$ simple topological excitations on the domain wall: the trivial one $\one$, only one $B_p = -1$ on the left of the domain wall (denoted by $X$), only one $B_q = -1$ on the right of the domain wall (denoted by $Y$), and two $B_p = -1$ on both sides (denoted by $Z$). It is clear that the fusion rules are given by
\[
X \otimes X = Y \otimes Y = Z \otimes Z = \one , \, X \otimes Y = Y \otimes X = Z .
\]

As depicted in Figure \ref{fig:domain_wall_e_m}, by using an $m$ string operator, an $m$ particle in the left bulk can be moved to the domain wall and becomes an $X$ particle. If we further apply a $\sigma_z^4$ operator, an $X$ particle at $p$ becomes an $e$ particle at $v$, which be moved into the right bulk. Therefore, the left-bulk-to-domain-wall map is
\[
\one \mapsto \one , \quad m \mapsto X , \quad e \mapsto Y , \quad f \mapsto Z ,
\]
and the right-bulk-to-domain-wall map is
\[
\one \mapsto \one , \quad e \mapsto X , \quad m \mapsto Y , \quad f \mapsto Z .
\]

This domain wall is also called the $e$-$m$-exchange domain wall, because if we move an $e$ particle across this domain wall, it becomes an $m$ particle. The exchange of $e$ and $m$ defines a braided autoequivalence $\phi \colon \Toric \to \Toric$, and the topological skeleton of this domain wall is precisely $\Toric_\phi$ (see Example \ref{expl:domain_wall_automorphism}).

\subsubsection{Fusion rules of gapped 1d domain walls in the toric code model} \label{sec:fusion_1d_domain_wall_toric_code}

Let us compute the fusion rules of gapped 1d domain walls in the toric code model.

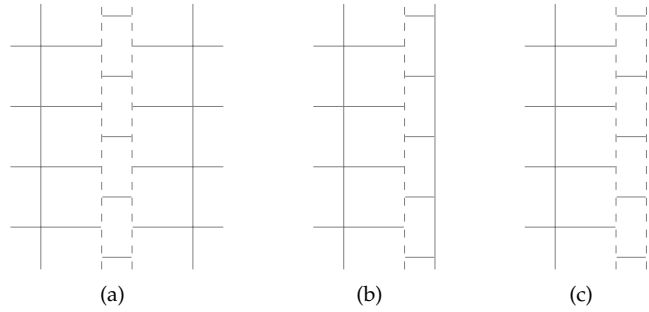
\begin{figure}[htbp]
\centering
\subfigure[]{
\begin{tikzpicture}[scale=0.8]
\draw[help lines,step=1.0] (-1.5,-1.7) grid (-0.01,2.7) ;
\draw[help lines,dashed] (0,-1.7)--(0,2.7) ;
\begin{scope}
\clip (0.01,-1.7) rectangle (0.49,2.7) ;
\draw[help lines,step=1.0,yshift=-0.5cm] (0.01,-1.2) grid (0.49,3.2) ;
\end{scope}
\draw[help lines,dashed] (0.5,-1.7)--(0.5,2.7) ;
\draw[help lines,step=1.0,xshift=-0.5cm] (1.01,-1.7) grid (2.5,2.7) ;
\end{tikzpicture}
}
\hspace{5ex}
\subfigure[]{
\begin{tikzpicture}[scale=0.8]
\draw[help lines,step=1.0] (-1.5,-1.7) grid (-0.01,2.7) ;
\draw[help lines,dashed] (0,-1.7)--(0,2.7) ;
\begin{scope}
\clip (0.01,-1.7) rectangle (0.49,2.7) ;
\draw[help lines,step=1.0,yshift=-0.5cm] (0.01,-1.2) grid (0.49,3.2) ;
\end{scope}
\draw[help lines] (0.5,-1.7)--(0.5,2.7) ;
\end{tikzpicture}
}
\hspace{5ex}
\subfigure[]{
\begin{tikzpicture}[scale=0.8]
\draw[help lines,step=1.0] (-1.5,-1.7) grid (-0.01,2.7) ;
\draw[help lines,dashed] (0,-1.7)--(0,2.7) ;
\begin{scope}
\clip (0.01,-1.7) rectangle (0.49,2.7) ;
\draw[help lines,step=1.0,yshift=-0.5cm] (0.01,-1.2) grid (0.49,3.2) ;
\end{scope}
\draw[help lines,dashed] (0.5,-1.7)--(0.5,2.7) ;
\end{tikzpicture}
}
\caption{the fusion of the $e$-$m$-exchange domain wall and other domain walls}
\label{fig:fusion_e_m_domain_wall}
\end{figure}

\medskip
First, let us consider the fusion rules of $e$-$m$-exchange domain wall with other domain walls. The fusion of two $e$-$m$-exchange domain walls is the trivial domain wall (see Figure \ref{fig:fusion_e_m_domain_wall} (a)). Indeed, if we move an $e$ particle across the domain wall, it changes twice and remains to be $e$. Similarly, any particle-like topological defects in the bulk can be moved across the domain wall and does not change. Mathematically, we have $\Toric_\phi \boxtimes_{\Toric} \Toric_\phi \simeq \Toric$.

The fusion rules of the $e$-$m$-exchange domain wall with four non-invertible domain walls reduce to the cases in Figure \ref{fig:fusion_e_m_domain_wall} (b) and (c). The fusion of the $e$-$m$-exchange domain wall with the smooth boundary is the rough boundary, because an $e$ particle becomes trivial after moving to the boundary, and an $m$ particle becomes an $E$ particle on the boundary. Similarly, the fusion of the $e$-$m$-exchange domain wall with the rough boundary is the smooth boundary. Mathematically, we have
\[
\Toric_\phi \boxtimes_{\Toric} \rep(\Zb_2) \simeq \vect_{\Zb_2} , \quad \Toric_\phi \boxtimes_{\Toric} \vect_{\Zb_2} \simeq \rep(\Zb_2) .
\]

The fusion rules of four non-invertible domain walls in the toric code model reduce to the fusion of two boundaries.

\begin{figure}[htbp]
\centering
\subfigure[]{
\begin{tikzpicture}[scale=0.8]
\draw[help lines,step=1.0] (-1,-1.7) grid (0.99,2.7) ;
\draw[help lines,dashed] (1,-1.7)--(1,2.7) ;

\draw[-stealth,e_str] (-1,2)--(1,2) ;
\fill[e_ext] (-1,2) circle (0.07) node[below right,black] {$E$} ;
\begin{scope}
\clip (0,0) rectangle (0.98,-1) ;
\draw[help lines,fill=m_ext] (0,0) rectangle (1,-1) node[midway,black] {$M$} ;
\end{scope}
\foreach \x in {-1,0}
	\draw[m_str] (\x,-1)--(\x,0) ;
\draw[-stealth,m_dual_str] (0.5,-0.5)--(-1.2,-0.5) ;
\end{tikzpicture}
}
\hspace{5ex}
\subfigure[]{
\begin{tikzpicture}[scale=0.8]
\draw[help lines,step=1.0] (-1,-1.7) grid (1,2.7) ;

\foreach \x/\xtext in {-1/1,0/2,1/3}
	\draw[m_str] (\x,0)--(\x,1) node[midway,right,black,scale=0.8] {$\xtext$} ;
\draw[m_dual_str] (-1.5,0.5)--(1.5,0.5) ;
\end{tikzpicture}
}
\hspace{5ex}
\subfigure[]{
\begin{tikzpicture}[scale=0.8]
\draw[help lines,step=1.0] (-1,-1.7) grid (1,2.7) ;

\foreach \x in {-1,0,1}
	\draw[m_str] (\x,0)--(\x,-1) ;
\foreach \x in {-1,0,1}
	\draw[m_str] (\x,1)--(\x,2) ;
\draw[m_dual_str] (-1.5,-0.5)--(1.5,-0.5) node[at end,above,black] {$X$} ;
\draw[m_dual_str] (-1.5,1.5)--(1.5,1.5) node[at end,above,black] {$Y$} ;
\fill[e_ext] (0,0) circle (0.07) node[above right,black] {$E$} ;
\end{tikzpicture}
}
\caption{(a): a narrow strip with a smooth and a rough boundary; (b): a narrow strip with two smooth boundaries; (c): the degeneracy of $E$.}
\label{fig:fusion_anomaly_free}
\end{figure}
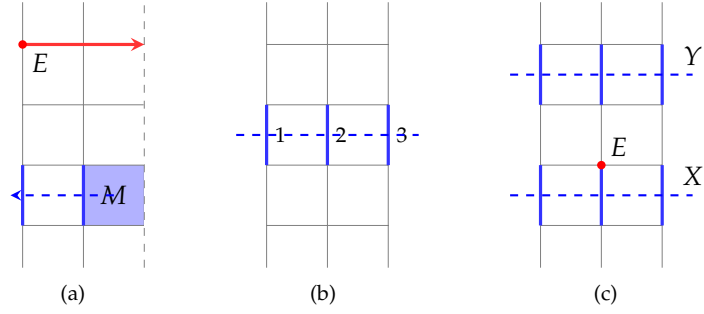

Figure \ref{fig:fusion_anomaly_free} (a) depicts a narrow strip of the toric code model with a smooth boundary and a rough boundary. This quasi-1d system can be viewed as an anomaly-free 1d topological order. There is no nontrivial topological excitations in this topological order. For example, an $e$ particle in the bulk can be annihilated on the rough boundary, and an $m$ particle can be annihilated on the smooth boundary. Similarly, an $E$ particle on the smooth boundary can be moved into the bulk and then annihilated on the rough boundary by a string operators. Such a string operator is a local operator because the strip is so narrow (with respect to the length scale). Hence the fusion of a smooth boundary with a rough boundary is the trivial 1d topological order:
\[
\rep(\Zb_2) \boxtimes_{\Toric} \vect_{\Zb_2} \simeq \vect .
\]
Since $\FZ_1(\vect) \simeq \vect$, the result is consistent with the boundary-bulk relation in 2d.

Figure \ref{fig:fusion_anomaly_free} (b) depicts another narrow strip of the toric code model with two smooth boundaries. In this case, an $m$ particle in the bulk can be annihilated on the boundary, but an $e$ particles (or $E$ particle) remains there. So we find two topological excitations $\one$ and $E$. Clearly we have $E \otimes E = \one$, as in the smooth boundary of the toric code model. It seems that the topological skeleton of this 1d topological order is $\rep(\Zb_2)$. However, the Drinfeld center $\FZ_1(\rep(\Zb_2))$ is not equivalent to $\vect$. So the result contradicts with the boundary-bulk relation in 2d. What is wrong?

\smallskip
The answer is that the fusion of two smooth boundaries is an unstable 1d topological order. Mathematically, $\rep(\Zb_2) \boxtimes_{\Toric} \rep(\Zb_2)$ is a multi-fusion category. Let us compute the GSD of this quasi-1d lattice model on a circle \cite{WW15}. Suppose there is an $L \times N$ square lattice and we impose a periodic boundary condition vertically. So the number of vertices is $V = (N+1)L$, the number of edges is $E = (2N+1)L$, and the number of plaquettes is $F = NL$. Moreover, we have $\prod_v A_v = 1$. Thus the GSD is $2^{E-(V-1)-F} = 2$.

We can also see this two-fold degeneracy in the following way \cite{KWZ15}. Consider an $m$ string operator $X \coloneqq \sigma_x^1 \sigma_x^2 \sigma_x^3$ (the dashed line in Figure \ref{fig:fusion_anomaly_free} (b)), which is a local operator because the strip is so narrow (with respect to the length scale). Clearly $X$ commutes with all $A_v$ and $B_p$ operators. It follows that $X$ acts on the ground state subspace invariantly, and two ground states correspond to two eigenspaces with $X = \pm 1$.

\begin{rem}
The existence of the operator $X$ implies that the ground state subspace is degenerate on an open interval (without boundary), not only on a circle. More generally, the GSD of an anomaly-free 1d topological order on a circle and on an open interval should be equal.
\end{rem}

The ground state subspace can be decomposed as the direct sum of two eigenspaces of $X$. Therefore, the trivial topological excitation $\one$ can be decomposed as the direct sum of simple ones $\one = \one_+ \oplus \one_-$, where $X$ acts on $\one_{\pm}$ as $\pm 1$.

The fusion rules are given by $\one_\pm \otimes \one_\pm = \one_\pm$ and $\one_\pm \otimes \one_\mp = 0$. Indeed, if the states on both the upper and lower half strip has eigenvalue $X = 1$, then so is the state on the whole strip; if the state on the lower half strip has $X = 1$ but the state on the upper half strip has $X = -1$, they can not be combined to a state on the whole strip unless there are some defects. This is because the string operator $X$ can be moved by applying $A_v$ operators. If the system is at the ground state thus all $A_v = 1$, then moving $X$ should not change the eigenvalue. The fusion rule $\one_+ \otimes \one_- = 0$ means this configuration is physically forbidden.

Similarly, the topological excitation $E$ can also be decomposed as the direct sum of simple ones. Suppose there is an $E$ particle located at a vertex $v$. The subspace of states with $A_v = -1$ also has dimension $2$ and can be decomposed by the eigenvalue of $X$ (see Figure \ref{fig:fusion_anomaly_free} (c)). But it can also be decomposed by the eigenvalues of $Y$. It can be verified that if a state in this subspace has eigenvalue $X = 1$, it must have eigenvalue $Y = -1$ because there is only one $A_v = -1$. If we view the state on the lower half strip as a topological excitation $\one_+$ and the state on the upper half strip as $\one_-$, this fact means $\one_+ \otimes E = E \otimes \one_- = \one_+ \otimes E \otimes \one_-$. We denote $E_{+-} \coloneqq \one_+ \otimes E \otimes \one_-$.

Also we define $E_{-+} \coloneqq \one_- \otimes E \otimes \one_+$, then clearly $E = E_{+-} \oplus E_{-+}$. It follows that the fusion rules are
\begin{gather*}
E_{+-} \otimes E_{-+} = \one_+ , \quad E_{-+} \otimes E_{+-} = \one_- , \quad E_{+-} \otimes E_{+-} = 0 , \quad E_{-+} \otimes E_{-+} = 0 , \\
\one_+ \otimes E_{+-} = E_{+-} = E_{+-} \otimes \one_- , \quad \one_- \otimes E_{-+} = E_{-+} = E_{-+} \otimes \one_+ , \\
\one_+ \otimes E_{-+} = 0 = E_{-+} \otimes \one_- , \quad \one_- \otimes E_{+-} = 0 = E_{+-} \otimes \one_+ .
\end{gather*}
These complicated fusion rules can be written in a more compact way. Define a multi-fusion category $\mathrm M_2(\vect)$ as follows:
\bit
\item As a category, $\mathrm M_2(\vect)$ is the direct sum of $4$ copies of $\vect$. Its objects are written as
\[
V =
\begin{pmatrix}
V_{11} & V_{12} \\ V_{21} & V_{22}
\end{pmatrix}
\]
where each $V_{ij} \in \vect$.
\item The tensor product of $\mathrm M_2(\vect)$ is given by the matrix product and the usual tensor product. More precisely, we have
\[
\begin{pmatrix}
V_{11} & V_{12} \\ V_{21} & V_{22}
\end{pmatrix}
\otimes
\begin{pmatrix}
W_{11} & W_{12} \\ W_{21} & W_{22}
\end{pmatrix}
=
\begin{pmatrix}
V_{11} \otimes W_{11} \oplus V_{12} \otimes W_{21} & V_{11} \otimes W_{12} \oplus V_{12} \oplus W_{22} \\ V_{21} \otimes W_{11} \oplus V_{22} \otimes W_{21} & V_{21} \otimes W_{12} \oplus V_{22} \otimes W_{22}
\end{pmatrix} .
\]
\item The tensor unit of $\mathrm M_2(\vect)$ is
\[
\begin{pmatrix}
\Cb & 0 \\ 0 & \Cb
\end{pmatrix}
=
\begin{pmatrix}
\Cb & 0 \\ 0 & 0
\end{pmatrix}
\oplus
\begin{pmatrix}
0 & 0 \\ 0 & \Cb
\end{pmatrix} ,
\]
which is not a simple object.
\eit
Then we have an equivalence $\rep(\Zb_2) \boxtimes_{\Toric} \rep(\Zb_2) \simeq \mathrm M_2(\vect)$ of multi-fusion categories:
\[
\one_+ \mapsto \begin{pmatrix}\Cb & 0 \\ 0 & 0\end{pmatrix} , \quad E_{+-} \mapsto \begin{pmatrix}0 & \Cb \\ 0 & 0\end{pmatrix} , \quad E_{-+} \mapsto \begin{pmatrix}0 & 0 \\ \Cb & 0\end{pmatrix} , \quad \one_- \mapsto \begin{pmatrix}0 & 0 \\ 0 & \Cb\end{pmatrix} .
\]
Moreover, by directly computing the half-braidings one can prove that $\FZ_1(\mathrm M_2(\vect)) \simeq \vect$, which gives the correct boundary-bulk relation in 2d (see Exercise \ref{exercise:Drinfeld_center_multi-fusion}).

In other words, the fusion of two smooth boundaries is the direct sum of two copies of the trivial 1d topological order.

\begin{rem} \label{rem:TO-neteq-TOSK}
A topological order is not the same as its topological skeleton. This distinction becomes manifest when we consider the category of topological orders \cite{KWZ15,KZ20a}. For example, the topological skeleton of the trivial 1d topological order $\tTO_1$ is $\vect$, but the topological skeleton of $\tTO_1 \oplus \tTO_1$ is $\mathrm M_2(\vect)$. This is similar to the fact that $\End(\Cb \oplus \Cb) \simeq \mathrm{M}_2(\Cb)$. 
\end{rem}

\begin{rem}
The construction of $\mathrm M_2(\vect)$ can be directly generalized to $\mathrm M_n(\CC)$ for any fusion category $\CC$ and $n \geq 1$. Similarly we have $\FZ_1(\mathrm M_n(\CC)) \simeq \FZ_1(\CC)$ as braided fusion categories. As an example, the relative tensor product of $\rep(\Zb_2) \boxtimes \vect_{\Zb_2}$ and $\vect_{\Zb_2} \boxtimes \vect_{\Zb_2}$ is
\[
\rep(\Zb_2) \boxtimes \vect_{\Zb_2} \boxtimes_{\Toric} \vect_{\Zb_2} \boxtimes \vect_{\Zb_2} \simeq \rep(\Zb_2) \boxtimes \mathrm M_2(\vect) \boxtimes \vect_{\Zb_2} \simeq \mathrm M_2(\rep(\Zb_2) \boxtimes \vect_{\Zb_2}) .
\]
So the fusion of 1d domain walls `sr' and `rr' is not a simple gapped 1d domain wall, but the superposition of two copies of `sr'. The complete fusion rules of simple 1d domain walls in the toric code model are listed in Table \ref{table:fusion_rule_1d_domain_wall_toric_code}.
\end{rem}

\begin{rem}
The multi-fusion category $\mathrm M_2(\vect)$ can also be realized by the 1+1D quantum Ising model (without transversal field):
\[
H = -\sum_i \sigma_z^i \sigma_z^{i+1} .
\]
There are two ground states with all $\sigma_z^i = +1$ or $\sigma_z^i = -1$. Two 0d domain walls between them are also clear. Fix a site $j$, consider the state with $\sigma_z^i = +1$ for $i < j$ and $\sigma_z^i = -1$ for $i \geq j$, then there is a 0d domain wall at the site $j$, which can not be annihilated by local operators. Similarly, the state with $\sigma_z^i = -1$ for $i < j$ and $\sigma_z^i = +1$ for $i \geq j$ gives another 0d domain wall.
\end{rem}

\begin{exercise} \label{exercise:Drinfeld_center_multi-fusion}
For all $n \geq 1$, prove that $\FZ_1(\mathrm M_n(\vect)) \simeq \vect$ as braided fusion categories. Hint: use Exercise \ref{exercise:linear_functor_preserve_direct_sum} (2) to determine a half-braiding in $\FZ_1(\mathrm M_n(\vect))$. Use the same method to prove that $\FZ_1(\mathrm M_n(\CC)) \simeq \FZ_1(\CC)$ for any fusion category $\CC$.
\end{exercise}

\begin{rem}
The readers might wonder why an unstable 1d topological order, often regarded as something `unnatural' or completely ignored in the most classifications, deserves a beautiful mathematical description of its topological skeleton. The reason is that the strip depicted in Figure \ref{fig:fusion_anomaly_free} is stable when it is sufficiently wide. It is important to include them when we consider the fusion of topological orders, especially in the study of the category of topological orders \cite{KWZ15,KZ20a,KZ22}. 
\end{rem}

\subsubsection{Witt equivalence}

\begin{defn}
We say that two anomaly-free stable 2d topological orders $\SC$ and $\SD$ are \emph{Witt equivalent} if there exists a gapped 1d domain $\SM$ wall between them.
\end{defn}

\begin{defn}[\cite{DMNO13}]
We say that two nondegenerate braided fusion categories $\CC$ and $\CD$ are \emph{Witt equivalent} if there exists a closed fusion $(\CC,\CD)$-bimodule $\CM$.
\end{defn}

Let us verify that the Witt equivalence of 2d topological orders is indeed an equivalence relation:
\bnu[(a)]
\item An anomaly-free stable 2d topological order $\SC$ is Witt equivalent to itself, because there is always the trivial domain wall $\SC_1$ in $\SC$ (see Remark \ref{expl:trivial_domain_wall}).
\item If there is a gapped 1d domain wall $\SM$ between $\SC$ and $\SD$, then we immediately get a domain wall $\overline{\SM}$ between $\SD$ and $\SC$ by reversing the orientation.
\item Suppose $\SM$ is a gapped 1d domain wall between $\SC$ and $\SD$, and $\SN$ is a gapped 1d domain wall between $\SD$ and $\SE$. Their fusion $\SM \boxtimes_\SD \SN$ is a gapped 1d domain wall between $\SC$ and $\SE$ (see Figure \ref{fig:relative_fusion_product}).
\enu
Similarly, the Witt equivalence of nondegenerate braided fusion categories is also an equivalence relation.

\begin{figure}[htbp]
\[
\begin{array}{c}
\begin{tikzpicture}[scale=0.8]
\fill[gray!30,opacity=0.7] (0,0)--(4,0)--(5,2)--(1,2)--cycle ;
\draw[->-,very thick] (3,2)--(2,0) node[at start,right] {$\SM_2$} ;
\node at (1.5,1) {$\SC_2$} ;
\node at (3.5,1) {$\SD_2$} ;
\fill[gray!30,opacity=0.5,yshift=0.5cm] (0,0)--(4,0)--(5,2)--(1,2)--cycle ;
\draw[->-,very thick,yshift=0.5cm] (3,2)--(2,0) node[at start,right] {$\SM_1$} ;
\node at (1.5,1.5) {$\SC_1$} ;
\node at (3.5,1.5) {$\SD_1$} ;
\end{tikzpicture}
\end{array}
\rightsquigarrow
\begin{array}{c}
\begin{tikzpicture}[scale=0.8]
\fill[gray!20] (0,0)--(4,0)--(5,2)--(1,2)--cycle ;
\draw[->-,very thick] (3,2)--(2,0) node[at start,above] {$\SM_1 \boxtimes \SM_2$} ;
\node at (1.5,1) {$\SC_1 \boxtimes \SC_2$} ;
\node at (3.5,1) {$\SD_1 \boxtimes \SD_2$} ;
\end{tikzpicture}
\end{array}
\]
\caption{stacking of 2d topological orders and 1d domain walls between them}
\label{fig:stacking_2d_1d}
\end{figure}
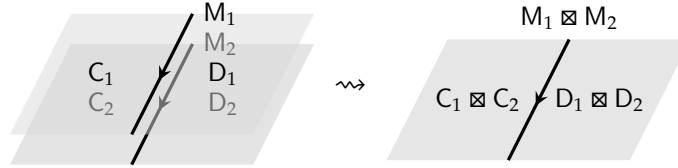

The stacking operation should preserve the Witt equivalence relation. More precisely, if $\SM$ is a domain wall between $\SC$ and $\SD$, and $\SM'$ is a domain wall between $\SC'$ and $\SD'$, then $\SM \boxtimes \SM'$ is a domain wall between $\SC \boxtimes \SC'$ and $\SD \boxtimes \SD'$ (see Figure \ref{fig:stacking_2d_1d}). So the stacking operation defines a multiplication on the set of Witt equivalence classes of anomaly-free stable 2d topological orders.
\bit
\item The multiplication is defined by $[\SC] \cdot [\SD] \coloneqq [\SC \boxtimes \SD]$.
\item There is an identity element given by the trivial 2d topological order $[\tTO_2]$. Indeed $[\tTO_2] \cdot [\SC] = [\tTO_2 \boxtimes \SC] = [\SC] = [\SC] \cdot [\tTO_2]$.
\item The inverse of $[\SC]$ is given by the time-reversal $[\overline{\SC}]$. Indeed, $[\SC] \cdot [\overline{\SC}] = [\SC \boxtimes \overline{\SC}] = [\tTO_2]$, where the gapped domain wall between $\SC \boxtimes \overline{\SC}$ and $\tTO_2$ is given by $\SC_1$ (see Figure \ref{fig:trivial_domain_wall}).
\item Moreover, this multiplication is commutative because $\SC \boxtimes \SD$ and $\SD \boxtimes \SC$ are the same topological order.
\eit
As a consequence, the Witt equivalence classes of 2d anomaly-free stable topological orders form an abelian group, where the multiplication is defined by the stacking. This abelian group is called the \emph{Witt group} of anomaly-free stable 2d topological orders. Similarly we can define the \emph{Witt group} of nondegenerate braided fusion categories \cite{DMNO13}.

All topological orders in a given Witt equivalence class must have the same chiral central charge. Thus taking the chiral central charge defines a group homomorphism from the Witt group of anomaly-free stable 2d topological orders to $\Qb$. Similarly, taking the additive central charge defines a group homomorphism from the Witt group of pseudo-unitary modular tensor categories (which is a subgroup of the Witt group of nondegenerate braided fusion categories) to $\Qb / 8 \Zb$.

\subsection{Gapped 0d domain walls}

In this subsection, we give a categorical description of 0d domain walls between gapped 1d domain walls and the boundary-bulk relation for 1d topological orders and their boundaries.

\subsubsection{The module category of 0d topological orders} \label{sec:0d_module_category}

Recall that 0+1D topological defects and instantons in an $n$d topological order form a category for $n \geq 1$ (see Theorem$^{\text{ph}}$ \ref{pthm:category_topological_defect}). For a 0d topological order $x$, we cannot talk about 0+1D topological defects of $x$, but 0+1D topological defects (or equivalently, 0d topological orders) living on the world line of $x$. Then Theorem$^{\text{ph}}$ \ref{pthm:category_topological_defect} still holds for a 0d topological order $x$ in the following sense: all 0d topological orders living on the world line of $x$, together with instantons between them, form a category $\CX$. By definition $x \in \CX$. We always assume that $\CX$ is finite semisimple.

\begin{rem}
Here we only consider gapped 0d topological defects. However, it is subtle to say that a 0d system is gapped. In practice, a stable 0d system is said to be gapped if its GSD is 1. In other words, a `gapless' 0d system has nontrivial GSD which is robust against perturbations. Usually this nontrivial GSD is protected by some symmetries.
\end{rem}

\begin{expl}
An anomaly-free 0d topological order is simply a quantum mechanic system because there is no thermodynamic limit. At zero temperature, the only observable is the ground state subspace. In other words, each 0d topological order is characterized by its ground state subspace, which is a finite-dimensional vector space (indeed, a Hilbert space). Thus the category of anomaly-free 0d topological order is equivalent to $\vect$.
\end{expl}

\begin{expl}
For an anomalous 0d topological order $x$, it is reasonable to talk about the thermodynamic limit if we choose an embedding of $x$ into an ambient higher-dimensional topological order. In this case the objects in the category $\CX$ are 0d topological orders which are able to live in the same ambient topological order. For example, let $\SC$ be an anomaly-free stable 2d topological order and $\CC$ be the modular tensor category of 0+1D topological defects of $\SC$. Then each $y \in \CC$ is an anomalous 0d topological order embedding in $\SC$. Clearly the category of 0d topological orders living on the world line of $y$, i.e., the category of 0d topological orders embedding in $\SC$, is $\CC$.
\end{expl}

\begin{figure}[htbp]
\centering
\begin{tikzpicture}[scale=1.0]
\draw[->-=0.7,ultra thick] (0,0)--(-5,0) node[near end,above] {$\SP$} ;
\fill (0,0) circle (0.1) node[above] {$x$} ;
\fill (-2,0) circle (0.07) node[above] {$a$} ;
\fill (-1,0) circle (0.07) node[above] {$b$} ;

\draw[-stealth,dashed] (-0.9,-0.1)--(-0.1,-0.1) ;
\draw[-stealth,dashed] (-1.9,-0.1)--(-1.1,-0.1) ;
\end{tikzpicture}
\caption{a 1d topological order $\SP$ and a 0d boundary $x$}
\label{fig:module}
\end{figure}
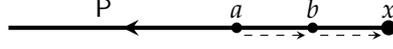

Now let us study the boundary-bulk relation of a 1d topological order and its 0d boundaries. Consider an anomaly-free 1d topological order $\SP$ and a 0d boundary $x$ (see Figure \ref{fig:module}). The topological skeleton of $\SP$ is a multi-fusion category $\CP$ and the category $\CX$ of 0d boundaries of $\SP$ (i.e., the category of 0d topological orders living on the boundary of $\SP$) is finite semisimple.

As in the 2d case, a bulk topological defect $a \in \CP$ can be moved to the boundary. This process may change the boundary topological order from $x$ to another one, which is denoted by $a \odot x$. We do not only move topological defects, but also the world lines and instantons attached to them. Hence $\odot$ is indeed a functor:
\begin{align*}
\odot \colon \CP \times \CX & \to \CX \\
(a,x) & \mapsto a \odot x .
\end{align*}
The physical intuition implies that two different ways of fusing two bulk topological defects $a,b$ with the boundary topological order $x$ should be the same (see Figure \ref{fig:module}). So we get an isomorphism (instanton) $s_{a,b,x} \colon (a \otimes b) \odot x \to a \odot (b \odot x)$. Finally, moving the trivial bulk topological defect $\one \in \CP$ to the boundary should not change the boundary topological order $x$. Thus we get an isomorphism (instanton) $l_x \colon \one \odot x \to x$ for each $x \in \CX$.

The above discussion leads to the following definition \cite{Ben65}.

\begin{defn}
Let $\CP$ be a monoidal category. A \emph{left $\CP$-module} or a \emph{left module over $\CP$} consists of the following data:
\bit
\item a category $\CX$;
\item a functor $\odot \colon \CP \times \CX \to \CX$, where $\odot(a,x)$ is also denoted by $a \odot x$;
\item a natural isomorphism $s_{a,b,x} \colon (a \otimes b) \odot x \to a \odot (b \odot x)$, called the \emph{associator};
\item a natural isomorphism $l_x \colon \one \odot x \to x$, called the \emph{left unitor};
\eit
and these data satisfy the following conditions:
\bnu
\item \textbf{(pentagon equation)} For any $a,b,c \in \CP$ and $x \in \CX$, the following diagram commutes:
\[
\xymatrix{
 & ((a \otimes b) \otimes c) \odot x \ar[dl]_{\alpha_{a,b,c} \odot \id_x} \ar[dr]^{\alpha_{a \otimes b,c,x}} \\
(a \otimes (b \otimes c)) \odot x \ar[d]_{s_{a,b \otimes c,x}} & & (a \otimes b) \odot (c \odot x) \ar[d]^{s_{a,b,c \otimes x}} \\
a \odot ((b \otimes c) \odot x) \ar[rr]^{\id_a \odot s_{b,c,x}} & & a \odot (b \odot (c \odot w))
}
\]
\item \textbf{(triangle equation)} For any $a \in \CP$ and $x \in \CX$, the following diagram commutes:
\[
\xymatrix{
(a \otimes \one) \odot x \ar[rr]^{s_{a,\one,x}} \ar[dr]_{\rho_a \odot \id_y} & & a \otimes (\one \odot x) \ar[dl]^{\id_a \odot l_y} \\
 & a \odot x
}
\]
\enu
When $\CP$ is a multi-fusion category, a left $\CP$-module is called \emph{finite semisimple} if it satisfies the following two conditions:
\bnu
\item $\CX$ is a finite semisimple category.
\item The left action functor $\odot \colon \CP \times \CX \to \CX$ is $\Cb$-bilinear (i.e., $\Cb$-linear in each variable).
\enu
\end{defn}

\begin{rem}
Similarly we can define a \emph{right module} over a monoidal category $\CQ$, which is essentially the same as a left module over $\CQ^\rev$. Moreover, for two monoidal categories $\CP,\CQ$, a \emph{$(\CP,\CQ)$-bimodule} is a category equipped with both a left $\CP$-module structure and a right $\CQ$-module structure, as well as the compatibility data/properties of these two structures. The details can be found in, for example, \cite[Definition 7.1.7]{EGNO15}.
\end{rem}

\begin{expl} \label{expl:regular_module}
Let $\CP$ be a monoidal category. Then $\CP$ itself is a left $\CP$-module with the left action defined by $a \odot b \coloneqq a \otimes b$ for $a,b \in \CP$. Similarly, $\CP$ is a $(\CP,\CP)$-bimodule.
\end{expl}

\begin{expl} \label{expl:monoidal_functor_pull_back}
Let $\CA,\CB$ be monoidal categories and $F \colon \CA \to \CB$ be a monoidal functor. Then $\CB$ is a $(\CA,\CA)$-bimodule with
\[
a \odot b \odot a' \coloneqq F(a) \otimes b \otimes F(a') , \quad a,a' \in \CA , \, b \in \CB .
\]
More generally, every left $\CB$-module $\CM$ can be viewed as a left $\CA$-module with the left $\CA$-action defined by $a \odot m \coloneqq F(a) \odot m$ for $a \in \CA$ and $m \in \CM$.
\end{expl}

\begin{expl}
Let $\CC$ be a braided fusion category and $\CM$ be a multi-fusion left $\CC$-module defined by a braided functor $F \colon \CC \to \FZ_1(\CM)$. Then $\CM$ is a left $\CC$-module with the left action defined by $a \odot x \coloneqq F(a) \otimes x$ for $a \in \CC$ and $x \in \CM$. Also, $\CM$ is a right $\CC$-module with the right action defined by $x \odot a \coloneqq x \otimes F(a)$ for $a \in \CC$ and $x \in \CM$. This can be viewed as a special case of Example \ref{expl:monoidal_functor_pull_back}.
\end{expl}

\begin{expl} \label{expl:opposite_module_category}
Let $\CP$ be a rigid monoidal category and $\CX$ be a left $\CP$-module. Then $\CX^\op$ is naturally a right $\CP$-module with
\[
x \odot a \coloneqq a^L \odot x , \quad a \in \CP , \, x \in \CX .
\]
There is also another right $\CP$-module structure defined by
\[
x \odot a \coloneqq a^R \odot x , \quad a \in \CP , \, x \in \CX .
\]
If $\CP$ is equipped with a pivotal structure, these two right $\CP$-module structures on $\CX$ are equivalent.
\end{expl}

\begin{rem}
Let $\CP$ be a multi-fusion category. Suppose $\CM,\CN$ are finite semisimple left $\CP$-modules. Then their direct sum $\CM \oplus \CN$ (see Definition \ref{defn:direct_sum_category}) is naturally a finite semisimple left $\CP$-module. A finite semisimple left $\CP$-module is called \emph{indecomposable} if it can not be written as a direct sum of two nonzero finite semisimple left $\CP$-modules. Given a multi-fusion category $\CP$, there are only finitely many indecomposable finite semisimple left $\CP$-modules (up to equivalence) \cite[Corollary 9.1.6]{EGNO15}.
\end{rem}

\begin{rem} \label{rem:module_functor_natural_transformation}
Let $\CP$ be a monoidal category and $\CM,\CN$ be left $\CP$-modules. A \emph{left $\CP$-module functor} from $\CM$ to $\CN$ is a functor $F \colon \CM \to \CN$ equipped with a natural isomorphism
\[
F^2_{x,m} \colon x \odot F(m) \to F(x \odot m) , \quad x \in \CP , \, m \in \CM ,
\]
that satisfies certain coherence conditions. Given two left $\CP$-module functors $F,G \colon \CM \to \CN$, a \emph{left $\CP$-module natural transformation} from $F$ to $G$ is a natural transformation $\alpha \colon F \Rightarrow G$ satisfying certain coherence conditions. Similarly one can define right module functors and right module natural transformations, as well as bimodule functors and bimodule natural transformations.
\end{rem}

In the language of module categories, we conclude that the category $\CX$ of 0d boundaries of a bulk 1d topological order $\SP$ is a finite semisimple left module over the topological skeleton $\CP$ (which is a multi-fusion category).

Note that the requirements that $\SP$ is anomaly-free and $x$ is a boundary are not necessary. For example, given a potentially anomalous 1d topological order $\SP$ and a gapped domain wall $x$ between $\SP$ and other 1d topological orders, $\CX$ is still a left module over $\CP$.

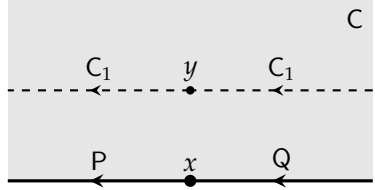
\begin{figure}[htbp]
\centering
\begin{tikzpicture}[scale=0.8]
\fill[gray!20] (-3,0) rectangle (3,3) node[at end,below left,black] {$\SC$} ;
\draw[->-,very thick] (3,0)--(0,0) node[midway,above] {$\SQ$} ;
\draw[->-,very thick] (0,0)--(-3,0) node[midway,above] {$\SP$} ;
\fill (0,0) circle (0.1) node[above] {$x$} ;
\draw[->-,,dashed,thick] (3,1.5)--(0,1.5) node[midway,above] {$\SC_1$} ;
\draw[->-,dashed,thick] (0,1.5)--(-3,1.5) node[midway,above] {$\SC_1$} ;
\fill (0,1.5) circle (0.07) node[above] {$y$} ;
\end{tikzpicture}
\caption{a typical configuration of 0d topological orders}
\label{fig:bimodule}
\end{figure}

\begin{expl}
Consider an anomaly-free 2d topological order $\SC$ with two 1d boundaries $\SP$ and $\SQ$, which are connected by a 0d domain wall $x$ (see Figure \ref{fig:bimodule}). Mathematically, the category $\CX$ of 0d topological orders living on the world line of $x$, i.e., the category of 0d domain walls between $\SP$ and $\SQ$, is a $(\CP,\CQ)$-bimodule, and both $\CP$ and $\CQ$ are closed multi-fusion left $\CC$-modules.

When $\SP = \SQ$, the category $\CX$ is equal to the category $\CP$. Both the left and right $\CP$-module action on $\CX$ are given by the tensor product of $\CP$ (see Example \ref{expl:regular_module}).
\end{expl}

\begin{expl}
Let $\SC$ be an anomaly-free 2d topological order. As depicted in Figure \ref{fig:bimodule}, a 0+1D topological defect $y \in \CC$ is a 0d domain wall between the trivial 1d domain wall $\SC_1$ (see Example \ref{expl:trivial_domain_wall}) and itself.
\end{expl}

\begin{expl}
Let $G$ be a finite group. A \emph{projective $G$-representation} is a vector space $V$ equipped with a group homomorphism $\bar{\rho} \colon G \to \mathrm{PGL}(V)$, where $\mathrm{PGL}(V) \coloneqq \mathrm{GL}(V) / \{\lambda \cdot \id_V \mid \lambda \in \Cb^\times\}$ is the group of projective linear transformations on $V$. It is also the automorphism group of the projective space $\mathbb P(V)$.

Given a projective $G$-representation $(V,\bar{\rho})$, since the canonical projection map $\pi \colon \mathrm{GL}(V) \to \mathrm{PGL}(V)$ is surjective, one can always find a lifting of $\bar{\rho}$, i.e., a map $\rho \colon G \to \mathrm{GL}(V)$ such that the following diagram commutes:
\[
\xymatrix{
G \ar[r]^{\rho} \ar[dr]_{\bar{\rho}} & \mathrm{GL}(V) \ar[d]^{\pi} \\
 & \mathrm{PGL}(V)
}
\]
We do not require $\rho$ to be a group homomorphism. Indeed, in general one can not find such a group homomorphism $\rho$. The equation $\bar{\rho}(g) \bar{\rho}(h) = \bar{\rho}(gh)$ implies that the map $\rho$ satisfies the equation
\[
\rho(g) \rho(h) = \beta(g,h) \cdot \rho(g,h)
\]
for some nonzero complex number $\beta(g,h) \in \Cb^\times$. The associativity $(\rho(g) \rho(h)) \rho(k) = \rho(g) (\rho(h) \rho(k))$ implies that the function $\beta \colon G \times G \to \Cb^\times$ satisfies the equation
\[
\beta(g,h) \beta(gh,k) = \beta(h,k) \beta(g,hk) ,
\]
or equivalently, $\mathrm d \beta = 1$ (see Example \ref{expl:vect_G_omega_monoidal_functor}). Such a function $\beta$ is called a \emph{2-cocycle} (valued in $\Cb^\times$). The space of 2-cocycles is denoted by $Z^2(G;\Cb^\times)$. We also call the pair $(V,\rho)$ a \emph{$\beta$-twisted $G$-representation}.

This 2-cocycle $\beta$ depends on the choice of the lifting $\rho$. If we choose a different lifting $\rho'$ of $\bar{\rho}$, the corresponding 2-cocycle $\beta'$ may be different from $\beta$. Assume that $\rho'(g) = \phi(g) \cdot \rho(g)$ for some nonzero complex number $\phi(g) \in \Cb^\times$, then
\[
\rho'(g) \rho'(h) = \phi(g) \phi(h) \cdot \rho(g) \rho(h) = \phi(g) \phi(h) \beta(g,h) \cdot \rho(gh) .
\]
It follows that
\be \label{eq:different_lifting_cohomology}
\beta'(g,h) \phi(gh) = \phi(g) \phi(h) \beta(g,h) .
\ee
Define
\[
(\mathrm d \phi)(g,h) \coloneqq \frac{\phi(h) \phi(g)}{\phi(gh)} , \quad g,h \in G.
\]
Then \eqref{eq:different_lifting_cohomology} can be rewritten as $\beta' = \beta \cdot \mathrm d \phi$. The operator $\mathrm d$ is usually called the \emph{differential}. It is straightforward to check that $\mathrm d \phi \in Z^2(G;\Cb^\times)$ is a 2-cocycle. A 2-cocycle obtained by the differential is called a \emph{2-coboundary}. The space of 2-coboundaries (valued in $\Cb^\times$) is denoted by $B^2(G;\Cb^\times)$. The quotient group $H^2(G;\Cb^\times) \coloneqq Z^2(G;\Cb^\times) / B^2(G;\Cb^\times)$ is called \emph{the second group cohomology group} of $G$ (with coefficient $\Cb^\times$). Then the above discussion implies that there is a well-defined cohomology class $[\beta] \in H^2(G;\Cb^\times)$ associated to each projective representation $(V,\bar{\rho})$.

Now we fix a 2-cocycle $\beta \in Z^2(G;\Cb^\times)$. Given two $\beta$-twisted $G$-representation $(V,\rho)$ and $(W,\sigma)$, a \emph{homomorphism} between them is a linear map $f \colon V \to W$ satisfying $f \circ \rho(g) = \sigma(g) \circ f$ for all $g \in G$. All finite-dimensional $\beta$-twisted $G$-representations and homomorphisms between them form a $\Cb$-linear category, denoted by $\rep(G,\beta)$. Indeed, $\rep(G,\beta)$ is finite semisimple.

Suppose $(V,\rho) \in \rep(G)$ and $(W,\sigma) \in \rep(G,\beta)$. It is easy to see that the tensor product space $V \otimes_\Cb W$ equipped with the $G$-action defined by
\[
g \mapsto \rho(g) \otimes_\Cb \sigma(g) , \quad g \in G ,
\]
is also a finite-dimensional $\beta$-twisted $G$-representation, denoted by $(V,\rho) \odot (W,\sigma)$. This defines a left $\rep(G)$-module structure on $\rep(G,\beta)$ with the left action $\odot \colon \rep(G) \times \rep(G,\beta) \to \rep(G,\beta)$.
\end{expl}

\begin{rem}
For each integer $n>1$ the cohomology group $H^2(\Zb_n;\Cb^\times)$ is trivial. Also we have $H^2(\Zb_2 \times \Zb_2;\Cb^\times) \simeq \Zb_2$, and the nontrivial cohomology class can be induced from the following projective $\Zb_2 \times \Zb_2$-representation on $\Cb^2$:
\[
(0,0) \mapsto \id_{\Cb_2} , \quad (1,0) \mapsto \sigma_x , \quad (0,1) \mapsto \sigma_y , \quad (1,1) \mapsto \sigma_z .
\]
Moreover, let $\beta$ be the 2-cocycle induced by the above projective representation. Then $\rep(\Zb_2 \times \Zb_2,\beta) \simeq \vect$, and the only simple object (up to isomorphism) is the above one.
\end{rem}

\begin{rem}
For every finite group $G$, the inclusion $\mathrm U(1) \hookrightarrow \Cb^\times$ induces a group isomorphism $H^2(G;\mathrm U(1)) \to H^2(G;\Cb^\times)$.
\end{rem}

\begin{rem}
More generally, suppose $H$ is a subgroup of $G$. Since the forgetful functor $F \colon \rep(G) \to \rep(H)$ is a monoidal functor (see Example \ref{expl:rep_G_forget_H}), every left $\rep(H)$-module is naturally a left $\rep(G)$-module (see Example \ref{expl:monoidal_functor_pull_back}). In particular, $\rep(H,\beta)$ is a left $\rep(G)$-module for each $\beta \in Z^2(H;\Cb^\times)$.

Indeed, every indecomposable finite semisimple left $\rep(G)$-module is equivalent to $\rep(H,\beta)$ for some subgroup $H$ of $G$ and 2-cocycle $\beta \in Z^2(H;\Cb^\times)$ \cite{Ost03}.
\end{rem}

\begin{rem}
Let $G$ be a finite group and $\beta \in Z^2(G;\Cb^\times)$. The \emph{$\beta$-twisted group algebra} $\Cb[G,\beta]$ is defined as follows. Its underlying vector space is freely generated by $G$, or equivalently, spanned by a family of symbols $\{v_g\}_{g \in G}$. The multiplication is defined by $v_g \cdot v_h \coloneqq \beta(g,h) \cdot v_{gh}$ for $g,h \in G$. Then $\rep(G,\beta)$ is equivalent to the category $\LMod_{\Cb[G,\beta]}(\vect)$ of finite-dimensional modules over $\Cb[G,\beta]$.
\end{rem}

\subsubsection{The boundary-bulk relation in 1d}

Let $\SP$ be an anomaly-free 1d topological order and $\CP$ be its topological skeleton. The category of 0d boundaries of $\SP$ is denoted by $\CX$. Moving a bulk topological excitation $a \in \CP$ to the boundary changes the boundary topological order and defines a functor from $\CX$ to itself. Mathematically, by fixing the first variable of the module action $\odot \colon \CP \times \CX \to \CX$ we get a $\Cb$-linear functor $a \odot - \colon \CX \to \CX$ for each $a \in \CP$.

We use $\FZ_0(\CX)$ to denote the category $\fun(\CX,\CX)$ of $\Cb$-linear functors from a finite semisimple category $\CX$ to itself (see Exercise \ref{exercise:functor_category}). It is a monoidal category where the tensor product is defined by the composition of functors. Then we see that the left $\CP$-module structure on $\CX$ induces a monoidal functor
\begin{align*}
\CP & \to \fun(\CX,\CX) \\
a & \mapsto a \odot - .
\end{align*}
Moreover, $\CX$ is a left module over $\FZ_0(\CX)$ with the left action defined by
\begin{align*}
\FZ_0(\CX) \times \CX & \to \CX \\
(F,x) & \mapsto F(x) .
\end{align*}

\begin{expl} \label{expl:matrix_as_functor_cat}
Let $\CX$ be a finite semisimple category with $n$ non-isomorphic simple objects, i.e., $\CX \simeq \vect^{\oplus n}$. Then $\FZ_0(\CX) \simeq \mathrm M_n(\vect)$ as monoidal categories. Hence $\FZ_0(\CX)$ is indeed a multi-fusion category.
\end{expl}

Given a multi-fusion category $\CP$ and a finite semisimple category $\CX$, a left $\CP$-module structure on $\CX$ is indeed equivalent to a $\Cb$-linear monoidal functor $\CP \to \fun(\CX,\CX)$. In other words, any left action of a monoidal category on $\CX$ factors through $\fun(\CX,\CX)$, in the sense that there exists a unique (up to isomorphism) $\Cb$-linear monoidal functor $\varphi \colon \CP \to \FZ_0(\CX)$ such that the following diagram commutes up to isomorphism:
\[
\xymatrix{
 & \FZ_0(\CX) \times \CX \ar[dr] \\
\CP \times \CX \ar[rr]^{\odot} \ar[ur]^{\varphi \times \id_\CX} & & \CX
}
\]


\begin{rem}
The functor category $\FZ_0(\CX)$ is a kind of $E_0$-center \cite{Lur17}.
\end{rem}

\begin{rem}
It is also convenient to write a 0d boundary $x \in \CX$ as a pair $(\CX,x)$ to emphasize the category $\CX$ of 0d boundaries. Mathematically, the pair $(\CX,x)$ is a kind of $E_0$-algebra.
\end{rem}

Similar to the 2d case, there is a boundary-bulk relation for 1d topological orders and their 0d boundaries.

\begin{pthm}[Boundary-bulk relation in 1d] \label{pthm:boundary_bulk_1d}
Suppose $\SP$ is an anomaly-free 1d topological order. Its topological skeleton is a (unitary) multi-fusion category $\CP$. The category of 0d boundaries of $\SP$ is denoted by $\CX$.
\bnu[(a)]
\item Moving bulk topological defects to boundaries induces a left $\CP$-module structure on $\CX$.
\item The monoidal functor $\CP \to \FZ_0(\CX)$ induced by the module action is an equivalence.
\enu
\end{pthm}

\begin{rem} \label{rem:1d_topological_order_classification}
Consider the direct sum of $n$ copies of the trivial 1d topological order $(\tTO_1)^{\oplus n}$. Clearly the category of its 0d boundaries is the direct sum $\vect^{\oplus n}$ of $n$ copies of $\vect$. By Example \ref{expl:matrix_as_functor_cat} and the boundary-bulk relation in 1d (Theorem$^{\text{ph}}$ \ref{pthm:boundary_bulk_1d}), the topological skeleton of $(\tTO_1)^{\oplus n}$ is $\mathrm M_n(\vect)$. This is consistent with the boundary-bulk relation in 2d (Theorem$^{\text{ph}}$ \ref{pthm:boundary_bulk}) because $\FZ_1(\mathrm M_n(\vect)) \simeq \vect$.

Conversely, a multi-fusion category $\CC$ satisfies $\FZ_1(\CC) \simeq \vect$ if and only if $\CC \simeq \mathrm M_n(\vect)$ for some positive integer $n$ (see for example \cite[Corollary 2.5.3]{KZ18}). Physically, an anomaly-free 1d topological order must be the direct sum of several copies of the trivial 1d topological order \cite{CGW11}. This also coincides with the classification of 1+1D fully extended TQFT's \cite{FHK94,SP11}.
\end{rem}

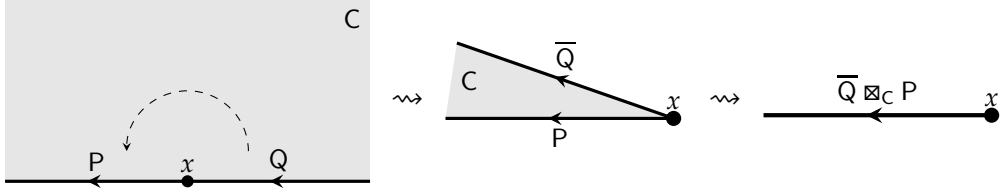
\begin{figure}[htbp]
\[
\begin{array}{c}
\begin{tikzpicture}[scale=0.8]
\fill[gray!20] (-3,0) rectangle (3,3) node[at end,below left,black] {$\SC$} ;
\draw[->-,very thick] (3,0)--(0,0) node[midway,above] {$\SQ$} ;
\draw[->-,very thick] (0,0)--(-3,0) node[midway,above] {$\SP$} ;
\fill (0,0) circle (0.1) node[above] {$x$} ;
\draw[-stealth,dashed] (1,0.5) arc (0:180:1) ;
\end{tikzpicture}
\end{array}
\rightsquigarrow
\begin{array}{c}
\begin{tikzpicture}[scale=1.0]
\fill[gray!20] (0,0)--(-3,0)--(-2.85,1) node[midway,right,black] {$\SC$}--cycle ;
\draw[->-,very thick] (0,0)--(-3,0) node[midway,below] {$\SP$} ;
\draw[->-,very thick] (0,0)--(-2.85,1) node[midway,above] {$\overline{\SQ}$} ;
\fill (0,0) circle (0.1) node[above] {$x$} ;
\end{tikzpicture}
\end{array}
\rightsquigarrow
\begin{array}{c}
\begin{tikzpicture}[scale=1.0]
\draw[->-,ultra thick] (0,0)--(-3,0) node[midway,above] {$\overline{\SQ} \boxtimes_\SC \SP$} ;
\fill (0,0) circle (0.1) node[above] {$x$} ;
\end{tikzpicture}
\end{array}
\]
\caption{the folding trick}
\label{fig:bimodule_folding}
\end{figure}

The boundary-bulk relation in 1d also helps us to find the relation between topological orders in a more general setting. For example, as depicted in Figure \ref{fig:bimodule_folding}, a 0d domain wall $x$ between two 1d boundaries $\SP$ and $\SQ$ of a 2d topological order $\SC$ can be viewed as a boundary of the anomaly-free 1d topological order $\overline{\SQ} \boxtimes_\SC \SP$, which is obtained by the relative fusion product. Mathematically, we have $\FZ_0(\CX) \simeq \CQ^\rev \boxtimes_\CC \CP$ as multi-fusion categories.

\begin{exercise} \label{exercise:bulk_of_a_particle}
Let $x$ be a particle-like topological defect in the toric code model. Its 1d bulk can be obtained via a dimensional reduction process (see Figure \ref{fig:dimensional_reduction}).
\bnu[(1)]
\item Put this 1d bulk on a circle. What is the GSD?
\item Find some local operators that commute with the Hamiltonian. These local operators should explain the nontrivial GSD of this 1d bulk (defined on an open interval). This is similar to the last example in Section \ref{sec:fusion_1d_domain_wall_toric_code}.
\item Determine the topological skeleton of this 1d bulk.
\item Check the boundary-bulk relation in 1d (Theorem$^{\text{ph}}$ \ref{pthm:boundary_bulk_1d}) in this example.
\enu
\end{exercise}

\subsubsection{Gapped 0d domain walls in the toric code model} \label{sec:0d_domain_wall_toric_code}

All nontrivial 0d domain walls between 1d domain walls in the toric code model can be reduced to the following two cases.

\begin{figure}[htbp]
\centering
\begin{tikzpicture}[scale=1.0]
\draw[help lines,step=1.0] (-3.5,0) grid (0,2.5) ;
\draw[help lines,step=1.0] (0,0.01) grid (3.5,2.5) ;
\draw[help lines,dashed] (0,0)--(3.5,0) ;

\draw[thick,fill=white] (0,0) circle (0.1) node[below] {$\CX
\simeq \vect$} ;
\end{tikzpicture}
\caption{a 0d domain wall between the smooth and rough boundaries of the toric code model}
\label{fig:domain_wall_smooth_rough}
\end{figure}
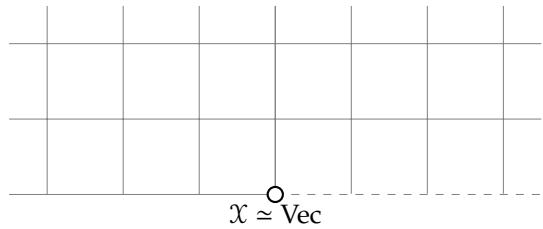

First we consider 0d domain walls between the smooth and rough boundaries as depicted in Figure \ref{fig:domain_wall_smooth_rough}. There is no $A_v$ operator on the vertex on the domain wall. Clearly both $e$ particles and $m$ particles can be created or annihilated by local operators near the domain wall, so there is only the trivial topological excitation $\one$ on the domain wall. In other words, the category $\CX$ of 0d domain walls between the smooth and rough boundaries has only one simple object $\one$ and thus is equivalent to $\vect$. This is consistent with the boundary-bulk relation in 1d and the relative fusion product $\rep(\Zb_2) \boxtimes_{\Toric} \vect_{\Zb_2} \simeq \vect \simeq \FZ_0(\vect)$ (see Figure \ref{fig:bimodule_folding}).

\begin{exercise}
The category of 0d domain walls between the smooth boundary and itself is $\rep(\Zb_2)$, which is clearly a $(\rep(\Zb_2),\rep(\Zb_2))$-bimodule. Find the bulk of such 0d domain walls by the folding trick (see Figure \ref{fig:bimodule_folding}) and check the boundary-bulk relation in 1d in this example.
\end{exercise}

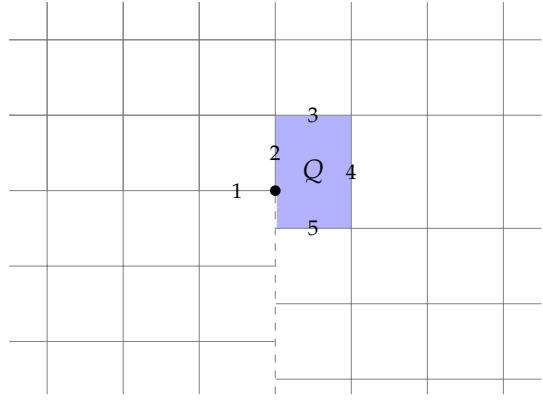
\begin{figure}[htbp]
\centering
\begin{tikzpicture}[scale=1.0]
\draw[help lines,step=1.0] (-3.5,0.01) grid (3.5,2.5) ;
\draw[help lines,step=1.0] (-0.01,2.5) grid (-3.5,-2.7) ;
\draw[help lines,step=1.0,yshift=-0.5cm] (0.01,0.51) grid (3.5,-2.2) ;
\draw[help lines,dashed] (0,-2.7)--(0,0) ;

\begin{scope}
\clip (0.02,-0.5)--(1,-0.5)--(1,1)--(0,1)--(0,0)--(0.02,0)--cycle ;
\draw[help lines,fill=m_ext] (0,-0.5) rectangle (1,1) node[midway,black] {$Q$} ;
\end{scope}
\fill (0,0) circle (0.07) ;

\node[link_label] at (-0.5,0) {$1$} ;
\node[link_label] at (0,0.5) {$2$} ;
\node[link_label] at (0.5,1) {$3$} ;
\node[link_label] at (1,0.25) {$4$} ;
\node[link_label] at (0.5,-0.5) {$5$} ;
\end{tikzpicture}
\caption{0d domain walls between the trivial 1d domain wall and the $e$-$m$-exchange domain wall}
\label{fig:0d_Majorana}
\end{figure}

Figure \ref{fig:0d_Majorana} depicts 0d domain walls between the trivial 1d domain wall and the $e$-$m$-exchange domain wall. The operator $Q$ is defined as \cite{Bom10,KK12}
\[
Q \coloneqq \sigma_x^1 \sigma_y^2 \sigma_z^3 \sigma_z^4 \sigma_z^5 .
\]
It can be easily verified that $Q$ commutes with all $A_v$ and $B_p$ operators, including those on the $e$-$m$-exchange domain wall.

If we take the Hamiltonian to be
\[
H = \sum_v (1-A_v) + \sum_p (1-B_p) - Q ,
\]
the ground state subspace is determined by eigenvalues $A_v = B_p = Q = +1$. It generates a topological excitation denoted by $\chi_+$, which is a 0d domain wall between the trivial 1d domain wall and the $e$-$m$-exchange domain wall. Similarly, if we take the Hamiltonian to be
\[
H = \sum_v (1-A_v) + \sum_p (1-B_p) + Q ,
\]
then the ground state is determined by the eigenvalues $A_v = B_p = +1$ and $Q = -1$. It also generates a topological excitation denoted by $\chi_-$. The category $\CX$ of 0d domain walls between the trivial 1d domain wall and the $e$-$m$-exchange domain wall has two simple objects $\chi_\pm$ and thus is equivalent to $\vect \oplus \vect$.

Now we choose the orientation such that $\CX$ is a $(\Toric,\Toric_\phi)$-bimodule. It is not hard to compute the module actions in the lattice model. The left $\Toric$-module action on $\CX$ is given by
\be \label{eq:Majorana_action_1}
e \odot \chi_\pm = \chi_\mp , \quad m \odot \chi_\pm = \chi_\mp , \quad f \odot \chi_\pm = \chi_\pm ,
\ee
and the right $\Toric_\phi$-module action on $\CX$ is given by
\be \label{eq:Majorana_action_2}
\chi_\pm \odot X = \chi_\mp , \quad \chi_\pm \odot Y = \chi_\mp , \quad \chi_\pm \odot Z = \chi_\pm .
\ee

\begin{exercise}
Prove \eqref{eq:Majorana_action_1} and \eqref{eq:Majorana_action_2} in the lattice model.
\end{exercise}

\begin{exercise}
Find the 1d bulk of $\chi_\pm$ and check the boundary-bulk relation in 1d in this example. Hint: use the same method in Exercise \ref{exercise:bulk_of_a_particle}.
\end{exercise}

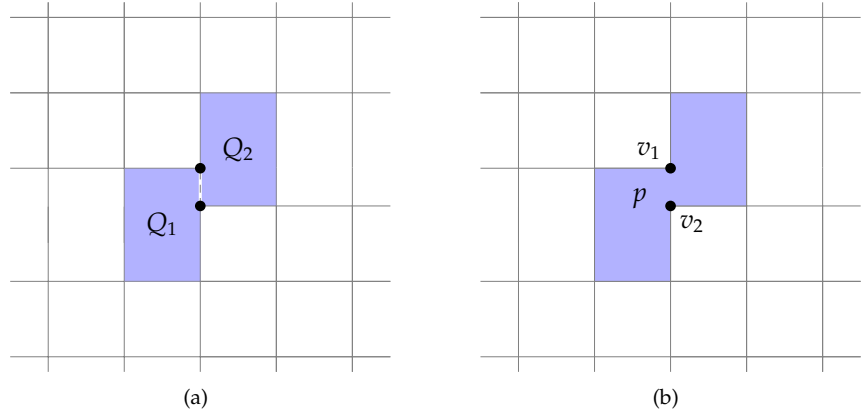
\begin{figure}[htbp]
\centering
\subfigure[]{
\begin{tikzpicture}[scale=1.0]
\draw[help lines,step=1.0] (-2.5,0.01) grid (2.5,2.2) ;
\draw[help lines,step=1.0] (-2.5,-0.99) grid (-0.01,2.2) ;
\draw[help lines,step=1.0,yshift=-0.5cm] (2.5,-0.01) grid (-2.5,-2.2) ;
\draw[help lines,step=1.0,yshift=-0.5cm] (0.01,0.51) grid (2.5,-2.2) ;
\draw[help lines,densely dashed] (0,-0.5)--(0,0) ;

\begin{scope}
\clip (0.02,-0.5)--(1,-0.5)--(1,1)--(0,1)--(0,0)--(0.02,0)--cycle ;
\draw[help lines,fill=m_ext] (0,-0.5) rectangle (1,1) node[midway,black] {$Q_2$} ;
\end{scope}
\begin{scope}
\clip (-0.02,-0.5)--(0,-0.5)--(0,-1.5)--(-1,-1.5)--(-1,0)--(-0.02,0)--cycle ;
\draw[help lines,fill=m_ext] (-1,-1.5) rectangle (0,0) node[midway,black] {$Q_1$} ;
\end{scope}
\fill (0,0) circle (0.07) ;
\fill (0,-0.5) circle (0.07) ;
\end{tikzpicture}
}
\hspace{5ex}
\subfigure[]{
\begin{tikzpicture}[scale=1.0]
\draw[help lines,step=1.0] (-2.5,0.01) grid (2.5,2.2) ;
\draw[help lines,step=1.0] (-2.5,-0.99) grid (-0.01,2.2) ;
\draw[help lines,step=1.0,yshift=-0.5cm] (2.5,-0.01) grid (-2.5,-2.2) ;
\draw[help lines,step=1.0,yshift=-0.5cm] (0.01,0.51) grid (2.5,-2.2) ;

\draw[help lines,fill=m_ext] (0,-0.5)--(1,-0.5)--(1,1)--(0,1)--(0,0)--(-1,0)--(-1,-1.5)--(0,-1.5)--cycle ;
\fill (0,0) circle (0.07) node[above left] {$v_1$} ;
\fill (0,-0.5) circle (0.07) node[below right] {$v_2$} ;
\node at (-0.4,-0.4) {$p$} ;
\end{tikzpicture}
}
\caption{the fusion of $\chi_\pm$ along the $e$-$m$-exchange domain wall}
\label{fig:fusion_chi_pm}
\end{figure}

\begin{exercise}
The 0d domain walls $\chi_\pm$ can be fused along the $e$-$m$-exchange domain wall. The results should be 0d topological defects in the toric code model.
\bnu[(1)]
\item Consider the `shortest' $e$-$m$-exchange domain wall and two end points as depicted in Figure \ref{fig:fusion_chi_pm} (a). Prove that
\[
Q_1 Q_2 = A_{v_1} A_{v_2} B_p , \qquad [Q_i,B_p] = 0 , \quad \{Q_i,A_{v_1}\} = \{Q_i,A_{v_2}\} = 0 , \, (i = 1,2)
\]
where the plaquette $p$ and vertices $v_1,v_2$ are depicted in Figure \ref{fig:fusion_chi_pm} (b).
\item Conclude that the fusion rules of $\chi_\pm$ along the $e$-$m$-exchange domain wall are (see \cite{Bom10,KK12})
\be \label{eq:fusion_chi_pm}
\chi_\pm \otimes \chi_\pm = \one \oplus f , \quad \chi_\pm \otimes \chi_\mp = e \oplus m .
\ee
\enu
\end{exercise}

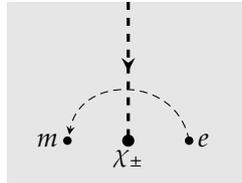
\begin{figure}[htbp]
\centering
\begin{tikzpicture}[scale=0.8]
\fill[gray!20] (-2,0) rectangle (2,3) ;
\draw[->-=0.5,very thick,dashed] (0,3)--(0,0.7) ;
\fill (0,0.7) circle (0.1) node[below] {$\chi_\pm$} ;
\fill (1,0.7) circle (0.07) node[right] {$e$} ;
\fill (-1,0.7) circle (0.07) node[left] {$m$} ;
\draw[-stealth,densely dashed] (1,0.7) .. controls (0.8,1.8) and (-0.8,1.8) .. (-0.97,0.8) ;
\end{tikzpicture}
\caption{The `braiding' of $e$ or $m$ with $\chi_\pm$}
\label{fig:crossed_braiding_chi_pm}
\end{figure}

\begin{rem} \label{rem:toric_chi_pm_crossed_braiding}
From \eqref{eq:Majorana_action_1} and \eqref{eq:fusion_chi_pm} we see that the set $\{\one,e,m,f,\chi_+,\chi_-\}$ is closed under the fusion. In other words, these topological defects generate a fusion category, denoted by $\Toric_+$. However, $\Toric_+$ is not a braided fusion category. For example, the `braiding' of $e$ with $\chi_\pm$ (see Figure \ref{fig:crossed_braiding_chi_pm}) is an instanton
\[
\chi_\pm \otimes e \to m \otimes \chi_\pm
\]
because moving an $e$ particle across the $e$-$m$-exchange domain wall gives an $m$ particle. These `braidings' do not form an ordinary braiding structure.

Roughly speaking, a $G$-crossed braided fusion category \cite{Tur00,Kir01,Mueg04,DGNO10} (where $G$ is a finite group) is a $G$-graded fusion category $\CC = \medoplus_{g \in G} \CC_g$ equipped with a compatible $G$-action and a $G$-crossed braiding
\[
c_{x,y} \colon x \otimes y \to g(y) \otimes x , \quad x \in \CC_g \, , y \in \CC .
\]
A typical example is $\vect_G^\omega$ for $\omega \in Z^3(G;\Cb^\times)$. It has a unique structure of a $G$-crossed braided fusion category (up to a unique isomorphism) with the obvious $G$-grading \cite{DGNO10}. In the above example, $\Toric_+$ admits a $\Zb_2$-grading $\Toric_+ = (\Toric_+)_0 \oplus (\Toric_+)_1$ with $(\Toric_+)_0 = \Toric$ and $(\Toric_+)_1$ generated by $\chi_\pm$. The $\Zb_2$-action on $\Toric_+$ is given by exchanging $e$ and $m$. Thus $\Toric_+$ is a $\Zb_2$-crossed braided fusion category \cite{BBCW19}.
\end{rem}

\subsubsection{Morita equivalence}

Suppose $\SP$ and $\SQ$ are gapped 1d boundaries of a 2d topological order $\SC$. Let $\CP$ and $\CQ$ be the topological skeletons of $\SP$ and $\SQ$, respectively. Assume that there exists at least one 0d domain wall $x$ between $\SP$ and $\SQ$. Then the category of 0d domain walls between $\SP$ and $\SQ$ is a nonzero finite semisimple $(\CP,\CQ)$-bimodule $\CX$ and $x \in \CX$. By reversing the orientation, the opposite category $\CX^\op$ is a finite semisimple $(\CQ,\CP)$-bimodule (see Example \ref{expl:opposite_module_category}), which is the category of 0d domain walls between $\SQ$ and $\SP$.

\begin{figure}[htbp]
\[
\begin{array}{c}
\begin{tikzpicture}[scale=0.8]
\fill[gray!20] (-3,0) rectangle (3,3) node[at end,below left,black] {$\SC$} ;
\draw[->-=0.58,very thick] (3,0)--(1,0) node[midway,below] {$\SP$} ;
\draw[->-=0.58,very thick] (1,0)--(-1,0) node[midway,below] {$\SQ$} ;
\draw[->-=0.58,very thick] (-1,0)--(-3,0) node[midway,below] {$\SP$} ;
\fill (-1,0) circle (0.1) node[above] {$x \in \CX$} ;
\fill (1,0) circle (0.1) node[above,xshift=0.15cm] {$y \in \CX^\op$} ;
\end{tikzpicture}
\end{array}
\rightsquigarrow
\begin{array}{c}
\begin{tikzpicture}[scale=0.8]
\fill[gray!20] (-2.5,0) rectangle (2.5,3) node[at end,below left,black] {$\SC$} ;
\draw[->-,very thick] (2.5,0)--(0,0) node[midway,below] {$\SP$} ;
\draw[->-,very thick] (0,0)--(-2.5,0) node[midway,below] {$\SP$} ;
\fill (0,0) circle (0.1) node[above] {$x \boxtimes_\SQ y \in \CX \boxtimes_\CQ \CX^\op$} ;
\end{tikzpicture}
\end{array}
\]
\caption{Morita equivalence}
\label{fig:Morita_equivalence}
\end{figure}
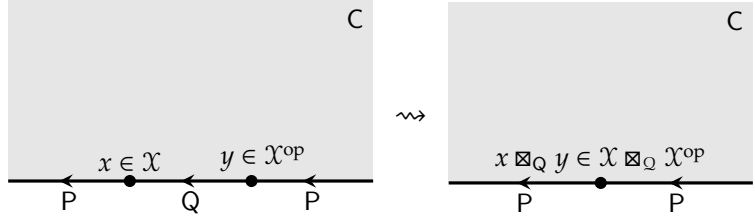

The 0d domain walls can be fused along $\SQ$ (see Figure \ref{fig:Morita_equivalence}). Thus the category of 0d domain walls between $\SP$ and $\SP$ can be obtained by the relative tensor product $\CX \boxtimes_\CQ \CX^\op$. On the other hand, the category of 0d domain walls between $\SP$ and itself is just the topological skeleton $\CP$ of $\SP$. Hence, there is an equivalence $\CX \boxtimes_\CQ \CX^\op \simeq \CP$ of $(\CP,\CP)$-bimodules. Similarly, there is also an equivalence $\CX^\op \boxtimes_\CP \CX \simeq \CQ$ of $(\CQ,\CQ)$-bimodules.

\begin{defn}[\cite{Mueg03,EO04,ENO10}]
Let $\CP,\CQ$ be multi-fusion categories. A finite semisimple $(\CP,\CQ)$-bimodule $\CX$ is called \emph{invertible} if there is an equivalence of $(\CP,\CP)$-bimodules:
\[
\CX \boxtimes_\CQ \CX^\op \simeq \CP
\]
and an equivalence of $(\CQ,\CQ)$-bimodules:
\[
\CX^\op \boxtimes_\CP \CX \simeq \CQ
\]
If there exists an invertible $(\CP,\CQ)$-bimodule, we say $\CP$ and $\CQ$ are \emph{Morita equivalent}.
\end{defn}


By the above discussion, the topological skeletons $\CP,\CQ$ of two gapped 1d boundaries $\SP,\SQ$ of a 2d topological order $\SC$ are Morita equivalent if there exists at least one 0d domain wall between $\SP$ and $\SQ$.

\begin{figure}[htbp]
\[
\begin{array}{c}
\begin{tikzpicture}[scale=0.8]
\fill[gray!20] (0,0) rectangle (1.5,3) node[midway,black] {$\SC$} ;
\draw[->-,very thick] (1.5,3)--(1.5,0) node[midway,right] {$\SP$} ;
\draw[->-,very thick] (0,0)--(0,3) node[midway,left] {$\SQ$} ;
\end{tikzpicture}
\end{array}
\rightsquigarrow
\begin{array}{c}
\begin{tikzpicture}[scale=0.8]
\draw[->-,ultra thick] (0,3)--(0,0) node[midway,right] {$\overline{\SQ} \boxtimes_\SC \SP$} ;
\fill (0,3) circle (0.1) node[right] {$x$} ;
\end{tikzpicture}
\end{array}
\rightsquigarrow
\begin{array}{c}
\begin{tikzpicture}[scale=0.8]
\fill[gray!20] (0,3)--(0.5,0)--(-0.5,0)--cycle ;
\draw[->-,very thick] (0,3)--(0.5,0) node[midway,right] {$\SP$} ;
\draw[->-,very thick] (-0.5,0)--(0,3) node[midway,left] {$\SQ$} ;
\fill (0,3) circle (0.1) node[right] {$x$} ;
\node at (0,1) {$\SC$} ;
\end{tikzpicture}
\end{array}
\rightsquigarrow
\begin{array}{c}
\begin{tikzpicture}[scale=0.8]
\fill[gray!20] (-2.5,0) rectangle (2.5,3) node[at end,below left,black] {$\SC$} ;
\draw[->-,very thick] (2.5,0)--(0,0) node[midway,above] {$\SQ$} ;
\draw[->-,very thick] (0,0)--(-2.5,0) node[midway,above] {$\SP$} ;
\fill (0,0) circle (0.1) node[above] {$x$} ;
\end{tikzpicture}
\end{array}
\]
\caption{the existence of a 0d domain wall between two boundaries}
\label{fig:existence_0d_domain_wall}
\end{figure}
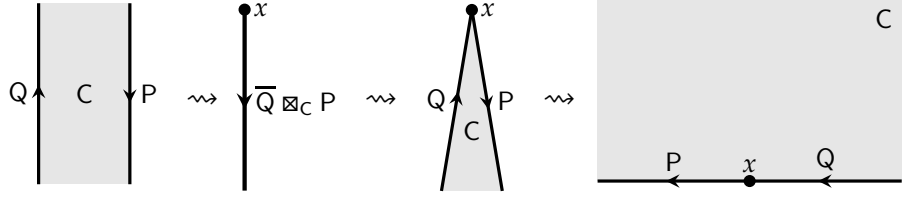

\begin{rem}
If $\SC$ is an anomaly-free stable 2d topological order, then for every gapped 1d boundaries $\SP,\SQ$ of $\SC$, there exists at least one 0d domain wall between them. Figure \ref{fig:existence_0d_domain_wall} depicts a proof. By the folding trick, a 0d domain wall between $\SP$ and $\SQ$ is the same as a 0d boundary of $\overline{\SQ} \boxtimes_\SC \SP$. Since every anomaly-free 1d topological order is the direct sum of several copies of the trivial 1d topological order (see Remark \ref{rem:1d_topological_order_classification}), it has at least one gapped boundary $x$.

However, if $\SC$ is not stable, it is possible that there is no 0d domain wall between two gapped 1d boundaries. Why does the above argument not apply to unstable case? The point is that the dimensional reduction $\overline{\SQ} \boxtimes_\SC \SP$ may be zero, and it means that such a dimensional reduction is physically forbidden.
\end{rem}

\begin{expl} \label{expl:domain_wall_Morita_equivalent}
Suppose $\SM$ is a gapped 1d domain walls in an anomaly-free stable 2d topological order $\SC$, then its topological skeleton $\CM$ is Morita equivalent to $\CC$ \cite{GJF19,KLWZZ20}. Indeed, by the folding trick, a gapped 1d domain wall in $\SC$ is the same as a gapped 1d boundary of $\SC \boxtimes \overline{\SC}$ (which is stable), and clearly $\SC_1$ is such a boundary (see Figure \ref{fig:trivial_domain_wall}).
\end{expl}

The above discussion, together with the boundary-bulk relation in 2d (Theorem$^{\text{ph}}$ \ref{pthm:boundary_bulk}), also lead to the following mathematical theorem \cite{Mueg03,NN08,ENO11}.

\begin{thm} \label{thm:Morita_equivalence_center}
Two indecomposable multi-fusion categories $\CP,\CQ$ are Morita equivalent if and only if their Drinfeld centers $\FZ_1(\CP),\FZ_1(\CQ)$ are equivalent as braided fusion categories.
\end{thm}

\begin{expl}
Let $G$ be a finite group. Recall Example \ref{expl:quantum_double_as_Drinfeld_center} that $\FZ_1(\rep(G))$ and $\FZ_1(\vect_G)$ are equivalent as braided fusion categories. Thus $\rep(G)$ and $\vect_G$ are Morita equivalent. Indeed, there is an invertible bimodule $\vect$, where the module action of $\rep(G)$ and $\vect_G$ on $\vect$ are induced by the forgetful functors $\rep(G) \to \vect$ and $\vect_G \to \vect$, respectively.
\end{expl}

\begin{rem} \label{rem:Morita_equivalence_module_dual}
Let $\CP$ be an indecomposable multi-fusion category and $\CX$ be a nonzero finite semisimple left $\CP$-module. Then all left $\CP$-module functors from $\CX$ to itself and left $\CP$-module natural transformations (see Remark \ref{rem:module_functor_natural_transformation}) also form a multi-fusion category, denoted by $\fun_\CP(\CX,\CX)$. Also $\CX$ is a finite semisimple left $\fun_\CP(\CX,\CX)$-module, or equivalently, a finite semisimple right $\fun_\CP(\CX,\CX)^\rev$-module. Then $\CP$ is Morita equivalent to $\fun_\CP(\CX,\CX)^\rev$ with an invertible $(\CP,\fun_\CP(\CX,\CX)^\rev)$-bimodule $\CX$.

Conversely, if $\CP,\CQ$ are multi-fusion categories with an invertible finite semisimple $(\CP,\CQ)$-bimodule $\CX$, then we have $\CQ \simeq \fun_\CP(\CX,\CX)^\rev$ and $\CP \simeq \fun_{\CQ^\rev}(\CX,\CX)$ as multi-fusion categories \cite{Mueg03,EO04,ENO10}.
\end{rem}

\begin{rem}
It is possible that there are different invertible bimodules between two multi-fusion categories. For example, let $G$ be a finite group. Then for each $\beta \in Z^2(G;\Cb^\times)$ the module category $\rep(G,\beta)$ is an invertible $(\rep(G),\rep(G))$-bimodule.
\end{rem}

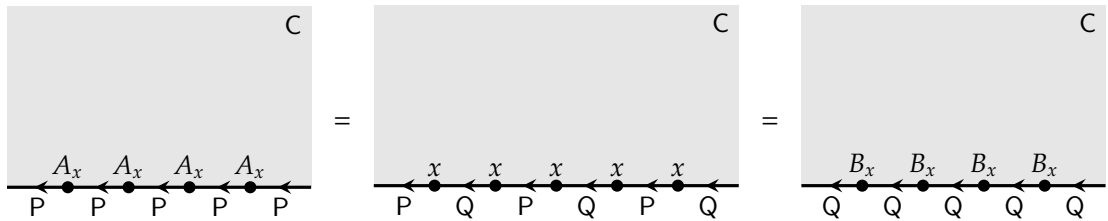
\begin{figure}[htbp]
\[
\begin{array}{c}
\begin{tikzpicture}[scale=0.8]
\fill[gray!20] (-2.5,0) rectangle (2.5,3) node[at end,below left,black] {$\SC$} ;
\draw[->-,very thick] (2.5,0)--(1.5,0) node[midway,below] {$\SP$} ;
\draw[->-,very thick] (1.5,0)--(0.5,0) node[midway,below] {$\SP$} ;
\draw[->-,very thick] (0.5,0)--(-0.5,0) node[midway,below] {$\SP$} ;
\draw[->-,very thick] (-0.5,0)--(-1.5,0) node[midway,below] {$\SP$} ;
\draw[->-,very thick] (-1.5,0)--(-2.5,0) node[midway,below] {$\SP$} ;
\fill (1.5,0) circle (0.1) node[above] {$A_x$} ;
\fill (0.5,0) circle (0.1) node[above] {$A_x$} ;
\fill (-0.5,0) circle (0.1) node[above] {$A_x$} ;
\fill (-1.5,0) circle (0.1) node[above] {$A_x$} ;
\end{tikzpicture}
\end{array}
=
\begin{array}{c}
\begin{tikzpicture}[scale=0.8]
\fill[gray!20] (-3,0) rectangle (3,3) node[at end,below left,black] {$\SC$} ;
\draw[->-,very thick] (3,0)--(2,0) node[midway,below] {$\SQ$} ;
\draw[->-,very thick] (2,0)--(1,0) node[midway,below] {$\SP$} ;
\draw[->-,very thick] (1,0)--(0,0) node[midway,below] {$\SQ$} ;
\draw[->-,very thick] (0,0)--(-1,0) node[midway,below] {$\SP$} ;
\draw[->-,very thick] (-1,0)--(-2,0) node[midway,below] {$\SQ$} ;
\draw[->-,very thick] (-2,0)--(-3,0) node[midway,below] {$\SP$} ;
\fill (2,0) circle (0.1) node[above] {$x$} ;
\fill (1,0) circle (0.1) node[above] {$x$} ;
\fill (0,0) circle (0.1) node[above] {$x$} ;
\fill (-1,0) circle (0.1) node[above] {$x$} ;
\fill (-2,0) circle (0.1) node[above] {$x$} ;
\end{tikzpicture}
\end{array}
=
\begin{array}{c}
\begin{tikzpicture}[scale=0.8]
\fill[gray!20] (-2.5,0) rectangle (2.5,3) node[at end,below left,black] {$\SC$} ;
\draw[->-,very thick] (2.5,0)--(1.5,0) node[midway,below] {$\SQ$} ;
\draw[->-,very thick] (1.5,0)--(0.5,0) node[midway,below] {$\SQ$} ;
\draw[->-,very thick] (0.5,0)--(-0.5,0) node[midway,below] {$\SQ$} ;
\draw[->-,very thick] (-0.5,0)--(-1.5,0) node[midway,below] {$\SQ$} ;
\draw[->-,very thick] (-1.5,0)--(-2.5,0) node[midway,below] {$\SQ$} ;
\fill (1.5,0) circle (0.1) node[above] {$B_x$} ;
\fill (0.5,0) circle (0.1) node[above] {$B_x$} ;
\fill (-0.5,0) circle (0.1) node[above] {$B_x$} ;
\fill (-1.5,0) circle (0.1) node[above] {$B_x$} ;
\end{tikzpicture}
\end{array}
\]
\caption{a critical point of the condensation from $\SP$ to $\SQ$}
\label{fig:Morita_equivalence_condensation}
\end{figure}

\begin{rem} \label{rem:Morita_equivalence_condensation}
Suppose $\SP,\SQ$ are gapped 1d boundaries of a 2d topological order $\SC$. Then $\SQ$ can be obtained from $\SP$ by condensing some topological defects of $\SP$. Indeed, suppose $x$ is a 0d domain wall between $\SP$ and $\SQ$. By reversing the orientation, $x$ is also a 0d domain wall between $\SQ$ and $\SP$ \cite{Kon14}. Then the fusion $A_x \coloneqq x \boxtimes_\SQ x$ is a 0d domain wall between $\SP$ and $\SP$, i.e., $A_x \in \CP$. Similarly, $B_x \coloneqq x \boxtimes_\SP x \in \CQ$ is a 0d domain wall between $\SQ$ and $\SQ$. The critical point of the condensation from $\SP$ to $\SQ$ driven by $A_x \in \CP$ is depicted in Figure \ref{fig:Morita_equivalence_condensation}. Conversely, it is also the critical point of the condensation from $\SQ$ to $\SP$ driven by $B_x \in \CQ$.

Mathematically, suppose $\CP,\CQ$ are fusion categories and $\CM$ is an invertible $(\CP,\CQ)$-bimodule. Then for any nonzero $x \in \CM$ the internal hom $A_x \coloneqq [x,x]_\CP$ is an algebra $A_x \in \CP$. Moreover, $\CQ$ is equivalent to the category $\BMod_{A_x|A_x}(\CP)$ of $(A_x,A_x)$-bimodules in $\CP$.
\end{rem}

\subsubsection{The structure of multi-fusion categories} \label{sec:multi_fusion}

Consider a (potentially anomalous) 1d topological order $\SP$ whose ground state subspace is $n$-fold degenerate. We only consider the case that the bulk of $\SP$ is stable. In other words, its topological skeleton $\CP$ is a multi-fusion category with $\one = \medoplus_{i=1}^n \one_i$, where each $\one_i$ is generated by a single ground state $\lvert \psi_i \rangle$ and thus is a simple object in $\CP$. The physical intuition that two different ground states $\lvert \psi_i \rangle$ and $\lvert \psi_j \rangle$ can not be smoothly connected implies that $\one_i \otimes \one_i = \one_i$ and $\one_i \otimes \one_j = 0$ for $i \neq j$.

Now we add a perturbation to $\SP$ such that the ground state falls into a single one $\lvert \psi_i \rangle$. Then we get a stable 1d topological order, denoted by $\SP_i$. Its topological skeleton is a fusion category, denoted by $\CP_{ii}$. Moreover, $\SP$ is the direct sum of these stable topological orders, i.e., $\SP = \medoplus_i \SP_i$. 

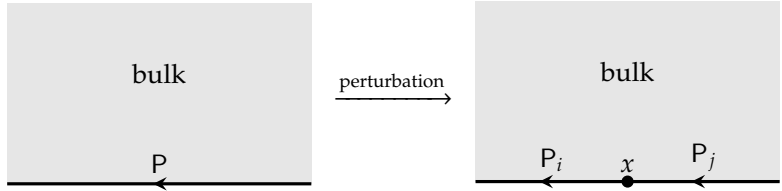
\begin{figure}[htbp]
\[
\begin{array}{c}
\begin{tikzpicture}[scale=0.8]
\fill[gray!20] (-2.5,0) rectangle (2.5,3) node[midway,above,black] {bulk} ;
\draw[->-=0.52,very thick] (2.5,0)--(-2.5,0) node[midway,above] {$\SP$} ;
\end{tikzpicture}
\end{array}
\xrightarrow{\text{perturbation}}
\begin{array}{c}
\begin{tikzpicture}[scale=0.8]
\fill[gray!20] (-2.5,0) rectangle (2.5,3) node[midway,above,black] {bulk} ;
\draw[->-=0.58,very thick] (2.5,0)--(0,0) node[midway,above] {$\SP_j$} ;
\draw[->-=0.58,very thick] (0,0)--(-2.5,0) node[midway,above] {$\SP_i$} ;
\fill (0,0) circle (0.1) node[above] {$x$} ;
\end{tikzpicture}
\end{array}
\]
\caption{a multi-fusion category $\CP$}
\label{fig:multi_fusion}
\end{figure}

If we add different perturbations in different regions, every choice of perturbations produces a 0d domain wall between these regions (see Figure \ref{fig:multi_fusion}). Clearly the category of 0d domain walls between $\SP_i$ and $\SP_j$ should be a full subcategory of $\CP$ defined by
\[
\CP_{ij} \coloneqq \one_i \otimes \CP \otimes \one_j \coloneqq \{\one_i \otimes x \otimes \one_j \mid x \in \CP\} .
\]
When $i=j$, it recovers the topological skeleton $\CP_{ii}$. In particular, the tensor unit of $\CP_{ii}$ is $\one_i$.

\begin{rem}
Since the perturbation is on the 1d boundary $\SP$ and does not affect the 2d bulk, we immediately know that all $\SP_i$'s share the same bulk with $\SP$. By the boundary-bulk relation in 2d, we have $\FZ_1(\CP_{ii}) \simeq \FZ_1(\CP)$ as braided fusion categories. Also, by Theorem \ref{thm:Morita_equivalence_center}, every $\CP_{ii}$ is Morita equivalent to $\CP$. Moreover, $\CP_{ij}$ is an invertible $(\CP_{ii},\CP_{jj})$-bimodule.
\end{rem}

It is clear that
\[
\CP = \one \otimes \CP \otimes \one = \bigl(\bigoplus_i \one_i \bigr) \otimes \CP \otimes \bigl(\bigoplus_j \one_j \bigr) = \bigoplus_{i,j} \one_i \otimes \CP \otimes \one_j = \bigoplus_{i,j} \CP_{ij} .
\]
Then we have the following structure theorem of a multi-fusion category \cite{ENO05,EO04,KZ18}.

\begin{thm}
Let $\CP$ be a multi-fusion category. Assume that $\one = \medoplus_i \one_i$ where each $\one_i$ is simple. Define $\CP_{ij} \coloneqq \one_i \otimes \CP \otimes \one_j$.
\bnu[(a)]
\item We have $\one_i \otimes \one_i = \one_i$ and $\one_i \otimes \one_j = 0$ for $i \neq j$.
\item $\CP = \medoplus_{i,j} \CP_{ij}$.
\item The tensor product of $\CP$ restricts to a functor $\CP_{ij} \times \CP_{jk} \simeq \CP_{ik}$ for every $i,j,k$. In other words, $\CP_{ij} \otimes \CP_{jk} \subseteq \CP_{ik}$.
\item When $\CP$ is indecomposable (see Remark \ref{rem:indecomposable_multi-fusion_category}), each component $\CP_{ij}$ is non-zero. Moreover, the tensor product induces an equivalence of $(\CP_{ii},\CP_{kk})$-bimodules for every $i,j,k$:
\[
\CP_{ij} \boxtimes_{\CP_{jj}} \CP_{jk} \simeq \CP_{ik} .
\]
In particular, $\CP_{ij}$ is an invertible $(\CP_{ii},\CP_{jj})$-bimodule for all $i,j$. Hence all fusion categories $\CP_{ii}$ are Morita equivalent.
\enu
\end{thm}

Hence, every multi-fusion category $\CP$ can be written as a square matrix $(\CP_{ij})$, in which diagonal components are fusion categories. It is indecomposable if and only if the matrix is not block-diagonal (up to permutation). The simplest example of an indecomposable multi-fusion category is $\mathrm M_n(\vect)$, in which every component is $\vect$.


\subsection{Condensation completion}

\subsubsection{Topological skeleton of a 2d topological order}

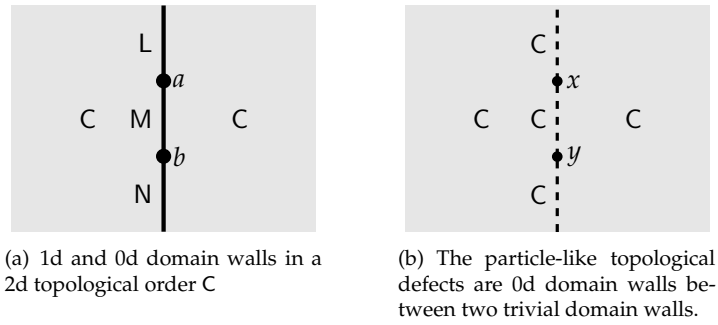
\begin{figure}[htbp]
\centering
\subfigure[1d and 0d domain walls in a 2d topological order $\SC$]{
\begin{tikzpicture}[scale=1.0]
\fill[gray!20] (-2,-1.5) rectangle (2,1.5) ;
\draw[ultra thick] (0,0.5)--(0,1.5) node[midway,left] {$\SL$} ;
\draw[ultra thick] (0,-0.5)--(0,0.5) node[midway,left] {$\SM$} ;
\draw[ultra thick] (0,-1.5)--(0,-0.5) node[midway,left] {$\SN$} ;
\fill (0,0.5) circle (0.1) node[right] {$a$} ;
\fill (0,-0.5) circle (0.1) node[right] {$b$} ;

\node at (-1,0) {$\SC$} ;
\node at (1,0) {$\SC$} ;
\end{tikzpicture}
}
\hspace{5ex}
\subfigure[The particle-like topological defects are 0d domain walls between two trivial domain walls.]{
\begin{tikzpicture}[scale=1.0]
\fill[gray!20] (-2,-1.5) rectangle (2,1.5) ;
\draw[dashed,very thick] (0,0.5)--(0,1.5) node[midway,left] {$\SC$} ;
\draw[dashed,very thick] (0,-0.50)--(0,0.5) node[midway,left] {$\SC$} ;
\draw[dashed,very thick] (0,-1.5)--(0,-0.5) node[midway,left] {$\SC$} ;
\fill (0,0.5) circle (0.07) node[right] {$x$} ;
\fill (0,-0.5) circle (0.07) node[right] {$y$} ;

\node at (-1,0) {$\SC$} ;
\node at (1,0) {$\SC$} ;
\end{tikzpicture}
}
\caption{The topological skeleton of a 2d topological order $\SC$}
\label{fig:top_skeleton_2d}
\end{figure}

As explained in Section \ref{sec:top_skeleton}, the topological skeleton of a 2d topological order $\SC$ consists of all topological defects in $\SC$. There are 1d domain walls in $\SC$, 0d domain walls between 1d domain walls and instantons (see Figure \ref{fig:top_skeleton_2d}). In particular, a particle-like topological defect is a 0d domain wall between two trivial 1d domain walls.

We have shown that the 0d domain walls between two 1d domain walls $\SM$ and $\SN$ (and instantons between them) form a category, denoted by $\Hom(\SM,\SN)$. The main structure of this category is the fusion of instantons on the world line. Moreover, these 0d domain walls can be fused along 1d domain walls. For example, two 0d domain walls $a$ and $b$ in Figure \ref{fig:top_skeleton_2d} (a) can be fused along $\SM$, and the result is a 0d domain wall between $\SL$ and $\SN$, denoted by $b \boxtimes_\SM a$ or simply $b \circ a$. In this process, the instantons on the world lines of 0d domain walls are also fused together. So the fusion of 0d domain walls defines a functor
\[
\circ \colon \Hom(\SM,\SN) \times \Hom(\SL,\SM) \to \Hom(\SL,\SN) .
\]

These defects and their relationship lead to the following structure.

\begin{defn}[not rigorous]
A \emph{2-category} $\BC$ consists of the following data:
\bit
\item a set $\ob(\BC)$, whose elements are called \emph{objects} of $\BC$;
\item a category $\Hom_\BC(x,y)$ for each $x,y \in \ob(\CC)$, called the \emph{hom category} from $x$ to $y$;
\item a functor $\circ \colon \Hom_\BC(y,z) \times \Hom_\BC(x,y) \to \Hom_\BC(x,z)$ for every $x,y,z \in \ob(\BC)$, called the \emph{composition functor};
\item an object $1_x \in \Hom_\BC(x,x)$ for every $x \in \ob(\BC)$, called the \emph{identity};
\item some higher coherence data;
\eit
These data should satisfy some conditions like (higher) associativity and unity axioms. The objects and morphisms in hom categories are also called \emph{1-morphisms} and \emph{2-morphisms} in $\BC$, respectively. In a fixed hom category $\Hom_\BC(x,y)$, the composition of two 2-morphisms $\alpha,\beta$ is denoted by $\beta \cdot \alpha$.
\end{defn}

\begin{notation}
In this work, 2-categories are labeled by \verb+\mathbf+ font: $\BC,\BD,\BE$\ldots.
\end{notation}

\begin{expl} \label{expl:2-cat}
We list some 2-categories:
\bnu
\item Categories, functors and natural transformations form a 2-category.
\item Finite semisimple categories, $\Cb$-linear functors and natural transformations form a 2-category, denoted by $2\vect$.
\item Monoidal categories, monoidal functors and monoidal natural transformations form a 2-category.
\item Fusion categories, $\Cb$-linear monoidal functors and monoidal natural transformations form a 2-category.
\item Braided monoidal categories, braided monoidal functors and monoidal natural transformations form a 2-category.
\item Let $\CC,\CD$ be a multi-fusion category. Finite semisimple left $\CC$-modules, left $\CC$-module functors and left $\CC$-module natural transformations (see Remark \ref{rem:module_functor_natural_transformation}) form a 2-category, denoted by $\LMod_\CC(2\vect)$. Similarly, finite semisimple right $\CD$-modules, $\CD$-module functors and $\CD$-module natural transformations form a 2-category $\RMod_\CD(2\vect)$, and finite semisimple $(\CC,\CD)$-bimodule, $(\CC,\CD)$-bimodule functors and $(\CC,\CD)$-bimodule natural transformations form a 2-category $\BMod_{\CC|\CD}(2\vect)$.
\enu
\end{expl}

\begin{rem} \label{rem:Morita_equivalent_module_2_cat}
Two multi-fusion categories $\CP$ and $\CQ$ are Morita equivalent if and only if $\RMod_\CP(2\vect)$ and $\RMod_\CQ(2\vect)$ are equivalent.
\end{rem}

The topological skeleton of a 2d topological order $\SC$, denoted by $\BC$, is a 2-category:
\bit
\item The objects are gapped 1d domain walls (i.e., string-like topological defects) in $\SC$.
\item The 1-morphisms between two objects (1d domain walls) are 0d domain walls between them.
\item The 2-morphisms between two 1-morphisms (0d domain walls) are instantons between them.
\item The composition functors are given by the fusion of 0d domain walls, and the identities are given by trivial 0d domain walls.
\eit
In particular, the hom category $\Hom_\BC(\one_\BC,\one_\BC)$ from the trivial 1d domain wall to itself is the category $\CC$ of particle-like topological defects of $\SC$.

\begin{figure}[htbp]
\centering
\begin{tikzpicture}[scale=0.8]
\fill[gray!20] (-1,0)--(5,0)--(6,2)--(0,2)--cycle node[black,above right] {$\SC$} ;

\draw[->-=0.6,very thick] (5.5,1)--(3.5,1) node[midway,above] {$\SL$} ;
\draw[->-=0.6,very thick] (3.5,1)--(1.5,1) node[midway,above] {$\SM$} ;
\draw[->-=0.6,very thick] (1.5,1)--(-0.5,1) node[midway,above] {$\SN$} ;

\fill (1.5,1) circle (0.1) node[below] {$x_1$} ;
\fill (3.5,1) circle (0.1) node[below] {$y_1$} ;

\draw (1.5,1)--(1.5,5) node[midway,left] {$x_2$} node[very near end,left] {$x_3$} ;
\draw[fill=white] (1.4,2.4) rectangle (1.6,2.6) node[midway,right] {$f_1$} ;
\draw[fill=white] (1.4,3.4) rectangle (1.6,3.6) node[midway,right] {$f_2$} ;

\draw (3.5,1)--(3.5,5) node[midway,right] {$y_2$} node[very near end,right] {$y_3$} ;
\draw[fill=white] (3.4,2.4) rectangle (3.6,2.6) node[midway,left] {$g_1$} ;
\draw[fill=white] (3.4,3.4) rectangle (3.6,3.6) node[midway,left] {$g_2$} ;
\end{tikzpicture}
\caption{the middle four exchange property}
\label{fig:middle_four_exchange}
\end{figure}
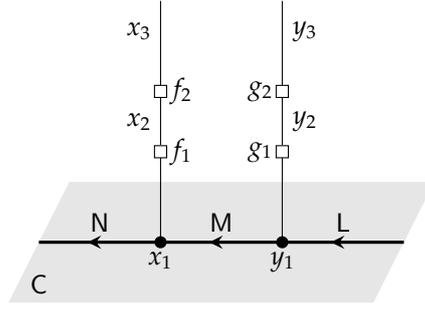

\begin{rem}
There are different fusion or composition of 2-morphisms in a 2-category. As depicted in Figure \ref{fig:middle_four_exchange}, instantons (2-morphisms) can be fused vertically. On the other hand, 0d domain walls (1-morphisms) can be fused horizontally along 1d domain walls (objects), and the instantons on the world lines are also horizontally fused together. So there are two ways to fuse four instantons: we can first fuse them vertically to get two instantons, and then do horizontal fusion; we can also horizontally fuse $x$ and $y$ and then do vertical fusion. These two processes should give the same result. In other words, we have
\[
(f_2 \cdot f_1) \circ (g_2 \cdot g_1) = (f_2 \circ g_2) \cdot (f_1 \circ g_1) .
\]
This equation means that the composition $\circ \colon \Hom_\BC(\SM,\SN) \times \Hom_\BC(\SL,\SM) \to \Hom_\BC(\SL,\SN)$ is a functor, and is also called the \emph{middle four exchange property} of a 2-category.
\end{rem}

Moreover, the topological skeleton $\BC$ of a 2d topological order $\SC$ has more structure than a 2-category because 1d domain walls (objects) can also be fused together. This intuition implies that $\BC$ is a \emph{monoidal 2-category}. We do not give the precise mathematical definition of a monoidal 2-category here.



\subsubsection{Topological skeleton of the toric code model}

As explained above, the topological skeleton of the toric code model is a monoidal 2-category, denoted by $\mathbf{TC}$. Its 2-category structure is depicted in the following diagram:
\[
\begin{tikzcd}[row sep=large]
 & \mbox{ss} \ar[looseness=8,out=120,in=60,"\rep(\Zb_2) \boxtimes \rep(\Zb_2)"] \ar[rr,harpoon,bend left=5,red] \ar[dl,harpoon,bend left=5,red] \ar[dd,harpoon,bend left=5,blue] \ar[ddrr,harpoon,bend left=5,red] \ar[drrr,harpoon,bend left=5,blue] & & \mbox{sr} \ar[looseness=8,out=120,in=60,"\rep(\Zb_2) \boxtimes \vect_{\Zb_2}"] \ar[ll,harpoon,bend left=5,red] \ar[dr,harpoon,bend left=5,red] \ar[dd,harpoon,bend left=5,blue] \ar[ddll,harpoon,bend left=5,red] \ar[dlll,harpoon,bend left=5,blue] \\
\mbox{unit} \ar[looseness=5,out=200,in=160,"\FZ_1(\rep(\Zb_2))"] \ar[ur,harpoon,bend left=5,red] \ar[dr,harpoon,bend left=5,red] \ar[urrr,harpoon,bend left=5,blue] \ar[drrr,harpoon,bend left=5,blue] \ar[rrrr,harpoon,bend left=5,red] & & & & \mbox{dual} \ar[looseness=4.8,out=20,in=-20,"\FZ_1(\rep(\Zb_2))"] \ar[ul,harpoon,bend left=5,red] \ar[dl,harpoon,bend left=5,red] \ar[ulll,harpoon,bend left=5,blue] \ar[dlll,harpoon,bend left=5,blue] \ar[llll,harpoon,bend left=5,red] \\
 & \mbox{rr} \ar[looseness=8,out=-60,in=-120,"\vect_{\Zb_2} \boxtimes \vect_{\Zb_2}"] \ar[rr,harpoon,bend left=5,red] \ar[ul,harpoon,bend left=5,red] \ar[uu,harpoon,bend left=5,blue] \ar[uurr,harpoon,bend left=5,red] \ar[urrr,harpoon,bend left=5,blue] & & \mbox{rs} \ar[looseness=8,out=-60,in=-120,"\vect_{\Zb_2} \boxtimes \rep(\Zb_2)"] \ar[ll,harpoon,bend left=5,red] \ar[ur,harpoon,bend left=5,red] \ar[uu,harpoon,bend left=5,blue] \ar[uull,harpoon,bend left=5,red] \ar[ulll,harpoon,bend left=5,blue]
\end{tikzcd}
\]
Here six vertices are simple objects of $\mathbf{TC}$, i.e., simple 1d domain walls in the toric code model. The vertex `unit' is the trivial domain wall, `dual' is the $e$-$m$-exchange domain wall, and the other four are non-invertible 1d domain walls (see Section \ref{sec:1d_domain_wall_toric_code}). The arrows between vertices denote hom categories between simple objects, i.e., the categories of 0d domain walls between 1d domain walls. Each black arrow denotes the fusion category of 0d domain walls between a 1d domain wall and itself, i.e., the topological skeleton of a 1d domain wall, which is computed in Section \ref{sec:1d_domain_wall_toric_code}. The other hom categories are essentially computed in Section \ref{sec:0d_domain_wall_toric_code}. A red line means that the corresponding hom category is equivalent to $\vect \oplus \vect$, and a blue line means that the corresponding hom category is equivalent to $\vect$.

Also, we list the fusion rules of simple objects of $\mathbf{TC}$ in Table \ref{table:fusion_rule_1d_domain_wall_toric_code}.

\begin{table}[htbp]
\centering
\renewcommand\arraystretch{1.3}
\begin{tabular}{c|c|c|c|c|c|c}
$\otimes$ & \mbox{unit} & \mbox{dual} & \mbox{ss} & \mbox{sr} & \mbox{rs} & \mbox{rr} \\
\hline
\mbox{unit} & \mbox{unit} & \mbox{dual} & \mbox{ss} & \mbox{sr} & \mbox{rs} & \mbox{rr} \\
\hline
\mbox{dual} & \mbox{dual} & \mbox{unit} & \mbox{rs} & \mbox{rr} & \mbox{ss} & \mbox{sr} \\
\hline
\mbox{ss} & \mbox{ss} & \mbox{sr} & 2\mbox{ss} & 2\mbox{sr} & \mbox{ss} & \mbox{sr} \\
\hline
\mbox{sr} & \mbox{sr} & \mbox{ss} & \mbox{ss} & \mbox{sr} & 2\mbox{ss} & 2\mbox{sr} \\
\hline
\mbox{rs} & \mbox{rs} & \mbox{rr} & 2\mbox{rs} & 2\mbox{rr} & \mbox{rs} & \mbox{rr} \\
\hline
\mbox{rr} & \mbox{rr} & \mbox{rs} & \mbox{rs} & \mbox{rr} & 2\mbox{rs} & 2\mbox{rr}
\end{tabular}
\caption{the fusion rules of simple string-like topological defects (gapped 1d domain walls) in the toric code model}
\label{table:fusion_rule_1d_domain_wall_toric_code}
\end{table}

\begin{exercise}
Explicitly calculate the composition functors of the 2-category $\mathbf{TC}$.
\end{exercise}

\begin{rem} \label{rem:toric_chi_pm_crossed_braided_monoidal_2}
For each $G$-crossed braided fusion category there is an associated monoidal 2-category \cite{Cui19}. The monoidal 2-category $\mathbf{TC}^\times$ associated to the $\Zb_2$-crossed braided fusion category $\Toric_+$ (see Remark \ref{rem:toric_chi_pm_crossed_braiding}) is the full sub-2-category of $\mathbf{TC}$ containing the trivial 1d domain wall and the $e$-$m$-exchange domain wall (i.e., two invertible objects in $\mathbf{TC}$).
\end{rem}

\subsubsection{Condensation completion} \label{sec:condensation_completion}

Moreover, the topological skeleton $\BC$ of a 2d topological order $\SC$ is a multi-fusion 2-category, a notion which is introduced by Douglas and Reutter \cite{DR18}. A multi-fusion 2-category is a monoidal 2-category satisfying some properties. The most important property is that a multi-fusion 2-category is \emph{idempotent complete} \cite{DR18}. The idempotent completeness of a 2-category is a generalization of the idempotent completeness of a 1-category defined in Definition \ref{defn:idempotent_complete}. The theory of idempotent completeness was later generalized to higher categories by Gaiotto and Johnson-Freyd \cite{GJF19} under the name of Karoubi completeness or condensation completeness (see also \cite{KLWZZ20}).

Physically the condensation completeness means that all condensation descendants are included. First we consider the modular tensor category $\CC$ of particle-like topological defects of an anomaly-free stable 2d topological order $\SC$. There is a 2-category $\mathrm B \CC$ defined by the following data:
\bit
\item There is only one object, i.e., $\ob(\mathrm B \CC) \coloneqq \{\ast\}$.
\item $\Hom_{\mathrm B \CC}(\ast,\ast) \coloneqq \CC$.
\item The composition and identity are defined by the tensor product and tensor unit of $\CC$.
\eit
Moreover, $\mathrm B \CC$ is a monoidal 2-category. This construction is similar to Example \ref{expl:delooping_group} and Example \ref{expl:delooping_abelian_group}. Since $\CC = \Hom_\BC(\one_\BC,\one_\BC)$, the monoidal 2-category $\mathrm B \CC$ is a sub-2-category of $\BC$, consisting of the trivial 1d domain wall in $\SC$ and 0d domain walls on the trivial 1d domain wall. Since all simple 1d domain walls in $\SC$ can be obtained from the trivial 1d domain wall by condensing some particle-like topological defects (see Remark \ref{rem:Morita_equivalence_condensation}), $\mathrm B \CC$ is not condensation complete and the condensation descendants form the 2-category $\BC$. We say that the condensation completion of $\mathrm B \CC$ is $\BC$ because $\BC$ is the minimal one among all condensation complete 2-categories containing $\mathrm B \CC$.

\begin{expl}
The condensation completion of $\mathrm B(\Toric)$ is $\mathbf{TC}$. Similarly, the condensation completion of $\mathbf{TC}^\times$ (see Remark \ref{rem:toric_chi_pm_crossed_braided_monoidal_2}) is also $\mathbf{TC}$ \cite{DR18}.
\end{expl}

Given a 2d topological order $\SC$ (which is not necessarily anomaly-free nor stable), its topological skeleton is a multi-fusion 2-category $\BC$, which can be viewed as a categorical description of $\SC$. Recall Remark \ref{rem:categorical_description_UMTC} that when $\SC$ is anomaly-free and stable, the unitary modular tensor category $\CC$ of particle-like topological defects of $\SC$ can also be viewed as a categorical description of $\SC$. What is the relation between these two descriptions? Ineed, when $\SC$ is anomaly-free and stable, the (unitary) modular tensor category $\CC$ and the fusion 2-category $\BC$ can determine each other \cite{JF20}. By definition we have $\Hom_\BC(1_\BC,1_\BC) = \CC$. Conversely, $\BC$ is equivalent to $\RMod_\CC(2\vect)$ (see Example \ref{expl:2-cat}) \cite{GJF19,KLWZZ20}. This is because that the topological skeleton of each 1d domain wall in $\SC$ is Morita equivalent to $\CC$ (see Example \ref{expl:domain_wall_Morita_equivalent} and Remark \ref{rem:Morita_equivalence_module_dual} \ref{rem:Morita_equivalent_module_2_cat}).

By describing a 2d topological order by its topological skeleton, there is another boundary-bulk relation for 2d topological orders and gapped 1d boundaries \cite{KLWZZ20,KZ20a}, similar to Theorem$^{\text{ph}}$ \ref{pthm:boundary_bulk_1d}.

\begin{pthm}[Boundary-bulk relation in 2d, 2nd version] \label{pthm:boundary_bulk_2d_version_2}
Let $\SC$ be an anomaly-free 2d topological order (which is not necessarily stable). Its topological skeleton is a multi-fusion 2-category $\BC$.
\bnu[(a)]
\item All gapped 1d boundaries of $\SC$ and 0d domain walls between them, as well as instantons, form a finite semisimple 2-category, denoted by $\BM$.
\item The bulk-to-boundary map induces an equivalence $\BC \simeq \FZ_0(\BM)$ of multi-fusion 2-categories, where $\FZ_0(\BM) \coloneqq \fun(\BM,\BM)$ is the 2-category of $\Cb$-linear 2-functors from $\BM$ to itself.
\enu
\end{pthm}

\begin{rem}
Suppose $\SC$ is an anomaly-free stable 2d topological order and $\SM$ is a gapped 1d boundary of $\SC$. By definition $\SM$ is an object of the 2-category $\BM$ of gapped boundaries of $\SC$, and $\Hom_\BM(\SM,\SM) = \CM$ is the topological skeleton of $\SM$. On the other hand, $\BM$ is the condensation completion of $\mathrm B \CM$, because the topological skeletons of all gapped 1d boundaries of $\SC$ are Morita equivalent. Also $\BM$ is equivalent to $\RMod_\CM(2\vect)$ (see Remark \ref{rem:Morita_equivalence_module_dual} \ref{rem:Morita_equivalent_module_2_cat}).
\end{rem}

\begin{rem}
When $\SC$ is an anomaly-free stable 2d topological order, Theorem$^{\text{ph}}$ \ref{pthm:boundary_bulk} is a direct corollary of Theorem$^{\text{ph}}$ \ref{pthm:boundary_bulk_2d_version_2} \cite{KLWZZ20} by taking the hom category of the tensor unit in the equivalence $\BC \simeq \FZ_0(\BM)$.
\end{rem}

\begin{rem} \label{rem:2d_TO_fusion_2_cat_trivial_center}
Recall Theorem$^{\text{ph}}$ \ref{pthm:2d_anomaly-free_nondegenerate} that a stable 2d topological order $\SC$ is anomaly-free if and only if the (unitary) braided fusion category $\CC$ of particle-like topological defects is nondegenerate, i.e., the M\"{u}ger center $\FZ_2(\CC) \simeq \vect$ is trivial. There is also a boundary-bulk relation for an anomaly-free stable 3d topological order $\SB$ and a 2d gapped boundary $\SC$ \cite{KWZ15,KWZ17}:
\bnu[(a)]
\item The bulk topological defects of codimension 2 and higher form a braided fusion 2-category $\BB$.
\item The topological skeleton of the boundary $\SC$ is a multi-fusion 2-category $\BC$.
\item The bulk-to-boundary map induces an equivalence $\BB \simeq \FZ_1(\BC)$ of braided fusion 2-categories.
\enu
This boundary-bulk relation has also been checked in a concrete lattice model \cite{KTZ20a}. In particular, a (not necessarily stable) 2d topological order $\SC$ is anomaly-free if and only if its topological skeleton $\BC$ satisfies $\FZ_1(\BC) \simeq 2\vect$.
\end{rem}

%% file: adv-topics.tex

\section{Advanced topics} \label{sec:adv-topics}

In this section, we recommend a few books on tensor categories and briefly outline a few advanced topics on the interaction between the physics of topological orders and the category theory.  We also provide some references so that intrigued readers can explore these topics by themselves. The physical literature on topological orders is gigantic. We only include those directly related to category theory. Many others can be found in Wen's reviews \cite{Wen17,Wen19} and references therein.


\subsection{Historical remarks and books on tensor categories} 

In previous sections, we have already seen how the mathematical theory of (braided) fusion categories emerges from the physics of 1d and 2d topological orders. Now we give brief historical remarks on fusion category and its application to physics. 

\smallskip
In mathematics, the notion of a (multi-)fusion category was studied long ago (see for example \cite{DM82}), but the name was coined in \cite{ENO05}. The subject was revived in 1990s under the influence of physics. In 1989, the mathematical structure of a modular tensor category, originally formulated non-categorically as the so-called Moore-Seiberg data,  was first discovered in the study of 1+1D rational conformal field theories by Moore and Seiberg \cite{MS89,MS90} (see \cite{Hua08a,Hua08} and references therein for the story on the mathematical side). It was later reformulated categorically by Reshetikhin and Turaev, and was used to construct the so-called Reshetikhin-Turaev TQFT's \cite{RT91,Tur20}. In 1992, Turaev and Viro introduced the state-sum construction of TQFT's based on modular tensor categories \cite{TV92}. Barrett and Westbury later generalized Turaev-Viro's construction to spherical fusion categories \cite{BW96,BW99}. In 2001, M\"{u}ger proved that the Drinfeld center of a spherical fusion category is a modular tensor category \cite{Mueg03a}, and Ostrik initiated the study of the representation theory of fusion categories \cite{Ost03,Ost03a}. Since then, the field of fusion categories has entered its golden age under the influence of many people: Drinfeld, M\"{u}ger, Ostrik, Etingof, Nikshych, Davydov and Gelaki, etc.  

In physics, around 1989, it was already noticed by physicists that modular tensor categories can be used to describe anyon systems (see for example \cite{MS89,FRS89,FG90,Reh90}). Many condensed matter physicists learned this fact from Kitaev's review in \cite[Appendix\ E]{Kit06}. In 2004, fusion categories were first used by Levin and Wen in their construction of string-net models (also called Levin-Wen models) \cite{LW05}. In 2011, the representation theory of fusion categories was later used by Kitaev and Kong to construct all gapped defects of codimension 1 and 2 in Levin-Wen models, including boundaries, walls and walls between walls \cite{KK12} (see also \cite{Kon13,LW14,HWW17,CCW17} for further developments). Around 2013-2014, the application of category theory in the study of topological orders (or SPT/SET orders) suddenly exploded. Some of the references will be provided in later subsections when we discuss advanced topics.

\smallskip
Now we recommend a few books and some references on tensor categories. For elementary notions in category theory, such as categories, functors, natural transformations, adjoint functors, limits/colimits, etc., readers can consult with standard textbooks on category theory (see for example \cite{Mac78,Awo10,Rie17}). For modular tensor categories, readers can consult with Turaev's book \cite{Tur20} and Bakalov and Kirillov Jr.'s book \cite{BK01}. Both books teach the so-called graphic calculus, which is a very useful tool for explicit calculation. They also discuss the relation between modular tensor categories and TQFT's. For further study of (braided) fusion categories, the standard references for mathematicians are the book ``Tensor categories'' written by Etingof, Gelaki, Nikshych and Ostrik \cite{EGNO15} and their influential papers \cite{Ost03,Ost03a,ENO05,EO04,ENO04,ENO11,ENO10}. The lecture notes \cite{Pen21} written by David Penneys are also friendly to physicists. Interestingly, \cite{EGNO15} does not use graphic calculus at all for good and important reasons. Physics oriented readers might find \cite{EGNO15} difficult to follow. In this case, we recommend Turaev and Virelizier's book \cite{TV17}, where the graphic calculus is used. But for those who really want to dig deeper in this subject, we would like to emphasize that non-graphic techniques in \cite{EGNO15} are unavoidable eventually. We also recommend readers to a few important developments \cite{DGNO07,DGNO10,DMNO13,DNO12,DN13,DSPS20,Shi16,KZ18,FSS19,BJS21,BJSS21,DN21} that are not covered in the book \cite{EGNO15}. There are other important developments of (braided) fusion (higher) categories related to physics. We choose to cover some of them in the later subsections.

\subsection{Factorization homology} \label{sec:factorization_homology}

We have emphasized in Section \ref{sec:Rn_observable} that the notion of a phase is a macroscopic notion defined on a disk-like region in the infinite size limit (i.e., on $\Rb^n$). In principle, it can be described by $\Rb^n$-observables in the long wave length limit. For example, in 2d, among all such observables, particle-like topological defects and their fusion-braiding properties form a unitary modular tensor category (UMTC) $\CA$, which characterizes the topological order up to invertible ones. The UMTC is an $\Rb^2$-observable but not a global observable defined on any 2-dimensional closed manifolds. 
For example, the double braiding can not be a global observable on a 2-sphere (see Remark \ref{rem:double_braiding_not_global}). 
We should ask the obvious question: what are the global observables on a closed 2-dimensional manifold $\Sigma$? 

It turns out that the global observables can be obtained by integrating the $\Rb^2$-observable $\CA$ over $\Sigma$. We denote the integral by $\int_\Sigma \CA$. It was explicitly computed in \cite{AKZ17}. The mathematical foundation of this integral is the mathematical theory of {\it factorization homology} (see \cite{Lur17} or a recent review \cite{AF20} and references therein). 

\begin{thm}[\cite{AKZ17}] \label{thm:FH}
Let $\Sigma$ be a closed 2-dimensional manifold of genus $g$ and $\CA$ be a UMTC. Then we have
\[
\int_\Sigma \CA = \biggl( \vect, \Hom_\CA \bigl(\one_\CA, (\medoplus_i i^* \otimes i)^{\otimes g} \bigr) \biggr) ,
\]
where the vector space $\Hom_\CA(\one_\CA, (\medoplus_i i^* \otimes i)^{\otimes g})$ is an object in $\vect$ and is precisely the ground state degeneracy (GSD) of the associated topological order on $\Sigma$. 
\end{thm}

\begin{rem}
Mathematical works on integrating braided monoidal categories on surfaces can also be found in David Ben-Zvi, Adrien Brochier, and David Jordan's works \cite{BZBJ18,BZBJ18a}. 
\end{rem}

\begin{rem}
The mathematical theory of factorization homology was originated from Beilinson and Drinfeld's theory of chiral homology \cite{BD04}, which was motivated by Geometric Langlands Correspondence and Conformal Field Theory. 
\end{rem}

\begin{rem}
By the mathematical theory of factorization homology, one can integrate any $E_n$-algebras, which can be viewed as the local observable algebras of certain quantum field theories, on any $n$-dimensional closed/open manifolds. 
For example, a braided monoidal category $\CA$ can be viewed as an $E_2$-algebra in the symmetric monoidal 2-category $\Cat$ of 1-categories. By integrating $\CA$ over a closed surface $\Sigma$, we obtain an $E_0$-algebra in $\Cat$, which is defined by a pair $(\CX, u)$, where $\CX$ is a 1-category and $u$ is a distinguished object in $\CX$. A special case of this general result is Theorem \ref{thm:FH}. This result can also be understood physically as follows. Integrating the UMTC $\CA$ over $\Sigma$ amounts to squeezing the associated topological order to a point, which can be viewed as a particle in the trivial 2d topological order, i.e., $(\vect, u)$. From this picture, it is easy to see that $\CX \neq \vect$ if $\CA$ has degenerate braidings. 
More discussion of the application of $E_n$-monoidal categories and $E_n$-centers in the study of topological orders can be found in \cite{JF20,KZ20a}. 
\end{rem}

\begin{rem}
In physics, it is also interesting to compute the GSD on 2-dimensional surface decorated by higher codimensional defects, such as gapped boundaries, domain walls and 0d defect junctions \cite{Kap14,HW15,LWW15} (see also \cite{HSW12,HWW13,WWH15,HWW17,HLPWW18,BHW17,WLHW18,WHW22} for the computation of the GSD in concrete lattice models in 2d or 3d). This amounts to compute the factorization homology on disk-stratified surfaces \cite{AKZ17}, the mathematical theory of which was developed in \cite{AF15,AFT17,AFT16,AFR18} (see also \cite{BZBJ18a}). 
\end{rem}

\subsection{String-net models}
Constructing lattice models to realize topological orders is an important direction in the study of topological orders. 
It turns out that category theory is also useful in the construction of lattice models and played an important role in the  introduction of category theory to the study of topological orders. 

\smallskip
It start from Kitaev's construction of quantum double models (including the toric code model) in \cite{Kit03}, where the idea of topological quantum computing was proposed. This work, together with Kitaev's honeycomb model \cite{Kit06}, played a crucial role in removing the doubts among theorists on the existence of quantum spin liquids (see \cite{ZKN17} for a review). Kitaev's quantum double models are built from the group algebras of finite groups and are the lattice model realizations of 2d Dijkgraaf-Witten TQFT's \cite{DW90}. The particle-like topological defects in these models can be identified with the UMTC $\FZ_1(\vect_G)$. Kitaev's quantum double model was later generalized to quantum double models based on (weak) $C^\ast$-Hopf algebras \cite{BMCA13,BK12,Cha14} and to twist quantum double models in \cite{HWW13}.

Partially motivated by Kitaev's work \cite{Kit03} and Turaev-Viro TQFT's \cite{TV92,BW96}, Levin and Wen introduced their string-net models, which are also called Levin-Wen models, in \cite{LW05}. Levin-Wen models can be viewed as the lattice model realizations of Turaev-Viro-Barrett-Westbury TQFT's \cite{KMR10,KKR10,Kir11,BK12}. A Levin-Wen model is built from the data of a unitary fusion category $\CC$. The category of particle-like topological defects was computed in a few examples and was conjectured to be the Drinfeld center of $\CC$ in \cite{LW05}. This conjecture was later proved in \cite{KK12} as a byproduct of a systematic construction of the topological defects of all codimensions in Levin-Wen models via the representation theory of unitary fusion categories (see also \cite{Kon13,LW14}). This work  further motivated the studies of gapped boundaries in quantum double models \cite{BSW11,BHW17,CCW17}. The relation between Kitaev's quantum double models and Levin-Wen models were clarified in \cite{BA09,BCKA13}. Roughly speaking, quantum double models based on $C^\ast$-Hopf algebras is equivalent to (via a Fourier transformation) the so-called extended string-net models constructed from the data of a unitary fusion category $\CC$, together with a fiber functor $\omega \colon \CC \to \vect$ \cite{BCKA13}. Another type of extended string-net models and their gapped boundaries (based on Frobenius algebras in unitary fusion categories) were extensively developed by Hu, et al. \cite{HGW18,HWW17,HLPWW18,WHW22}. 

\medskip
String-net models in 2d were further generalized to constructions based on mathematical structures different from unitary fusion categories \cite{Run20,HW20,LLB21} and to non-unitary cases in \cite{FGHTTW12,LVHV20,CLYH22} and to symmetry enriched topological orders \cite{GWW15,CCC+15,GK16,BGK17,HBFL16}. String-net models in 3d were considered in Levin and Wen's original paper \cite{LW05}, and were further developed in the so-called Walker-Wang models \cite{WW11}, and in a generalization based on $G$-crossed braided fusion categories \cite{Cui19}, and in (higher) gauge theories \cite{BCKMM17,BCKMM19,ZLW19,BD19,BD20,BD21,BD21a}, and in its full generality based on fusion 2-categories \cite{DR18,XLLC21}.

\subsection{Anyon condensation and higher algebras} \label{sec:anyon-condensation}

A 2d topological order can be transformed into another one via the so-called anyon condensations (or boson condensations). In physics, a systematic study of anyon condensations was initiated by Bais, Schroers, Slingerland  in 2002 based on the idea of Hopf symmetry breaking \cite{BSS02,BSS03}, and was further developed by Bais and Slingerland in an influential work \cite{BS09}, and was followed by many others (see for example \cite{BMD08,BSH09,BW10,KS11,BSS11,BR12,Lev13,BJQ13,BJLP19}). In mathematics, without knowing its relation to anyon condensations, the corresponding mathematical theory was developed earlier and independently by B\"{o}ckenhauer, Evans and Kawahigashi \cite{BEK99,BEK00,BEK01} in 1999-2001 in terms of the language of subfactors and by Kirillov, Jr.~and Ostrik \cite{KO02} in 2001 in terms of the language of algebras in modular tensor categories (see also \cite{FFRS06})\footnote{A dictionary between two languages provided by Kawahigashi can be found in the Introduction Section in \cite{Kon14}.}, and by Davydov, M\"{u}ger, Nikshych and Ostrik in \cite{DMNO13}.
Ironically, the connection between the mathematical results and the physical problem was unnoticed for some years before it was fully established in 2013 \cite{Kon14} (see also \cite{ERB14,HW15a} and see \cite{Bur18} for a review and references therein).

\begin{figure}[t] 
\[
\begin{tikzpicture}
\fill[blue!30] (-2,0) rectangle (2,3) ;
\fill[green!30] (0,1.5) circle (1);
\node at (-1.5,2.5) {$\SC$} ;
\node at (0,1.5) {$\SD$} ;
\draw [blue, ultra thick] (0,1.5) circle [radius=1] ;
\end{tikzpicture}
\]
\caption{We consider an anyon condensation occurring in a 2d region in a 2d topological phase $\SC$. It produces a gapped domain wall between $\SC$ and the new phase $\SD$.}
\label{fig:set-up}
\end{figure}
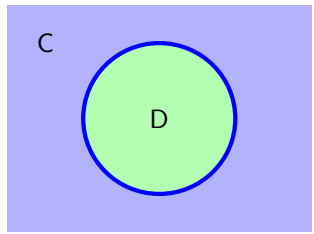

The mathematical theory of anyon condensation demands a lot of new mathematical notions. But we decide to summarize this theory below without explaining these notions not only because readers can get some glimpse of this theory and an impression that the theory is rich, precise and completely computable, but also because it reveals its relation to higher algebras and their higher representation theories (see Remark \ref{rem:higher-algebra}). 
\begin{pthm}[\cite{Kon14}] \label{pthm:anyon-condensation}
If a 2d topological order $\SD=(\CD,c)$ is obtained from another one $\SC=(\CC,c)$ via a 2d anyon condensation, and if a 1d gapped domain wall is produced by this condensation (see Figure \ref{fig:set-up}), then we must have
\bnu
\item The vacuum particle in $\SD$ can be identified with a connected commutative symmetric normalized-special $\dagger$-Frobenius algebra $A$ in $\CC$. Moreover, $\CD$ consists of all deconfined particles and can be identified with the category $\CC_A^{\mathrm{loc}}$ of local $A$-modules in $\CC$, i.e., $\CD = \CC_A^{\mathrm{loc}}$. 

\item Excitations on the wall include all confined and deconfined particles, and can be identified with the unitary fusion category $\CC_A$, which denotes the category of right $A$-modules in $\CC$.

\item Anyons in $\SC$ move onto the wall according to the central functor $- \otimes A \colon \CC \to \CC_A$ defined by $x \mapsto x \otimes A$ for all $x\in \CC$.
 
\item Anyons in $\SD$ move onto the wall according to the embedding $\CC_A^{\mathrm{loc}} \hookrightarrow \CC_A$, then can move out to the $\SD$-side freely. 

\enu
\end{pthm}

\begin{rem} \label{rem:higher-algebra}
In Theorem$^{\text{ph}}$ \ref{pthm:anyon-condensation}, both of the modular tensor category $\CC$ and the commutative algebra $A$ in $\CC$ are examples of $E_2$-algebras \cite{Lur17} or 2-disk algebras \cite{AF20}, which are the 2-dimensional cases of higher dimensional algebras (or simply higher algebras). A local $A$-module is an $A$-module equipped with a 2-dimensional $A$-action. It is an example of higher dimensional representations of an higher algebra. One can also condense particles along a line. This process was called a 1d condensation in \cite{Kon14}, and is controlled by an $E_1$-algebra and its bimodules (see Remark \ref{rem:Morita_equivalence_condensation}). Higher dimensional condensation theory has not yet been developed. However, it is clear that it relies on the higher representation theories of higher algebras (see for example \cite{Lur17,Hau17} and references therein).
\end{rem}

\begin{rem}
More general type of condensations between any two Witt equivalent 2d topological orders was developed in \cite{CRS19}.
\end{rem}

\subsection{2d SPT/SET orders} \label{sec:SPT/SET_2d}


For a quantum many-body system that realizes a trivial topological order, Gu and Wen discovered that, if we impose a symmetry onto the system, then it can realize different and non-trivial phases (e.g. the Haldane phase) that are protected by the symmetry \cite{GW09}. They named such a new phase a symmetry protected topological (or trivial) (SPT) order, which is the trivial topological order if we ignore the symmetry. Chen, Gu and Wen later introduced the notion of a symmetry enriched topological (SET) order \cite{CGW10}, which is a non-trivial topological order if we ignore the symmetry. Note that a topological order is simply an SET order with a trivial symmetry, and an SPT order is an SET order with a trivial topological order. 

\medskip
There are a few non-categorical approaches towards SPT orders, such as K-theory \cite{Kit09}, (super) group cohomology \cite{CGLW13,GW14,WG18}, cobordism groups \cite{Kap14b,KTTW15} and stable homotopy theory \cite{Kit11,Kit13,Kit15,FH21,GJF19a}. We only focus on the categorical approach here. We have seen that the categorical description of a 2d topological order is given by a UMTC. Therefore, it is natural to expect that the categorical description of a 2d SPT/SET order is deeply related to UMTC. For a finite onsite symmetry, there are two approaches.  
\bnu
\item One approach is based on the idea of gauging the symmetry by introducing invertible defects of codimension 1 \cite{BBCW19,TLF16}. For a bosonic SPT/SET order, it is shown that the associated mathematical structure is a $G$-crossed braided fusion category \cite{Tur00,Kir01,Mueg04,DGNO10} (see Remark \ref{rem:toric_chi_pm_crossed_braiding}). This approach can also be applied to fermionic SET orders \cite{FVM18,ABK21,BB22,BB22a}.

\item The other approach is also based on the idea of gauging the symmetry but in a different way. In \cite{LG12}, Levin and Gu gauged the symmetry in a 2d $\Zb_2$ SPT order and obtained a gauge theory (i.e., 2d topological order) as the result of the gauging. This idea was soon formulated 
mathematically by Lan, Kong and Wen in \cite{LKW16a,LKW17} as a categorical gauging process, in which additional particles are introduced the system so that each of them has non-trivial double braidings with at least one symmetry charges. After the gauging, we obtain a UMTC. This categorical gauging process has a mathematical name called \emph{minimal modular extension} first appeared in M\"{u}ger's work \cite{Mueg00}, and it works for all bosonic and fermionic 2d SPT/SET orders. Moreover, these minimal modular extensions form a finite abelian group \cite{LKW16a}, which recovers the classification of bosonic 2d SPT orders by $H^3(G;\mathrm U(1))$ \cite{CGLW13} (see Example \ref{expl:Drinfeld_center_vect_G_omega}) and Kitaev's 16-fold way \cite{Kit06} (see Remark \ref{rem:Ising_16-fold_way}). Further developments of minimal modular extensions can be found in \cite{BGHNPRW17,GVR17,VR19,DN21,OY21,JFR21,KZ21a,Nik22}.

\enu
In the bosonic cases, the equivalence between above two approaches were known \cite{Bru00,Kir01,Mueg04}. 
The idea of minimal modular extension can be generalized to higher dimensions (see \cite{KTZ20,KLWZZ20,KZ21a}).

For SPT orders with spacetime symmetries, there are a lot of physical and non-categorical works (see for example \cite{FK11,SHFH17,WL17,JR17,TE18,BBCJW19,Xio18,SXG18,SHQFH19,ET19,SFQ20,SXH20,GOPWW20,FH20,Deb21} and references therein). There are also works on the real space construction of SET orders with space symmetries (see for example \cite{Lak16,QJW19,MW20,NMLW21} and references therein). Unfortunately, the categorical description of SPT/SET orders with spacetimes symmetries has not yet been developed. This is an interesting direction for future development.

\subsection{Boundary-bulk relation and gapless boundaries} \label{sec:bbr}

We have seen the boundary-bulk relation in the toric code model, which can be stated as the bulk UMTC is the center of the boundary unitary fusion category. We have also mentioned in Theorem$^{\text{ph}}$ \ref{pthm:boundary_bulk} that the boundary-bulk relation holds for all anomaly-free stable 2d topological orders and their gapped 1d boundaries. It turns out that it is not an isolated phenomenon. It was known earlier that the boundary-bulk duality (i.e., open-closed duality) in rational conformal field theories (RCFT) also states that the bulk CFT is the center of a compatible boundary CFT \cite{FFRS08,KR08,KR09,Dav10} (see \cite{Kon11} for a review and references therein). If we view the same phenomena occur in different theories as a robust phenomena, it must have a simple reason.

In mathematics, the notion of center can be defined for any algebraic object in any monoidal higher category by its universal property, which is entirely the same for all different notions of center, such as the ordinary center of a group or an algebra, Drinfeld center, M\"{u}ger center, Hochschild cohomology and $E_n$-centers, etc. The universal property of the center only depends on how we define the morphisms between objects instead of the set-theoretical definition of an object. Therefore, a formal proof of the boundary-bulk relation for general $n$d topological orders can be obtained by checking the universal property if we know how to define the right notion of a morphism between two potentially anomalous $n$d topological orders. This notion was introduced in \cite{KWZ15,KWZ17}. Based on this notion, it is almost tautological to prove the following rather strong result. 

\begin{pthm}[Boundary-Bulk Relation \cite{KWZ15,KWZ17}] \label{pthm:bbr}
For any anomaly-free $n$d topological order $\SC$ and one of its gapped/gaples boundary $\SX$, the bulk $\SC$ is the center of the boundary $\SX$, and we denote this relation by $\SC = \FZ(\SX)$. 
\end{pthm}

\begin{rem}
The proof of the above physical Theorem actually works for even more general systems. More precisely, we can replace the anomaly-free $n$d topological order by an anomaly-free $n$d quantum liquid-like gapped/gapless phase \cite{KZ20a}. By `liquid-like', we mean that the phase (and all higher codimensional defects) is `soft' enough so that it does not depend rigidly on the local geometry (e.g., metric) of the phase. In the gapped cases, the phase is topological thus completely independent of the metric. In the gapless cases, the phase is of CFT-type in the sense that it only depends covariantly on the deformations of the metric in a finite and controllable way similar to 2D rational CFT's. We call such a `liquid-like' quantum phase a quantum liquid. Quantum liquids include all topological orders, SPT/SET orders, gapped symmetry-breaking orders and certain CFT-type gapless phases. 
\end{rem}

The significance of Theorem$^{\mathrm{ph}}$ \ref{pthm:bbr} is two-fold. On the one hand, if we know the categorical description $\CX$ of the boundary $\SX$, by computing the center, we immediately obtain that of the bulk. On the other hand, if we know the categorical description $\CC$ of $\SC$, we can obtain that of $\SX$ by solving the equation $\CC=\FZ_1(?)$. For example, if $\CC$ is the UMTC of a 2d topological order $\SC$, then the solutions to the equation $\CC=\FZ_1(?)$ provide the mathematical descriptions of its 1d boundaries. When $?$ is a unitary fusion category, then it describes a gapped boundary. When the equation does not have a solution in unitary fusion categories, the boundaries are necessarily gapless. The mathematical theory of gapless boundaries of 2d topological orders was developed in \cite{KZ18b,KZ20,KZ21}. A gapless boundary can be mathematically described by a unitary fusion category $\CXs$ enriched in a UMTC such that $\CC=\FZ_1(\CXs)$ \cite{KZ18a,KYZZ21}. This result led to a new and systematic approach towards 1+1D topological phase transitions \cite{CJKYZ20}. The mathematical theory of enriched fusion categories was developed only recently (see  \cite{MP17,KZ18a,Zhe17,MPP18,JMPP21,KZ21,KYZZ21}). 
The gapless boundaries of higher dimensional topological orders demand us to develop the mathematical theory of enriched higher categories (see \cite{KZ21} for more discussions).


\begin{rem}
The statement of boundary-bulk relation in Theorem$^{\mathrm{ph}}$ \ref{pthm:bbr} can be further generalized to include domain walls between boundaries and higher codimensional walls between walls. For 2d topological orders with gapped boundaries, this generalized boundary-bulk relation amounts to the fact that the Drinfeld center defines a fully faithful functor \cite{KZ18}. Generalizing this result to a pointed Drinfeld center functor, we obtain the boundary-bulk relation including domain walls in 1+1D rational CFT's \cite{KYZ21}. For 2d topological orders with gapped/gapless boundaries, it amounts to the fact that the $E_1$-center of certain braided-enriched fusion categories defines a monoidal equivalence \cite{KZ21}. For higher dimensional topological orders with gapped boundaries and gapped higher codimensional walls, the boundary-bulk relation can also be formulated as a center functor \cite{KZ21a}.
\end{rem}

\subsection{Higher dimensional topological orders and SPT/SET orders}

We have seen that an anomaly-free 1d topological order can be described by a unitary fusion 1-category with a trivial $E_1$-center (see Section \ref{sec:boundary_bulk}). We have also seen that an anomaly-free 2d topological order can be described either by a UMTC or by a unitary fusion 2-category with a trivial $E_1$-center (see Remark \ref{rem:2d_TO_fusion_2_cat_trivial_center}). In the study of higher dimensional topological orders, we are forced to enter the mathematical world of higher categories.

In recent years, higher category becomes a very hot topics in mathematics. There are a lot of exciting developments of $(\infty,n)$-categories (see for example \cite{Lur09,Cis19} and references therein). Unfortunately, it is not clear to us how to apply such heavy machinery to the study of topological orders. 

Without knowing how to rigorously define a fusion higher category, guided by the so-called remote-detectable principle \ref{principle:remote-detectable} and boundary-bulk relation, Kong, Wen and Zheng proposed in \cite{KW14,KWZ15} that topological defects of codimension 2 or higher in an anomaly-free stable $n$d topological order form a unitary braided fusion ($n$-1)-category with a trivial $E_2$-center, or equivalently, its defects of codimension 1 or higher form a unitary fusion $n$-category with a trivial $E_1$-center. Combining with useful physical intuitions, this proposal led Lan, Kong and Wen to a classification theory of 3d topological orders with only bosonic particles \cite{LKW18}, which was later rigorously proved by Johnson-Freyd \cite{JF20a}, and led Lan and Wen to that of  3d topological orders with fermionic particles \cite{LW19}, which was later slightly corrected and rigorously proved by Johnson-Freyd \cite{JF20a} (see also \cite{FHH21} for physical realizations). Further categorical studies of 3d and 4d topological orders can be found in \cite{JFY21,JFY21a}.

Rigorous categorical results related to $n$d topological orders for $n\geq 3$ start from Douglas and Reutter's definition of a multi-fusion 2-category and their state-sum construction of 3+1D TQFT's based on spherical fusion 2-categories \cite{DR18}. Their definition of a multi-fusion 2-category is based on the idea of the 2-category version of idempotent completion, which was briefly introduced in \cite{CR16} and thoroughly developed in \cite{DR18}. It turns out that this idempotent completion precisely amounts to adding condensation descendants to the categorical description of a topological order \cite{GJF19,KLWZZ20}. The necessity of this idempotent completion in the categorical description of topological orders was further confirmed by an explicit computation of $\FZ_1(2\mathrm{Vec}_G^\omega)$ for a finite group and $\omega \in H^4(G;\mathrm U(1))$, i.e., the $E_1$-center of $2\mathrm{Vec}_G^\omega$, which describes the topological defects of codimension 2 or higher in 3+1D Dijkgraaf-Witten Theory \cite{KTZ20}. Douglas and Reutter's idempotent completion was later generalized to higher categories by Gaiotto and Johnson-Freyd \cite{GJF19} under the name of Karoubi completion or condensation completion (see Remark \ref{rem:cc}). This generalization is not mathematically complete in the senes that it did not choose any concrete model of higher categories and assumed the not-yet-developed theory of colimits in higher categories. However, it seems that Gaiotto and Johnson-Freyd's work solved the problems that are orthogonal to the higher coherence problems, and should work in any reasonable models of higher categories. This work led to an incomplete-but-workable definition of a multi-fusion $n$-category \cite{JF20}. Based on this definition, Johnson-Freyd developed a mathematical theory of multi-fusion $n$-categories and answered many questions in \cite{KW14,KWZ15} (see Remark \ref{rem:cc}). This theory was further developed in \cite{KZ20a,KZ21a} from a slightly different perspective emphasizing more on separable $n$-categories. 

\begin{rem} \label{rem:cc}
All defects of all codimensions in a 2d non-chiral topological order, including all condensation descendants, were explicitly constructed via Levin-Wen models \cite{KK12}.  A categorical description of the complete set of these defects was known (see \cite[Example 16]{KW14}). The study of condensation completion for topological orders in higher dimensions was initiated in a physical context in \cite[Section XI.H]{KW14}. 
\end{rem}

Based on the idea of boundary-bulk relation \cite{KWZ15,KWZ17}, a unified mathematical theory of all SPT/SET orders in all dimensions was developed in \cite{KLWZZ20,KLWZZ20a}.

\begin{rem}
It was noticed in \cite{KZ20a} that a small modification of the main result in \cite{KLWZZ20} automatically provides the classification of all gapped symmetry-breaking phases. Applying the topological Wick rotation introduced in \cite{KZ20,KZ21} to the results in \cite{KLWZZ20}, Kong and Zheng proposed to use enriched higher categories to give a grand unification of the categorical description of all quantum liquids, including topological orders, SPT/SET orders (with a finite onsite symmetry), gapped symmetry-breaking phases and CFT-type gapless phases \cite{KZ20a}. The validity of this categorical description of all 1d gapped liquids with a finite onsite symmetry was checked for Ising chain and Kitaev chain in \cite{KWZ22}. Moreover, the earlier classification of all 1d gapped quantum phases with a bosonic/fermionic finite onsite symmetry by studying ground state wave functions \cite{CGW11,SPGC11,CGW11a} was rediscovered in this new categorical approach \cite{KZ20a,KWZ22}. A more complete mathematical theory of all quantum liquids was developed in \cite{KZ22}.
\end{rem}

